\documentclass[aps,prb,twocolumn,amsmath,amssymb,superscriptaddress,scrartcl,eqsecnum,longbibliography]{revtex4-1}
\usepackage{extarrows}
\usepackage{slashed}
\usepackage{amsmath,amsfonts,amssymb,graphics,graphicx,epsfig,color,times,indentfirst,layout}
\usepackage{lipsum}
\usepackage{mathrsfs,bbold}
\usepackage[unicode=true,pdfusetitle,bookmarks=false,colorlinks=true,citecolor=black,urlcolor=black,linkcolor=black]{hyperref}
\usepackage{mathtools}

\usepackage{tikz}
\usepackage{tikz-cd}
\usetikzlibrary{arrows}
\usetikzlibrary{intersections}
\usetikzlibrary{shapes.geometric}
\usepackage{float}
\usetikzlibrary{shapes, positioning}

\def\be{\begin{equation}}
\def\ee{\end{equation}}

\begin{document}

\title{Distinguish modular categories and 2+1D topological orders beyond modular data:\\
Mapping class group of higher genus manifold
}

\date{\today}

\author{Xueda Wen}

\author{Xiao-Gang Wen}

\affiliation{Department of Physics, Massachusetts Institute of Technology, Cambridge, MA 02139, USA}

\begin{abstract}
It was believed that modular data are enough to distinguish different modular
categories (and topological orders in 2+1-dimensions). Then counterexamples to
this conjecture were found by Mignard and Schauenburg in 2017.  In this work,
we show that the simplest counterexamples can be distinguished by studying the
representations of mapping class groups of a punctured torus or a
genus-2 closed manifold.

\end{abstract}
\maketitle

\tableofcontents

\section{Introduction}

Landau pointed out the reason that two states of matter belong to different
phases is that they have different symmetries.\cite{LL58} In last 30 years, we
started to realize that two quantum states of matter with identical symmetry
can still belong to different phases.\cite{W8987} Those quantum states are
complicated many-body states.  Beside characterize them as messy and complex,
it is hard to describe their internal structure, not to mention to distinguish
them as different phases of matter. In 1989, a method to probe the internal
structures of those messy complex many-body states was discovered: we put the
many-body system on closed spatial manifold $M^d$ with different topologies,
and then measure the ground state degeneracy $GSD(M^d)$.\cite{W8987,WN9077}
Such topology dependent ground state degeneracy $GSD(M^d)$ reveal the universal
internal structures of the many-body state.  A concept of \emph{topological
order} was introduced to describe such an internal structure which is
beyond the Landau symmetry breaking order.

If one accepts such topological order as a new kind of order, one may ask if
the ground state degeneracy $GSD(M^d)$ for different spatial topologies fully
characterize topological order or not?  It turns what that the answer is no.
So finding physical quantities that can fully characterize topological order is
a key central question in developing a comprehensive theory of topological
order.  In early days, it was proposed to use non-Abelian geometric
phases\cite{WZ8411} of the degenerate ground states induced by deforming the
space, to further characterize topological order.\cite{W9039,KW9327} Since the
non-Abelian geometric phases lead to a  representations of
mapping-class-group (MCG) of the space $M^d$: $R(g),\ g\in \text{MCG}(M^d)$, it
was proposed\cite{W9039} to use those  representations $R(g)$ of
$\text{MCG}(M^d)$ to fully characterize topological order, and to develop a
comprehensive theory of topological order.

For 2+1D topological order, the MCG for a torus is $\text{MCG}(S^1\times
S^1)\cong \text{SL}(2,\mathbb{Z})$.  Also there is natural way to choose a
canonical basis for the degenerate ground states on torus.\cite{ZV1313,ZSH1355}
The representation of $\text{SL}(2,\mathbb{Z})$ in the canonical basis is
generated by $S$ and $T$ matrices, which are called the modular data.  It turns
out that $S$ and $T$ matrices in the canonical basis contain a lot of
information about the 2+1D topological order.  For a long time people speculate
that  modular data ({\it i.e.}  representations of
$\text{MCG}(T^2)$), plus the central charge of the edge states,\cite{W9505} can
fully characterize 2+1D topological order.\cite{Kitaev2006,RSW0777,W150605768}
However, recently in Ref. \onlinecite{MS1708}, Mignard and Schauenburg (MS)
found some topological orders that have the same modular data and central
charge.  Thus, modular data and central charge are not enough to fully
characterize topological order.  In this paper, we will show that
representations of genus-2 mapping class group $\text{MCG}(\Sigma_2)$ can
distinguish those topological orders discovered in Ref. \onlinecite{MS1708}.

To understand why some topological orders cannot be distinguished by modular
data, let us introduce another way to characterize topological orders.  Instead
of using the general ground states on closed spatial manifold $M^d$ with
various topologies, we may also consider the ground states on punctured sphere
$S^d_\text{punc}$ with various punctures to characterize 2+1D topological
orders.  This leads to the unitary modular tensor category (MTC) theory for
2+1D topological orders, where the punctures correspond to the objects in the
MTC.  MTC are algebraic models of anyons (\textit{i.e.} the punctures) in 2+1D
topological phases of matter.\cite{Kitaev2006,RSW0777}  It is suggested that
the data of a modular category should be supported on the punctured sphere up
to 4 punctures, the torus, and the once-punctured torus, with consistency
relations supported on the 5-punctured sphere and twice-punctured
torus\cite{luo1999grothendieck,Wang1805} (See also the related discussions in
the content of conformal field theory
\cite{Moore1989,Friedan1986,Vafa1987,Vafa1988,Sonoda1988} ).

The modular data of a modular tensor category $\mathcal{C}$, also named $S$ and
$T$ matrices, are square matrices indexed by the simple objects or anyons.
They can be viewed as the  representation of the modular group of a
genus-one closed surface $T^2$, realized by the modular action on the
quasi-particle basis.  In fact, this is how the name of `modular data' comes.
Alternatively, these two matrices can be considered as topological invariants
in the framework of topological quantum field theory (TQFT): $S$ matrix is
related to the Hopf-link invariants colored by two simple objects, and $T$
matrix can be viewed as a single loop with a twist.  Practically, $S$ matrix
determines the fusion rules of anyons through Verlinde formula.  With a
properly chosen basis, the modular $T$ matrix is a diagonal matrix with entries
$\theta_a$, where $\theta_a$ is the topological spin for anyon $a$.  The
modular $T$ matrix is of finite order due to Vafa's theorem.\cite{Vafa1988}
Physicists proposed to use $S$ and $T$ matrices as order parameters for the
classification of topological phases of matter.\cite{Wen1303}  It was
believed/conjectured that a modular category is fully determined by the modular
data.\cite{W9039,KW9327,Wang1310}  In a computer-based
work,\cite{MS1708_Computer32} it is found that the modular data are a complete
invariant for $D^{\omega}(G)$, which are twisted Drinfeld doubles of finite
groups $G$, when the group order $|G|$ is smaller than $32$.



Later in 2017, in the work by Mignard and Schauenburg,\cite{MS1708} a
family of counterexamples were discovered showing that arbitrarily many
inequivalent modular categories can share the same modular data.  These
counterexamples are found among the quite accessible class of MTCs,
\textit{i.e.}, twisted quantum doubles of finite groups.  These counterexamples
are defined over the same non-abelian finite group but twisted with different
cocycles.

Then one question arises naturally: How to distinguish these counterexamples?
In the original paper by MS,\cite{MS1708}  the authors found these categories
to be distinct by proving the inexistence of suitable equivalences, but not by
finding extra topological invariants to distinguish them. It is desirable to
find certain physical quantities to distinguish these different categories.

There may be two ways to study this problem.  One way is based on the fact that
modular $S$ and $T$ matrices are associated with the Hopf-link and a twsted
unknot invariants, respectively.  We know there are infinite types of links
(with two components or multi components) and knots. It is natural to search
for other link or knot invariants beyond the Hopf-link and the twisted unknot.
This is the method used in Refs.\onlinecite{Wang1805,DT1806,MS1806}.  In
Ref.\onlinecite{Wang1805}, it is found that all the knots that are two-braid
closures colored by $a\in \Pi_{\mathcal{C}}$ and all the links that are
two-braid closures with the two components colored by $a, b\in \Pi_c$ are
determined by the modular data, where $\Pi_\mathcal{C}$ represents the
collection of anyons in the modular tensor category $\mathcal{C}$.  Therefore,
the next candidates to distinguish the MS MTCs may be found among the links and
knots which are at least three-braid closures.

Then in Ref.\onlinecite{Wang1805}, it is found that the whitehead link
invariant together with the $T$ matrix can be used to distinguish the simplest
counterexamples in MS MTCs. Later in Ref.\onlinecite{DT1806}, based on a
computer search, it is found that many link and knot invariants together with
$T$ matrix can fulfill the same aim.  Remarkably, in Ref.\onlinecite{MS1806},
it is \textit{proved} that the Borromean ring together with $T$ matrix are
enough to distinguish \textit{all} the counterexamples in MS MTCs.  One common
feature of these nontrivial link/knot invariants that go beyond modular data is
that they are the closure of braids with more than two strands. In fact, all
the nontrivial link/knot invariants mentioned above are the closure of
three-strand braids.  Note that the inverse is not true, \textit{i.e.}, a
closure of braids with more than two strands may not be able to distinguish the
counterexamples in MS MTCs.

The other way is based on the fact that modular $S$ and $T$ matrices correspond
to the  representations of mapping class group of a torus. Since the
modular data on a torus is not enough, it is natural to study the MCG
representations of higher-genus surface\cite{W9039} with or
without punctures. With the  representations of the MCG of higher
genus manifold, one may define topological invariants to distinguish the
counterexamples in MS-MTCs. This is the main aim of this work.

One may wonder if we can distinguish different counterexamples in MS MTCs by
using the full algebraic data $\{N_{ab}^c, R_a^{bc}, F^{abc}_d\}$, which
represent the fusion multiplicities, $R$-matrix, and $F$-matrix,
respectively.\cite{Moore1989,Kitaev2006,bonderson2007} The problem is that one
needs to solve the pentagon and hexagon equations, which is a hard problem even
for MTCs with a small number of simple objects.  As will be seen shortly, the
simplest counterexamples in MS MTCs are non-abelian theories with 49 anyons,
and it is a formidable task to solve for the corresponding $F$ and $R$
matrices.  In addition, there is too much information in $F$ and $R$ matrices
including the gauge redundancy.  It is desirable for us to find a minimal set
of gauge invariant data to fully determine different MS MTCs.

\subsection{Main results}

We have seen that the modular data from a torus is not enough to characterize
2+1D topological orders.  To find more data, it is natural to consider MCG
representations for a genus-2 surface.  We find that MCG representations for a
genus-2 surface indeed add extra data, beyond that of representations for the
genus-1 surface.  This points to a direction to build a quantitative theory of
topological order in any dimensions based on MCG representations.

In this paper, we concentrate on 2+1D topological orders.  We study the
 representations of the MCG of a punctured torus $\Sigma_{1,1}$ and a
closed manifold of genus two $\Sigma_{2,0}$ for a MTC.  This is used to
distinguish the simplest counterexamples in MS MTCs with
$G=\mathbb{Z}_{11}\rtimes \mathbb{Z}_5$ twisted by the 3-cocycles $\omega\in
H^3(G,U(1))$.  There are in total 5 different categories $\mathcal{C}_u$ with
$u=0,\, 1,\, 2,\, 3,\, 4$, but only 3 inequivalent modular data.  In
particular, the categories $\mathcal{C}_{u=1}$ and $\mathcal{C}_{u=4}$ share
the same modular data, $\mathcal{C}_{u=2}$ and $\mathcal{C}_{u=3}$ share the
same modular data, and $u=5$ has another set of modular data.

For a closed genus-2 surface $\Sigma_{2,0}$, the five generators of
MCG$(\Sigma_{2,0})$ may be considered as the Dehn twists along the five closed
curves $a_1$, $a_2$, $b_1$, $b_2$, and $c$ as follows:
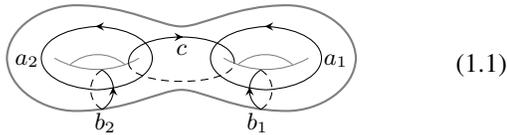
\begin{eqnarray}\label{Genus2_5Dehntwists}
\small
\begin{tikzpicture}[baseline={(current bounding box.center)}]

\draw  [gray](-60*0.8pt,0pt) arc (-120:-60:40*0.8pt);
\draw  [gray](-52.5*0.8pt,-3*0.8pt) arc (130:50:20*0.8pt);
\draw  [gray](60*0.8pt,0pt) arc (-60:-120:40*0.8pt);
\draw  [gray](52.5*0.8pt,-3*0.8pt) arc (50:130:20*0.8pt);

\draw [thick][gray]
(50*0.8pt,25*0.8pt)..controls (65*0.8pt,25*0.8pt) and (82*0.8pt,15*0.8pt)..
(82.5*0.8pt,0pt).. controls (82*0.8pt,-15*0.8pt) and (65*0.8pt,-25*0.8pt).. 
(50*0.8pt,-25*0.8pt)..controls (25*0.8pt,-25*0.8pt) and (10*0.8pt,-15*0.8pt)..
(0pt,-15*0.8pt)..controls (-10*0.8pt,-15*0.8pt) and (-25*0.8pt,-25*0.8pt)..
(-50*0.8pt,-25*0.8pt)..controls (-65*0.8pt,-25*0.8pt) and (-82*0.8pt,-15*0.8pt)..(-82.5*0.8pt,0pt)..controls (-82*0.8pt,15*0.8pt) 
and (-65*0.8pt,25*0.8pt)..(-50*0.8pt,25*0.8pt)..controls (-25*0.8pt,25*0.8pt) 
and (-10*0.8pt,15*0.8pt)..(0pt,15*0.8pt).. controls (10*0.8pt,15*0.8pt) and (25*0.8pt,25*0.8pt)..(50*0.8pt,25*0.8pt);

\draw (30pt,-4.5pt) arc (120:240:8.6pt);
\draw[densely dashed] (30pt,-4.5pt) arc (60:-60:8.6pt);

\draw [densely dashed] (-30pt,-4.5pt) arc (120:240:8.6pt);
\draw (-30pt,-4.5pt) arc (60:-60:8.6pt);

\draw (40*0.8pt,15*0.8pt)..controls (75*0.8pt,13*0.8pt) and (75*0.8pt,-13*0.8pt)..(40*0.8pt,-15*0.8pt);
\draw (40*0.8pt,15*0.8pt)..controls (5*0.8pt,13*0.8pt) and (5*0.8pt,-13*0.8pt)..(40*0.8pt,-15*0.8pt);

\draw (-40*0.8pt,15*0.8pt)..controls (-75*0.8pt,13*0.8pt) and (-75*0.8pt,-13*0.8pt)..(-40*0.8pt,-15*0.8pt);
\draw (-40*0.8pt,15*0.8pt)..controls (-5*0.8pt,13*0.8pt) and (-5*0.8pt,-13*0.8pt)..(-40*0.8pt,-15*0.8pt);

\draw (-25*0.8pt,-2*0.8pt)..controls (-25*0.8pt,14*0.8pt) and (25*0.8pt,14*0.8pt)..(25*0.8pt,-2*0.8pt);
\draw [densely dashed] (-25*0.8pt,-2*0.8pt)..controls (-25*0.8pt,-14*0.8pt) and (25*0.8pt,-14*0.8pt)..(25*0.8pt,-2*0.8pt);

\node at (0pt,3pt){$c$};

\node at (-36*0.8pt, -33*0.8+2pt){$b_2$};
\node at (36*0.8pt, -33*0.8+2pt){$b_1$};

\node at (-73*0.8pt, 0*0.8-1pt){$a_2$};
\node at (73*0.8pt, 0*0.8-1pt){$a_1$};

\draw[>=stealth,->] (0.5pt,8pt)--(0.52pt,8pt);
\draw[>=stealth,->] (39.52*0.8pt,15*0.8pt)--(39.5*0.8pt,15*0.8pt);
\draw[>=stealth,->] (-40.5*0.8pt,15*0.8pt)--(-40.52*0.8pt,15*0.8pt);

\draw[>=stealth,->] (-25.8pt,-11.2pt)--(-25.8pt,-11pt);
\draw[>=stealth,->] ( 25.8pt,-11.2pt)--( 25.8pt,-11pt);

\end{tikzpicture}
\end{eqnarray}
The basis vectors for the degenerate ground states of a topological order on a genus-2 surface 
$\Sigma_{2,0}$ can be chosen as 
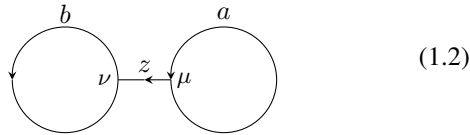
\begin{eqnarray}\label{BasisI_introduction}
\begin{tikzpicture}[baseline={(current bounding box.center)}]
\draw[>=stealth,<-] (0pt,0pt) arc (180:-180:20pt) ;
\draw[>=stealth,<-] (60pt,0pt) arc (180:-180:20pt) ;
\draw [>=stealth,->] (60pt,0pt)--(50pt,0pt);
\draw (50pt,0pt)--(40pt,0pt);
\node at (20pt, 25pt){$b$};
\node at (80pt, 25pt){$a$};
\node at (50pt, 5pt){$z$};
\node at (35pt, 0pt){\small $\nu$};
\node at (65pt, 0pt){\small $\mu$};
\end{tikzpicture}
\end{eqnarray}
where $a$, $b$, and $z$ denote anyon types, and $\mu$ and $\nu$ denote the fusion channels. 
By studying how the modular transformations act on the basis vectors, one can define the 
 representations of MCG($\Sigma_{2,0}$) for the five generators (Dehn twists) as 
$T_{a_1}$, $T_{b_1}$, $T_{a_2}$, $T_{b_2}$, and $T_c$. Alternatively, we can define the 
$S$ matrix as $S_i=T_{b_i}\cdot T_{a_i}^{-1}\cdot T_{b_i}=T_{a_i}^{-1}\cdot T_{b_i}\cdot T_{a_i}^{-1}$. 
Then by rewriting $T_i:=T_{a_i}$ ($i=1,\, 2$), the  representations of the five generators of MCG$(\Sigma_{2,0})$
are denoted as $T_1$, $S_1$, $T_2$, $S_2$, and $T_c$.
It is found that $T_i$ and $S_i$ ($i=1,\,2$) are also the  representations of the generators 
of MCG$(\Sigma_{1,1})$, where $\Sigma_{1,1}$ is a punctured torus. For this reason, we may 
call $S_i$ the punctured $S$ matrix, in comparison to the modular $S$ matrix for a torus without puncture.

In this work, we solve the punctured matrices $S_i$ and $T_i$ ($i=1,2$) explicitly.
With these data, we can distinguish the 5 different categories in MS MTCs in the following different ways:

\textit{ (i) Punctured $S$ matrix together with modular $T$ matrix.}

There are gauge freedoms in the punctured $S$ matrix. However, the diagonal parts 
of a punctured $S$-matrix are gauge invariant, and can be used to distinguish the counterexamples in MS MTCs.
More explicitly, 
by relabeling anyons, the modular $T$ matrix for $\mathcal{C}_{u=1\,(2)}$ 
is sent to $T$ for $\mathcal{C}_{u=4\,(3)}$.  
Then one can find that with the same way of 
relabeling anyons, one can not send the diagonal parts of punctured $S$ matrix for $\mathcal{C}_{u=1\,(2)}$  to those for $\mathcal{C}_{u=4\,(3)}$.
That is, the diagonal parts of punctured $S$ matrix together with the modular $T$ matrix can be used to distinguish the 5 different categories.

\textit{ (ii) Topological invariants.}

We propose an efficient and simple way to distinguish the MS MTCs. 
We construct the topological invariants based on the trace of \textit{words} as follows:
\be\label{TopInvariant_MCG}
W_{\Sigma_{2,0}}=\text{Tr}\left( D_1^{n_1}\cdots D_i^{n_i}\cdots \right)
\ee 
where the \textit{letters} $D_i$ can be arbitrarily chosen among the  representations 
of the five generators in MCG$(\Sigma_{2,0})$, and $n_i$ is an arbitrary integer.
Here the trace is taken over the Hilbert space of degenerate ground states on genus-2 surface $\Sigma_{2,0}$.

There are infinite number of words that can be used to distinguish the MS MTCs. 
As a simple example, we can choose $\small W_{\Sigma_{2,0}}=\text{Tr}\left[ (T_1)^7 (S_1)^{-7}\right]$, and the 
result of topological invariants for the 5 different MS MTCs is:
\be\label{Topo_invariants_intro}
W_{\Sigma_{2,0}}=
\left\{
\begin{split}
&a_0+b \cdot  e^{\frac{i 2\pi}{5}\times 0},\quad u=0,\\
&a_1+b \cdot  e^{\frac{i 2\pi}{5}\times 1},\quad u=1,\\
&a_2+b \cdot  e^{\frac{i 2\pi}{5}\times 2},\quad u=2,\\
&a_2+b \cdot  e^{\frac{i 2\pi}{5}\times 3},\quad u=3,\\
&a_1+b \cdot  e^{\frac{i 2\pi}{5}\times 4},\quad u=4,\\
\end{split}
\right.
\ee
where $a_i$ and $b$ are certain constants (see Table \ref{W1_Genus2_QP}).
It is found a single topological invariant can be used to distinguish different categories. 
This is in contrast to Refs.\onlinecite{Wang1805,DT1806,MS1806} where one needs to 
track how the link/knot invariant changes by permuting anyons. The topological 
invariant defined above is independent of the permutation or relabeling of anyons.
Furthermore, we show that each topological invariant in \eqref{TopInvariant_MCG}
is related to a specific link invariant.

As an alternative approach, we study the  representation of MCG$(\Sigma_{2,0})$
for a general Dijkgraaf-Witten theory with a finite group $G$ and three-cocyle $\omega\in H^3(G,U(1))$
on the lattice. We consider the minimal triangulation of a genus-2 surface as follows:
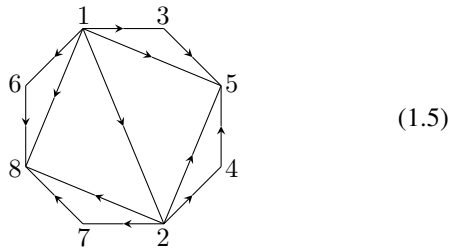
\begin{eqnarray}\label{Genus2Triangle_introduction}
\begin{tikzpicture}[baseline={(current bounding box.center)}]
\draw [>=stealth,->](-15.3073pt,36.9552pt)--(0pt,36.9552pt); \draw (0pt,36.9552pt)--(15.3073pt,36.9552pt);
\draw [>=stealth,->](15.3073pt,36.9552pt)--(26.1312pt,26.1312pt); \draw (26.1312pt,26.1312pt)--(36.9552pt,15.3073pt);
\draw [>=stealth,->](15.3073pt,-36.9552pt)--(0pt,-36.9552pt); \draw (0pt,-36.9552pt)--(-15.3073pt,-36.9552pt);
\draw [>=stealth,->](-15.3073pt,-36.9552pt)--(-26.1312pt,-26.1312pt); \draw (-26.1312pt,-26.1312pt)--(-36.9552pt,-15.3073pt);
\draw [>=stealth,->](15.3073pt,-36.9552pt)--(26.1312pt,-26.1312pt); \draw (26.1312pt,-26.1312pt)--(36.9552pt,-15.3073pt);
\draw [>=stealth,->](-15.3073pt,36.9552pt)--(-26.1312pt,26.1312pt); \draw (-26.1312pt,26.1312pt)--(-36.9552pt,15.3073pt);
\draw [>=stealth,->](36.9552pt,-15.3073pt)--(36.9552pt,0pt); \draw (36.9552pt,0pt)--(36.9552pt,15.3073pt);
\draw [>=stealth,->](-36.9552pt,15.3073pt)--(-36.9552pt,0pt); \draw (-36.9552pt,0pt)--(-36.9552pt,-15.3073pt);
\node at (-15pt,42pt){$1$};
\node at (15pt,42pt){$3$};

\node at (-15pt,-42pt){$7$};
\node at (15pt,-42pt){$2$};

\node at (41pt,16pt){$5$};
\node at (41pt,-16pt){$4$};

\node at (-41pt,16pt){$6$};
\node at (-41pt,-16pt){$8$};

\draw [>=stealth,->](-15.3073pt,36.9552pt)--(0pt,0pt); \draw (0pt,0pt)--(15.3073pt,-36.9552pt);
\draw [>=stealth,->](-15.3073pt,36.9552pt)--(10.8239pt,26.1312pt); \draw (10.8239pt,26.1312pt)--(36.9552pt,15.3073pt); 
\draw [>=stealth,->](15.3073pt,-36.9552pt)--(-10.8239pt,-26.1312pt); \draw (-10.8239pt,-26.1312pt)--(-36.9552pt,-15.3073pt); 
\draw [>=stealth,->](15.3073pt,-36.9552pt)--(26.1312pt,-10.8239pt); \draw (26.1312pt,-10.8239pt)--(36.9552pt,15.3073pt);
\draw [>=stealth,->](-15.3073pt,36.9552pt)--(-26.1312pt,10.8239pt); \draw (-26.1312pt,10.8239pt)--(-36.9552pt,-15.3073pt);
\end{tikzpicture}
\end{eqnarray}
where we identify the edges 
$[24]$ with $[35]$, $[45]$ with $[13]$, $[27]$ with $[68]$, and $[78]$ with $[16]$.
This triangulation has six triangle faces and only one vertex.
By defining the ground state wavefunction on the above triangulated lattice, we can perform Dehn twists on this lattice,
and study how the basis vectors transform into each other. Then we can obtain the  representations of these 
generators. The only input are the finite group $G$ and the 3-cocycle $\omega$.

We apply this lattice gauge theory approach to MS MTCs with $G=\mathbb{Z}_{11}\rtimes \mathbb{Z}_5$.
Since the ground state bases we choose are different from the quasi-particle bases, the matrices $S_i$ and $T_i$
(which are basis dependent) look totally different from those obtained in the quasi-particle bases. 
But the topological invariants obtained from these two different approaches are the same, as expressed in \eqref{Topo_invariants_intro}. 
More topological invariants are presented in Sec.\ref{Sec: Lattice}.

The methods discussed above also apply to the punctured torus 
(see more detail in the main text) and a higher-genus surface.

In short, by considering the MS MTCs on the genus-2 surface and a punctured torus, 
we use both the quasi-particle basis calculation and the lattice gauge theory 
approach to obtain the  representations $S_i$ and $T_i$ in MCG$(\Sigma_{2,0})$ 
and MCG$(\Sigma_{1,1})$. 
With these  representations $S_i$ and $T_i$, we 
construct the topological invariants to distinguish the simplest counterexamples in MS MTCs
that cannot be distinguished by the modular data. 


The rest of this work is organized as follows.
We introduce the basic properties of mapping class group (MCG) of a surface of genus $g$ 
with $n$ punctures, and then introduce how to construct the topological invariants based on 
the  representations of MCG
in the rest of the introduction. 
In Sec.\ref{Sec: MS MTC}, we give a brief review of the 
MS modular categories, including the types of anyons and the modular data. Then in Sec.\ref{Sec: Quasi-particle}, by focusing on the simplest counterexamples in MS MTCs, we study the  representations of 
the mapping class group of a punctured torus and a genus-two surface.
With these  representations, we construct topological invariants to distinguish different MS MTCs.
In Sec.\ref{Sec: Lattice}, with an independent method on the topological lattice gauge theory,
we study the  representations of 
MCG on a genus-2 manifold and a punctured torus for a general Dijkgraaf-Witten theory. 
We apply the general results to the MS MTCs, and obtain the same results of topological invariants 
as those in Sec.\ref{Sec: Quasi-particle}.  
We give some discussions and conclude in Sec.\ref{Sec: DiscussAndConclusion}.
We also give several appendices on the properties of MS MTCs, algebraic theory of anyons and its application 
in the MCG representations, and further details on the lattice gauge theory approach and so on.

\subsection{Mapping class group}

In this subsection, we give a brief introduction to the mapping class group (MCG) of a connected and orientable 
two-dimensional surface with and without punctures.

\begin{figure}[h]
\begin{tikzpicture}[baseline={(current bounding box.center)}]
\draw  (-60pt,0pt) arc (-120:-60:20pt);
\draw  (-56pt,-1.5pt) arc (130:50:10pt);
\draw (-65pt,0pt)..controls (-63pt,10pt) and (-37pt,10pt)..(-35pt,0pt);
\draw [>=stealth,->](-65pt,0pt)..controls (-63pt,-10pt) and (-37pt,-10pt)..(-35pt,0pt);

\draw (-50pt,-2.5pt)..controls (-45pt,-5pt) and (-45pt,-20pt)..(-50pt,-22.5pt);
\draw [dashed](-50pt,-2.5pt)..controls (-55pt,-5pt) and (-55pt,-20pt)..(-50pt,-22.5pt);

\draw [>=stealth,->](-46.2pt,-12.5pt)--(-46.2pt,-12.4pt);
\draw  (-60+40pt,0pt) arc (-120:-60:20pt);
\draw  (-56+40pt,-1.5pt) arc (130:50:10pt);

\draw (-65+40pt,0pt)..controls (-63+40pt,10pt) and (-37+40pt,10pt)..(-35+40pt,0pt);
\draw [>=stealth,->](-65+40pt,0pt)..controls (-63+40pt,-10pt) and (-37+40pt,-10pt)..(-35+40pt,0pt);

\draw (-50+40pt,-2.5pt)..controls (-45+40pt,-5pt) and (-45+40pt,-20pt)..(-50+40pt,-22.5pt);
\draw [dashed](-50+40pt,-2.5pt)..controls (-55+40pt,-5pt) and (-55+40pt,-20pt)..(-50+40pt,-22.5pt);
\draw [>=stealth,->](-46.2+40pt,-12.5pt)--(-46.2+40pt,-12.4pt);

\draw  (-60+80pt,0pt) arc (-120:-60:20pt);
\draw  (-56+80pt,-1.5pt) arc (130:50:10pt);

\draw (-65+80pt,0pt)..controls (-63+80pt,10pt) and (-37+80pt,10pt)..(-35+80pt,0pt);
\draw [>=stealth,->](-65+80pt,0pt)..controls (-63+80pt,-10pt) and (-37+80pt,-10pt)..(-35+80pt,0pt);

\draw (-50+80pt,-2.5pt)..controls (-45+80pt,-5pt) and (-45+80pt,-20pt)..(-50+80pt,-22.5pt);
\draw [dashed](-50+80pt,-2.5pt)..controls (-55+80pt,-5pt) and (-55+80pt,-20pt)..(-50+80pt,-22.5pt);
\draw [>=stealth,->](-46.2+80pt,-12.5pt)--(-46.2+80pt,-12.4pt);

\draw  (-60+140pt,0pt) arc (-120:-60:20pt);
\draw  (-56+140pt,-1.5pt) arc (130:50:10pt);

\draw (-65+140pt,0pt)..controls (-63+140pt,10pt) and (-37+140pt,10pt)..(-35+140pt,0pt);
\draw [>=stealth,->](-65+140pt,0pt)..controls (-63+140pt,-10pt) and (-37+140pt,-10pt)..(-35+140pt,0pt);

\draw (-50+140pt,-2.5pt)..controls (-45+140pt,-5pt) and (-45+140pt,-20pt)..(-50+140pt,-22.5pt);
\draw [dashed](-50+140pt,-2.5pt)..controls (-55+140pt,-5pt) and (-55+140pt,-20pt)..(-50+140pt,-22.5pt);
\draw [>=stealth,->](-46.2+140pt,-12.5pt)--(-46.2+140pt,-12.4pt);


\draw [thick](-50pt,-22.5pt)..controls (-60pt,-25pt) and (-78pt,-20pt)..(-79pt,0pt);
\draw [thick](-50pt,22.5pt)..controls (-60pt,23pt) and (-78pt,20pt)..(-79pt,0pt);

\draw [thick](-50+40pt,-21-1.5pt)..controls(-50+35pt,-21-1.5pt) and (-50+30pt,-18-1.5pt) ..(-50+20pt,-18-1.5pt)..
controls (-50+10pt,-18-1.5pt) and (-50+5pt,-21-1.5pt)..(-50pt,-21-1.5pt);

\draw [thick](-50+40pt,21+1.5pt)..controls(-50+32pt,21+1.5pt) and (-50+30pt,18+1.5pt) ..(-50+20pt,18+1.5pt)..
controls (-50+10pt,18+1.5pt) and (-50+8pt,21+1.5pt)..(-50pt,21+1.5pt);

\draw [thick](-50+40+40pt,-21-1.5pt)..controls(-50+32+40pt,-21-1.5pt) and (-50+30+40pt,-18-1.5pt) ..(-50+20+40pt,-18-1.5pt)..
controls (-50+10+40pt,-18-1.5pt) and (-50+8+40pt,-21-1.5pt)..(-50+40pt,-21-1.5pt);
\draw [thick](-50+40+40pt,21+1.5pt)..controls(-50+32+40pt,21+1.5pt) and (-50+30+40pt,18+1.5pt) ..(-50+20+40pt,18+1.5pt)..
controls (-50+10+40pt,18+1.5pt) and (-50+8+40pt,21+1.5pt)..(-50+40pt,21+1.5pt);

\draw [thick](-50+40+100pt,-22.5pt)..controls(-50+130pt,-22.5pt) and (-50+120pt,-18pt) ..(-50+20+90pt,-18pt)..
controls (-50+100pt,-18pt) and (-50+90pt,-22.5pt)..(-50+80pt,-22.5pt);

\draw [thick](-50+40+100pt,22.5pt)..controls(-50+130pt,22.5pt) and (-50+120pt,18pt) ..(-50+20+90pt,18pt)..
controls (-50+100pt,18pt) and (-50+90pt,22.5pt)..(-50+80pt,22.5pt);

\draw [thick] (90pt,22.5pt)..controls (100pt,23pt) and (119pt,20pt)..(120pt,0pt);
\draw [thick] (90pt,-22.5pt)..controls (100pt,-23pt) and (119pt,-20pt)..(120pt,0pt);

\node at (60pt,-2pt){$\cdots$};
\node at (-50pt,12pt){\small $a_1$};
\node at (-10pt,12pt){\small $a_2$};
\node at (30pt,12pt){\small $a_3$};
\node at (90pt,12pt){\small $a_g$};
\node at (-40pt,-14pt){\small $b_1$};
\node at (0pt,-14pt){\small $b_2$};
\node at (0+40pt,-14pt){\small $b_3$};
\node at (0+100pt,-14pt){\small $b_g$};
\end{tikzpicture}
\caption{Homology basis for a surface of genus $g$.}
\label{Homology_g_Fig}
\end{figure}
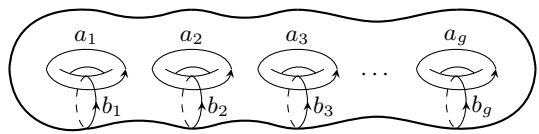

We denote the connected and orientable surface of genus $g$ with $n$ punctures as $\Sigma_{g,n}$, 
where the genus $g$ is the number of handles as shown in Fig.\ref{Homology_g_Fig}, and the
$n$ punctures are obtained by removing $n$ individual points from the surface.
It is usually convenient to think of the punctures as marked points on the surface.

For simplicity, we start from a closed oriented manifold $\Sigma_{g,0}$ of genus $g$ without punctures. 
Its topology is completely classified by the Euler number
$\chi(\Sigma_{g,0})=2-2g$. The first homology group $H_1(\Sigma_{g,0})$ has the dimension $\text{dim}\,H_1(\Sigma_{g,0})=2g$, and 
we can choose a canonical homology basis $\{a_i,\, b_i\}$ ($i=1,\cdots, g$) as shown in Fig.\ref{Homology_g_Fig}.
That is, the first homology group $H_1(\Sigma_{g,0})$ is generated by the $2g$ loops $a_i$ and $b_i$ in Fig.\ref{Homology_g_Fig}.

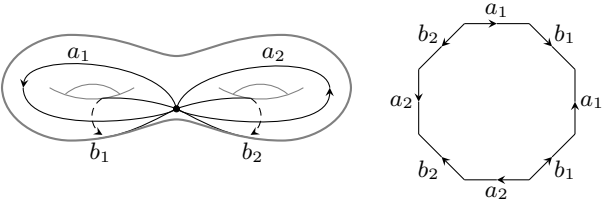
\begin{figure}[h]
\small
\begin{tikzpicture}[baseline={(current bounding box.center)}]

\draw  [gray](-60*0.8pt,0pt) arc (-120:-60:40*0.8pt);
\draw  [gray](-52.5*0.8pt,-3*0.8pt) arc (130:50:20*0.8pt);
\draw  [gray](60*0.8pt,0pt) arc (-60:-120:40*0.8pt);
\draw  [gray](52.5*0.8pt,-3*0.8pt) arc (50:130:20*0.8pt);

\draw (-35*0.8pt,-5*0.8pt)..controls (-20*0.8pt,-4*0.8pt) and (-5*0.8pt,-8*0.8pt)..(0pt,-10*0.8pt);
\draw (-37*0.8pt,-24*0.8pt)..controls (-20*0.8pt,-22*0.8pt) and (-5*0.8pt,-12*0.8pt)..(0pt,-10*0.8pt);

\draw  [>=stealth,->][densely dashed] (35*0.8pt,-5*0.8pt) arc (60:-60:10*0.8pt);
\draw [>=stealth,->][densely dashed](-35*0.8pt,-5*0.8pt) arc (120:240:10*0.8pt);

\draw  (0pt,-10*0.8pt)..controls (5*0.8pt,-8*0.8pt) and (20*0.8pt,-4*0.8pt)..(35*0.8pt,-5*0.8pt);
\draw  (0pt,-10*0.8pt)..controls (5*0.8pt,-12*0.8pt) and (20*0.8pt,-22*0.8pt)..(37*0.8pt,-24*0.8pt);

\draw  (0pt,-10*0.8pt)..controls (-5*0.8pt,-15*0.8pt) and (-72*0.8pt,-24*0.8pt)..(-72.5*0.8pt,0pt);
\draw [>=stealth,->](0pt,-10*0.8pt)..controls(-5*0.8pt,13*0.8pt) and (-70*0.8pt,18*0.8pt)..(-72.5*0.8pt,0pt);



\draw  [>=stealth,->](0pt,-10*0.8pt)..controls (5*0.8pt,-15*0.8pt) and (72*0.8pt,-24*0.8pt)..(72.5*0.8pt,0pt);
\draw (72.5*0.8pt,0pt)..controls(70*0.8pt,18*0.8pt) and (5*0.8pt,11*0.8pt)..(0pt,-10*0.8pt);

\draw [thick][gray]
(50*0.8pt,25*0.8pt)..controls (65*0.8pt,25*0.8pt) and (82*0.8pt,15*0.8pt)..
(82.5*0.8pt,0pt).. controls (82*0.8pt,-15*0.8pt) and (65*0.8pt,-25*0.8pt).. 
(50*0.8pt,-25*0.8pt)..controls (25*0.8pt,-25*0.8pt) and (10*0.8pt,-15*0.8pt)..
(0pt,-15*0.8pt)..controls (-10*0.8pt,-15*0.8pt) and (-25*0.8pt,-25*0.8pt)..
(-50*0.8pt,-25*0.8pt)..controls (-65*0.8pt,-25*0.8pt) and (-82*0.8pt,-15*0.8pt)..(-82.5*0.8pt,0pt)..controls (-82*0.8pt,15*0.8pt) 
and (-65*0.8pt,25*0.8pt)..(-50*0.8pt,25*0.8pt)..controls (-25*0.8pt,25*0.8pt) 
and (-10*0.8pt,15*0.8pt)..(0pt,15*0.8pt).. controls (10*0.8pt,15*0.8pt) and (25*0.8pt,25*0.8pt)..(50*0.8pt,25*0.8pt);

\node at (-36*0.8pt, -33*0.8+2pt){$b_1$};
\node at (36*0.8pt, -33*0.8+2pt){$b_2$};

\node at (-46*0.8pt, 17*0.8-1pt){$a_1$};
\node at (46*0.8pt, 17*0.8-1pt){$a_2$};


\draw [fill] (0pt,-10*0.8pt) circle [radius=0.04]; 
\end{tikzpicture}
\quad
\begin{tikzpicture}[baseline={(current bounding box.center)}]
\draw [>=stealth,->](-15.3073*0.8pt,36.9552*0.8pt)--(0pt,36.9552*0.8pt); \draw (0pt,36.9552*0.8pt)--(15.3073*0.8pt,36.9552*0.8pt);
\draw [>=stealth,->](15.3073*0.8pt,36.9552*0.8pt)--(26.1312*0.8pt,26.1312*0.8pt); \draw (26.1312*0.8pt,26.1312*0.8pt)--(36.9552*0.8pt,15.3073*0.8pt);
\draw [>=stealth,->](15.3073*0.8pt,-36.9552*0.8pt)--(0pt,-36.9552*0.8pt); \draw (0pt,-36.9552*0.8pt)--(-15.3073*0.8pt,-36.9552*0.8pt);
\draw [>=stealth,->](-15.3073*0.8pt,-36.9552*0.8pt)--(-26.1312*0.8pt,-26.1312*0.8pt); \draw (-26.1312*0.8pt,-26.1312*0.8pt)--(-36.9552*0.8pt,-15.3073*0.8pt);
\draw [>=stealth,->](15.3073*0.8pt,-36.9552*0.8pt)--(26.1312*0.8pt,-26.1312*0.8pt); \draw (26.1312*0.8pt,-26.1312*0.8pt)--(36.9552*0.8pt,-15.3073*0.8pt);
\draw [>=stealth,->](-15.3073*0.8pt,36.9552*0.8pt)--(-26.1312*0.8pt,26.1312*0.8pt); \draw (-26.1312*0.8pt,26.1312*0.8pt)--(-36.9552*0.8pt,15.3073*0.8pt);
\draw [>=stealth,->](36.9552*0.8pt,-15.3073*0.8pt)--(36.9552*0.8pt,0pt); \draw (36.9552*0.8pt,0pt)--(36.9552*0.8pt,15.3073*0.8pt);
\draw [>=stealth,->](-36.9552*0.8pt,15.3073*0.8pt)--(-36.9552*0.8pt,0pt); \draw (-36.9552*0.8pt,0pt)--(-36.9552*0.8pt,-15.3073*0.8pt);

\node at (0pt,42*0.8+1pt){$a_1$};
\node at (0pt,-42*0.8-1pt){$a_2$};

\node at (45*0.8pt,0pt){$a_1$};
\node at (-45*0.8pt,0pt){$a_2$};

\node at (31*0.8+1pt,31*0.8+1pt){$b_1$};
\node at (31*0.8+1pt,-31*0.8-1pt){$b_1$};

\node at (-31*0.8-1pt,31*0.8+1pt){$b_2$};
\node at (-31*0.8-1pt,-31*0.8-1pt){$b_2$};




\end{tikzpicture}
\caption{A manifold $\Sigma_{2,0}$ of genus $g=2$ (left).
We cut the Riemann surface along the canonical curves ($a_i$ and $b_i$ cycles) to obtain a connected $4g$-gon, which is an
octagon here (right).}
\label{Genus2_homology}
\end{figure}

Once the homology basis $\{a_i,\, b_i\}$ ($i=1,\cdots, g$) is chosen, 
it is useful to represent the manifold $\Sigma_{g,0}$ in terms of a $4g$-side polygon in the following way.
We choose a basepoint $x_0$ on $\Sigma_{g,0}$, and cut the surface along the
$2g$ curves that are homologous to the canonical basis. Then the Riemann surface unfolds into a $4g$ polygon 
(see Fig.\ref{Genus2_homology} for the example of $\Sigma_{2,0}$).
Note that that all the $4g$ vertices of the polygon are identified to a single 
point (which is the basepoint $x_0$ here) on the Riemann surface.
Equivalently, a Riemann surface $\Sigma_{g,0}$ can be considered as the quotient space of a polygon
with the edges identified in pairs.
To indicate which paired edges are to be identified,
we start at a definite vertex, proceed around the boundary of the polygon, and record the letters
assigned to the different sides in succession. 
If the arrow on the side points in the same (opposite) direction that we go around the boundary, then 
we write the letter for that side with exponent $1$ ($-1$).
For the case of $g=2$ in Fig.\ref{Genus2_homology}, the identifications of edges are indicated by the symbols
$b_2^{-1}a_2^{-1}b_2a_2b_1^{-1}a_1^{-1}b_1a_1$, where the symbols are read from right to left.
Similarly, for $\Sigma_{g,0}$ in Fig.\ref{Homology_g_Fig}, the symbols representing the identification of 
edges for a $4g$-sided polygon is $b_g^{-1}a_g^{-1}b_ga_g\cdots b_2^{-1}a_2^{-1}b_2a_2b_1^{-1}a_1^{-1}b_1a_1$.
Representing a Riemann surface $\Sigma$ by a polygon will be useful in studying a topological lattice gauge theory in Sec.\ref{Sec: Lattice}.

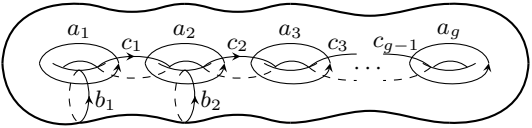
\begin{figure}[h]
\begin{tikzpicture}[baseline={(current bounding box.center)}]
\draw  (-60pt,0pt) arc (-120:-60:20pt);
\draw  (-56pt,-1.5pt) arc (130:50:10pt);

\draw (-65pt,0pt)..controls (-63pt,10pt) and (-37pt,10pt)..(-35pt,0pt);
\draw [>=stealth,->](-65pt,0pt)..controls (-63pt,-10pt) and (-37pt,-10pt)..(-35pt,0pt);

\draw (-50pt,-2.5pt)..controls (-45pt,-5pt) and (-45pt,-20pt)..(-50pt,-22.5pt);
\draw [dashed](-50pt,-2.5pt)..controls (-55pt,-5pt) and (-55pt,-20pt)..(-50pt,-22.5pt);
\draw [>=stealth,->](-46.2pt,-12.5pt)--(-46.2pt,-12.4pt);
\draw  (-60+40pt,0pt) arc (-120:-60:20pt);
\draw  (-56+40pt,-1.5pt) arc (130:50:10pt);

\draw (-65+40pt,0pt)..controls (-63+40pt,10pt) and (-37+40pt,10pt)..(-35+40pt,0pt);
\draw [>=stealth,->](-65+40pt,0pt)..controls (-63+40pt,-10pt) and (-37+40pt,-10pt)..(-35+40pt,0pt);

\draw (-50+40pt,-2.5pt)..controls (-45+40pt,-5pt) and (-45+40pt,-20pt)..(-50+40pt,-22.5pt);
\draw [dashed](-50+40pt,-2.5pt)..controls (-55+40pt,-5pt) and (-55+40pt,-20pt)..(-50+40pt,-22.5pt);
\draw [>=stealth,->](-46.2+40pt,-12.5pt)--(-46.2+40pt,-12.4pt);

\draw  (-60+80pt,0pt) arc (-120:-60:20pt);
\draw  (-56+80pt,-1.5pt) arc (130:50:10pt);

\draw (-65+80pt,0pt)..controls (-63+80pt,10pt) and (-37+80pt,10pt)..(-35+80pt,0pt);
\draw [>=stealth,->](-65+80pt,0pt)..controls (-63+80pt,-10pt) and (-37+80pt,-10pt)..(-35+80pt,0pt);

\draw  (-60+140pt,0pt) arc (-120:-60:20pt);
\draw  (-56+140pt,-1.5pt) arc (130:50:10pt);

\draw (-65+140pt,0pt)..controls (-63+140pt,10pt) and (-37+140pt,10pt)..(-35+140pt,0pt);
\draw [>=stealth,->](-65+140pt,0pt)..controls (-63+140pt,-10pt) and (-37+140pt,-10pt)..(-35+140pt,0pt);

\draw (-43pt,-1.5pt)..controls (-38pt,4pt) and (-20.5pt,4pt)..(-15.5pt,-1.5pt);
\draw [dashed](-43pt,-1.5pt)..controls (-38pt,-7pt) and (-20.5pt,-7pt)..(-15.5pt,-1.5pt);

\draw [>=stealth,->](-29.1pt,2.6pt)--(-29pt,2.6pt);

\draw (-43+40pt,-1.5pt)..controls (-38+40pt,4pt) and (-20.5+40pt,4pt)..(-15.5+40pt,-1.5pt);
\draw [dashed](-43+40pt,-1.5pt)..controls (-38+40pt,-7pt) and (-20.5+40pt,-7pt)..(-15.5+40pt,-1.5pt);
\draw [>=stealth,->](-29.1+40pt,2.6pt)--(-29+40pt,2.6pt);

\draw (-43+80pt,-1.5pt)..controls (-38+80pt,3.0pt) and (-30+80pt,3.5pt)..(-25+80pt,3.5pt);
\draw [dashed](-43+80pt,-1.5pt)..controls (-38+80pt,-6pt) and (-30+80pt,-6.5pt)..(-25+80pt,-6.5pt);

\draw (43+40pt,-1.5pt)..controls (38+40pt,3.0pt) and (30+40pt,3.5pt)..(25+40pt,3.5pt);
\draw [dashed](43+40pt,-1.5pt)..controls (38+40pt,-6pt) and (30+40pt,-6.5pt)..(25+40pt,-6.5pt);


\draw [thick](-50pt,-22.5pt)..controls (-60pt,-25pt) and (-78pt,-20pt)..(-79pt,0pt);
\draw [thick](-50pt,22.5pt)..controls (-60pt,23pt) and (-78pt,20pt)..(-79pt,0pt);

\draw [thick](-50+40pt,-21-1.5pt)..controls(-50+35pt,-21-1.5pt) and (-50+30pt,-18-1.5pt) ..(-50+20pt,-18-1.5pt)..
controls (-50+10pt,-18-1.5pt) and (-50+5pt,-21-1.5pt)..(-50pt,-21-1.5pt);

\draw [thick](-50+40pt,21+1.5pt)..controls(-50+32pt,21+1.5pt) and (-50+30pt,18+1.5pt) ..(-50+20pt,18+1.5pt)..
controls (-50+10pt,18+1.5pt) and (-50+8pt,21+1.5pt)..(-50pt,21+1.5pt);

\draw [thick](-50+40+40pt,-21-1.5pt)..controls(-50+32+40pt,-21-1.5pt) and (-50+30+40pt,-18-1.5pt) ..(-50+20+40pt,-18-1.5pt)..
controls (-50+10+40pt,-18-1.5pt) and (-50+8+40pt,-21-1.5pt)..(-50+40pt,-21-1.5pt);
\draw [thick](-50+40+40pt,21+1.5pt)..controls(-50+32+40pt,21+1.5pt) and (-50+30+40pt,18+1.5pt) ..(-50+20+40pt,18+1.5pt)..
controls (-50+10+40pt,18+1.5pt) and (-50+8+40pt,21+1.5pt)..(-50+40pt,21+1.5pt);

\draw [thick](-50+40+100pt,-22.5pt)..controls(-50+130pt,-22.5pt) and (-50+120pt,-18pt) ..(-50+20+90pt,-18pt)..
controls (-50+100pt,-18pt) and (-50+90pt,-22.5pt)..(-50+80pt,-22.5pt);

\draw [thick](-50+40+100pt,22.5pt)..controls(-50+130pt,22.5pt) and (-50+120pt,18pt) ..(-50+20+90pt,18pt)..
controls (-50+100pt,18pt) and (-50+90pt,22.5pt)..(-50+80pt,22.5pt);

\draw [thick] (90pt,22.5pt)..controls (100pt,23pt) and (119pt,20pt)..(120pt,0pt);
\draw [thick] (90pt,-22.5pt)..controls (100pt,-23pt) and (119pt,-20pt)..(120pt,0pt);

\node at (60pt,-2pt){$\cdots$};

\node at (-50pt,12pt){\small $a_1$};
\node at (-10pt,12pt){\small $a_2$};
\node at (30pt,12pt){\small $a_3$};

\node at (-30pt,7pt){\small $c_1$};
\node at (10pt,7pt){\small $c_2$};
\node at (48pt,7pt){\small $c_3$};
\node at (70pt,7pt){\small $c_{g-1}$};

\node at (90pt,12pt){\small $a_g$};

\node at (-40pt,-14pt){\small $b_1$};
\node at (0pt,-14pt){\small $b_2$};

\end{tikzpicture}
\caption{Generators of $\text{MCG}(\Sigma_{g,0})$.
Dehn twists around the $(2g+1)$ simple closed curves generate $\text{MCG}(\Sigma_{g,0}$).
}
\label{MCG_g}
\end{figure}

Now let us introduce the mapping class group of a two dimensional manifold.
Let $\text{Diff}(\Sigma)$ be the group of orientation-preserving diffeomorphisms of $\Sigma$, and let 
$\text{Diff}_0(\Sigma)$ be the normal subgroup of diffeomorphisms homotopic to
the identity. Then the mapping class group, sometimes called the modular group, is defined by
\be
\text{MCG}(\Sigma)=\text{Diff}(\Sigma)/\text{Diff}_0(\Sigma).
\ee
For a closed manifold of genus $g$ (with $g>1$), the mapping class group can be generated by 
$(2g+1)$ Dehn twists along the $(2g+1)$ simple closed curves as depicted in Fig.\ref{MCG_g}.\cite{humphries1979generators}
In general, a Dehn twist along a closed curve means that we cut out an annulus around the curve
and rotate the two boundaries of the annulus relative to each other by $2\pi$, and then glue it back.
It is noted that if one is not limited to Dehn twists, the number of generators needs not to grow
with the complexity of the surface. For an orientable surface of genus $g$, either closed or with 
one puncture, its mapping class group can be generated by only two elements.\cite{wajnryb1996mapping}

The relations of the $(2g+1)$ Dehn twists that generate $\text{MCG}(\Sigma_{g,0})$ have been well studied,
and one may refer to Ref.\onlinecite{Wajnryb1983} for more details.
It is useful to consider how a nontrivial diffeomorphism acts on the homology basis in Fig.\ref{Homology_g_Fig}.
For example, the Dehn twist around $a_1$ induces the transformation on the homology basis
$b_1\to b_1+a_1$, with other homology bases unaffected. The Dehn twist along $c_1$ induces the transformation $a_1\to a_1-b_1+b_2$ and
$a_2\to a_2+b_1-b_2$, with other bases unaffected.

One basic example of $\text{MCG}(\Sigma_{g,0})$ is for $g=1$, \textit{i.e.}, a torus without any puncture/boundary. 
One can find that $\text{MCG}(\Sigma_{1,0})\cong \text{SL}(2,\mathbb{Z})$, with the explicit expression:
\be
\text{MCG}(\Sigma_{1,0})\cong \langle\mathfrak{s}, \mathfrak{t}| \mathfrak{s}^4=1, (\mathfrak{s}\mathfrak{t})^3=\mathfrak{s}^2 \rangle,
\ee
where $\mathfrak{t}$ can be considered as the Dehn twist $\mathfrak{t}_{a_1}$ around $a_1$ in Fig.\ref{Homology_g_Fig}, 
and $\mathfrak{s}$, the so-called $S$ transformation, is a proper combination of the two Dehn twists around $a_1$ and $b_1$ as 
$\mathfrak{s}=\mathfrak{t}_{b_1}\cdot\mathfrak{t}_{a_1}^{-1}\cdot\mathfrak{t}_{b_1}=\mathfrak{t}_{a_1}^{-1}\cdot\mathfrak{t}_{b_1}\cdot \mathfrak{t}_{a_1}^{-1}$. Apparently, $\text{MCG}(\Sigma_{1,0})$ can be alternatively generated by the two Dehn twists 
$\mathfrak{t}_{a_1}$ and $\mathfrak{t}_{b_1}$.

Now we consider adding punctures on the manifold. 
For the surface $\Sigma_{g,n}$ as introduced at the beginning of the introduction, it 
can be viewed as removing $n$ distinct points from the compact manifold $\Sigma_{g,0}$.
$\text{MCG}(\Sigma_{g,n})$ is generated by Denh twists along the curves as in the closed manifold, except that now we have more curves.
For example, we consider a closed curve that includes a puncture in its interior, then doing a Dehn twist along this curve is nontrivial now.

\begin{figure}[ht]
\begin{tikzpicture}[baseline={(current bounding box.center)}]
\small
\draw [gray](0+12pt,0pt) arc (-60:-120:25pt);
\draw [gray](-3+12pt,-1.6pt) arc (50:130:15pt);

\draw [gray][thick]
(0pt,20pt)..controls (15pt,20pt) and (34pt,12pt)..
(34.5pt,0pt)..controls (34pt,-12pt) and (15pt,-20pt)..(0pt,-20pt)..
controls (-15pt,-20pt) and (-34pt,-12pt)..(-34.5pt,0pt)..controls(-34pt,12pt) and (-15pt,20pt)..(0pt,20pt);

\draw  [>=stealth,->](-20pt,-10*0.8pt)..controls (-16pt,-15pt) and (20pt,-24*0.8pt)..(22pt,0pt);
\draw (22pt,0pt)..controls(20pt,18pt) and (-16pt,15pt)..(-20pt,-10*0.8pt);
\node at (28pt,0pt){${a_1}$};

\draw  (-20pt,-8pt)..controls (-18pt,-6pt) and (-10pt,-4pt)..(-2pt,-3.3pt);
\draw  (-20pt,-8pt)..controls (-18pt,-14pt) and (-10pt,-18pt)..(-4pt,-20pt);
\draw [>=stealth,->][dashed](-2pt,-3.3pt)..controls (3pt,-4pt) and (3pt,-19pt)..(-4pt,-20pt);
\node at (-4pt,-24pt){${b_1}$};

\draw [>=stealth,->](-20pt,-8pt)..controls (-22pt,-8pt) and (-34pt,-2pt)..(-26pt,5pt);
\draw (-26pt,5pt)..controls (-22.5pt,7pt) and (-22pt,2pt)..(-20pt,-8pt);
\node at (-24pt,9pt){${c}$};

\draw [fill] (-20pt, -8pt) circle [radius=0.04];

\draw [black] (-25pt,0pt) circle [radius=0.05]; 
\end{tikzpicture}
\quad\quad
\begin{tikzpicture}[baseline={(current bounding box.center)}]
\draw (-20pt,20pt)--(-20pt,0pt); \draw [>=stealth,<-] (-20pt, 0pt) -- (-20pt,-20pt);
\draw (20pt,20pt)--(20pt,0pt); \draw [>=stealth,<-] (20pt, 0pt) -- (20pt,-20pt);
\draw [>=stealth,->] (-20pt,20pt)--(0pt,20pt); \draw (0pt,20pt)--(20pt,20pt);
\draw [>=stealth,->] (-20pt,-20pt)--(0pt,-20pt);  \draw (0pt, -20pt) -- (20pt,-20pt);
\draw [black] (-10pt,-9pt) circle [radius=0.05]; 
\node at (28pt,0pt){${b_1}$};
\node at (0pt,-27pt){${a_1}$};
\node at (0pt,27pt){${a_1}$};
\node at (-28pt,0pt){${b_1}$};
\draw [>=stealth,->](-20pt,-20pt)..controls (0pt,15pt) and (0pt,-10pt)..(-20pt,-20pt);
\node at (0pt,0pt){${c}$};
\end{tikzpicture}
\caption{A punctured torus $\Sigma_{1,1}$(left), where $\circ$ denotes the puncture where a point is removed,
 and its unfolding by cutting along $a_1$ and $b_1$ (right).
}
\label{PuncturedTorus}
\end{figure}
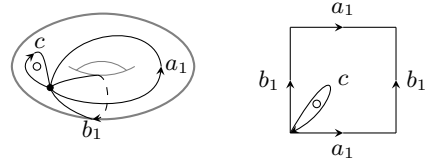

For later use, let us consider the punctured torus $\Sigma_{1,1}$, as depicted in Fig.\ref{PuncturedTorus}.
One can find that MCG$(\Sigma_{1,1})$ is a central extension of SL($2,\mathbb{Z}$), with the explicit expression:
\be\label{MCG_punctuedTorus}
\text{MCG}(\Sigma_{1,1})\cong \langle\mathfrak{s}, \mathfrak{t}| \mathfrak{s}^4=\mathfrak{r}^{-1}, (\mathfrak{s}\mathfrak{t})^3=\mathfrak{s}^2 \rangle,
\ee
where $\mathfrak{r}$ represents the $2\pi$ rotation around the puncture or equivalently the Dehn twist around the closed curve $c$
in Fig.\ref{PuncturedTorus}.
That is, $\text{MCG}(\Sigma_{1,1})$ is still generated by $\mathfrak{s}$ and $\mathfrak{t}$, or equivalently the Dehn twists 
$\mathfrak{t}_{a_1}$ and $\mathfrak{t}_{a_2}$ around the closed curves $a_1$ and $a_2$ in Fig.\ref{PuncturedTorus}. But the 
modular relation is modified because of the introduction of the puncture.

For a manifold $\Sigma_{g,n}$ with $n>1$, $\text{MCG}(\Sigma_{g,n})$ is allowed to permute punctures, and one needs to
introduce the operation of `half twist' which exchanges two punctures. It can be proved that for any $g,\, n\ge 0$, the group $\text{MCG}(\Sigma_{g,n})$ 
is generated by a finite number of Dehn twists and half-twists.
One may refer to, \textit{e.g.}, Ref.\onlinecite{MCG2011primer}, for more details on $\text{MCG}(\Sigma_{g,n})$.

\subsection{ Representation of mapping class group}
\label{Sec: Introduction_PojectiveRep_MCG}

Now given a unitary modular category, we introduce the construction of Hilbert space of states and the  representations of mapping class group.
Let us give a general picture first. 
It is known that a modular category gives rise to a (2+1) dimensional topological field theories (TQFTs).
For a (2+1) dimensional TQFT, the two dimensional surface $\Sigma$ bounds
a three dimensional open manifold. The path integral on this 3-manifold  can be viewed as a wavefunction in the Hilbert space $\mathcal{H}(\Sigma)$
associated with $\Sigma$.\cite{witten1989quantum}  
Considering a diffeomorphism $f:\, \Sigma\to \Sigma$, then there is a corresponding automorphism 
$f_{\ast}:\, \mathcal{H}(\Sigma)\to \mathcal{H}(\Sigma)$, so that composition of diffeomorphisms of $\Sigma$ corresponds to composition of 
vector-space isomorphisms.
That is, the mapping class group of $\Sigma$ acts as automorphisms of $\mathcal{H}(\Sigma)$. It therefore provides a (possibly )
representation of MCG.

Now we give a more precise description. 
Let us start from a two dimensional closed oriented surfaces $\Sigma_{g,0}$, which can be considered as the boundaries of three dimensional 
manifold $H_g$ in $\mathbb{R}^3$. For example, the two dimensional surface $\Sigma_{1,0}=S_1\times S_1$ bounds a three dimensional solid torus $D_2\times S_1$ or $S^1\times D^2$.
Following \onlinecite{turaev2016quantum} (see also Refs.\onlinecite{reshetikhin1991invariants,reshetikhin1990ribbon,bloomquist2018topological}), 
we assign a trivalent graph at the core of the 3-manifold $H_g$ (More rigorously, the trivalent graph here is called a ribbon graph in literature\cite{turaev2016quantum}).  
That is, the neighborhood of this trivalent graph is $H_g$.
Then, we color this trivalent graph by different anyons $a\in \Pi_{\mathcal{C}}$ in the modular category $\mathcal{C}$.
Here $\Pi_\mathcal{C}$ represents the collection of anyons in the modular category. 
One canonical choice of the colored trivalent graph for $\Sigma_{g,0}$ is:
\begin{eqnarray}\label{Genus_g_basis}
\footnotesize
\begin{tikzpicture}[baseline={(current bounding box.center)}]
\draw [>=stealth,->](12pt,-12pt) arc (-90:90:12pt) ;
\draw [>=stealth,->](12pt,12pt) arc (90:270:12pt) ;
\node at (12pt,18pt){{$b_1$}};
\node at (12pt,-17pt){{$c_1=b_1$}};

\draw [>=stealth,->](12+35pt,-12pt) arc (-90:90:12pt) ;
\draw [>=stealth,->](12+35pt,12pt) arc (90:270:12pt) ;
\node at (12+35pt,18pt){{$b_2$}};
\node at (12+35pt,-17pt){{$c_2$}};
\draw [>=stealth,->](24pt,0pt)--(31pt,0pt); \draw(31pt,0pt)--(35pt,0pt);
\node at (30pt,5pt){{$a_1$}};

\draw [>=stealth,->](12+70pt,-12pt) arc (-90:90:12pt) ;
\draw [>=stealth,->](12+70pt,12pt) arc (90:270:12pt) ;
\node at (12+70pt,18pt){{$b_3$}};
\node at (12+70pt,-17pt){{$c_3$}};
\draw [>=stealth,->](59pt,0pt)--(66pt,0pt);\draw(66pt,0pt)--(70pt,0pt);
\node at (65pt,5pt){{$a_2$}};

\draw [>=stealth,->](94pt,0pt)--(101pt,0pt);\draw(101pt,0pt)--(105pt,0pt);
\node at (100pt,5pt){{$a_3$}};

\node at (113pt,0pt){{\scriptsize $\cdots$}};
\draw [>=stealth,->](120pt,0pt)--(127pt,0pt);\draw(127pt,0pt)--(131pt,0pt);
\node at (122pt,5pt){{$a_{g-1}$}};

\draw [>=stealth,->](12+131pt,-12pt) arc (-90:90:12pt) ;
\draw [>=stealth,->](12+131pt,12pt) arc (90:270:12pt) ;
\node at (12+131pt,18pt){{$b_g$}};
\node at (12+131pt,-17pt){{$c_g=b_{g+1}$}};
\quad
\end{tikzpicture}
\end{eqnarray}
where $a_i$, $b_i$, and $c_i$ denote the anyon types.  
Note that at the vertices we need to specify the fusion channel $u_i$
 where $a_i$ and $b_{i+1}$ fuse into $c_{i+1}$ , and $\nu_i$ where $\bar{b}_i$ and $c_{i}$ fuse into $a_i$.
Then the Hilbert space $\mathcal{H}(\Sigma_{g,0})$ is spanned by the basis 
$|a_i,b_i,c_i; \mu_i, \nu_i\rangle$ with $i=1,\cdots, g$,
corresponding to different ways of coloring of the trivalent graph in \eqref{Genus_g_basis}.
For an Abelian theory, it can be found that $a_1,\cdots, a_{g-1}$ all become identity anyons, and the configuration in \eqref{Genus_g_basis}
decomposes into $g$ isolated circles.
The dimension of the Hilbert space $\mathcal{H}(\Sigma_{g,0})$, or the ground state degeneracy (GSD) on a closed manifold $\Sigma_{g,0}$, 
can be expressed in terms of the modular $S$ matrix as\cite{verlinde1988fusion}
\be\label{GSD_genus_g}
\small
\text{dim} \mathcal{H}(\Sigma_{g,0})=\sum_{a\in \Pi_{\mathcal{C}}}\left(\frac{1}{S_{0a} }\right)^{2(g-1)}.
\ee

Now if we consider an oriented manifold $\Sigma_{g,n}$  of genus $g$ with $n$ punctures, then the basis in \eqref{Genus_g_basis}
is modified as
\begin{eqnarray}\label{Genus_g_Puncture_n_basis}
\footnotesize
\begin{tikzpicture}[baseline={(current bounding box.center)}]
\draw [>=stealth,->](12pt,-12pt) arc (-90:90:12pt) ;
\draw [>=stealth,->](12pt,12pt) arc (90:270:12pt) ;
\node at (12pt,18pt){{$b_1$}};

\draw [>=stealth,->](12+35pt,-12pt) arc (-90:90:12pt) ;
\draw [>=stealth,->](12+35pt,12pt) arc (90:270:12pt) ;
\node at (12+35pt,18pt){{$b_2$}};
\node at (12+35pt,-17pt){{$c_2$}};
\draw [>=stealth,->](24pt,0pt)--(31pt,0pt); \draw(31pt,0pt)--(35pt,0pt);
\node at (30pt,5pt){{$a_1$}};

\draw [>=stealth,->](12+70pt,-12pt) arc (-90:90:12pt) ;
\draw [>=stealth,->](12+70pt,12pt) arc (90:270:12pt) ;
\node at (12+70pt,18pt){{$b_3$}};
\node at (12+70pt,-17pt){{$c_3$}};
\draw [>=stealth,->](59pt,0pt)--(66pt,0pt);\draw(66pt,0pt)--(70pt,0pt);
\node at (65pt,5pt){{$a_2$}};

\draw [>=stealth,->](94pt,0pt)--(101pt,0pt);\draw(101pt,0pt)--(105pt,0pt);
\node at (100pt,5pt){{$a_3$}};

\node at (113pt,0pt){{\scriptsize $\cdots$}};
\draw [>=stealth,->](120pt,0pt)--(127pt,0pt);\draw(127pt,0pt)--(131pt,0pt);
\node at (122pt,5pt){{$a_{g-1}$}};

\draw [>=stealth,->](12+131pt,-12pt) arc (-90:90:12pt) ;
\draw [>=stealth,->](12+131pt,12pt) arc (90:270:12pt) ;
\node at (12+131pt,18pt){{$b_g$}};
\node at (12+131pt,-17pt){{$c_g$}};
\draw [>=stealth,->](155pt,0pt)--(162pt,0pt);
\draw (162pt,0pt)--(170pt,0pt);
\draw [>=stealth,->](170pt,0pt)--(170pt,8pt); \draw(170pt,8pt)--(170pt,12pt);

\node at (162pt,-5pt){{$e_1$}};
\node at (170pt,18pt){{$i_1$}};

\draw [>=stealth,->](170pt,0pt)--(177pt,0pt);\draw (177pt,0pt)--(182pt,0pt);
\draw [>=stealth,->](182pt,0pt)--(182pt,8pt); \draw(182pt,8pt)--(182pt,12pt);
\node at (177pt,-5pt){{$e_2$}};
\node at (182pt,18pt){{$i_2$}};

\draw [>=stealth,->](182pt,0pt)--(189pt,0pt);
\node at (195pt,0pt) {{$\cdots$}};
\draw (199pt,0pt)--(205pt,0pt);
\draw [>=stealth,->](205pt,0pt)--(205pt,8pt); \draw(205pt,8pt)--(205pt,12pt);
\node at (205pt,18pt){{$i_n$}};
\node at (215pt,0pt) {{$ $}};
\end{tikzpicture}
\end{eqnarray}
which span the Hilbert space $\mathcal{H}(\Sigma_{g,n})$.
Here $i_1, \cdots, i_n$ in \eqref{Genus_g_Puncture_n_basis} denote the anyon types at the punctures. 
The dimension of Hilbert space with \textit{fixed} anyon types $i_1, \cdots, i_n$ at the punctures
is expressed in terms of the modular $S$ matrix as\cite{Moore1989}
\be\label{HibertSpace_Punctures_introduction}
\small
\begin{split}
&\text{dim}\mathcal{H}\left(\Sigma_{g,n}; i_1,\cdots, i_n \right)
=\sum_{a\in \Pi_{\mathcal{C}}} \left(\frac{1}{S_{0a}}\right)^{2(g-1)}\frac{S_{i_1 a}}{S_{0a}}\cdots\frac{S_{i_n a}}{S_{0a}},
\end{split}
\ee
which reduces to the result in Eq.\eqref{GSD_genus_g} when $i_1,\,\cdots,\,i_n$ 
at the punctures are all identity anyons.
It is remarked that by gluing the punctured-torus bases (see the left plot in \eqref{PunctureTorusBasis}) to the 
punctures $i_m$ ($m=1,\cdots, n$) in \eqref{Genus_g_Puncture_n_basis}, 
one can obtain the basis in a closed manifold of higher genus.

Now we construct the  representations of mapping class groups with the basis vectors 
$|v_{\alpha}\rangle\in\mathcal{H}(\Sigma_{g,n})$,
where $\alpha=1,\cdots, \text{dim}\,\mathcal{H}(\Sigma_{g,n})$. 
The general procedures are as follows. For an orientation preserving diffeomorphism $f:\, \Sigma_{g,n}\to \Sigma_{g,n}$, we consider 
a mapping cylinder $\Sigma_{g,n}\times [0, 1]$, with
$\Sigma_{g,n}\times \{0\}$ parametrized by identity, and $\Sigma_{g,n}\times \{1\}$ parametrized by $f$. 
Now we glue the 3-manifold $H_g$ with a specified coloring $\alpha$ in \eqref{Genus_g_Puncture_n_basis}
to the surface $\Sigma_{g,n}\times \{0\}$ with the identity operation, and glue another 3-manifold $H_g$ with coloring $\beta$
to the surface $\Sigma_{g,n}\times \{1\}$ with $f$. Then we obtain a 3 dimensional manifold $M$ with a certain ribbon graph
denoted by $\Omega$. The path integral on the manifold $M$ with coloring $\Omega$ corresponds to the amplitude 
$\langle v_{\beta}|f_{\ast}|v_{\alpha}\rangle$, 
which defines the action of the mapping class group on the Hilbert space,  $f_{\ast}: \mathcal{H}(\Sigma_{g,n})\to \mathcal{H}(\Sigma_{g,n})$. 
Then we obtain the  representation of $\text{MCG}(\Sigma_{g,n})$. 
 
Note that the manifold $M$ and the coloring $\Omega$ depend on both $f$ and the choice of $\alpha$ ($\beta$). As an illustration, we consider
the example of $\Sigma_{1,0}$.\cite{witten1989quantum}
The basis $|v_{\alpha(\beta)}\rangle$ in $\mathcal{H}(\Sigma_{1,0})$ can be viewed as an anyon loop carrying 
anyon charge $\alpha$ ($\beta$) threading through a solid torus. Repeating the gluing procedure above, one can find that if $f=\mathfrak{s}$, 
then we obtain a three manifold $M=S^3$, with $\Omega$ being a Hopf link of two anyon loops carrying anyon charges 
$\alpha$ and $\beta$, respectively.
On the other hand, if $f=\mathfrak{t}$, we have $M=S^2\times S^1$, and $\Omega$ corresponds to two parallel anyon loops carrying anyon
charge $\alpha$ and $\beta$ threading along $S^1$ direction, with the loop $\beta$ twisted by $2\pi$.
Evaluating the path integral on the closed manifold $M$, one can obtain the  
modular data $S$ and $T$ matrices,  corresponding to the  representation of $\mathfrak{s}$ and $\mathfrak{t}$, respectively.\cite{witten1989quantum}
For a unitary modular category, the modular data $(S,T)$ satisfy the following conditions:
\be\label{ST_relation_genus1}
\begin{split}
&(ST)^3=S^2, \quad S^2=C,\quad C^2=I,
\end{split}
\ee
where 
$d_a$ is the quantum dimension of anyon $a$, and $\mathcal{D}$
is the total quantum dimension, and
$C=(\delta_{a\bar{b}} )_{a,b\in\Pi_{\mathcal{C}} }$ is called the charge conjugation matrix of $\mathcal{C}$.
With the quasi-particle basis chosen in \eqref{Genus_g_basis} and $g=1$, $T$ is a diagonal matrix
with the diagonal elements related to the 
topological spin $\theta_a$ of anyon $a$ as follows
\be
e^{i\theta_a}=e^{i 2\pi \frac{c}{24}}\, T_{aa},
\ee 
where $c$ is the chiral central charge, which is $0$ in the MS MTCs.
Moreover, $S$ is a unitary matrix and has the following symmetries $S_{ab}=S_{ba}=S_{\bar{a}b}^{\ast}
=S^{\ast}_{a\bar{b}}=S_{\bar{a}\bar{b}}=S_{\bar{b}\bar{a}}$.

For a punctured torus $\Sigma_{1,1}$, recall that $\text{MCG}(\Sigma_{1,1})$ in Eq.\eqref{MCG_punctuedTorus} is still 
generated by $\mathfrak{s}$ and $\mathfrak{t}$. We denote the corresponding  representations as $S^{(z)}$
and $T^{(z)}$, where $z$ denotes the anyon type at the puncture. Then $S^{(z)}$ and $T^{(z)}$ satisfy the following modular relations\cite{Kitaev2006}
\be\label{ST_relation_puncture}
\small
(S^{(z)}T^{(z)})^3= (S^{(z)})^2, \quad (S^{(z)})^2=C^{(z)},\quad (C^{(z)})^2=\theta_{z}^{\ast},
\small
\ee
where 
$C^{(z)}$ may be viewed as the punctured charge conjugation, and $\theta_{z}$ is the topological spin of anyon $z$ at the puncture.
More details on the modular relations in \eqref{ST_relation_puncture} will be discussed in Sec.\ref{Sec: punctured S and T}.

For $\Sigma_{g,0}$, a general computation of the  representations of the $(2g+1)$ generators
in Fig.\ref{MCG_g} is performed in Ref.\onlinecite{bloomquist2018topological}. 
Therein, the  representations are expressed in terms of $F$ and $R$ matrices. 
The challenging problem is that, as we mentioned in the introduction, it is hard to calculate the 
$F$ and $R$ matrices even for a modular category with a small number of anyons.

\subsection{Topological invariants }

Based on the  representations of mapping class groups on a two dimensional 
manifold $\Sigma_{g,n}$ of genus $g$ with $n$ punctures,
one may construct different \textit{words}: $w:=D_{i_1}^{n_1}\cdot D_{i_2}^{n_2}\cdots 
D_{i_N}^{n_N}$, where the \textit{letters} $D_{i_m}$ denote the 
representations of $\text{MCG}(\Sigma_{g,n})$, and the power $n_m$
in $D_{i_m}^{n_m}$ is an arbitrary (positive or negative) integer. 
Then the topological invariants we construct are of the form
\be\label{Word_Def}
W:=\text{Tr}(w)=\text{Tr}\left(D_{i_1}^{n_1}\cdot D_{i_2}^{n_2}\cdots 
D_{i_j}^{n_j}\cdots\right).
\ee
Here the trace is over $\mathcal{H}(\Sigma_{g,n})$ as introduced in the previous subsection.
The physical meaning of the topological invariant in Eq.\eqref{Word_Def} can be understood as the partition function
on a closed 3 dimensional manifold of certain (possibly complicate) topology.
A well known example is that for the torus $\Sigma_{1,0}$, $\text{Tr}\big[(TST)^n\big]$ can be understood as the partition function 
on the special Lens space $L_{n,1}=S^3/\mathbb{Z}_n$, where $S$ and $T$ are the modular matrices.

We want to emphasize that it is possible to choose a specific basis $|v_a\rangle$ in $\mathcal{H}(\Sigma_{g,n})$, 
and define the topological invariant as $\langle v_a| D_{i_1}^{n_1}\cdot D_{i_2}^{n_2}\cdots 
D_{i_N}^{n_N} |v_a\rangle$. For example, for the torus $\Sigma_{1,0}$, $\langle v_a|S|v_a\rangle$ corresponds to a Hopf link 
invariants of two loops labeled by anyons $a$ in $S^3$. 
This kind of topological invariant, however, depends on the choice of anyon charge $a$. In MS MTCs, by permuting the anyons,
the modular data can be mapped to each other among different categories. To distinguish the MS MTCs, we want to design 
a topological invariant which is independent of the permutation of anyons. This is one of the underlying reasons why we 
use ``Tr''  over the whole Hilbert space $\mathcal{H}(\Sigma_{g,n})$ in the definition in Eq.\eqref{Word_Def}.

\section{MS modular categories}
\label{Sec: MS MTC}

In this section we give a brief review of the basic properties of MS modular categories.\cite{MS1708}
The modular data of MS MTCs will be used in constructing the representations of 
modular groups of higher-genus manifolds in Sec.\ref{Sec: Quasi-particle}.

The MS modular categories (MS-MCs) are modular categories that go beyond the modular data. 
They are representation categories of quantum doubles $D^{\omega}(G)$
of $G=\mathbb{Z}_q\rtimes_n \mathbb{Z}_p$, twisted by the three-cocyles 
$\omega\in H^3(\mathbb{Z}_q\rtimes_n \mathbb{Z}_p,U(1))$.
In the following we will always consider the normalized cocycles, \textit{i.e.},
the value of cocycle is 1 when one of arguments equals to identity.
Here $\omega$ satisfy the 3-cocyle condition
\be\label{3cocycle_condition}
\begin{split}
&\omega(g_1,g_2,g_3)\cdot \omega(g_0\cdot g_1,g_2,g_3)^{-1}\cdot \omega(g_0,g_1\cdot g_2,g_3)\\
&\cdot \omega(g_0,g_1,g_2\cdot g_3)^{-1}\cdot \omega(g_0,g_1,g_2)=1.
\end{split}
\ee

For the finite nonabelian gauge group $G=\mathbb{Z}_q\rtimes_n \mathbb{Z}_p$, 
$p$ and $q$ are primes with $p|(q-1)$.
$\mathbb{Z}_p$ acts on $\mathbb{Z}_q$ as multiplication by an element $n$ of multiplicative order $p$ in $\mathbb{Z}_q$.
To be concrete, the group $G$ has the presentation
\be
G=\{(a^l,b^m)|a^q=b^p=1,\,bab^{-1}=a^n \},
\ee
with $l\in\{0,1,\cdots,q-1\}$,  $m\in \{0,1,\cdots,p-1\}$, $n^p=1$ mod $q$, and $n\neq 1$.
\footnote{
It is known that there exist exactly 
two non-isomorphic groups of order $pq$. One is the cyclic group $\mathbb{Z}_{pq}$ which is Abelian, 
and the other is a non-Abelian group, which is a 
semidirect product $\mathbb{Z}_q\rtimes \mathbb{Z}_p=\mathbb{Z}_q\rtimes_n \mathbb{Z}_p$.} 
All such choices of $n$ give rise to isomorphic groups, \textit{i.e.}, the group does not depend on the choice of $n$.
The multiplication of two group elements are 
$(a^l,b^m)\cdot (a^{l'}, b^{m'})=(a^l (b^m a^{l'} b^{-m}),b^{m+m'})
=(a^{l+n^m l'},b^{m+m'})$.

For the cohomology group of $G$, one can consider the following short exact sequence:\cite{mac1995homology,MS1708}
\be
\small
0\to H^3(\mathbb{Z}_p,U(1))\to H^3(G,U(1))\to H^3(\mathbb{Z}_q, U(1))^{\mathbb{Z}_p}\to 0.
\ee
Denote $\kappa: \mathbb{Z}_p^3\to U(1)$ as the generator of the cohomology group $H^3(\mathbb{Z}_p,U(1))\cong \mathbb{Z}_p$.
We can take $\kappa$ as the $p$-th roots of unity. Then $\omega\in H^3(G,U(1))$ can be considered as the inflation of $\kappa$ to $G$,
\textit{i.e.}, $\omega=\text{Inf}_{\mathbb{Z}_p}^G \kappa$.
Then we can define $p$ modular categories 
of twisted quantum double of $G$ by $\mathcal{C}_u=D^{\omega^u}(G)$, with
$u=0, 1, \cdots, p-1$, and $\omega\in H^3(\mathbb{Z}_q\rtimes_n \mathbb{Z}_p,U(1))\cong \mathbb{Z}_p$.
To be concrete, the $3$-cocyles for MS MTCs are
\be\label{3cocycle}
\small
\omega^{u}(g,h,k)=\exp\big[\frac{2\pi i}{p^2}u\cdot [k_b]_p\cdot
\big(
[g_b]_p+[h_b]_p-[g_b+h_b]_p
\big)\big],
\ee
with $u=0,\cdots,p-1$ characterizing inequivalent cocycles. 
where we have denoted the group elements as
\be
g:=(a^{g_a},b^{g_b}),
\ee
and $[x]_p:=x$ mod $p$. 
In general, different equivalent classes of cocycles give rise to different modular data, but this is not the case for MS MTCs.
For this family of modular categories with $p>3$, MS found that there are $p$ inequivalent modular categories.
However, there are only three sets of distinct modular data.

Before introducing the modular data for the $p$ inequivalent modular categories, let us emphasize the meaning of `equivalence' (or 'inequivalence') of two modular categories of $D^{\omega}(G)$.
Two cocycles $\omega$ and $\nu$ on a group $G$ give the same category not only if they are cohomologous, but also if their 
cohomology classes are mapped to each other under the action of the automorphism group of $G$.
Alternatively, $D^{\omega}(G)$ and $D^{\nu}(G)$ are equivalent as modular categories if and only if there is 
an automorphism $f$ of $G$ such that $\nu$ and $f^{\ast}\omega$ are cohomologous.
For the MS MTCs with $G=\mathbb{Z}_q\rtimes \mathbb{Z}_p$, one can find that any automorphisms of $G$ fixes the unique subgroup 
$\mathbb{Z}_q$ of order $q$ and induces the identity
on the quotient $\mathbb{Z}_p$. Recall that the 3-cocycles $\omega$ are inflated from $H^3(\mathbb{Z}_p, U(1))$, 
then the $p$ modular categories are pairwise inequivalent.

\subsection{Anyons}
\label{MS MTCs: anyons}

In MS MTCs, there are in total $\frac{q^2-1+p^3}{p}$ anyons,
including $\big(\frac{q-1}{p}+p\big)$ type-$I$ anyons,
 $\big(\frac{q-1}{p}\times q\big)$ type-$A$ anyons, and $p(p-1)$ type-$B$ anyons. 
The anyons in $D^{\omega}(G)$ are parametrized by pairs  
\be\label{AnyonData}
\big ( g, \tilde{\chi} \big),
\ee
or alternatively by pairs $\big ( [g], \tilde{\chi} \big)$,
where $g\in G$ is a representative of a conjugacy class $[g]$ in $G$. 
To distinguish with the character $\chi$ of a linear representation,
$\tilde{\chi}$ is denoted as the character of 
$\alpha_g$- irreducible representation of the centralizer $C_G(g):=\{x|gx=xg\}$.
Here $\alpha_g$ is a two cocycle on $C_G(g)$ obtained from the three cocycle 
$\omega$ as follows
\be\label{2cocycle}
\alpha_g(x,y):=\frac{\omega(g,x,y)\cdot \omega(x,y,g)}{\omega(x,g,y)}.
\ee
Each $\alpha_g$ determines a class of  representations called $\alpha_g$- representation $\tilde{\rho}:$
$C_G(g)\to \text{GL}(C_G(g))$ obeying
\be\label{Rep}
\tilde{\rho}^g(x) \tilde{\rho}^g(y)=\alpha_g(x,y)\tilde{\rho}^g(x,y).
\ee
The two-cocycle condition for $\alpha_g$ corresponds to the associativity 
$\tilde{\rho}^g(x)[\tilde{\rho}^g(y) \tilde{\rho}(z)]=[\tilde{\rho}^g(x) \tilde{\rho}^g(y)] \tilde{\rho}^g(z)$.
Physically, for the pairs in \eqref{AnyonData}, we may call $g$ or $[g]$ the flux (magnetic charge), and $\tilde{\chi}$
the charge (electric charge). Moreover, the (charge) conjugation of $\big ( g, \tilde{\chi} \big)$ is isomorphic to 
$\big ( g^{-1}, \tilde{\chi}^{\ast} \big)$.

One can check there are $\frac{q-1}{p}+p$ conjugacy classes of $G$.
Depending on the conjugacy classes, we divide the anyons into three types (See Appendix \ref{Appendix_ZqZp} for more details):

(i) type-$I$ anyons:  $I_i:=(1,\chi_i)$.

The conjugacy class is $\{1\}$, where $1=(a^0,b^0)$ is the identity element of $G$. 
Since the two-cocyle $\alpha_g$ is trivial for $g=1$, here $\chi$ in Eq.\eqref{AnyonData} is the irreducible character of $G$.
The number of irreducible representations of $G$ are also $\frac{q-1}{p}+p$, equal to the number of conjugacy classes in $G$. 
This is the number of type-$I$ anyons.
In particular, $p$ of them are one dimensional, and $\frac{q-1}{p}$ of them are $p$ dimensional. 
Denoting the dimension of irreducible representation $\rho_{\mu}$ as $d_{\mu}$, then
one can check that
$\sum_{\mu} d_{\mu}^2=1^2\cdot p+p^2\cdot \frac{q-1}{p}=pq=|G|$.

(ii) type-$A$ anyons: $A_{l,m}:=\big(a^l,\, \chi_m\big)$.

Here we denote $a^l:=(a^l,b^0)$, where the integer $l$ has the value 
\be\label{ConjugacyA}
l\in \mathbb{Z}_q^{\times}/\langle n\rangle,
\ee
where $\mathbb{Z}_q^{\times}$ is the multiplicative group of integers coprime to $q$.
The conjugacy class that contains the group element $(a^l,b^0)$ is $\{(a^{l'},b^0)| l'=l\cdot n^k\,\text{mod}\, q;\, k=0, 1,\cdots, p-1\}$.
Note that the size of the conjugacy class $\big|[(a^l,b^0)]\big|=p$. The number of conjugacy classes is $|\mathbb{Z}^{\times}|/p=(q-1)/p$.
The centralizer subgroup of the representative $a^l$ is $C_G(a^l)=\{ (a^{l'},b^0)| l'=0, 1,\cdots, q-1 \}=\mathbb{Z}_q$.
One can find the 3-cocycle in Eq.\eqref{3cocycle} is trivial for the centralizer subgroup $\mathbb{Z}_q$. 
Therefore, $\tilde{\chi}$ reduce to $\chi$, the linear irreducible
representations in $\mathbb{Z}_q$, which are all one dimensional.
The quantum dimensions of a type-$A$ anyons are $p\times 1=p$.

(iii) type-$B$ anyons: $B_{k,n}:=\big( b^k,\tilde{\chi}_n\big)$.

Here we denote $b^k:=(a^0, b^k)$, where $k\in \{1,2,\cdots, p-1\}$.
The conjugacy class containing the group element $b^k$ is $\{(a^{l}, b^k)| l=0,\cdots,q-1\}$.
Then the size of conjugacy class is $q$. The number of conjugacy classes is $p-1$.
The centralizer subgroup of $(a^l,b^k)$ is $C_G((a^l,b^k))=\{ (a^0,b^{k'})| k'=0,1,\cdots, p-1\}=\mathbb{Z}_p$.
Different from the type-$I$ and type-$A$ anyons, now the 3-cocyle in Eq.\eqref{3cocycle} plays a role.
One can check that the associated 2-cocyle in Eq.\eqref{2cocycle} now becomes
\be
\alpha_{b^k}(b^{k_1},b^{k_2})=\omega(b^{k_1}, b^{k_2}, b^k),
\ee
where $k_1, k_2\in\{0, 1, \cdots, p-1\}$.
It can be found that the two cocycle $\alpha_{b^k}$ is actually a two coboundary. 
This kind of twisting is sometimes called ``cohomology trivial" (CT).\cite{CGR00}
For the two-coboundary $\alpha_{b^k}$, one can find a one-cochain $\epsilon_{b^k}$ such that 
\be\label{2cocycle_CT}
\alpha_{b^k}(g,h)=\delta \epsilon_{b^k}(g,h)=\epsilon_{b^k}(g)\, \epsilon_{b^k}(h)\, \epsilon_{b^k}^{-1}(gh).
\ee

Based on the definition of $\alpha_g$- representations in Eq.\eqref{Rep},
one can find that the  representation and the linear representation are related by a phase:
\be\label{LinearRep}
\tilde{\rho}^{b^k}_{n}(x) =\epsilon_{b^k}(x) \rho^{b^k}_{n}(x).
\ee
Here $n$ labels the $n$-th  representation. That is, the  representation is obtained 
from a usual representation by adding a twist $\epsilon_{b^k}$, which is a $U(1)$ phase.
Both the number and dimensions of the irreducible presentations stay the same by adding this twist, \textit{i.e.},
a CT twisted theory has the same number  and quantum dimensions of anyons as the untwisted theory.
Since the irreducible presentations of the centralizer subgroup $\mathbb{Z}_p$ are all one dimensional, 
the quantum dimensions of type-$B$ anyons are $q\times 1=q$.

A short summary for the three types of anyons: for type-$I$ and type-$A$ anyons, the 3-cocyle $\omega \in H^3(G,U(1))$
does not enter the definition of anyons, and only the group data of $G$ plays a role. 
For type-$B$ anyons, however, the 3-cocyle $\omega$ will twist the irreducible presentation of the centralizer subgroup.
Strictly speaking, the irreducible presentation is twisted by the two-cocyle obtained from $\omega$. 
This will affect the modular data as will be discussed below.  
In total, there are $\frac{q-1}{p}+p$ type-$I$ anyons, with $\frac{q-1}{p}$ of them $p$ dimensioanl, and $p$ of them one dimensional, 
$\frac{q-1}{p}\cdot q$ type-$A$ anyons with quantum dimension $p$, and $(p-1)\cdot p$ type-$B$ anyons with quantum dimensions $q$.
One can check explicitly that the total quantum dimension is indeed $\sum_i d_i^2=p^2q^2=|G|^2$.

\subsection{Modular data}
\label{SubSec:Modular data}

Having defined the types of anyons, now we discuss the modular data of MS MTCs,
which will be used in constructing the  representations of MCG of higher-genus manifold in 
Sec.\ref{Sec: Quasi-particle}. In addition, we will check explicitly there are only three sets 
of $S$ and $T$ matrices in MS MTCs, up to anyon permutation. 

A general construction of modular data for a twisted quantum double of a finite group $G$ can be found in 
Ref.\onlinecite{CGR00}. 
Let us consider the $T$ matrix first.
The topological spin of anyon $\big ( g, \tilde{\chi} \big)$ is
\be\label{TopologicalSpin0}
\theta_{(g,\chi)}=\frac{\text{tr} \tilde{\chi}(g)}{\text{tr}\tilde{\chi}(1)}.
\ee
For type-$I$ and type-$A$ anyons, the irreducible representations are all linear. Then one simply has 
$\theta_{(g,\chi)}=\frac{\text{tr} \chi(g)}{\text{tr}\chi(1)}$.
In particular, for type-$I$ anyons $(1,\chi_i)$, the topological spin is trivial, \textit{i.e.}, 
\be
\theta(1,\chi_i)=\frac{\text{tr} \chi_i(1)}{\text{tr}\chi_i(1)}=1.
\ee
For type-$A$ anyon $\big(a^l,\, \chi_m\big)$, $\chi_m$ is the $m$-th linear irreducible character of $\mathbb{Z}_q$ and is generated by 
$\chi(a)=e^{\frac{2\pi i}{q}}$, where we have defined $a^l:=(a^l,b^0)$. 
Then the topological spin of anyon $(a^l,\chi_m)$ is
\be\label{TopologicalSpin_A}
\theta_{(a^l, \chi_m)}=\exp\Big(\frac{2\pi i}{q} lm\Big).
\ee
Now let us check type-$B$ anyons $\big( b^k,\tilde{\chi}_n\big)$.
The  irreducible representation is twisted by $\alpha_{b^k}^{u}$, with
\be 
\begin{split}
&\alpha^{u}_{b^k}(g,h)=
\exp\big[\frac{2\pi i}{p^2}uk\cdot
\big(
[g_b]_p+[h_b]_p-[g_b+h_b]_p
\big)\big],
\end{split}
\ee
where we have defined $g:=(a^{g_a},b^{g_b})$, and $[x]_p=x$ mod $p$.
Recall that $\alpha_{b^k}^{u}$ is a two-coboundary and can be expressed as a function of 
one-cochain in Eq.\eqref{2cocycle_CT}, with
\be\label{epsilon_bk}
\epsilon_{b^k}^{u}(g)=\exp\big(\frac{2\pi i}{p^2}ku\cdot [g_b]_p\big).
\ee
Based on Eq.\eqref{LinearRep}, the $m$-th $\alpha_{b^k}$- character is related with the linear character as follows:
\be\label{Character_B}
\tilde{\chi}^{b^k}_m(x)=\chi_m(x)\cdot \epsilon_{b^k}(x)=(\chi(x))^m \cdot \epsilon_{b^k}(x),
\ee
where $m\in\{0,1,2,\cdots,p-1\}$, and $\chi$ is the generator of linear characters, with the expression
\be
\chi(b^k):=\exp\big(\frac{2i\pi}{p}k\big).
\ee
Then the topological spin of a type-$B$ anyons $(b^k, \tilde{\chi}_m)$ is
\be\label{TopologicalSpin_typeB}
\theta^{(u)}_{(b^k,\tilde{\chi}_m)}=\frac{\text{tr}\chi_m(b^k)}{\text{tr}\chi_m(1)}\epsilon^{u}_{b^k}(b^k)=\exp\big(\frac{2\pi i}{p^2}(p\cdot mk+k^2 u)\big),
\ee
where $k\in \{1,2,\cdots, p-1\}$.

As a short summary of the moduar $T$ matrix, or the topological spins, we have

(i) type-$I$ anyons: $\theta_{(1,\chi)}=1$;

(ii) type-$A$ anyons: $\theta_{(a^l,\chi^m)}=e^{\frac{2\pi i}{q} lm}$;

(iii) type-$B$ anyons: $\theta^{(u)}_{(b^k,\tilde{\chi}^m)}=e^{\frac{2\pi i}{p^2} \left(pmk+k^2 u\right)}$.

That is, only the topological spins of type-$B$ anyons depend on the equivalence class of 3-cocycle $\omega^{u}$, where 
$u=0,\cdots, p-1$.

Now let us check the modular $S$ matrix, which are indexed by two simples $(g,\tilde{\chi}_{\mu})$ and $(h,\tilde{\chi}_{\nu})$, as follows:\cite{CGR00}
\be\label{Sab1}
S_{(g,\tilde{\chi}_{m}),(h,\tilde{\chi}_{n})}=\frac{1}{|G|}\sum_{\substack{g'\in [g], h'\in [h] \\g'h'=h'g'}}  \big[\tilde{\chi}^{g'}_{m}(h')\big]^{\ast}\, \big[\tilde{\chi}^{h'}_{n}(g') \big]^{\ast}.
\ee
One can find that $S_{a,b}$ depends on the 3-cocyle $\omega$ only when there is at least one type-$B$ anyon in the indices.
In fact, by further looking into the details of $S$-matrix as shown in Appendix \ref{Appendix: Smatrix}, one can find that
$S_{a,b}$ depends on the $3$-cocycle only when both $a$ and $b$ are 
type-$B$ anyons. In this case, the modular $S$ matrix has a simple expression (see Appendix \ref{Appendix: Smatrix})
\be\label{S_BB}
\small
S^{(u)}_{(b^k,\tilde{\chi}_m),(b^{k'},\tilde{\chi}_n)}=\frac{1}{p}\exp\Big( -\frac{2\pi i}{p^2} \big[2u k k'+p(kn + k'm)\big]\Big),
\ee
where $u=0,\cdots,p-1$ denotes inequivalent classes of 3-cocycle $\omega$.

One remark here: 
If we focus on the modular data that depend on the 3-cocycle $\omega^u$, they are topological spins
$\theta^{(u)}_{(b^k,\tilde{\chi}^m)}=e^{\frac{2\pi i}{p^2} \left(pmk+k^2 u\right)}$, and $S^{(u)}_{(b^k,\tilde{\chi}_m),(b^{k'},\tilde{\chi}_n)}$ 
as expressed in Eq.\eqref{S_BB}. 
One can find they are the same as the modular data for a twisted quantum double of $G=\mathbb{Z}_p$ (see appendix \ref{SubSec: Zp}).
For the later case, however, there are only three inequivalent categories for $p>3$,\footnote{We thank P. Schauenburg for comments on this point.} 
 while for MS MTCs there are $p$ inequivalent categories.\cite{MS1708}
This difference may be intuitively understood by considering that 
the twisted quantum double of $G=\mathbb{Z}_p$ is an abelian theory, and there is no more new information in the  
representations of MCG of higher genus manifold (Recall that for an abelian theory, the basis vector in \eqref{Genus_g_basis}
decomposes into disconnected loops with $a_1=\cdots=a_{g-1}=1$, where $1$ denotes the identity anyon).
But MS MTCs are non-abelian, and there are new structures in the ground states on higher-genus 
manifolds, which introduces extra information compared to the genus-1 case.

\subsubsection{Equivalence of modular data}
\label{Sec: proof_Equiv_ModularData}

Now we are ready to see how the modular data of different categories can be mapped to each other.\cite{MS1708}
Mathematically, this is based on the application of Galois group actions.
There are two steps. Step 1 is to check how the modular data transform as we change the 3-cocyle from $\omega^{\nu}$
to $\omega^u$. Step 2 is to consider how the modular data transform as we permute anyons within the same category.
One can refer to Appendix \ref{Sec: Galois} for an introduction of the Galois symmetry in the modular data.

Step 1: For the modular category of $D^{\omega^u}(G)$, $\omega^u$ (here $u\neq 0$) is related to $\omega$ by $\omega^u=\sigma^r \omega$, where $r=0, 1, \cdots, p-2$. 
Mathematically, $\sigma$ is an element
of the absolute Galois group of abelian extensions $\Gamma:=\text{Gal}(\mathbb{Q}^{ab})/\mathbb{Q}$.
One can choose $\sigma$ such that $\sigma(\zeta_p)=\zeta_p^m$, where $m$ is the primitive root of the prime number $p$, with
 $m^{p-1}=1$ mod $p$.
Then by considering the mapping $\omega\xrightarrow{\sigma^r} \omega^u$ where $u=m^r$, 
there is a bijection $p_r$ between the simple objects (anyons) in $D^{\omega}(G)$ and $D^{\omega^u}(G)$ as follows
\be
p_r:\quad (g, \tilde{\chi}) \,\,\text{ in } \,\, D^{\omega}(G) \longrightarrow (g, \sigma^r\tilde{\chi}) \,\,\text{ in } \,\, D^{\omega^u}(G).
\ee
Hereafter, in some cases we may call the bijection $p_{r}$ of anyons in two categories 
as `permutation' of anyons in two categories.
This bijection can be understood in the following way. 
The 3-cocyle $\omega^u$ affects the definition/property of anyons through its  representation of the centralizer $C_G(g)$ 
in \eqref{Rep}. By replacing $\omega$ with $\omega^u=\sigma^r \omega=\omega^{m^r}$, $\alpha_g$ in Eq.\eqref{2cocycle} is replaced by $\sigma^r \alpha_g$.
Then to obtain a $\sigma^r \alpha_g$- representation,
every matrix element in the  representation $\tilde{\rho}$ of $C_G(g)$  should be acted by $\sigma^r$, 
which results in the replacement of $\tilde{\chi}$ by $\sigma^r\tilde{\chi}$.
Then from the definition of modular $S$ and $T$ matrices in Eqs.\eqref{Sab1} 
and \eqref{TopologicalSpin0}, one can find that the modular data of $D^{\omega^u}(G)$
are related to those of $D^{\omega}(G)$ as
\be
S^{(u)}_{p_r(i), p_r(j)}=\sigma^r(S^{(u=1)}_{ij}), \quad T^{(u)}_{p_r(i), p_r(i)}=\sigma^r(T^{(u=1)}_{ii}),
\ee
where $u=m^r$ mod $p$. More generally, the modular data of $D^{\omega^u}(G)$ and those of $D^{\omega^v}(G)$ are related by
\be
S^{(u)}_{p_{r,s}(i), p_{r,s}(j)}=\sigma^{r-s}(S^{(v)}_{ij}), \quad T^{(u)}_{p_{r,s}(i), p_{r,s}(i)}=\sigma^{r-s}(T^{(v)}_{ii}),
\ee
where, without loss of generality, we have assumed that $r>s$ with $\omega^u=\sigma^r \omega$ 
and $\omega^{v}=\sigma^s \omega$ (so that $u=m^r$ mod $p$, and $v=m^s$ mod $p$),
and $p_{r,s}$ denotes the corresponding bijection between the simple objects in $D^{\omega^u}(G)$ and $D^{\omega^v}(G)$.

Step 2: Now let us check how the modular data transform by permuting anyons within the category.
It is known that for the twisted quantum double $D^{\omega^u}(G)$, for each $\sigma\in \Gamma$ (see Step 1) there 
exists a unique permutation $\hat{\sigma}$ of simple objects in $D^{\omega^u}(G)$, so that 
$S_{\hat{\sigma}(i), \hat{\sigma}(j) }=\sigma^2(S_{i,j})$ and $T_{\hat{\sigma}(i), \hat{\sigma}(i) }=\sigma^2(T_{i,i})$
(See also Appendix \ref{Sec: Galois}).\cite{deBoer1991,Coste1994,etingof2005fusion,CGR00}

Comparing the two steps above, one can find that if $(r-s)$ is an even number, then the modular categories $D^{\omega^u}(G)$ and $D^{\omega^{v}}(G)$ share the same modular data.
That is, the modular data only depends on the parity of $r$. In short, there are at most three distinct sets of modular data, corresponding to $D^{\omega^u}(G)$ with $u=0$, $u=m^{2n}$, and $u=m^{2n+1}$, respectively.

From this point of view, as suggested in Ref.\onlinecite{MS1708}, the different categories 
that share the same modular data in MS MTCs are `Galois twists' of each other.

\section{ Representations of mapping class group: quasi-particle basis}
\label{Sec: Quasi-particle}

In this section, we study the  representations of mapping class group for the MS MTCs on a genus-2 manifold.  
All the calculation is based on the simplest examples of MS MTCs with $G=\mathbb{Z}_{11}\rtimes \mathbb{Z}_5$.
We will use the quasi-particle basis, \textit{i.e.}, the basis colored by anyons.
One merit of the quasi-particle basis is that we can relate the topological invariants constructed from 
the MCG representations to various link invariants.

For a genus-2 manifold (without punctures), there are two choices of canonical basis as 
\begin{eqnarray}\label{BasisI}
\text{basis I:}\quad\quad
\begin{tikzpicture}[baseline={(current bounding box.center)}]
\draw[>=stealth,<-] (0pt,0pt) arc (180:-180:20pt) ;
\draw[>=stealth,<-] (60pt,0pt) arc (180:-180:20pt) ;
\draw [>=stealth,->] (60pt,0pt)--(50pt,0pt);
\draw (50pt,0pt)--(40pt,0pt);
\node at (20pt, 25pt){$b$};
\node at (80pt, 25pt){$a$};
\node at (50pt, 5pt){$z$};
\node at (35pt, 0pt){\small $\nu$};
\node at (65pt, 0pt){\small $\mu$};
\end{tikzpicture}
\end{eqnarray}
and
\begin{eqnarray}\label{BasisII}
\text{basis II:}\quad\quad
\begin{tikzpicture}[baseline={(current bounding box.center)}]
\draw[>=stealth,->] (10*0.7pt,17*0.7pt) arc (40:180:30*0.7pt);
\draw  (-43*0.7pt,-2*0.7pt) arc (180:320:30*0.7pt);
\draw (10*0.7pt,17*0.7pt) arc (140:0:30*0.7pt);
\draw [>=stealth,<-] (63*0.7pt,-2*0.7pt) arc (0:-140:30*0.7pt);
\draw (10*0.7pt,17*0.7pt) -- (10*0.7pt, -2*0.7pt);
\draw [>=stealth,<-] (10*0.7pt, -2*0.7pt) -- (10*0.7pt,-22*0.7pt);
\node at (-35*0.7pt,0pt){$b$};
\node at (55*0.7pt,0pt){$a$};
\node at (15*0.7pt,0pt){{ $z$}};
\node at (7.5pt, 18pt){\small $\mu$};
\node at (7.5pt, -20pt){\small $\nu$};
\end{tikzpicture}
\end{eqnarray}
where $a$, $b$, and $z$ denote anyons, and $u$ ($\nu$) denotes the fusion channel. 
For example, in basis I, there are two loops colored by anyons $a$ and $b$, respectively. 
These two loops can be connected by a third line colored by anyon $z$ if $N_{a\bar{a}}^z, N_{b\bar{b}}^z>0$. 
The two different bases in \eqref{BasisI} and \eqref{BasisI} can be transformed into each other based on a $F$ transformation (see Sec.\ref{Appendix: Other basis}).
As the number of genus $g$ increases, there are more choices of canonical basis. See, \textit{e.g.}, the case of $g=3$ in \eqref{g=2TrivalentGraph}.

If there are punctures on the manifold, the quasiparticle basis will have have open ends corresponding to the anyons at the punctures.
For example, the quasi-particle basis for a torus with one puncture and two punctures can be expressed as follows:
\begin{eqnarray}\label{PunctureTorusBasis}
\begin{tikzpicture}[baseline={(current bounding box.center)}]
\draw[>=stealth,<-] (0pt,0pt) arc (180:-180:20pt) ;
\draw [>=stealth,->] (0pt,0pt)--(-7pt,0pt);
\draw (-7pt,0pt)--(-14pt,0pt);
\node at (20pt, 25pt){$a$};
\node at (-10pt, 5pt){$z$};
\node at (5pt, 0pt){$\mu$};
\end{tikzpicture}
\quad\quad
\begin{tikzpicture}[baseline={(current bounding box.center)}]
\draw[>=stealth,->] (0pt,0pt) arc (-180:180:20pt);
\draw[>=stealth,->] (40pt,0pt) arc (0:5:20pt) ;
\draw [>=stealth,->] (40-37.3pt,10pt)--(40-44.43pt,13.3pt); \draw (40-44.43pt,13.3pt)--(40-48pt,15pt);
\draw [>=stealth,->] (40-37.3pt,-10pt)--(40-44.43pt,-13.3pt); \draw (40-44.43pt,-13.3pt)--(40-48pt,-15pt);
\node at (20pt, 25pt){$a$};
\node at (40-53pt, 15pt){$z_1$};
\node at (40-32pt, 8pt){$\mu$};
\node at (40-53pt, -15pt){$z_2$};
\node at (40-32pt, -8pt){ $\nu$};
\node at (40-44pt, 0pt){$b$};
\end{tikzpicture}
\end{eqnarray}
It is noted that by gluing punctured-torus bases, one may obtain the basis for a higher-genus manifold.
For example, by gluing two copies of bases for the once-punctured torus
(the left in \eqref{PunctureTorusBasis}) along the 
puncture, one can obtain the genus-2 basis in \eqref{BasisI}.
Hereafter we will mainly focus on genus-2 manifold and use basis I in \eqref{BasisI}. It is noted that the basic structure in basis I is the vertex structure:
\begin{eqnarray}\label{vertex_basis}
\begin{tikzpicture}[baseline={(current bounding box.center)}]
\draw [>=stealth,->] (0pt,-15pt)--(0pt,-6pt);\draw (0pt,-6pt)--(0pt,0pt);
\draw [>=stealth,->] (0pt,0pt)--(-6pt,8pt); \draw(-6pt,8pt)--(-7.5pt,10pt);
\draw [>=stealth,->] (0pt,0pt)--(6pt,8pt); \draw(6pt,8pt)--(7.5pt,10pt);
\node at (-8pt,15pt){$a$};
\node at (8pt,15pt){$z$};
\node at (0pt,-20pt){$a$};
\node at (4pt,-2pt){\small$\mu$};
\end{tikzpicture}
\end{eqnarray}
which is actually the basis in the so-called splitting space $V_a^{az}$ (whose dual space is the fusion vector space $V_{az}^a$) 
in algebraic anyon theory (see Appendix \ref{Sec: Appendix_anyon}). In the following, we will first specify the fusion rules of $a\times \bar{a}$ 
in \eqref{vertex_basis}, and then study how the modular transformations act on this vertex basis.

\subsection{MS modular categories with $G=\mathbb{Z}_{11}\rtimes \mathbb{Z}_5$}
\label{Sec: Z11Z5_data}

The twisted quantum double $D^{\omega}(G)$ of $G=\mathbb{Z}_{11}\rtimes \mathbb{Z}_5$
are the simplest examples of MS MTCs, with group order $55$, and $49$ simple objects.
Throughout this work, all the concrete calculation is based on these examples.
In the following, we will review the basic data that are necessary for our later study
on the MCG representations.

Recall that the simple objects of a twisted quantum double of a finite group
are labeled by the `flux' (conjugacy class) and `charge' (the irreducible
representation of the centralizer of conjugacy class).  There are
$p+\frac{q-1}{p}=7$ conjugacy classes of $G=\mathbb{Z}_{11}\rtimes
\mathbb{Z}_5$, which we label as $[1]$, $[a^1]$, $[a^2]$, $[b^1]$, $[b^2]$,
$[b^3]$, and $[b^4]$, as described in Table \ref{ConjugacyClass}.

\begin{table}[h]
\centering
\small
\begin{tabular}{cccccccc}
&  anyons &\vline  &Conjugacy class  &\vline     &Centralizer  \\ \hline
& type-$I$&\vline  &$[1]=\{1\}$  &\vline       &$G=\mathbb{Z}_{11}\rtimes\mathbb{Z}_5$   \\ \hline
& type-$A$&\vline  &$[a^1]=\{a,a^3,a^4,a^5,a^9\}$  &\vline         &$\mathbb{Z}_{11}$  \\ 
&        	&\vline  &$[a^2]=\{a^2,a^6,a^7,a^8,a^{10}\}$  &\vline       &$\mathbb{Z}_{11}$  \\ \hline
& 		&\vline  &$[b^1]=\{a^0b^1,a^1b^1, a^2b^1,\cdots, a^{10}b^1\}$  &\vline          &$\mathbb{Z}_{5}$   \\ 
&type-$B$	&\vline  &$[b^2]=\{a^0b^2,a^1b^2, a^2b^2,\cdots, a^{10}b^2\}$  &\vline       &$\mathbb{Z}_{5}$  \\ 
&		&\vline  &$[b^3]=\{a^0b^3,a^1b^3, a^2b^3,\cdots, a^{10}b^3\}$  &\vline          &$\mathbb{Z}_{5}$    \\ 
&		&\vline  &$[b^4]=\{a^0b^4,a^1b^4, a^2b^4,\cdots, a^{10}b^4\}$  &\vline          &$\mathbb{Z}_{5}$    \\ \hline
\end{tabular}
\caption{Conjugacy classes and the corresponding centralizers of $G=\mathbb{Z}_{11}\rtimes \mathbb{Z}_5$. 
}
\label{ConjugacyClass}
\end{table}

As discussed in the previous section, depending on the conjugacy classes, there are three types of anyons. 
Here we denote them as type-$I$, type-$A$, and type-$B$ anyons:
\be\label{Z11Z5Simples}
\begin{split}
I_i:=&\left(1,\chi_i\right),\\
A_{l,m}=&\left(a^l,\omega_{11}^m\right),\\
B_{k,n}=&\left(b^k,\tilde{\omega}_5^n\right),
\end{split}
\ee
where $\chi_i$, $\omega_{11}^m$, and $\omega_{5}^n$ are the corresponding  character of centralizer subgroup (see the following). 

-- type-$I$ anyons (pure charge): $(1,\chi_i)$

In this case, the flux is trivial, \textit{i.e.}, the conjugacy class is $\{1\}$, with $1$ representing the identity group element in $G$,
and therefore type-$I$ anyons are all pure charges.
The centralizer of $1$ is the 
total group $G=\mathbb{Z}_{11}\rtimes \mathbb{Z}_5$. 
The character table of $G$ is \cite{Wang1805}
\begin{eqnarray}
\centering
\small
\begin{tabular}{cccccccccc}
$G$              &\vline     &$[1]$  &$[a^1]$  &$[a^2]$ &$[b^1]$  &$[b^2]$  &$[b^3]$ &$[b^4]$\\ \hline
$\chi_0$  &\vline      &1  &1  			&1				&1  			&1  			&1 			&1\\
$\chi_1$  &\vline      &1  &1  			&1 				&$\xi_5$  		&$\xi_5^2$  	&$\xi_5^3$ 	&$\xi_5^4$\\
$\chi_2$  &\vline      &1  &1 			&1 				&$\xi_5^2$  	&$\xi_5^4$  	&$\xi_5$ 		&$\xi_5^3$\\
$\chi_3$  &\vline      &1  &1  			&1 				&$\xi_5^3$   	&$\xi_5$  		&$\xi_5^4$ 	&$\xi_5^2$\\
$\chi_4$  &\vline      &1  &1  			&1 				&$\xi_5^4$   	&$\xi_5^3$  	&$\xi_5^2$ 	&$\xi_5$\\
$\chi_5$  &\vline      &5  &$\sigma$  		&$\sigma^{\ast}$ 	&0  			&$0$  		&0 			&$0$\\
$\chi_6$  &\vline      &5  &$\sigma^{\ast}$ 	&$\sigma$ 		&0  			&$0$  		&0			&$0$\\ \hline
\end{tabular}
\label{Gcharacter}
\end{eqnarray}
where $\xi_m=e^{\frac{2\pi i}{5}}$, and $\sigma=e^{\frac{2\pi i}{11}1}+e^{\frac{2\pi i}{11}3}+e^{\frac{2\pi i}{11}4}+e^{\frac{2\pi i}{11}5}+e^{\frac{2\pi i}{11}9}$.
One can find there are in total $7$ irreducible representations for $G$, with five of them 1
dimensional, \textit{i.e.}, $\chi_i(1)=1$ and two of them 5 dimensional, \textit{i.e.}, $\chi_i(1)=5$.
One can check that $1^2\cdot 5+5^2\cdot 2=55=|G|$.
Corresponding to the characters $\chi_i$ in \eqref{Gcharacter}, 
we denote these pure charges as
\be
\small
\begin{split}
I_0, I_1, I_2, I_3, I_4, \quad &d=1,\\
I_5, I_6, \quad &d=5.
\end{split}
\ee
The topological spins of all type-$I$ anyons are trivial, \textit{i.e.}, $\theta_{I_i}=1$.

\begin{table}
\centering
\footnotesize
\begin{tabular}{cccc}
Label  &\vline     &$d$ &$\theta$\\ \hline\hline
$I_0$  &\vline       &$1$  &$1$  \\ 
$I_1$  &\vline          &$1$  &$1$  \\ 
$I_2$  &\vline       &$1$  &$1$  \\ 
$I_3$  &\vline          &$1$  &$1$  \\ 
$I_4$  &\vline       &$1$  &$1$  \\ 
$I_5$  &\vline          &$5$  &$1$  \\ 
$I_6$  &\vline          &$5$  &$1$  \\ \hline
$A_{1,0}$  &\vline       &$5$  &$1$  \\ 
$A_{1,1}$  &\vline       &$5$  &$\exp\left(\frac{i2\pi}{11}\right)$  \\ 
$A_{1,2}$  &\vline       &$5$  &$\exp\left(\frac{i2\pi}{11}2\right)$  \\ 
$A_{1,3}$  &\vline       &$5$  &$\exp\left(\frac{i2\pi}{11}3\right)$  \\ 
$A_{1,4}$  &\vline       &$5$  &$\exp\left(\frac{i2\pi}{11}4\right)$  \\ 
$A_{1,5}$  &\vline        &$5$  &$\exp\left(\frac{i2\pi}{11}5\right)$  \\ 
$A_{1,6}$  &\vline       &$5$  &$\exp\left(\frac{i2\pi}{11}6\right)$ \\
$A_{1,7}$  &\vline       &$5$  &$\exp\left(\frac{i2\pi}{11}7\right)$  \\ 
$A_{1,8}$  &\vline       &$5$  &$\exp\left(\frac{i2\pi}{11}8\right)$  \\ 
$A_{1,9}$  &\vline        &$5$  &$\exp\left(\frac{i2\pi}{11}9\right)$  \\ 
$A_{1,10}$  &\vline       &$5$  &$\exp\left(\frac{i2\pi}{11}10\right)$ \\ \hline
$A_{2,0}$  &\vline       &$5$  &$1$  \\ 
$A_{2,1}$  &\vline       &$5$  &$\exp\left(\frac{i4\pi}{11}\right)$  \\ 
$A_{2,2}$  &\vline       &$5$  &$\exp\left(\frac{i4\pi}{11}2\right)$  \\ 
$A_{2,3}$  &\vline       &$5$  &$\exp\left(\frac{i4\pi}{11}3\right)$  \\ 
$A_{2,4}$  &\vline       &$5$  &$\exp\left(\frac{i4\pi}{11}4\right)$  \\ 
$A_{2,5}$  &\vline        &$5$  &$\exp\left(\frac{i4\pi}{11}5\right)$  \\ 
$A_{2,6}$  &\vline       &$5$  &$\exp\left(\frac{i4\pi}{11}6\right)$ \\
$A_{2,7}$  &\vline       &$5$  &$\exp\left(\frac{i4\pi}{11}7\right)$  \\ 
$A_{2,8}$  &\vline       &$5$  &$\exp\left(\frac{i4\pi}{11}8\right)$  \\ 
$A_{2,9}$  &\vline        &$5$  &$\exp\left(\frac{i4\pi}{11}9\right)$  \\ 
$A_{2,10}$  &\vline       &$5$  &$\exp\left(\frac{i4\pi}{11}10\right)$ \\ \hline
$B_{1,0}$  &\vline       &$11$  &$\exp\left(\frac{i2\pi}{25}1^2 u\right)$ \\
$B_{1,1}$  &\vline       &$11$  &$\exp\left(\frac{i2\pi}{25}(5\cdot 1\cdot 1+1^2u)\right)$  \\ 
$B_{1.2}$  &\vline       &$11$  &$\exp\left(\frac{i2\pi}{25}(5\cdot 1\cdot 2+1^2u)\right)$  \\ 
$B_{1.3}$  &\vline        &$11$  &$\exp\left(\frac{i2\pi}{25}(5\cdot 1\cdot 3+1^2u)\right)$  \\ 
$B_{1,4}$  &\vline       &$11$  &$\exp\left(\frac{i2\pi}{25}(5\cdot 1\cdot 4+1^2u)\right)$ \\ \hline
$B_{2,0}$  &\vline       &$11$  &$\exp\left(\frac{i2\pi}{25}2^2 u\right)$ \\
$B_{2,1}$  &\vline       &$11$  &$\exp\left(\frac{i2\pi}{25}(5\cdot 2\cdot 1+2^2u)\right)$  \\ 
$B_{2.2}$  &\vline       &$11$  &$\exp\left(\frac{i2\pi}{25}(5\cdot 2\cdot 2+2^2u)\right)$  \\ 
$B_{2.3}$  &\vline        &$11$  &$\exp\left(\frac{i2\pi}{25}(5\cdot 2\cdot 3+2^2u)\right)$  \\ 
$B_{2,4}$  &\vline       &$11$  &$\exp\left(\frac{i2\pi}{25}(5\cdot 2\cdot 4+2^2u)\right)$ \\ \hline
$B_{3,0}$  &\vline       &$11$  &$\exp\left(\frac{i2\pi}{25}3^2 u\right)$ \\
$B_{3,1}$  &\vline       &$11$  &$\exp\left(\frac{i2\pi}{25}(5\cdot 3\cdot 1+3^2u)\right)$  \\ 
$B_{3.2}$  &\vline       &$11$  &$\exp\left(\frac{i2\pi}{25}(5\cdot 3\cdot 2+3^2u)\right)$  \\ 
$B_{3.3}$  &\vline        &$11$  &$\exp\left(\frac{i2\pi}{25}(5\cdot 3\cdot 3+3^2u)\right)$  \\ 
$B_{3,4}$  &\vline       &$11$  &$\exp\left(\frac{i2\pi}{25}(5\cdot 3\cdot 4+3^2u)\right)$ \\ \hline
$B_{4,0}$  &\vline       &$11$  &$\exp\left(\frac{i2\pi}{25}4^2 u\right)$ \\
$B_{4,1}$  &\vline       &$11$  &$\exp\left(\frac{i2\pi}{25}(5\cdot 4\cdot 1+4^2u)\right)$  \\ 
$B_{4.2}$  &\vline       &$11$  &$\exp\left(\frac{i2\pi}{25}(5\cdot 4\cdot 2+4^2u)\right)$  \\ 
$B_{4.3}$  &\vline        &$11$  &$\exp\left(\frac{i2\pi}{25}(5\cdot 4\cdot 3+4^2u)\right)$  \\ 
$B_{4,4}$  &\vline       &$11$  &$\exp\left(\frac{i2\pi}{25}(5\cdot 4\cdot 4+4^2u)\right)$ \\ \hline
\end{tabular}
\caption{Quantum dimensions and topological spins for twisted quantum double of $G=\mathbb{Z}_{11}\rtimes \mathbb{Z}_5$. 
Here $u=0, 1, 2, 3$, and $4$.}
\label{49Anyons}
\end{table}

-- type-$A$ anyons: $(a^l, \omega_{11}^m)$

These anyons correspond to conjugacy classes $[a^1]$ and $[a^2]$.
As seen from Table \ref{ConjugacyClass}, the size of the conjugacy class is $5$, and the centralizer is $\mathbb{Z}_{11}$.
From Eq.\eqref{3cocycle}, one can find the 3-cocycle in this case is trivial, and the irreducible presentations are 
all linear, and one dimensional. 
The generator of characters for $\mathbb{Z}_{11}$ is $\omega_{11}(a)=e^{\frac{2\pi i}{11}}$, so that $\omega^m_{11}(a^l)=e^{\frac{2\pi i}{11}\cdot lm}$.
There are in total $2\times 11=22$ type-$A$ anyons, with the quantum dimension $d_i=5\times 1=5$. 
As seen from Eq.\eqref{TopologicalSpin_A}, the topological spin for  anyon $(a^l, \omega_{11}^m)$ is $e^{\frac{2\pi i}{11} lm}$, which is
independent of the 3-cocycle.

-- type-$B$ anyons: $\left(b^k,\tilde{\omega}_5^n\right)$

The corresponding conjugacy classes are $[b^1]$, $[b^2]$, $[b^3]$, and $[b^4]$.
As seen from Table \ref{ConjugacyClass}, the size of conjugacy class is $11$. The centralizer subgroup
is $\mathbb{Z}_{5}$.
The irreducible  representations
 of $\mathbb{Z}_{5}$ are all one dimensional (see Eq.\eqref{Character_B}).
There are in total $4\times 5=20$ type-$B$ anyons, with quantum dimensions $d_i=11\times 1=11$.
The topological spin of anyon $(b^k, \tilde{\chi}^m)$ is
$\theta^{(u)}_{(b^k,\tilde{\chi}^m)}=e^{\frac{2\pi i}{25} \left(5mk+k^2 u\right)}$,
where $u=0, \cdots, 4$.

In short, there are in total $49$ anyons.
The quantum dimensions and topological spins of these anyons are summarized in Table \ref{49Anyons}, 
where $u=0, 1, 2, 3, 4$ denote different categories.

For the modular $S$ matrix, the explicit expression has been given in Appendix \ref{Appendix: Smatrix}.
The elements that depend on the 3-cocycle $\omega^u$ are 

\be\label{SBBZ11Z5}
\small
S^{(u)}_{(b^k,\tilde{\chi}_m),(b^{k'},\tilde{\chi}_n)}
=\frac{1}{5}\exp\Big( -\frac{2\pi i}{25} \big[2u k k'+5(kn + k'm)\big]\Big).
\ee

According to the previous section, $m=2$ is the primitive root of the prime number $p=5$. Then the five categories $D^{\omega^u}(G)$ are 
divided into three sets with $u=0$, $u=m^{2n}$, and $u=m^{2n+1}$ mod $p$. That is,   
one can find there are three distinct sets of modular data for (1) $u=0$, (2) $u=1$ and $u=4$, and (3) $u=2$ and $u=3$, respectively.

As mentioned at the beginning of this section, to specify the fusion/splitting basis in Eq.\eqref{BasisI},
we need to know the fusion rules of $a\times \bar{a}$. 
With the modular $S$ matrix in Eq.\eqref{Sab1}, this can be obtained via Verlinde's formula
\be
N_{ab}^c=\sum_i \frac{S_{ai} \, S_{bi} \, S^{\ast}_{ci} }{ S_{0i} }.
\ee
The explicit results are listed as follows:

(1) $a\times \bar{a}=\sum_c N_{a\bar{a}}^c c$, with $a=I_i$
\begin{small}
\be\label{fusion1}
\begin{split}
I_0\times I_0=&I_0,\\
I_1\times I_4=&I_0,\\
I_2\times I_3=&I_0,\\
I_5\times I_6=&I_0+I_1+I_2+I_3+I_4+2\, I_5 +2\,I_6.
\end{split}
\ee
\end{small}

(2) $a\times \bar{a}=\sum_c N_{a\bar{a}}^c c$, with $a=A_{l,m}$
\be\label{fusion2}
\small
\begin{split}
A_{1,0}\times A_{2,0}=&I_0+I_1+I_2+I_3+I_4\\
&+2\, A_{1,0} +2\, A_{2,0},\\
A_{1,1}\times A_{2,6}=&I_0+I_1+I_2+I_3+I_4\\
&+A_{1,6} +A_{1,10}+A_{2,3} +A_{2,5} ,\\
A_{1,2}\times A_{2,1}=&I_0+I_1+I_2+I_3+I_4\\
&+A_{1,1} +A_{1,9}+A_{2,6} +A_{2,10} ,\\
A_{1,3}\times A_{2,7}=&I_0+I_1+I_2+I_3+I_4\\
&+A_{1,7} +A_{1,8}+A_{2,4} +A_{2,9} ,\\
A_{1,4}\times A_{2,2}=&I_0+I_1+I_2+I_3+I_4\\
&+A_{1,2} +A_{1,7}+A_{2,1} +A_{2,9} ,\\
A_{1,5}\times A_{2,8}=&I_0+I_1+I_2+I_3+I_4\\
&+A_{1,6} +A_{1,8}+A_{2,3} +A_{2,4} ,\\
A_{1,6}\times A_{2,3}=&I_0+I_1+I_2+I_3+I_4\\
&+A_{1,3} +A_{1,5}+A_{2,7} +A_{2,8} ,\\
A_{1,7}\times A_{2,9}=&I_0+I_1+I_2+I_3+I_4\\
&+A_{1,4} +A_{1,9}+A_{2,2} +A_{2,10} ,\\
A_{1,8}\times A_{2,4}=&I_0+I_1+I_2+I_3+I_4\\
&+A_{1,3} +A_{1,4}+A_{2,2} +A_{2,7} ,\\
A_{1,9}\times A_{2,10}=&I_0+I_1+I_2+I_3+I_4\\
&+A_{1,2} +A_{1,10}+A_{2,1} +A_{2,5} ,\\
A_{1,10}\times A_{2,5}=&I_0+I_1+I_2+I_3+I_4\\
&+A_{1,1} +A_{1,5}+A_{2,6} +A_{2,8} ,\\
\end{split}
\ee
Note that the above fusion results in \eqref{fusion1} and \eqref{fusion2} are independent of $u$.
This can be understood based on the fact that $S_{I_i, x}$ and $S_{A_{l,m}, x}$
are independent of $u$ for arbitrary anyon type $x$.
Or more essentially, the definitions of type-$I$ and type-$A$ anyons are independent of $u$.

(3) $a\times \bar{a}=\sum_c N_{a\bar{a}}^c c$, with $a=B_{k,n}$

Now the fusion rules depend on $u$ as follows.
For different choices of $u$, the fusion results are the same, but the dual anyons of $B_{k,n}$ are different.

-- $u=0$:
\be\label{fusion3u0}
\small
\begin{split}
B_{1,0}\times B_{4,0}=&I_0+
I_5+I_6\\
&+
A_{1,0}+A_{1,1}+A_{1,2}+A_{1,3}+A_{1,4}+A_{1,5}\\
&+A_{1,6}+A_{1,7}+A_{1,8}+A_{1,9}+A_{1,10}\\
&+
A_{2,0}+A_{2,1}+A_{2,2}+A_{2,3}+A_{2,4}+A_{2,5}\\
&+A_{2,6}+A_{2,7}+A_{2,8}+A_{2,9}+A_{2,10}.\\
\end{split}
\ee
The quantum dimensions on the two sides of fusion rules satisfy the relation $11\times 11=1+5\times 24$.
Interestingly, one can find that the fusion results for the other $9$ pairs (anyons and their dual anyons) 
are the same as $B_{1,0}\times B_{4,0}$, \textit{i.e.}, 
\be
\small
\begin{split}
&B_{1,0}\times B_{4,0}=B_{1,1}\times B_{4,4}=B_{1,2}\times B_{4,3}\\
=&B_{1,3}\times B_{4,2}=B_{1,4}\times B_{4,1}\\
=&B_{2,0}\times B_{3,0}=B_{2,1}\times B_{3,4}=B_{2,2}\times B_{3,3}\\
=&B_{2,3}\times B_{3,2}=B_{2,4}\times B_{3,1}.
\end{split}
\ee
It is emphasized that the ``$=$" here means the fusion results (but not the fusion rules) are the same.

-- $u=1$:

One can find the fusion results of the following anyons and their dual anyons are the same as that in Eq.\eqref{fusion3u0}.
\be
\small
\begin{split}
&B_{3,0}\times B_{2,3}=B_{3,1}\times B_{2,2}=B_{3,2}\times B_{2,1}\\
=&B_{3,3}\times B_{2,0}=B_{3,4}\times B_{2,4}\\
=&B_{4,0}\times B_{1,3}=B_{4,1}\times B_{1,2}=B_{4,2}\times B_{1,1} \\
=&B_{4,3}\times B_{1,0}=B_{4,4}\times B_{1,4}.
\end{split}
\ee
Similarly, for $u=2, 3$, and $4$, we have

-- $u=2$:
\be
\small
\begin{split}
&B_{1,0}\times B_{4,1}=B_{1,1}\times B_{4,0}=B_{1,2}\times B_{4,4}\\
=&B_{1,3}\times B_{4,3}=B_{1,4}\times B_{4,2}\\
=&B_{2,0}\times B_{3,1}=B_{2,1}\times B_{3,0}=B_{2,2}\times B_{3,4}\\
=&B_{2,3}\times B_{3,3}=B_{2,4}\times B_{3,2}.
\end{split}
\ee

-- $u=3$:
\be
\small
\begin{split}
&B_{1,0}\times B_{4,4}=B_{1,1}\times B_{4,3}=B_{1,2}\times B_{4,2}\\
=&B_{1,3}\times B_{4,1}=B_{1,4}\times B_{4,0}\\
=&B_{2,0}\times B_{3,4}=B_{2,1}\times B_{3,3}=B_{2,2}\times B_{3,2}\\
=&B_{2,3}\times B_{3,1}=B_{2,4}\times B_{3,0}.
\end{split}
\ee

-- $u=4:$
\be\label{Fusion_u4}
\small
\begin{split}
&B_{1,0}\times B_{4,2}=B_{1,1}\times B_{4,1}=B_{1,2}\times B_{4,0}\\
=&B_{1,3}\times B_{4,4}=B_{1,4}\times B_{4,3}\\
=&B_{2,0}\times B_{3,2}=B_{2,1}\times B_{3,1}=B_{2,2}\times B_{3,0}\\
=&B_{2,3}\times B_{3,4}=B_{2,4}\times B_{3,3}.
\end{split}
\ee
One can find that for different 3-cocycle $\omega^u$, the dual anyons of $B_{k,n}$
are different. This is also straightforwardly seen by looking at the topological spins 
in Table \ref{49Anyons}, noting that $\theta_a=\theta_{\bar{a}}$.
Based on these fusion rules, we can fix the basis in Eq.\eqref{vertex_basis}.
For example, if we fix $z=A_{1,1}$, it is found that $a$ can only be chosen as 
$A_{1,2}$, $A_{2,1}$, $A_{1,10}$, $A_{2,5}$, and all the $20$ type-$B$ anyons.
In addition, there is only one fusion channel for these choices of $a$, \textit{i.e.}, $N_{a\bar{a}}^z=1$.
Then the Hilbert space spanned by the basis in Eq.\eqref{vertex_basis}
with fixed $z=A_{1,1}$ is 24 dimensional.
It is noted that in certain cases, \textit{e.g.}, $z=A_{1,0}$, we may have more than one fusion channels with $N_{a\bar{a}}^z>1$.
The dimension of Hilbert space for $\oplus_a V_a^{az}$ with fixed $z$ is $\sum_a N_{a\bar{a}}^z$.

\subsubsection{Simple currents}
\label{Sec: SimpleCurrents}

There are some fine structures in the modular data due to those anyons with quantum dimension $1$, 
which are called `simple currents' in literature.\cite{schellekens1989extended,bernard1987string, kreuzer1994simple} 
Understanding such fine structures will help us in analyzing the patterns of a punctured $S$ matrix in the next subsection. 

The simple currents, which will be denoted as $j$ here, can be defined as any $j \in \Pi_{\mathcal{C}}$, with quantum dimension 
$d_j=S_{0j}/S_{00}=1$. The fusion rules of $j$ with any other simple objects $a\in \Pi_{\mathcal{C}}$ are simply
$j\times a=a'$, where $a'\in \Pi_{\mathcal{C}}$. That is, there is only one anyon appearing in the fusion results.
The effect of simple currents on modular $S$ and $T$ matrices have been studied in, \textit{e.g.}, 
Refs.\onlinecite{schellekens1989extended,bernard1987string, kreuzer1994simple,CGR00}. 
Here we focus on the simplest counterexamples in MS MTCs with $G=\mathbb{Z}_{11}\rtimes \mathbb{Z}_5$.
From Table \ref{49Anyons}, one can find there are in total five simple currents $I_i$, with $i=0,\cdots,4$.
Since $I_0$ is the identity anyon, its fusion rules with other anyons are trivial, and we will not write them down.
Note that there are five different categories with $u=0, 1, 2, 3, 4, 5$, but 
interestingly, the fusion rules of the simple currents with other simple objects are independent of $u$ as follows:
\be
\small
\begin{split}
I_1\times I_1=&I_2,\, I_1\times I_2=I_3, \,I_1\times I_3=I_4, I_1\times I_4=I_0,\\
I_2\times I_2=&I_4,\, I_2\times I_3=I_0, \, I_2\times I_4=I_1,\\
I_3\times I_3=&I_1,\,I_3\times I_4=I_2,\\
I_4\times I_4=&I_3,\\
\end{split}
\ee
\be
\small
I_k\times A_{l,m}=A_{l,m}, \quad k=1,2,3,4,\\ 
\ee
and
\be\label{Fusion_SimpleCurrent_Banyon}
\small
\begin{split}
I_1\times B_{i,j}=&B_{i,j+1}, \\
I_4\times B_{i,j}=&B_{i,j-1}, \\
I_2\times B_{i,j}=&B_{i,j+2}, \\
I_3\times B_{i,j}=&B_{i,j-2},
\end{split}
\ee
where $i=1,2,3,4$ and $j=j\,\,\text{mod}\,\,5$. 
It is noted that in the above fusion rules $B_{i,j'}=B_{i,j}\times I_k$, the index `$i$' which labels the flux
keeps the same, but only the index `$j$' which labels the charge changes. This is because
the simple currents $I_k$ are pure charges, and fusing $I_k$ with any other anyon $a\in\Pi_{\mathcal{C}}$
can at most change the charge of anyon $a$.

Diagrammatically, Eq.\eqref{Fusion_SimpleCurrent_Banyon} can be expressed as

\begin{eqnarray}\label{split}
\small
B_{ij'}=B_{ij}\times I_k:\quad
\begin{tikzpicture}[baseline={(current bounding box.center)}]
\draw [>=stealth,->] (0pt,-15pt)--(0pt,2pt);\draw (0pt,2pt)--(0pt,16pt);
\node at (0pt,-20pt){$B_{ij'}$};
\end{tikzpicture}
=
\begin{tikzpicture}[baseline={(current bounding box.center)}]
\draw [>=stealth,->] (0pt,-15pt)--(0pt,2pt);\draw (0pt,2pt)--(0pt,16pt);
\draw [dashed][>=stealth,->] (3.5pt,-15pt)--(3.5pt,2pt);\draw [dashed](3.5pt,2pt)--(3.5pt,16pt);
\node at (-1pt,-20pt){$B_{ij}$};
\node at (11pt,0pt){$I_{k}$};
\end{tikzpicture}
\end{eqnarray}
where the dashed line represents the world line of $I_k$ anyons.
We are interested in the effect of simple currents $I_k$ ($k=1,2,3,4$) on $S_{B_{i,j}, B_{i',j'}}$ and $T_{B_{i,j}, B_{i,j}}$.
From the fusion rules \eqref{Fusion_SimpleCurrent_Banyon} and the modular data in Appendix \eqref{Appendix: Smatrix}, 
it is found that
\be\label{ST_pattern}
\small
\left\{
\begin{split}
S_{(I_k \times B_{i,j}),B_{m,n}}=&\,S_{B_{i,j},B_{m,n}} \cdot e^{2\pi i Q_k(B_{m,n})},\\
T_{(I_k\times B_{m,n}), (I_k\times B_{m,n})}=&\,T_{B_{i,j}, B_{i,j}} \cdot e^{-2\pi i Q_k(B_{m,n})}.
\end{split}
\right.
\ee
where $e^{2\pi i Q_k(B_{m,n})}$ is a pure $U(1)$ phase with the expression
\be\label{Phase_factor}
\small
e^{2\pi i Q_k(B_{m,n})}=\frac{S_{I_k, B_{m,n}} }{ S_{I_0, B_{m,n}} }
=\frac{S_{B_{m,n},I_k} }{ S_{B_{m,n},I_0}}= e^{-\frac{2\pi i}{5}\cdot k\cdot m}.
\ee 
This phase can be viewed as the Aharonov–Bohm phase introduced by dragging 
an $I_k$ anyon ($k=0,1,2,3,4$) around anyon $B_{m,n}$.
Diagrammatically, the relations in Eq.\eqref{ST_pattern} can be depicted as follows:
\begin{eqnarray}\label{S_Ik}
\small
\begin{tikzpicture}[baseline={(current bounding box.center)}]
\draw (20pt,0pt) circle (20pt);
\draw[line width=6pt, draw=white] (0pt,0pt) circle (20pt);
\draw (0pt,0pt) circle (20pt);
\draw [>=stealth,->] (20pt,0.1pt)--(20pt,0.11pt);
\draw [>=stealth,->] (40pt,0.1pt)--(40pt,0.11pt);
\draw[line width=6pt, draw=white]  (0pt,0pt) arc (-180:-270:20pt);
\draw (0pt,0pt) arc (-180:-270:20pt);
\node at (0pt,-27pt){$B_{i,j'}$};
\node at (25pt,-27pt){$B_{m,n}$};
\end{tikzpicture}
=
\begin{tikzpicture}[baseline={(current bounding box.center)}]
\draw (20pt,0pt) circle (20pt);
\draw[line width=6pt, draw=white] (0pt,0pt) circle (20pt);
\draw (0pt,0pt) circle (20pt);
\draw [>=stealth,->][dashed](18pt,0pt) arc (0:360:18pt);
\draw [>=stealth,->] (20pt,0.1pt)--(20pt,0.11pt);
\draw [>=stealth,->] (40pt,0.1pt)--(40pt,0.11pt);
\draw[line width=6pt, draw=white]  (0pt,0pt) arc (-180:-270:20pt);
\draw (0pt,0pt) arc (-180:-270:20pt);
\node at (0pt,-27pt){$B_{i,j}$};
\node at (-12pt,0pt){$I_k$};
\node at (25pt,-27pt){$B_{m,n}$};
\end{tikzpicture}
\end{eqnarray}
where $B_{ij'}=B_{ij}\times I_k$, and
\begin{eqnarray}\label{T_Ik}
\small
\begin{tikzpicture}[baseline={(current bounding box.center)}]
\draw [>=stealth,->](-2pt,-20pt)--(-2pt,-10pt); \draw (-2pt,-10pt)--(-2pt,-2pt);
\draw (-4pt,8pt)..controls (-10pt,15pt) and (-18pt,0pt)..(-10pt,-4pt)..controls (-3pt,-6pt) and (-2pt,4pt)..(-2pt,20pt);
\node at (-15pt,-14pt){$B_{m,n'}$};
\end{tikzpicture}
=
\begin{tikzpicture}[baseline={(current bounding box.center)}]
\draw [>=stealth,->](0pt,-20pt)--(0pt,-10pt); \draw (0pt,-10pt)--(0pt,0pt);
\draw [dashed][>=stealth,->](2.5pt,-20pt)--(2.5pt,-6pt); \draw [dashed] (2.5pt,-6pt)--(2.5pt,1pt);
\draw (-4pt,8pt)..controls (-10pt,15pt) and (-18pt,0pt)..(-10pt,-4pt)..controls (-3pt,-6pt) and (-2pt,4pt)..(-2pt,20pt);
\draw [dashed] (-4pt,11pt)..controls (-10pt,18pt) and (-24pt,0pt)..(-10pt,-7pt)..controls (-1pt,-10pt) and (1pt, 4pt)..(1pt,22pt);
\node at (-12pt,-14pt){$B_{m,n}$};
\node at (8pt,-8pt){$I_k$};
\end{tikzpicture}
\end{eqnarray}
where $B_{m,n'}=B_{m,n}\times I_k$. 
Then by considering the local operation\cite{Kitaev2006}
\be\label{local_shrink}
\small
\begin{tikzpicture}[baseline={(current bounding box.center)}]
\draw [>=stealth,->](0pt,0pt) arc (110:360:8pt);
\draw (8+8*0.3420pt,-8*0.9397pt) arc (0:70:8pt);
\draw [>=stealth,->](3pt,-12pt)--(3pt,-2pt); \draw (3pt,-2pt)--(3pt,10pt);
\draw (3pt,-25pt)--(3pt,-18pt);
\node at (3pt,-31pt){$\bar{x}$};
\node at (15pt,-7pt){$a$};
\end{tikzpicture}
=\frac{S_{ax}}{S_{0x}}
\begin{tikzpicture}[baseline={(current bounding box.center)}]
\draw [>=stealth,->](3pt,-24pt)--(3pt,-5pt); \draw (3pt,-5pt)--(3pt,10pt);
\node at (3pt,-30pt){$\bar{x}$};
\end{tikzpicture}
\ee
one can remove the $I_k$ anyon in \eqref{S_Ik} and \eqref{T_Ik} by introducing extra phases 
$\frac{S_{I_k, B_{m,n}} }{ S_{I_0, B_{m,n}} }$ in $S$-matrix, and
$\frac{S_{I_k, \overline{B_{m,n}}} }{ S_{I_0, \overline{B_{m,n}}} }=\frac{S^{\ast}_{I_k, B_{m,n}} }{ S^{\ast}_{I_0, B_{m,n}} }$ in $T$ matrix,
as expressed in Eq.\eqref{ST_pattern}.

\subsection{Punctured $S$ and $T$ matrices}
\label{Sec: punctured S and T}

In this subsection, we study the properties of punctured $S$ and $T$ matrices, 
and in particular present the results for the MS MTCs with $G=\mathbb{Z}_{11}\rtimes \mathbb{Z}_5$.

First, we give an intuitive picture on how to obtain the punctured $S$ and $T$ matrices 
based on the modular transformation of a punctured torus.
As introduced in Sec.\ref{Sec: Introduction_PojectiveRep_MCG},
the canonical basis on a punctured torus can be considered as the path integral over
a solid torus $D^2\times S^1$ as follows:
\begin{eqnarray}\label{PuncturedTorusBasis}
|\psi_{a\bar{a},\mu}^{(z)}\rangle:=
\begin{tikzpicture}[baseline={(current bounding box.center)}]

\draw  [gray](0+12pt,0pt) arc (-60:-120:25pt);
\draw  [gray](-3+12pt,-1.6pt) arc (50:130:15pt);

\draw (-30pt,5pt)..controls(-27pt,4pt) and (-26pt,3pt)..(-25pt,3pt);
\draw [>=stealth,->] (-19.5pt,0pt)..controls (-21pt,1pt) and (-22pt,2pt)..(-25pt,3pt);

\node at (24pt, 0pt){$a$};
\node at (-30pt, 0pt){$z$};
\begin{small}
\node at (-16pt, -1pt){$\mu$};
\end{small}

\draw [gray]
(0pt,20pt)..controls (15pt,20pt) and (34pt,12pt)..
(34.5pt,0pt)..controls (34pt,-12pt) and (15pt,-20pt)..(0pt,-20pt)..
controls (-15pt,-20pt) and (-34pt,-12pt)..(-34.5pt,0pt)..controls(-34pt,12pt) and (-15pt,20pt)..(0pt,20pt);

\draw
(0pt,10pt)..controls (10pt,10pt) and (19pt,5pt)..
(19.5pt,0pt)..controls (19pt,-5pt) and (10pt,-10pt)..(0pt,-10pt);

\draw [>=stealth,->]
(0pt,10pt)..controls (-10pt,10pt) and (-19pt,5pt)..
(-19.5pt,0pt)..controls (-19pt,-5pt) and (-10pt,-10pt)..(0pt,-10pt);
\draw (-30pt,5pt) circle [radius=0.05]; 
\end{tikzpicture}
\quad \text{on } D^2\times S^1 
\nonumber
\end{eqnarray}
where $\circ$ denotes the puncture on the two dimensional surface
$\partial (D^2\times S^1)=S^1\times S^1$, $z$ denotes the anyon at the puncture,
and $\mu$ denotes the channel that $a$ and $\bar{a}$ fuse into $z$.
Note that when $z=1$, $|\psi_{a\bar{a},\mu}^{(z)}\rangle$ 
reduces to the canonical basis on a torus $T^2$ without any puncture.

Then we can define the matrix element of the punctured $S$-matrix as 
\be\label{PuncturedS_WF}
S^{(z)}_{a,\mu;b,\nu}:= \frac{\langle \psi_{a\bar{a},\mu}^{(z)} | \,\hat{S}\,| \psi_{b\bar{b},\nu}^{(z)}\rangle}
{\sqrt{\langle \psi_{a\bar{a},\mu}^{(z)} | \psi_{a\bar{a},\mu}^{(z)} \rangle \cdot \langle \psi_{b\bar{b},\nu}^{(z)} | \psi_{b\bar{b},\nu}^{(z)}\rangle}},
\ee
where we have

\begin{eqnarray}\label{PuncturedTorusOverlap}
\langle
\psi_{a\bar{a},\mu}^{(z)}
|\psi_{a\bar{a},\mu}^{(z)}\rangle=
\begin{tikzpicture}[baseline={(current bounding box.center)}]

\draw  [gray](0+12pt,0pt) arc (-60:-120:25pt);
\draw  [gray](-3+12pt,-1.6pt) arc (50:130:15pt);

\draw  (-27.2pt,5pt)..controls(-27pt,4pt) and (-26pt,3pt)..(-25pt,3pt);
\draw  [>=stealth,->] (-19.5pt,0pt)..controls (-21pt,1pt) and (-22pt,2pt)..(-25pt,3pt);

\node at (16pt, 0pt){$a$};
\node at (25pt,1pt){$\bar{a}$};
\node at (-24pt, -2pt){$z$};
\begin{small}
\node at (-16pt, -1pt){$\mu$};
\end{small}

\begin{small}
\node at (-30pt, 3pt){$\mu$};
\end{small}

\draw [gray]
(0pt,20pt)..controls (15pt,20pt) and (34pt,12pt)..
(34.5pt,0pt)..controls (34pt,-12pt) and (15pt,-20pt)..(0pt,-20pt)..
controls (-15pt,-20pt) and (-34pt,-12pt)..(-34.5pt,0pt)..controls(-34pt,12pt) and (-15pt,20pt)..(0pt,20pt);

\draw 
(0pt,10pt)..controls (10pt,10pt) and (19pt,5pt)..
(19.5pt,0pt)..controls (19pt,-5pt) and (10pt,-10pt)..(0pt,-10pt);

\draw [>=stealth,->]
(0pt,10pt)..controls (-10pt,10pt) and (-19pt,5pt)..
(-19.5pt,0pt)..controls (-19pt,-5pt) and (-10pt,-10pt)..(0pt,-10pt);

\draw [>=stealth,->]
(0pt,10*1.5pt)..controls (10*1.5pt,10*1.5pt) and (19*1.5pt,5*1.5pt)..
(19.5*1.5pt,0pt)..controls (19*1.5pt,-5*1.5pt) and (10*1.5pt,-10*1.5pt)..(0pt,-10*1.5pt);

\draw 
(0pt,10*1.5pt)..controls (-10*1.5pt,10*1.5pt) and (-19*1.5pt,5*1.5pt)..
(-19.5*1.5pt,0pt)..controls (-19*1.5pt,-5*1.5pt) and (-10*1.5pt,-10*1.5pt)..(0pt,-10*1.5pt);
\end{tikzpicture}
\quad \text{on } S^2\times S^1,
\nonumber
\end{eqnarray}
and similarly for $\langle\psi_{b\bar{b},\nu}^{(z)}
|\psi_{b\bar{b},\nu}^{(z)}\rangle$.
They can be evaluated based on the surgery approach\cite{witten1989quantum}, and the result is
\be
\langle
\psi_{a\bar{a},\mu}^{(z)}
|\psi_{a\bar{a},\mu}^{(z)}\rangle=
\langle\psi_{b\bar{b},\nu}^{(z)}
|\psi_{b\bar{b},\nu}^{(z)}\rangle=\sqrt{d_z}.
\ee
The numerator in \eqref{PuncturedS_WF} is
\begin{eqnarray}\label{Numerator}
\small
\langle \psi_{a\bar{a},\mu}^{(z)} | \,\hat{S}\,| \psi_{b\bar{b},\nu}^{(z)}\rangle=
\begin{tikzpicture}[baseline={(current bounding box.center)}]
\draw (20*0.8pt,0pt) circle (20*0.8pt);
\draw [line width=6*0.8pt, draw=white] (0pt,0pt) circle (20*0.8pt);
\draw (0pt,0pt) circle (20*0.8pt);
\draw [line width=6*0.8pt, draw=white]  (0pt,0pt) arc (-180:-270:20*0.8pt);
\draw (0pt,0pt) arc (-180:-270:20*0.8pt);
\node at (-15*0.8+40*0.8pt,0pt){$a$};
\node at (5*0.8+39*0.8pt,0pt){$b$};
\node at (-4pt,-19pt){$\mu$};
\node at (19pt,-19pt){$\nu$};
\draw [>=stealth,->] (20*0.8pt,0.1*0.8pt)--(20*0.8pt,0.11*0.8pt);
\draw [>=stealth,->] (40*0.8pt,0.1*0.8pt)--(40*0.8pt,0.11*0.8pt);

\draw [>=stealth,->]  (20*0.8pt,-20*0.8pt)..controls (15*0.8pt,-30*0.8pt) and (5*0.8pt,-30*0.8pt)..(0pt,-20*0.8pt);
\node at (10*0.8pt,-35*0.8+2pt){$z$};
\draw [gray](7.8pt,-3pt) circle (30pt);
\end{tikzpicture}
\quad \text{on }S^3.
\end{eqnarray}
That is, after the $S$ transformation, two solid torus are glued together as a $S^3$, with the two punctures identified.

Hereafter, without specific explanation, all the link/knot invariants can be
considered as embedded in the three manifold $S^3$, and for convenience we will
remove $\bigcirc$ which represents $S^3$ in Eq.\eqref{Numerator}.

Based on the above analysis, one can find that the punctured $S$-matrix can be presented as
(note that an extra normalization factor $1/\mathcal{D}$ is introduced):
\begin{eqnarray}\label{PuncturedSmatrix}
S_{a,\mu;b,\nu}^{(z)}&=
\frac{1}{\mathcal{D}}
\cdot 
\frac{1}{\sqrt{d_z}}
\cdot 
\begin{tikzpicture}[baseline={(current bounding box.center)}]
\draw (20pt,0pt) circle (20pt);
\draw[line width=6pt, draw=white] (0pt,0pt) circle (20pt);
\draw (0pt,0pt) circle (20pt);
\draw [>=stealth,->] (20pt,0.1pt)--(20pt,0.11pt);
\draw [>=stealth,->] (40pt,0.1pt)--(40pt,0.11pt);
\draw[line width=6pt, draw=white]  (0pt,0pt) arc (-180:-270:20pt);
\draw (0pt,0pt) arc (-180:-270:20pt);
\node at (-15+40pt,0pt){$a$};
\node at (5+40pt,0pt){$b$};
\draw[>=stealth,->]  (20pt,-20pt)..controls (15pt,-30pt) and (5pt,-30pt)..(0pt,-20pt);
\node at (10pt,-32pt){$z$};
\node at (-5pt,-25pt){\small$\mu$};
\node at (22pt,-25pt){\small$\nu$};
\end{tikzpicture}
\end{eqnarray}
One can see clearly that for the case of $z=1$, the presentation above reduces to the conventional modular $S$ matrix.
It is noted that $S_{a,\mu; b,\nu}^{(z)}$ can also be expressed in terms of $F$ and $R$ symbols 
(also called $F$ and $R$ matrices)\cite{Moore1989,Kitaev2006}.
 
Similarly, one can define the punctured $T$ matrix as
\be
T^{(z)}_{a,\mu;b,\nu}:=\frac{\langle \psi_{a\bar{a},\mu}^{(z)} | \,\hat{T}\,| \psi_{b\bar{b},\nu}^{(z)}\rangle}
{\sqrt{\langle \psi_{a\bar{a},\mu}^{(z)} | \psi_{a\bar{a},\mu}^{(z)} \rangle \cdot \langle \psi_{b\bar{b},\nu}^{(z)} | \psi_{b\bar{b},\nu}^{(z)}\rangle}},
\ee
where the numerator $\langle \psi_{a\bar{a},\mu}^{(z)} | \,\hat{T}\,| \psi_{b\bar{b},\nu}^{(z)}\rangle$ corresponds to a path integral over $S^2\times S^1$.
With the straightforward surgery approach\cite{witten1989quantum}, one can find that
\be\label{PunctureT}
T^{(z)}_{a,\mu;b,\nu}=\delta_{a,b}\, \delta_{\mu,\nu}\,\theta_a,
\ee
with $N_{a\bar{a}}^z>0$.
For the twisted quantum double $D^{\omega}(G)$ of $G=\mathbb{Z}_q\rtimes_n \mathbb{Z}_p$, the topological 
spin $\theta_a$ can be straightforwardly obtained based on Eq.\eqref{TopologicalSpin0}.
That is, based on the modular data and the fusion rules (which are also determined by the modular data due to Verlinde formula),
we can obtain the punctured $T$-matrix in \eqref{PunctureT}.

To obtain the punctured $S$ matrix, our strategy is to solve the modular relations for
punctured $S$ and $T$ as follows:\cite{Moore1989,Kitaev2006}
\be\label{ModularRelationSzTz}
\left\{
\begin{split}
& \big(S^{(z)}\big)^2=C^{(z)},\\
& \big(C^{(z)}\big)^2=\theta_z^{\ast},\\
& \big(S^{(z)}T^{(z)}\big)^3= \big(S^{(z)}\big)^2,\\
\end{split}
\right.
\ee
$C^{(z)}$ can be understood as the `punctured' charge conjugation, which will reduce to the conventional charge conjugation
$C$ when $z=1$.
In addition, $S^{(z)}$ and $T^{(z)}$ are unitary matrices.
This is apparent for $T^{(z)}$ based on the expression in \eqref{PunctureT}.
The unitarity property of $S^{(z)}$ is related to the braiding non-degeneracy of modular categories.\cite{Kitaev2006} 
Intuitively, it means that a nontrivial anyon (which is not an identity) can be detected by Aharonov-Bohm measurement
by dragging a test particle around it.
With the unitary property of $S^{(z)}$ and $T^{(z)}$, the last equation in Eq.\eqref{ModularRelationSzTz} is equivalent to 
$\big(S^{(z)}\big)^{\dag}T^{(z)}\big(S^{(z)}\big)=\big(T^{(z)}\big)^{\dag}\big(S^{(z)}\big)^{\dag}\big(T^{(z)}\big)^{\dag}$,
which is useful in solving $S^{(z)}$.
 Furthermore, from Eq.\eqref{ModularRelationSzTz}, one can also observe that
$\big(S^{(z)}\big)^4=\theta_z^{\ast}$.
This is related to $\text{MCG}(\Sigma_{1,1})$ in \eqref{MCG_punctuedTorus}, 
where $ \mathfrak{s}^4=\mathfrak{r}^{-1}$, with $\mathfrak{r}$ representing the Dehn twist around the puncture.

We will give further details on the modular relations in \eqref{ModularRelationSzTz} in the rest of this subsection.
It is convenient to study the modular relations in \eqref{ModularRelationSzTz} by acting the
operators on the basis vectors in the Hilbert space $\mathcal{H}(\Sigma_{1,1})$. We denote the Hilbert space
on a punctured torus as $\mathcal{H}(\Sigma_{1,1})=\oplus_b V_b^{bz}$, with $z$ representing the anyonic charge at the puncture.
Then one has\cite{Moore1989,Kitaev2006} 
\begin{eqnarray}\label{Sz}
S^{(z)}
\begin{tikzpicture}[baseline={(current bounding box.center)}]
\draw [>=stealth,->] (0pt,-15pt)--(0pt,-6pt);\draw (0pt,-6pt)--(0pt,0pt);
\draw [>=stealth,->] (0pt,0pt)--(-6pt,8pt); \draw(-6pt,8pt)--(-7.5pt,10pt);
\draw [>=stealth,->] (0pt,0pt)--(6pt,8pt); \draw(6pt,8pt)--(7.5pt,10pt);
\node at (-8pt,15pt){$b$};
\node at (8pt,15pt){$z$};
\node at (0pt,-20pt){$b$};
\node at (6pt,-2pt){\small$\mu$};
\end{tikzpicture}
:=\frac{1}{\mathcal{D}} \sum_a d_a \,\,
\begin{tikzpicture}[baseline={(current bounding box.center)}]
\draw (10pt,0pt) arc (0:180:10pt);
\draw [>=stealth,->](-10pt,0pt) arc (-180:0:10pt);
\draw[line width=4pt, draw=white] (-5pt,-20pt)--(-5pt,5pt);
\draw [>=stealth,->](-5pt,-18pt)--(-5pt,0pt); \draw(-5pt,0pt)--(-5pt,5pt);
\draw (-5pt, 12pt)--(-5pt,18pt);
\draw [>=stealth,->](5pt, 8.66 pt)--(8pt,14.264pt);\draw(8pt,14.264pt)--(10pt,18pt);
\node at (-5pt,-23pt){$a$};
\node at (-5pt,23pt){$a$};
\node at (15pt,0pt){$b$};
\node at (10pt,23pt){$z$};
\node at (11pt,10pt){\small$\mu$};
\end{tikzpicture},
\end{eqnarray}
\begin{eqnarray}\label{Tz}
T^{(z)}
\begin{tikzpicture}[baseline={(current bounding box.center)}]
\draw [>=stealth,->] (0pt,-15pt)--(0pt,-6pt);\draw (0pt,-6pt)--(0pt,0pt);
\draw [>=stealth,->] (0pt,0pt)--(-6pt,8pt); \draw(-6pt,8pt)--(-7.5pt,10pt);
\draw [>=stealth,->] (0pt,0pt)--(6pt,8pt); \draw(6pt,8pt)--(7.5pt,10pt);
\node at (-8pt,15pt){$b$};
\node at (8pt,15pt){$z$};
\node at (0pt,-20pt){$b$};
\node at (6pt,0pt){\small$\mu$};
\end{tikzpicture}
:=\theta_b\,\,
\begin{tikzpicture}[baseline={(current bounding box.center)}]
\draw [>=stealth,->] (0pt,-15pt)--(0pt,-6pt);\draw (0pt,-6pt)--(0pt,0pt);
\draw [>=stealth,->] (0pt,0pt)--(-6pt,8pt); \draw(-6pt,8pt)--(-7.5pt,10pt);
\draw [>=stealth,->] (0pt,0pt)--(6pt,8pt); \draw(6pt,8pt)--(7.5pt,10pt);
\node at (-8pt,15pt){$b$};
\node at (8pt,15pt){$z$};
\node at (0pt,-20pt){$b$};
\node at (6pt,0pt){\small$\mu$};
\end{tikzpicture},
\end{eqnarray}
and
\begin{eqnarray}\label{Cz}
C^{(z)}
\begin{tikzpicture}[baseline={(current bounding box.center)}]
\draw [>=stealth,->] (0pt,-15pt)--(0pt,-6pt);\draw (0pt,-6pt)--(0pt,0pt);
\draw [>=stealth,->] (0pt,0pt)--(-6pt,8pt); \draw(-6pt,8pt)--(-7.5pt,10pt);
\draw [>=stealth,->] (0pt,0pt)--(6pt,8pt); \draw(6pt,8pt)--(7.5pt,10pt);
\node at (-8pt,15pt){$b$};
\node at (8pt,15pt){$z$};
\node at (0pt,-20pt){$b$};
\node at (6pt,0pt){\small$\mu$};
\end{tikzpicture}
:=\theta_b^{\ast}\,\,
\begin{tikzpicture}[baseline={(current bounding box.center)}]
\draw [>=stealth,->](0pt,0pt)..controls (-4pt,-15pt) and (-10pt,-15pt)..(-20pt,10pt);
\draw[line width=6pt, draw=white] (0pt,0pt)..controls (-4pt,15pt) and (-10pt,15pt)..(-20pt,-14pt);
\draw (0pt,0pt)..controls (-4pt,15pt) and (-10pt,15pt)..(-20pt,-14pt);
\draw [>=stealth,->](-16pt,-3.5pt)--(-16+0.2pt,-3.5+0.5pt);
\draw [>=stealth,->] (0pt,0pt)--(6pt,8pt); \draw(6pt,8pt)--(7.5pt,10pt);
\node at (-20pt,16pt){$\bar{b}$};
\node at (8pt,15pt){$z$};
\node at (-18pt,-20pt){$\bar{b}$};
\node at (6pt,0pt){\small$\mu$};
\end{tikzpicture}.
\end{eqnarray}
In this basis, $T^{(z)}$ is a diagonal matrix with entries $\theta_b$, which is the topological spin of anyon $b$ that satisfies
the fusion rule $b\times \bar{b}=I+N_{b\bar{b}}^z z+\cdots$, with $N_{b\bar{b}}^z>0$. 
Note that $N_{ij}^k=\text{dim} V_{ij}^k$.
Based on the definitions in Eqs.\eqref{Sz}, \eqref{Tz}, and\eqref{Cz}, one can prove 
the modular relations in Eq.\eqref{ModularRelationSzTz}.
For example,  the first relation $\left(S^{(z)}\right)^2=C^{(z)}$ in Eq.\eqref{ModularRelationSzTz} can be proved as follows:
\be\label{SzSquare}
\small
\begin{split}
&\big(S^{(z)}\big)^2
\begin{tikzpicture}[baseline={(current bounding box.center)}]
\draw [>=stealth,->] (0pt,-15pt)--(0pt,-6pt);\draw (0pt,-6pt)--(0pt,0pt);
\draw [>=stealth,->] (0pt,0pt)--(-6pt,8pt); \draw(-6pt,8pt)--(-7.5pt,10pt);
\draw [>=stealth,->] (0pt,0pt)--(6pt,8pt); \draw(6pt,8pt)--(7.5pt,10pt);
\node at (-8pt,15pt){$b$};
\node at (8pt,15pt){$z$};
\node at (0pt,-20pt){$b$};
\node at (6pt,-2pt){\small$\mu$};
\end{tikzpicture}
=
\frac{1}{\mathcal{D}^2} \sum_{x,a} d_x d_a  \
\begin{tikzpicture}[baseline={(current bounding box.center)}]
\draw (0pt,0pt) circle (10pt);
\draw (15pt, 8.66 pt)--(20pt,18pt);
\draw (10pt,0pt) circle (10pt);
\draw [line width=4pt, draw=white](10pt,10pt) arc (90:180:10pt);
\draw (10pt,10pt) arc (90:180:10pt);
\draw [line width=4pt, draw=white](10pt,0pt) arc (0:-90:10pt);
\draw (10pt,0pt) arc (0:-90:10pt);
\draw[line width=4pt, draw=white] (-5pt,-18pt)--(-5pt,-5pt);
\draw (-5pt,18pt)--(-5pt,10pt);
\draw [>=stealth,->] (-5pt,-18pt)--(-5pt, 0pt); \draw (-5pt,0pt)--(-5pt,5pt);
\draw [>=stealth,->] (10pt,1.0pt)--(10pt,1.01pt);
\draw [>=stealth,->] (20pt,1.0pt)--(20pt,1.01pt);
\node at (-5pt,-23pt){$x$};
\node at (-5pt,23pt){$x$};
\node at (24pt,0pt){$b$};
\node at (20pt,23pt){$z$};
\begin{small}
\node at (23pt,10pt){$\mu$};
\end{small}
\node at (14pt,0pt){$a$};
\end{tikzpicture}\\
=&
\frac{1}{\mathcal{D}^2} \sum_{x,a} d_x d_a  \theta_x^{\ast}\
\begin{tikzpicture}[baseline={(current bounding box.center)}]
\draw (0pt,0pt) circle (10pt);
\draw (15pt, 8.66 pt)--(20pt,18pt);
\draw (10pt,0pt) circle (10pt);
\draw [line width=4pt, draw=white](10pt,10pt) arc (90:180:10pt);
\draw (10pt,10pt) arc (90:180:10pt);
\draw [line width=4pt, draw=white](10pt,0pt) arc (0:-90:10pt);
\draw (10pt,0pt) arc (0:-90:10pt);
\draw (-5pt,0pt)..controls (-6pt,-15pt) and (-15pt,-15pt)..(-25pt,15pt);
\draw [line width=4pt, draw=white](-5pt,0pt)..controls (-6pt,15pt) and (-15pt,15pt)..(-25pt,-15pt);
\draw (-5pt,0pt)..controls (-6pt,15pt) and (-15pt,15pt)..(-25pt,-15pt);
\draw [line width=4pt, draw=white](-10pt,0pt) arc (-180:-90:10pt);
\draw [>=stealth,->] (-5pt,0.1pt)--(-5pt,-0.1pt);
\draw [>=stealth,->] (10pt,1.0pt)--(10pt,1.01pt);
\draw [>=stealth,->] (20pt,1.0pt)--(20pt,1.01pt);
\draw (-10pt,0pt) arc (-180:-90:10pt);
\node at (-25pt,-20pt){$x$};
\node at (-25pt,20pt){$x$};
\node at (24pt,0pt){$b$};
\node at (20pt,23pt){$z$};
\begin{small}
\node at (23pt,10pt){$\mu$};
\end{small}
\node at (14pt,0pt){$a$};
\end{tikzpicture}\\
=&\sum_x \delta_{x,\bar{b}}\,\theta_x^{\ast}
\begin{tikzpicture}[baseline={(current bounding box.center)}]
\draw [>=stealth,->](0pt,0pt)..controls (-4pt,-15pt) and (-10pt,-15pt)..(-20pt,10pt);
\draw[line width=6pt, draw=white] (0pt,0pt)..controls (-4pt,15pt) and (-10pt,15pt)..(-20pt,-14pt);
\draw (0pt,0pt)..controls (-4pt,15pt) and (-10pt,15pt)..(-20pt,-14pt);
\draw [>=stealth,->](-16pt,-3.5pt)--(-16+0.2pt,-3.5+0.5pt);
\draw [>=stealth,->] (0pt,0pt)--(6pt,8pt); \draw(6pt,8pt)--(7.5pt,10pt);
\node at (-20pt,16pt){$x$};
\node at (8pt,15pt){$z$};
\node at (-18pt,-20pt){$x$};
\node at (6pt,0pt){\small$\mu$};
\end{tikzpicture}
=
\theta_b^{\ast}
\begin{tikzpicture}[baseline={(current bounding box.center)}]
\draw [>=stealth,->](0pt,0pt)..controls (-4pt,-15pt) and (-10pt,-15pt)..(-20pt,10pt);
\draw[line width=6pt, draw=white] (0pt,0pt)..controls (-4pt,15pt) and (-10pt,15pt)..(-20pt,-14pt);
\draw (0pt,0pt)..controls (-4pt,15pt) and (-10pt,15pt)..(-20pt,-14pt);
\draw [>=stealth,->](-16pt,-3.5pt)--(-16+0.2pt,-3.5+0.5pt);
\draw [>=stealth,->] (0pt,0pt)--(6pt,8pt); \draw(6pt,8pt)--(7.5pt,10pt);
\node at (-20pt,16pt){$\bar{b}$};
\node at (8pt,15pt){$z$};
\node at (-18pt,-20pt){$\bar{b}$};
\node at (6pt,0pt){\small$\mu$};
\end{tikzpicture},
\end{split}
\ee
where we have considered the fact $\theta_b=\theta_{\bar{b}}$ in the last step. By comparing Eq.\eqref{SzSquare} 
with the definition of $C^{(z)}$ in Eq.\eqref{Cz}, one can find that $\big(S^{(z)}\big)^2=C^{(z)}$.
Some further details in \eqref{SzSquare} and 
the proof of the other two relations in Eq.\eqref{ModularRelationSzTz}  can be found in the appendix \ref{Sec: Appendix_anyon}.

Before solving Eqs.\eqref{ModularRelationSzTz} for $S^{(z)}$, it is helpful to understand the property of matrix elements of $C^{(z)}$, 
which are expressed as follows:
\be\label{CzMatrixElement}
\begin{split}
C^{(z)}_{a,\mu;b,\nu}=&\delta_{a,\bar{b}}\,\delta_{\mu,\nu}\,\frac{\theta_b^{\ast}}{\sqrt{d_a\,d_b\,d_z}}
\begin{tikzpicture}[baseline={(current bounding box.center)}]
\draw (-20pt,0pt)..controls (-16pt,10pt) and (-4pt,10pt)..(0pt,0pt)..controls (4pt,-10pt) and (16pt,-10pt)..(20pt,0pt);
\draw [line width=8pt, draw=white](-20pt,0pt)..controls (-16pt,-10pt) and (-4pt,-10pt)..(0pt,0pt)..controls (4pt,10pt) and (16pt,10pt)..(20pt,0pt);
\draw [>=stealth,->](-20pt,0pt)..controls (-16pt,-10pt) and (-4pt,-10pt)..(0pt,0pt);
\draw (0pt,0pt)..controls (4pt,10pt) and (16pt,10pt)..(20pt,0pt);
\draw [>=stealth,->](20pt,0pt)..controls (28pt,5pt) and (15pt,25pt)..(0pt,24pt);
\draw (0pt,24pt)..controls (-15pt,25pt) and (-28pt,5pt)..(-20pt,0pt);
\node at (0pt,18pt){$z$};
\node at (0pt,-12pt){$\bar{b}$};
\node at (-24pt,-4pt){\small$\mu$};
\node at ( 24pt,-4pt){\small$\nu$};
\end{tikzpicture}\\
=&\delta_{a,\bar{b}}\, \big[R_z^{\bar{b}b}\big]^{-1}_{\mu\nu} \cdot \theta_b^{\ast},
\end{split}
\ee
where we have used the definition of $R$ matrix:
\be
\begin{tikzpicture}[baseline={(current bounding box.center)}]
\draw [>=stealth,->] (0pt,-15pt)--(0pt,-6pt);\draw (0pt,-6pt)--(0pt,0pt);
\draw [>=stealth,->] (0pt,0pt)..controls (7pt,2pt) and (7pt,4pt)..(-8pt,12pt);
\draw [line width=4pt, draw=white](0pt,0pt)..controls (-7pt,2pt) and (-7pt,4pt)..(8pt,12pt);
\draw [>=stealth,->] (0pt,0pt)..controls (-7pt,2pt) and (-7pt,4pt)..(8pt,12pt);
\node at (-12pt,15pt){$b$};
\node at (12pt,15pt){$a$};
\node at (0pt,-20pt){$c$};
\node at (6pt,-2pt){\small$\mu$};
\end{tikzpicture}
=R_{ab}
\begin{tikzpicture}[baseline={(current bounding box.center)}]
\draw [>=stealth,->] (0pt,-15pt)--(0pt,-6pt);\draw (0pt,-6pt)--(0pt,0pt);
\draw [>=stealth,->] (0pt,0pt)--(-6pt,8pt); \draw(-6pt,8pt)--(-7.5pt,10pt);
\draw [>=stealth,->] (0pt,0pt)--(6pt,8pt); \draw(6pt,8pt)--(7.5pt,10pt);
\node at (-8pt,15pt){$a$};
\node at (8pt,15pt){$b$};
\node at (0pt,-20pt){$c$};
\node at (6pt,-2pt){\small$\mu$};
\end{tikzpicture}
=\sum_{\nu} (R^{ab}_c)_{\mu\nu}
\begin{tikzpicture}[baseline={(current bounding box.center)}]
\draw [>=stealth,->] (0pt,-15pt)--(0pt,-6pt);\draw (0pt,-6pt)--(0pt,0pt);
\draw [>=stealth,->] (0pt,0pt)--(-6pt,8pt); \draw(-6pt,8pt)--(-7.5pt,10pt);
\draw [>=stealth,->] (0pt,0pt)--(6pt,8pt); \draw(6pt,8pt)--(7.5pt,10pt);
\node at (-8pt,15pt){$b$};
\node at (8pt,15pt){$a$};
\node at (0pt,-20pt){$c$};
\node at (6pt,-2pt){\small$\nu$};
\end{tikzpicture}.
\ee
One can check that for the specific case of $z=1$, $C^{(z)}_{a,\mu;b,\nu}$ reduces to
the conventional charge conjugation:
\be
C_{a,b}^{(z=1)}=\delta_{a,\bar{b}}\frac{\theta_b^{\ast}}{\sqrt{d_a\,d_b}}
\begin{tikzpicture}[baseline={(current bounding box.center)}]
\draw (-20pt,0pt)..controls (-16pt,10pt) and (-4pt,10pt)..(0pt,0pt)..controls (4pt,-10pt) and (16pt,-10pt)..(20pt,0pt);
\draw [line width=8pt, draw=white](-20pt,0pt)..controls (-16pt,-10pt) and (-4pt,-10pt)..(0pt,0pt)..controls (4pt,10pt) and (16pt,10pt)..(20pt,0pt);
\draw [>=stealth,->](-20pt,0pt)..controls (-16pt,-10pt) and (-4pt,-10pt)..(0pt,0pt);
\draw (0pt,0pt)..controls (4pt,10pt) and (16pt,10pt)..(20pt,0pt);
\draw [dashed][>=stealth,->](20pt,0pt)..controls (28pt,5pt) and (15pt,25pt)..(0pt,24pt);
\draw [dashed](0pt,24pt)..controls (-15pt,25pt) and (-28pt,5pt)..(-20pt,0pt);
\node at (0pt,-12pt){$\bar{b}$};
\node at (0pt,18pt){\small $z=1$};
\end{tikzpicture}
=\delta_{a,\bar{b}}.
\ee
In addition, from the result in Eq.\eqref{CzMatrixElement}, one can prove 
the relation $\big(C^{(z)}\big)^2=\theta_z^{\ast}$ in \eqref{ModularRelationSzTz} straightforwardly as
\be
\small
\begin{split}
\big[C^{(z)}\big]^2_{a,\mu;b,\nu}
=&\sum_{c,\lambda} C^{(z)}_{a,\mu;c,\lambda} C^{(z)}_{c,\lambda;b,\nu}\\
=&\sum_{c,\lambda} \delta_{a,\bar{c}}\delta_{c,\bar{b}} (R^{c\bar{c}}_z)^{-1}_{\mu,\lambda}\cdot \theta_c^{\ast} \cdot (R^{b\bar{b}}_z)^{-1}_{\lambda,\nu}\cdot \theta_b^{\ast} \\
=& \delta_{a,b} \,\delta_{\mu,\nu}\,\theta_z^{\ast},
\end{split}
\ee
where in the last step we have used the ribbon property $\sum_{\lambda}[R^{ab}_c]_{\mu\lambda}
[R^{ba}_c]_{\lambda \nu}=\frac{\theta_c}{\theta_a\theta_b}\delta_{\mu,\nu}$, and the properties of topological spins
$\theta_a=\theta_{\bar{a}}$ and $\theta^{\ast}_a=\theta_a^{-1}$.

One remark on \eqref{CzMatrixElement}:

The term $[R^{\bar{b}b}_z]^{-1}_{\mu \nu}$ in the expression of $C^{(z)}_{a,\mu;b,\nu}$ in 
Eq.\eqref{CzMatrixElement} is in general not gauge invariant, even for the multiplicity-free case.
As a comparison, $R^{bb}_z$ (and $R^{\bar{b}\bar{b}}_z$)
for the multiplicity-free case are gauge invariant quantities, with $R^{bb}_z=\pm\frac{\theta_z^{1/2}}{\theta_b}$, 
where the sign $\pm$ is determined by the modular data.\cite{Wang1805}
One can find that $R^{\bar{b}b}_z$ and $R^{\bar{b}b}_z$ (for the multiplicity-free case) are gauge invariant
only when $b=\bar{b}$. In MS MTCs, except for the identity anyon $I_0$, none of the anyons are self-dual,
and $R^{\bar{b}b}_z$ ($R^{\bar{b}b}_z$) are not gauge invariant. We give more discussions on the gauge 
freedom in $S^{(z)}$ and $C^{(z)}$ in the following subsection.

\subsubsection{Gauge freedom and patterns in the solutions}

The solutions to the modular relation in Eq.\eqref{ModularRelationSzTz} are not unique,
because they have a gauge freedom associated with each distinct vertex that amounts to the choice of
basis vectors. More explicitly, we can consider the basis transformation in the splitting space $V_c^{ab}$
as follows
\be\label{gauge_basis}
\begin{tikzpicture}[baseline={(current bounding box.center)}]
\draw [>=stealth,->] (0pt,-15pt)--(0pt,-6pt);\draw (0pt,-6pt)--(0pt,0pt);
\draw [>=stealth,->] (0pt,0pt)--(-6pt,8pt); \draw(-6pt,8pt)--(-7.5pt,10pt);
\draw [>=stealth,->] (0pt,0pt)--(6pt,8pt); \draw(6pt,8pt)--(7.5pt,10pt);
\node at (-8pt,15pt){$a$};
\node at (8pt,15pt){$b$};
\node at (0pt,-20pt){$c$};
\node at (6pt,-2pt){\small$\mu$};
\end{tikzpicture}
=\sum_{\mu'} [u^{ab}_c]_{\mu \mu'}
\begin{tikzpicture}[baseline={(current bounding box.center)}]
\draw [>=stealth,->] (0pt,-15pt)--(0pt,-6pt);\draw (0pt,-6pt)--(0pt,0pt);
\draw [>=stealth,->] (0pt,0pt)--(-6pt,8pt); \draw(-6pt,8pt)--(-7.5pt,10pt);
\draw [>=stealth,->] (0pt,0pt)--(6pt,8pt); \draw(6pt,8pt)--(7.5pt,10pt);
\node at (-8pt,15pt){$a$};
\node at (8pt,15pt){$b$};
\node at (0pt,-20pt){$c$};
\node at (6pt,-2pt){\small$\mu'$};
\end{tikzpicture}
\ee
or equivalently, $|a,b;c,\mu\rangle=\sum_{\mu'} [u^{ab}_c]_{\mu \mu'}|a,b;c,\mu'\rangle$.
If one requires that the $F$ and $R$ matrices in the anyon theory are presented by unitary matrices, the basis transformation 
above should also be unitary.\cite{bonderson2007}
For the multiplicity-free case, such that $u=u'$ denotes the unique fusion channel, 
then $u^{ab}_c$ in Eq.\eqref{gauge_basis} is a $U(1)$ phase. 

The gauge freedom in Eq.\eqref{gauge_basis} means there are a family of solutions 
to the modular relations in Eq.\eqref{ModularRelationSzTz}, and also to the more general relations such as Pentagon and Hexagon equations.
This can be intuitively seen from the diagrammatic representation of $S^{(z)}_{a,\mu;b,\nu}$ in \eqref{PuncturedSmatrix}
and $C^{(z)}_{a,\mu;b,\nu}$ in \eqref{CzMatrixElement}. There are a pair of vertices in both $S^{(z)}_{a,\mu;b,\nu}$ and $C^{(z)}_{a,\mu;b,\nu}$.
If we change the gauge choice associated with the vertex structures, then the concrete value of 
$S^{(z)}_{a,\mu;b,\nu}$ and $C^{(z)}_{a,\mu;b,\nu}$ will change accordingly.
More explicitly, with the basis transformation in \eqref{gauge_basis}, one can find that different quantities transform as:
\be\label{GaugeTransform_SCT}
\small
\begin{split}
&[S']^{(z)}_{a,\mu';b,\nu'}=\sum_{\mu,\nu} (u^{bz}_b)^{-1}_{\nu'\nu}\cdot S^{(z)}_{a,\mu;b,\nu}\cdot (u^{az}_a)_{\mu\mu'},\\
&[C']^{(z)}_{a,\mu';\bar{a},\nu'}=\sum_{\mu,\nu} (u^{\bar{a}z}_{\bar{a}})^{-1}_{\nu'\nu}\cdot C^{(z)}_{a,\mu;\bar{a},\nu}\cdot (u^{az}_a)_{\mu\mu'},\\
&[T']^{(z)}_{a,\mu'; b,\nu'}=T^{(z)}_{a,\mu';b,\nu'}.
\end{split}
\ee
where we have considered the expressions of $C^{(z)}$ and $T^{(z)}$
in Eqs.\eqref{CzMatrixElement} and \eqref{PunctureT}, respectively.
Interestingly, from Eq.\eqref{GaugeTransform_SCT}, one can find that 
\be\label{SumOverZ_Sz}
\sum_{\mu'}[S']^{(z)}_{a,\mu';a,\mu'}=\sum_{\mu} S^{(z)}_{a,\mu; a,\mu}.
\ee
That is, although $S^{(z)}_{a,\mu;b,\nu}$ is not gauge invariant with respect to the basis transformation in \eqref{gauge_basis}, 
$\sum_{\mu} S^{(z)}_{a,\mu; a,\mu}$ is a gauge invariant quantity. The result in Eq.\eqref{SumOverZ_Sz} is also
studied in the diagrammatical representation of anyons in Ref.\onlinecite{Wang1805}.

For simplicity, we will focus on the multiplicity-free case in the following discussions. This is the case
we are interested in for the MS MTCs (See, e.g., the fusion rules in \eqref{fusion3u0}). 
Generalization to the case that is not multiplicity-free is straightforward.
In the multiplicity-free case, Eq.\eqref{GaugeTransform_SCT} can be simplified as
\be\label{GaugeTransform_SCT_free}
\begin{split}
&[S']^{(z)}_{a,b}=S^{(z)}_{a,b}\cdot (u^{bz}_b)^{-1}\cdot u^{az}_a,\\
&[C']^{(z)}_{a,\bar{a}}=C^{(z)}_{a,\bar{a}}\cdot (u^{\bar{a}z}_{\bar{a}})^{-1}\cdot u^{az}_a,\\
&[T']^{(z)}_{a,b}=T^{(z)}_{a,b}.
\end{split}
\ee
where $u^{ab}_c$ is simply a phase factor. We give several remarks on Eq.\eqref{GaugeTransform_SCT_free} as follows:

-- One can find that 
\be
[S']^{(z)}_{a,a}=S^{(z)}_{a,a},
\ee 
which is a simplified version of Eq.\eqref{SumOverZ_Sz}. That is, $S^{(z)}_{a,a}$ is a gauge invariant quantity in the multiplicity-free case.
As will be discussed in Sec.\eqref{PuncturedS_distinguish_MS_MC}, we will use this quantity to distinguish different MS MTCs.

-- The punctured $S$ matrix is in general not symmetric, \textit{i.e.}, $S^{(z)}_{a,b}\neq S^{(z)}_{b,a}$. 
This can be understood based on the first equation in \eqref{GaugeTransform_SCT_free}. One can find that
\be
\small
\frac{[S']^{(z)}_{a,b}}{[S']^{(z)}_{b,a}}=\frac{S_{a,b}^{(z)}}{S_{b,a}^{(z)}}\cdot\frac{(u^{az}_a)^2}{(u^{bz}_b)}.
\ee
If $S^{(z)}_{a,b}= S^{(z)}_{b,a}$, then after a basis transformation in \eqref{gauge_basis}, one can obtain 
$[S']^{(z)}_{a,b}/[S']^{(z)}_{b,a}=(u^{az}_a)^2/(u^{bz}_b)^2$, which is in general not equal to $1$, and therefore
 $[S']^{(z)}_{a,b}$ is asymmetric.
This is in contrast to the modular $S$ matrix, which is symmetric, \textit{i.e.}, $S_{a,b}=S_{b,a}$.

-- $C^{(z)}_{a,\bar{a}}$ is gauge invariant if $\bar{a}=a$, but not gauge invariant if $\bar{a}\neq a$.
This is straightforward based on the second equation in \eqref{GaugeTransform_SCT_free}.
It is also related to the fact that $R^{\bar{b}b}_z$ in \eqref{CzMatrixElement} is not gauge invariant for $\bar{b}\neq b$.
Note that for the counterexamples in MS MTCs, for all anyons but $I_0$, we have $\bar{a}\neq a$. 

-- $T^{(z)}_{a,b}$ is gauge invariant, with the explicit expression in Eq.\eqref{PunctureT}.

In the above we have discussed the gauge freedom in $S^{(z)}$, $C^{(z)}$, and $T^{(z)}$ in Eq.\eqref{ModularRelationSzTz}.
Now we need to fix a simple gauge choice to solve Eq.\eqref{ModularRelationSzTz}.
We will consider the simple currents $I_k$ ($k=1,2,3,4$) as introduced in Sec.\ref{Sec: SimpleCurrents}.
The gauge choice we make is 
\begin{eqnarray}\label{GaugeChoice}
\small
\begin{tikzpicture}[baseline={(current bounding box.center)}]
\draw [>=stealth,->] (0pt,-16pt)--(0pt,-6pt);\draw (0pt,-6pt)--(0pt,0pt);
\draw [>=stealth,->] (0pt,0pt)--(-6pt,8pt); \draw(-6pt,8pt)--(-7.5pt,10pt);
\draw [>=stealth,->] (0pt,0pt)--(6pt,8pt); \draw(6pt,8pt)--(7.5pt,10pt);
\node at (-8pt,16pt){$B_{i,j'}$};
\node at (8pt,15pt){$z$};
\node at (0pt,-22pt){$B_{i,j'}$};
\end{tikzpicture}
=
\begin{tikzpicture}[baseline={(current bounding box.center)}]
\draw [>=stealth,->] (0pt,-16pt)--(0pt,-6pt);\draw (0pt,-6pt)--(0pt,0pt);
\draw [dashed][>=stealth,->] (-3pt,-16pt)--(-3pt,-6pt);\draw [dashed](-3pt,-6pt)--(-3pt,0pt);
\draw [>=stealth,->] (0pt,0pt)--(-6pt,8pt); \draw(-6pt,8pt)--(-7.5pt,10pt);
\draw [dashed][>=stealth,->] (-3pt,0-2pt)--(-9pt,8-2pt); \draw[dashed](-9pt,8-2pt)--(-10.5pt,10-2pt);
\draw [>=stealth,->] (0pt,0pt)--(6pt,8pt); \draw(6pt,8pt)--(7.5pt,10pt);
\node at (-8pt,16pt){$B_{i,j}$};
\node at (8pt,15pt){$z$};
\node at (3pt,-22pt){$B_{i,j}$};
\node at (-8pt,-10pt){$I_k$};
\end{tikzpicture}
\end{eqnarray}
where $B_{i,j'}=B_{i,j}\times I_k$.
This can be viewed as a generalization of \eqref{split} to the vertex structure. We emphasize that 
there is no gauge freedom in \eqref{split}. For \eqref{GaugeChoice}, however, there can be a phase-factor
difference on the two sides. For example, under the basis transformation in \eqref{gauge_basis}, 
a phase factor $u^{B_{i,j}, z}_{B_{i,j}}/u^{B_{i,j'}, z}_{B_{i,j'}}$ will be introduced in Eq.\eqref{GaugeChoice}.
Here we make the simple gauge choice such that Eq.\eqref{GaugeChoice} holds.
Once we find the solutions to the modular relations in Eq.\eqref{ModularRelationSzTz}, we can obtain 
a family of gauge-equivalent solutions by considering the basis transformations in \eqref{gauge_basis}.

With the gauge choice in \eqref{GaugeChoice}, one can find the following patterns in $S^{(z)}$:
\begin{eqnarray}
\small
\begin{tikzpicture}[baseline={(current bounding box.center)}]
\draw (20pt,0pt) circle (20pt);
\draw[line width=6pt, draw=white] (0pt,0pt) circle (20pt);
\draw (0pt,0pt) circle (20pt);
\draw [>=stealth,->] (20pt,0.1pt)--(20pt,0.11pt);
\draw [>=stealth,->] (40pt,0.1pt)--(40pt,0.11pt);
\draw[line width=6pt, draw=white]  (0pt,0pt) arc (-180:-270:20pt);
\draw (0pt,0pt) arc (-180:-270:20pt);
\node at (0pt,25pt){$B_{i,j'}$};
\node at (25pt,25pt){$B_{m,n}$};
\draw[>=stealth,->]  (20pt,-20pt)..controls (15pt,-30pt) and (5pt,-30pt)..(0pt,-20pt);
\node at (10pt,-32pt){$z$};
\end{tikzpicture}
=
\begin{tikzpicture}[baseline={(current bounding box.center)}]
\draw (20pt,0pt) circle (20pt);
\draw[line width=6pt, draw=white] (0pt,0pt) circle (20pt);
\draw (0pt,0pt) circle (20pt);
\draw [dashed](0pt,0pt) circle (18pt);
\draw [>=stealth,->] (20pt,0.1pt)--(20pt,0.11pt);
\draw [>=stealth,->] (40pt,0.1pt)--(40pt,0.11pt);
\draw [>=stealth,->] (18pt,0.1pt)--(18pt,0.11pt);

\draw[line width=6pt, draw=white]  (0pt,0pt) arc (-180:-270:20pt);
\draw (0pt,0pt) arc (-180:-270:20pt);
\node at (0pt,25pt){$B_{i,j}$};
\node at (-12pt,0pt){$I_k$};
\node at (25pt,25pt){$B_{m,n}$};
\draw[>=stealth,->]  (20pt,-20pt)..controls (15pt,-30pt) and (5pt,-30pt)..(0pt,-20pt);
\node at (10pt,-32pt){$z$};
\end{tikzpicture}
\end{eqnarray}
where $B_{i,j'}=B_{i,j}\times I_k$, as seen in Eq.\eqref{Fusion_SimpleCurrent_Banyon}.
By shrinking the $I_k$ anyon, and using the relation in Eq.\eqref{local_shrink}, one can obtain the following results:
\be\label{Sz_pattern}
\begin{split}
S^{(z)}_{(I_k \times B_{i,j}),B_{m,n}}=&\frac{S_{I_k,B_{m,n}}}{S_{I_0,B_{m,n}}}\cdot S^{(z)}_{B_{i,j},B_{m,n}},\\
S^{(z)}_{B_{m,n}, (I_k\times B_{i,j})}=&\frac{S_{B_{m,n},I_k} }{S_{B_{m,n},I_0}}\cdot S^{(z)}_{B_{m,n},B_{i,j}},\\
\end{split}
\ee
where the phase factors $\frac{S_{I_k,B_{m,n}}}{S_{I_0,B_{m,n}}}$ and $\frac{S_{B_{m,n},I_k} }{S_{B_{m,n},I_0}}$ 
are expressed in Eq.\eqref{Phase_factor}.
This pattern holds for all the five categories with $u=0, \cdots, 4$ in the simplest counterexamples in MS MTCs with 
$G=\mathbb{Z}_{11}\rtimes \mathbb{Z}_5$.

The concrete solutions of punctured $S$ matrix for MS MTCs with 
$G=\mathbb{Z}_{11}\rtimes \mathbb{Z}_5$
can be found in online materials [\onlinecite{OnlineData}].
Recall that in MS MTCs, only type-$B$ anyons contain the information of the 3-cocycle $\omega^u$, with $u=0,1,2,3,4$.
This indicates the nontrivial $S^{(z)}$ that are candidates to distinguish different MS MTCs are those containing type-$B$ anyons. 
Based on the fusion rules in Eqs.\eqref{fusion1}, \eqref{fusion2}, and \eqref{fusion3u0}, the nontrivial $S^{(z)}$ correspond to those 
with $z=I_5$, $I_6$, and all type-$A$ anyons.

On the other hand, for $z=I_k$ ($k=1,2,3,4$), only type-$A$ anyons participate in the fusion $a\times \bar{a}=I_0+N_{a\bar{a}}^z z+\cdots$, 
\textit{i.e.}, we have $N_{a\bar{a}}^z>0$ only if $a$ are type-$A$ anyons, as seen from \eqref{fusion2}.
In this case, since both $I_k$ and type-$A$ anyons are defined independent of the 3-cocyle $\omega^{u}$,
it is expected that $S^{(z)}$ with $z=I_k$ ($k=1,2,3,4$) are independent of the 3-cocyle $\omega^{u}$, 
and therefore are trivial for our motivations. 
This is verified in the lattice calculation in Sec.\ref{Sec: Lattice}. 
We did not solve for these trivial $S^{(z)}$ which are independent of $u$.

It is emphasized that the solutions in [\onlinecite{OnlineData}] are not unique, and one can obtain a family of gauge-equivalent solutions by performing the 
basis transformation in Eq.\eqref{gauge_basis}. For the multiplicity-free case, this corresponds to the gauge transformation in Eq.\eqref{GaugeTransform_SCT_free}.


\subsection{Distinguish MS MTCs with $S^{(z)}$ and $T$}
\label{PuncturedS_distinguish_MS_MC}

As discussed in the previous subsections, in general the punctured $S$ matrix depends on the gauge 
freedom in the choice of basis vectors in the fusion/splitting space.
To distinguish different MS MTCs, we hope to find some topological invariants, which is our main aim in the next subsections.
But before that, it is noted that certain matrix elements in the punctured 
$S$ matrix are topological invariants themselves, such as $\sum_{\mu=1}^{N_{a\bar{a}}^z}S^{(z)}_{a\mu,a\mu}$,
as seen from Eq.\eqref{SumOverZ_Sz}.

In this subsection, we show that the simplest
 counterexamples in MS MTCs can be distinguished based on the diagonal parts of punctured $S$-matrix 
together with the modular $T$ matrix.
For example, one can find that the permutations of anyons taking $T^{u=1(2)}$ to $T^{u=4(3)}$ cannot take $S^{(z),u=1(2)}_{B,B}$ 
to $S^{(z),u=4(3)}_{B,B}$ simultaneously, indicating that the punctured $S$ matrix together with $T$ matrix can 
distinguish the two categories with $u=1(2)$ and $u=4(3)$. 
This method is similar to those used in Ref.\onlinecite{Wang1805,DT1806,MS1806} in spirit, where one needs to track whether two sets
of topological invariants transform in a consistent way.

More explicitly, let us consider the categories $\mathcal{C}_u$ with $u=1$ and $u=4$.
From the modular $T$-matrix in Table \ref{49Anyons}, one can find that, for example, 
$A_{1,1}$ in $\mathcal{C}_{u=1}$  can only be sent to $A_{1,1}$ or $A_{2,6}$
in $\mathcal{C}_{u=4}$  to keep the topological spin unchanged 
(Note that $A_{1,1}$ and $A_{2,6}$ are dual anyons 
in both $\mathcal{C}_{u=1}$ and $\mathcal{C}_{u=4}$ as seen from the fusion rules in \eqref{fusion2} with $\theta_a=\theta_{\bar{a}}$).
Here the meaning of ``sending $A_{1,1}$ in $\mathcal{C}_{u=1}$ to $A_{1,1}$ or $A_{2,6}$
in $\mathcal{C}_{u=1}$ " is the bijection of anyons as introduced in Sec.\ref{Sec: proof_Equiv_ModularData}.
That is, we consider a bijection of anyons between two categories that share the same modular data, such that
$T^{(u)}_{p_r(a),p_r(b)}=T^{(u')}_{a,b}$.
For the example above, one can find that $T^{(u=4)}_{A_{1,1}, A_{1,1}}=T^{(u=4)}_{A_{2,6}, A_{2,6}}=T^{(u=1)}_{A_{1,1}, A_{1,1}}$.

\begin{table}[h]
\centering
\footnotesize
\begin{tabular}{cccccccccccc}
$a$  		 &\vline        &$u=1$  &\vline         &$u=4$\\ \hline\hline
$B_{1,0}$  &\vline            &$\frac{1}{5}e^{-\frac{144}{275} i\pi}$  &\vline             &$\frac{1}{5}e^{\frac{274}{275} i\pi}$ \\
$B_{1,1}$  &\vline             &$\frac{1}{5}e^{\frac{186}{275} i\pi}$  &\vline             &$\frac{1}{5}e^{\frac{54}{275} i\pi}$  \\ 
$B_{1.2}$  &\vline           &$\frac{1}{5}e^{-\frac{34}{275} i\pi}$  &\vline             &$\frac{1}{5}e^{-\frac{166}{275} i\pi}$  \\ 
$B_{1.3}$  &\vline            &$\frac{1}{5}e^{-\frac{254}{275} i\pi}$  &\vline              &$\frac{1}{5}e^{\frac{164}{275} i\pi}$  \\ 
$B_{1,4}$  &\vline            &$\frac{1}{5}e^{\frac{76}{275} i\pi}$  &\vline              &$\frac{1}{5}e^{-\frac{56}{275} i\pi}$   \\ \hline

$B_{2,0}$  &\vline            &$\frac{1}{5}e^{-\frac{126}{275} i\pi}$  &\vline             &$\frac{1}{5}e^{-\frac{104}{275} i\pi}$   \\
$B_{2,1}$  &\vline             &$\frac{1}{5}e^{-\frac{16}{275} i\pi}$ &\vline              &$\frac{1}{5}e^{\frac{6}{275} i\pi}$   \\ 
$B_{2.2}$  &\vline            &$\frac{1}{5}e^{\frac{94}{275} i\pi}$  &\vline              &$\frac{1}{5}e^{\frac{116}{275} i\pi}$   \\ 
$B_{2.3}$  &\vline             &$\frac{1}{5}e^{\frac{204}{275} i\pi}$  &\vline              &$\frac{1}{5}e^{\frac{226}{275} i\pi}$    \\ 
$B_{2,4}$  &\vline             &$\frac{1}{5}e^{-\frac{236}{275} i\pi}$  &\vline              &$\frac{1}{5}e^{-\frac{214}{275} i\pi}$  \\ \hline

$B_{3,0}$  &\vline             &$\frac{1}{5}e^{\frac{204}{275} i\pi}$  &\vline              &$\frac{1}{5}e^{\frac{116}{275} i\pi}$   \\
$B_{3,1}$  &\vline            &$\frac{1}{5}e^{\frac{94}{275} i\pi}$  &\vline             &$\frac{1}{5}e^{\frac{6}{275} i\pi}$    \\ 
$B_{3.2}$  &\vline             &$\frac{1}{5}e^{-\frac{16}{275} i\pi}$  &\vline              &$\frac{1}{5}e^{-\frac{104}{275} i\pi}$    \\ 
$B_{3.3}$  &\vline              &$\frac{1}{5}e^{-\frac{126}{275} i\pi}$  &\vline              &$\frac{1}{5}e^{-\frac{214}{275} i\pi}$    \\ 
$B_{3,4}$  &\vline             &$\frac{1}{5}e^{-\frac{236}{275} i\pi}$  &\vline            &$\frac{1}{5}e^{\frac{226}{275} i\pi}$  \\ \hline

$B_{4,0}$  &\vline              &$\frac{1}{5}e^{-\frac{254}{275} i\pi}$   &\vline             &$\frac{1}{5}e^{-\frac{166}{275} i\pi}$   \\
$B_{4,1}$  &\vline             &$\frac{1}{5}e^{-\frac{34}{275} i\pi}$   &\vline             &$\frac{1}{5}e^{\frac{54}{275} i\pi}$   \\ 
$B_{4.2}$  &\vline              &$\frac{1}{5}e^{\frac{186}{275} i\pi}$  &\vline              &$\frac{1}{5}e^{\frac{274}{275} i\pi}$   \\ 
$B_{4.3}$  &\vline               &$\frac{1}{5}e^{-\frac{144}{275} i\pi}$   &\vline              &$\frac{1}{5}e^{-\frac{56}{275} i\pi}$   \\ 
$B_{4,4}$  &\vline             &$\frac{1}{5}e^{\frac{76}{275} i\pi}$   &\vline              &$\frac{1}{5}e^{\frac{164}{275} i\pi}$   \\ \hline
\end{tabular}
\caption{The diagonal elements $S^{(z)}_{B_{k,n}B_{k,n}}$ of the punctured $S$ matrix with 
$z=A_{1,1}$ and $A_{2,6}$ ($A_{1,1}$ and $A_{2,6}$ are dual anyons)
for $\mathcal{C}_{u=1}$ and $\mathcal{C}_{u=4}$, respectively. 
The complete data for $S^{(z)}_{a,\mu;b,\nu}$ can be found in materials online.\cite{OnlineData}}
\label{SaaSampleA11}
\end{table}

After fixing how $z$ are sent from $\mathcal{C}_{u=1}$ to $\mathcal{C}_{u=4}$ to keep the modular $T$ matrix unchanged, 
now let us consider the anyons $B_{k,n}$ in $S^{(z)}_{B_{k,n}B_{k,n}}$.
From Table \ref{49Anyons}, the topological spins for type-$B$ anyons have the form 
\be\label{TopologicalSpin}
\small
\theta=\exp\left(\frac{i2\pi}{25}s\right),
\ee
with
\be
\small
\begin{split}
u=1,\quad s\in\{&1,6,11,16,21;\quad 4, 14, 24, 9, 19;\\
& 9, 24,14,4,19;\quad 16,11,6, 1, 21\},\\
u=4,\quad s\in\{&4,9,14,19,24;\quad 16,1,11,21,6;\\
&11,1,16,6,21;\quad 14,9,4,24,19\},
\end{split}
\ee
where $s$ is listed with the order in Table \ref{49Anyons}.
One can find that, for example, $B_{1,0}$ in $\mathcal{C}_{u=1}$ can only be sent to $B_{2,1}$ or $B_{3,1}$ in $\mathcal{C}_{u=4}$ 
(Note that $B_{2,1}$ and $B_{3,1}$ in $\mathcal{C}_{u=4}$
are dual anyons as seen from \eqref{Fusion_u4}, with $\theta_a=\theta_{\bar{a}}$).
Till now, based on the modular $T$ matrix, we just observed how different anyons can be sent/permuted from $\mathcal{C}_{u=1}$
 to $\mathcal{C}_{u=4}$ categories. Next, we need to check if $S^{(z)}_{B_{k,n}B_{k,n}}$ is mapped in a consistent way.

In Table \ref{SaaSampleA11}, we show the results for the diagonal elements $S_{B_{k,n}, B_{k,n}}^{(z)}$ with $z=A_{1,1}$ and $A_{2,6}$,
for both $\mathcal{C}_{u=1}$ and $\mathcal{C}_{u=4}$ categories. 
One can find that $S^{(z=A_{1,1})}_{B_{1,0},B_{1,0}}=\frac{1}{5}e^{-\frac{144}{275} i\pi}$ in $\mathcal{C}_{u=1}$, 
but $S^{(z=A_{1,1})}_{B_{2,1},B_{2,1}}=S^{(z=A_{1,1})}_{B_{3,1},B_{3,1}}
=S^{(z=A_{2,6})}_{B_{2,1},B_{2,1}}=S^{(z=A_{2,6})}_{B_{3,1},B_{3,1}}
=\frac{1}{5}e^{\frac{6}{275} i\pi}$ in $\mathcal{C}_{u=4}$. This means $S^{(z)}_{B_{k,n}B_{k,n}}$ is not mapped in the same way as 
the modular $T$ matrix by permuting anyons, indicating that $\mathcal{C}_{u=1}$ and $\mathcal{C}_{u=4}$ are different categories.

Then curious readers may ask whether $S^{(z=A_{1,1})}_{B_{1,0},B_{1,0}}$ for 
$\mathcal{C}_{u=1}$ can be sent to any other quantity? By looking through the data for $S^{(z)}_{B_{k,n}B_{k,n}}$,
we found that $S^{(z=A_{1,1})}_{B_{1,0},B_{1,0}}$ in $\mathcal{C}_{u=1}$ is sent to $S^{(z=A_{1,9})}_{B_{2,1},B_{2,1}}
=S^{(z=A_{2,10})}_{B_{2,1},B_{2,1}}=S^{(z=A_{1,9})}_{B_{3,1},B_{3,1}}=S^{(z=A_{2,10})}_{B_{3,1},B_{3,1}}$ in $\mathcal{C}_{u=4}$, 
as seen from Table \ref{SaaSampleA19A20}.
That is, to fix $S_{B_{k,n}, B_{k,n}}^{(z)}$, we need to send $A_{1,1}$ in $\mathcal{C}_{u=1}$ to $A_{1,9}$ or $A_{2,10}$ 
in $\mathcal{C}_{u=4}$. Apparently, 
$A_{1,1}$ and $A_{1,9}$ ($A_{2,10}$) have different topological spins and the corresponding $T$ matrix elements
cannot be mapped consistently from $\mathcal{C}_{u=1}$ to $\mathcal{C}_{u=4}$.

A summary of how $S^{(z)}_{B_{i,j},B_{i,j}}$ in $\mathcal{C}_{u=1}$ are mapped to 
$S^{(z')}_{B_{i',j'},B_{i',j'}}$ in $\mathcal{C}_{u=4}$ can be found in appendix \ref{Sec: Other_Sz}.

\begin{table}[h]
\centering
\footnotesize
\begin{tabular}{cccccccccccc}
$a$  		 &\vline        &$u=1$  &\vline         &$u=4$\\ \hline\hline
$B_{1,0}$  &\vline            &$\frac{1}{5}e^{\frac{156}{275} i\pi}$  &\vline             &$\frac{1}{5}e^{\frac{24}{275} i\pi}$ \\
$B_{1,1}$  &\vline             &$\frac{1}{5}e^{-\frac{64}{275} i\pi}$  &\vline             &$\frac{1}{5}e^{-\frac{196}{275} i\pi}$  \\ 
$B_{1.2}$  &\vline           &$\frac{1}{5}e^{\frac{266}{275} i\pi}$  &\vline             &$\frac{1}{5}e^{\frac{134}{275} i\pi}$  \\ 
$B_{1.3}$  &\vline            &$\frac{1}{5}e^{\frac{46}{275} i\pi}$  &\vline              &$\frac{1}{5}e^{-\frac{86}{275} i\pi}$  \\ 
$B_{1,4}$  &\vline            &$\frac{1}{5}e^{-\frac{174}{275} i\pi}$  &\vline              &$\frac{1}{5}e^{\frac{244}{275} i\pi}$   \\ \hline

$B_{2,0}$  &\vline            &$\frac{1}{5}e^{\frac{274}{275} i\pi}$  &\vline             &$\frac{1}{5}e^{-\frac{254}{275} i\pi}$   \\
$B_{2,1}$  &\vline             &$\frac{1}{5}e^{-\frac{166}{275} i\pi}$ &\vline              &$\frac{1}{5}e^{-\frac{144}{275} i\pi}$   \\ 
$B_{2.2}$  &\vline            &$\frac{1}{5}e^{-\frac{56}{275} i\pi}$  &\vline              &$\frac{1}{5}e^{-\frac{34}{275} i\pi}$   \\ 
$B_{2.3}$  &\vline             &$\frac{1}{5}e^{\frac{54}{275} i\pi}$  &\vline              &$\frac{1}{5}e^{\frac{76}{275} i\pi}$    \\ 
$B_{2,4}$  &\vline             &$\frac{1}{5}e^{\frac{164}{275} i\pi}$  &\vline              &$\frac{1}{5}e^{\frac{186}{275} i\pi}$  \\ \hline

$B_{3,0}$  &\vline             &$\frac{1}{5}e^{\frac{54}{275} i\pi}$  &\vline              &$\frac{1}{5}e^{-\frac{34}{275} i\pi}$   \\
$B_{3,1}$  &\vline            &$\frac{1}{5}e^{-\frac{56}{275} i\pi}$  &\vline             &$\frac{1}{5}e^{-\frac{144}{275} i\pi}$    \\ 
$B_{3.2}$  &\vline             &$\frac{1}{5}e^{-\frac{166}{275} i\pi}$  &\vline              &$\frac{1}{5}e^{-\frac{254}{275} i\pi}$    \\ 
$B_{3.3}$  &\vline              &$\frac{1}{5}e^{\frac{274}{275} i\pi}$  &\vline              &$\frac{1}{5}e^{\frac{186}{275} i\pi}$    \\ 
$B_{3,4}$  &\vline             &$\frac{1}{5}e^{\frac{164}{275} i\pi}$  &\vline            &$\frac{1}{5}e^{\frac{76}{275} i\pi}$  \\ \hline

$B_{4,0}$  &\vline              &$\frac{1}{5}e^{\frac{46}{275} i\pi}$   &\vline             &$\frac{1}{5}e^{\frac{134}{275} i\pi}$   \\
$B_{4,1}$  &\vline             &$\frac{1}{5}e^{\frac{266}{275} i\pi}$   &\vline             &$\frac{1}{5}e^{-\frac{196}{275} i\pi}$   \\ 
$B_{4.2}$  &\vline              &$\frac{1}{5}e^{-\frac{64}{275} i\pi}$  &\vline              &$\frac{1}{5}e^{\frac{24}{275} i\pi}$   \\ 
$B_{4.3}$  &\vline               &$\frac{1}{5}e^{\frac{156}{275} i\pi}$   &\vline              &$\frac{1}{5}e^{\frac{244}{275} i\pi}$   \\ 
$B_{4,4}$  &\vline             &$\frac{1}{5}e^{-\frac{174}{275} i\pi}$   &\vline              &$\frac{1}{5}e^{-\frac{86}{275} i\pi}$   \\ \hline
\end{tabular}
\caption{The diagonal elements $S^{(z)}_{B_{k,n}B_{k,n}}$ of the punctured $S$ matrix with $z=A_{1,9}$ and $A_{2,10}$ ($A_{1,9}$ 
and $A_{2,10}$ are dual anyons)
for $u=1$ and $u=4$, respectively. The complete data for $S^{(z)}_{a,\mu;b,\nu}$ can be found in materials online.}
\label{SaaSampleA19A20}
\end{table}

In short, by permuting anyons, the topological invariants $S^{(z)}_{B_{k,n}B_{k,n}}$ and $T$ matrix are mapped from $\mathcal{C}_{u=1}$ to
$\mathcal{C}_{u=4}$ categories in different ways. Therefore, we conclude that the punctured $S$ matrix together with the modular $T$ matrix can distinguish 
$\mathcal{C}_{u=1}$ and $\mathcal{C}_{u=4}$ categories. One can check that $\mathcal{C}_{u=2}$ and $\mathcal{C}_{u=3}$ categories can be distinguished in the same way.

Several remarks before we leave this subsection: 

-- Not all the punctured $S$ matrices can be used to distinguish different categories. This is true even if $z$ are type-$A$ anyons.
For example, when $z=A_{1,0}$ or $A_{2,0}$, which has trivial topological spin, the punctured $S$ matrix $S^{(z)}$ together with 
$T$ matrix cannot distinguish different categories. This is explicitly shown in Appendix.\ref{Sec: Other_Sz}. 
We observed that only for $z=A_{1,i}$ and $A_{2,i}$ with $i=1,\cdots, 10$, which have nontrivial topological spins, \textit{i.e.}, $\theta_z\neq 1$,
the corresponding punctured $S$ matrix $S^{(z)}$ together with $T$ can distinguish different categories.

-- As discussed in Sec.\ref{SubSec:Modular data},
because of Galois symmetry in the modular data, it is well understood about the transformation property of modular $S$ and $T$ matrices
by permuting anyons in the same category as well as among different categories. It is interesting to study how the punctured $S$
matrix transforms under the permutation of anyons.

-- Although the method above can be used to distinguish MS MTCs, but the procedure is complicate.
We need to track how the bijections/permutations of anyons map the data of 
$T$ and $S^{(z)}_{a,a}$ between two different categories. It will be nice if we can distinguish the MS MTCs with a single number.
In addition, if we work on a lattice gauge theory (see Sec.\ref{Sec: Lattice}), since it is not clear to us how to write down the 
quasi-particle basis, it is difficult to do anyon permutation as what we have done in this subsection. For these reasons, it is desired 
to define topological invariants which are independent of anyon types, as studied in the following subsection.

-- In Ref.\onlinecite{Wang1805}, the link invariant of Whitehead link was proposed to distinguish the most
simplest counterexamples in MS MTCs. It is proved that the Whitehead link invariant is determined by 
the diagonal elements of punctured $S$ matrix together with the modular data.
Here we obtain the punctured $S$ matrix with both the diagonal and off-diagonal elements, which provide more 
information than the Whitehead link invariant.

\subsection{Topological invariants: Trace of words and link invariants}
\label{Sec: Topological_invariants}

Based on the punctured $S$ and $T$ matrices obtained in the previous subsections, we can 
construct topological invariants on a punctured torus or a genus-two closed manifold.
More precisely, we calculate these topological invariants in the Hilbert space 
$\mathcal{H}(\Sigma_{1,1})$ or $\mathcal{H}(\Sigma_{2,0})$.

The merit of topological invariants constructed in this subsection is that
\textit{we only need a single number to distinguish those different MS MTCs.}

We illustrate this idea in the cases of (i) a punctured torus and (ii) a genus-two manifold, respectively.

\subsubsection{Punctured torus}
\label{Sec: Punctured Torus_QuasiParticleBasis}

With the punctured $S$ and $T$ matrices, we can construct an arbitrary `word' of the following form:
\be\label{Wz}
w^{(z)}:=
\big(
S^{(z)}\big)^{n_1}\big(T^{(z)}\big)^{n_2}\big(S^{(z)}\big)^{n_3}\big(T^{(z)}\big)^{n_4}\cdots 
\ee
where $n_i$ are arbitrary integers.
The topological invariant can be expressed as a trace of words as
\be\label{Wz_PuncturedTorus}
W_{\Sigma_{1,1}}^{(z)}=\text{Tr}(w^{(z)}),
\ee
where the trace is over the Hilbert space $\mathcal{H}(\Sigma_{1,1}; z)$ with a fixed anyon charge $z$
(see the basis of $\mathcal{H}(\Sigma_{1,1}; z)$ in the left plot of \eqref{PunctureTorusBasis}).

There are infinite number of nontrivial words which may be used to distinguish the MS MTCs. Here we list a few of them:
\be\label{Words}
\small
\left\{
\begin{split}
w_1^{(z)}=&(\theta^{(z)})^2\cdot (T^{(z)})^7S^{(z)},\\
w_2^{(z)}=&(\theta^{(z)})^3\cdot \big[(T^{(z)})^3S^{(z)}\big]^2,\\
w_3^{(z)}=&(\theta^{(z)})^2\cdot \big[ (T^{(z)})^4S^{(z)}\big]^3,\\
w_4^{(z)}=& \big[(T^{(z)})^5S^{(z)}\big]^4,\\
w_5^{(z)}=&(\theta^{(z)})^3\cdot \big[(T^{(z)})^3S^{(z)}\big]^2\cdot \big[(T^{(z)})^2 S^{(z)}\big]^5,\\
w_6^{(z)}=&(\theta^{(z)})^5\cdot \big[ (T^{(z)})^3S^{(z)}\big]^4,\\
w_7^{(z)}=&(\theta^{(z)})^8\cdot \big[ (T^{(z)})^3S^{(z)}\big]^2,
\end{split}
\right.
\ee
where we have $\theta^{(z)}=\big(S^{(z)}\big)^{-4}$.

The reason why the trace of words in Eqs.\eqref{Wz_PuncturedTorus} are topological 
invariants is that they are related to the link invariants. 
In the following, we give two examples on $\text{Tr}(w_1^{(z)})$ and $\text{Tr}(w_2^{(z)})$,
and then give some general remarks.

Let us present some preliminaries first, by looking at the expression of  
$\text{Tr}(T^{(z)})$ and $\text{Tr}(S^{(z)})$. For $\text{Tr}(T^{(z)})$, it is straightforward to check that 
$\text{Tr}(T^{(z)})=\sum_a N_{a\bar{a}}^z \,\theta_a$.
For $\text{Tr}(S^{(z)})$, we have
\be\label{TrS}
\small
\begin{split}
\text{Tr}[S^{(z)}]=\sum_{a,\mu}S_{a\mu,a\mu}^{(z)}=
\sum_{a,\mu}
\frac{1}{\mathcal{D}}
\cdot 
\frac{1}{\sqrt{d_z}}\cdot
\begin{tikzpicture}[baseline={(current bounding box.center)}]
\draw (20/1.5pt,0pt) circle (20/1.5pt);
\draw[line width=6/1.5pt, draw=white] (0pt,0pt) circle (20/1.5pt);
\draw (0pt,0pt) circle (20/1.5pt);
\draw[line width=6/1.5pt, draw=white]  (0pt,0pt) arc (-180:-270:20/1.5pt);
\draw (0pt,0pt) arc (-180:-270:20/1.5pt);
\node at (-15/1.5+28/1.5pt,0pt){$a$};
\node at (5/1.5+28/1.5pt,0pt){$a$};
\node at (-5/1.5pt,-25/1.5pt){\small$\mu$};
\node at (22/1.5+1pt,-25/1.5pt){\small$\mu$};
\draw[>=stealth]  (20/1.5pt,-20/1.5pt)..controls (15/1.5pt,-30/1.5pt) and (5/1.5pt,-30/1.5pt)..(0/1.5pt,-20/1.5pt);
\node at (10/1.5pt,-35/1.5pt){$z$};
\draw [>=stealth,->] (20/1.5pt,0.5pt)--(20/1.5pt,0.51pt);
\draw [>=stealth,->] (40/1.5pt,0.5pt)--(40/1.5pt,0.51pt);
\draw [>=stealth,->] (8/1.5pt,-18.2pt)--(8/1.5-0.1pt,-18.2pt);
\end{tikzpicture}
\small
\end{split}
\ee
By using the following relation (see also Eq.\eqref{omega_loop_2} in the appendix):
\be\label{SimplifySz_maintext}
\small
\sum_x S_{0z}S^{\ast}_{zx}\,
\begin{tikzpicture}[baseline={(current bounding box.center)}]
\draw (-10pt,0pt) arc (180:0:10pt);
\draw [>=stealth,->](-10pt,0pt) arc (-180:0:10pt);
\draw[line width=4pt, draw=white] (-4pt,0pt)--(-4pt,18pt);
\draw [>=stealth,->](-4pt,-7pt)--(-4pt,5pt);\draw(-4pt,5pt)--(-4pt,18pt);
\draw (-4pt,-13pt)--(-4pt,-20pt);
\draw[line width=4pt, draw=white] (4pt,0pt)--(4pt,18pt);
\draw [>=stealth,->](4pt,-7pt)--(4pt,5pt); \draw(4pt,5pt)--(4pt,18pt);
\draw (4pt,-13pt)--(4pt,-20pt);
\node at (15pt,0pt){$x$};
\node at (-4pt,-25pt){${a}$};
\node at (-4pt,23pt){${a}$};
\node at (4pt,-25pt){${b}$};
\node at (4pt,23pt){${b}$};
\end{tikzpicture}
\,
=
\,
\sum_{\mu}\sqrt{\frac{d_z}{d_a d_b}}
\begin{tikzpicture}[baseline={(current bounding box.center)}]
\draw [>=stealth,->](-10pt,-18pt)--(-5pt,-13pt);\draw(-5pt,-13pt)--(0pt,-8pt);
\draw [>=stealth,->](10pt,-18pt)--(5pt,-13pt);\draw(5pt,-13pt)--(0pt,-8pt);
\draw [>=stealth,->](0pt,-8pt)--(0pt,2pt); \draw(0pt,2pt)--(0pt,8pt);
\draw [>=stealth,->](0pt,8pt)--(-7pt,15pt);\draw(-7pt,15pt)--(-10pt,18pt);
\draw [>=stealth,->](0pt,8pt)--(7pt,15pt);\draw(7pt,15pt)--(10pt,18pt);
\node at (-13pt,-18pt){$a$};
\node at (-13pt,18pt){$a$};
\node at (13pt,-18pt){$b$};
\node at (13pt,18pt){$b$};
\node[right] at (0pt,6pt){$\mu$};
\node[right] at (-10pt,0pt){$z$};
\node[right] at (0pt,-6pt){$\mu$};
\end{tikzpicture}
\ee
Eq.\eqref{TrS} can be rewritten as

\be\label{TrS2}
\begin{split}
\text{Tr}[S^{(z)}]
=&\frac{1}{\mathcal{D}} \sum_{a,x} \frac{d_a}{d_z}\cdot S_{0\bar{z}}\cdot S_{\bar{z}x}^{\ast}
\begin{small}
\begin{tikzpicture}[baseline={(current bounding box.center)}]
\draw (20/1.5pt,0pt) circle (20/1.5pt);
\draw[line width=6/1.5pt, draw=white] (0pt,0pt) circle (20/1.5pt);
\draw (0pt,0pt) circle (20/1.5pt);
\draw[line width=6/1.5pt, draw=white]  (0pt,0pt) arc (-180:-270:20/1.5pt);
\draw (0pt,0pt) arc (-180:-270:20/1.5pt);
\node at (-15/1.5+28/1.5pt,0pt){$a$};
\node at (5/1.5+28/1.5pt,0pt){$a$};
\draw[>=stealth]  (20/1.5pt,-20/1.5pt)..controls (20/1.5pt,-48/1.5pt) and (0pt,-48/1.5pt)..(0pt,-20/1.5pt);
\draw[>=stealth]  (27/1.5pt,-19/1.5pt)..controls (27/1.5pt,-57/1.5pt) and (-7/1.5pt,-57/1.5pt)..(-7/1.5pt,-19/1.5pt);
\draw[line width=2/1.5pt, draw=white]  (26.6/1.5pt,-19/1.5pt) arc (-72:-91:20/1.5pt);
\draw[line width=2/1.5pt, draw=white]  (-6.6/1.5pt,-19/1.5pt) arc (-108:-89:20/1.5pt);
\draw[>=stealth, line width=3.5/1.5pt,draw=white]  (10/1.5pt,-40/1.5pt)..controls (16/1.5pt,-25/1.5pt) and (16/1.5pt,-64/1.5pt)..(10/1.5pt,-49/1.5pt);
\draw[>=stealth]  (10/1.5pt,-40/1.5pt)..controls (16/1.5pt,-25/1.5pt) and (16/1.5pt,-64/1.5pt)..(10/1.5pt,-49/1.5pt);
\node at (12/1.5pt,-32/1.5pt){$x$};
\draw [>=stealth,->] (20/1.5pt,0.5pt)--(20/1.5pt,0.51pt);
\draw [>=stealth,->] (40/1.5pt,0.5pt)--(40/1.5pt,0.51pt);
\draw [>=stealth,->] (9.6pt,-30.19pt)--(9.6pt,-30.2pt);
\end{tikzpicture}
\end{small}
\end{split}
\ee
which is related to the whitehead link as studied in Ref.\onlinecite{Wang1805}.
Then for $\text{Tr}(w_1^{(z)})$, since both $\theta^{(z)}$ and $T^{(z)}$ are diagonal matrices, 
it is straightforward to check that

\be
\small
\text{Tr}(w_1^{(z)})=
\frac{1}{\mathcal{D}} \sum_{a,x} \theta_z^2 \cdot \theta_a^7\cdot  \frac{d_a}{d_z}\cdot S_{0\bar{z}}\cdot S_{\bar{z}x}^{\ast}
\begin{tikzpicture}[baseline={(current bounding box.center)}]
\draw (20/1.5pt,0pt) circle (20/1.5pt);
\draw[line width=6/1.5pt, draw=white] (0pt,0pt) circle (20/1.5pt);
\draw (0pt,0pt) circle (20/1.5pt);
\draw[line width=6/1.5pt, draw=white]  (0pt,0pt) arc (-180:-270:20/1.5pt);
\draw (0pt,0pt) arc (-180:-270:20/1.5pt);
\node at (-15/1.5+28/1.5pt,0pt){$a$};
\node at (5/1.5+28/1.5pt,0pt){$a$};
\draw[>=stealth]  (20/1.5pt,-20/1.5pt)..controls (20/1.5pt,-48/1.5pt) and (0pt,-48/1.5pt)..(0pt,-20/1.5pt);
\draw[>=stealth]  (27/1.5pt,-19/1.5pt)..controls (27/1.5pt,-57/1.5pt) and (-7/1.5pt,-57/1.5pt)..(-7/1.5pt,-19/1.5pt);
\draw[line width=2/1.5pt, draw=white]  (26.6/1.5pt,-19/1.5pt) arc (-72:-91:20/1.5pt);
\draw[line width=2/1.5pt, draw=white]  (-6.6/1.5pt,-19/1.5pt) arc (-108:-89:20/1.5pt);
\draw[>=stealth, line width=3.5/1.5pt,draw=white]  (10/1.5pt,-40/1.5pt)..controls (16/1.5pt,-25/1.5pt) and (16/1.5pt,-64/1.5pt)..(10/1.5pt,-49/1.5pt);
\draw[>=stealth]  (10/1.5pt,-40/1.5pt)..controls (16/1.5pt,-25/1.5pt) and (16/1.5pt,-64/1.5pt)..(10/1.5pt,-49/1.5pt);
\node at (12/1.5pt,-32/1.5pt){$x$};
\draw [>=stealth,->] (20/1.5pt,0.5pt)--(20/1.5pt,0.51pt);
\draw [>=stealth,->] (40/1.5pt,0.5pt)--(40/1.5pt,0.51pt);
\draw [>=stealth,->] (9.6pt,-30.19pt)--(9.6pt,-30.2pt);
\end{tikzpicture}.
\ee

As the second example, now let us check $W_{\Sigma_{1,1}}^{z}:=\text{Tr}(w_2^{(z)})$. 
First, let us look at how $\omega_2^{(z)}$ acts on the basis as follows:
\be
\small
\begin{split}
&(\theta^{(z)})^3\big(T^{(z)}\big)^3S^{(z)}
\big(T^{(z)}\big)^3S^{(z)}
\begin{tikzpicture}[baseline={(current bounding box.center)}]
\draw [>=stealth,->] (0pt,-15pt)--(0pt,-6pt);\draw (0pt,-6pt)--(0pt,0pt);
\draw [>=stealth,->] (0pt,0pt)--(-6pt,8pt); \draw(-6pt,8pt)--(-7.5pt,10pt);
\draw [>=stealth,->] (0pt,0pt)--(6pt,8pt); \draw(6pt,8pt)--(7.5pt,10pt);
\node at (-8pt,15pt){$b$};
\node at (8pt,15pt){$z$};
\node at (0pt,-20pt){$b$};
\node at (6pt,-2pt){\small$\mu$};
\end{tikzpicture}\\
=&(\theta^{(z)})^3\big(T^{(z)}\big)^3S^{(z)}\,\,
\frac{1}{\mathcal{D}} \sum_a d_a (\theta_a)^3\,\,
\begin{tikzpicture}[baseline={(current bounding box.center)}]
\draw (0pt,0pt) circle (10pt);
\draw[line width=4pt, draw=white] (-5pt,-20pt)--(-5pt,5pt);
\draw [>=stealth,->](-5pt,-18pt)--(-5pt,0pt); \draw(-5pt,0pt)--(-5pt,5pt);
\draw (-5pt, 12pt)--(-5pt,18pt);
\draw [>=stealth,->](5pt, 8.66 pt)--(10pt,18pt);
\node at (-5pt,-23pt){$a$};
\node at (-5pt,23pt){$a$};
\node at (15pt,0pt){$b$};
\node at (10pt,23pt){$z$};
\draw [>=stealth,->] (10pt,0.5pt)--(10pt,0.51pt);
\node at (13pt,10pt){\small$\mu$};
\end{tikzpicture}\\
=&\frac{1}{\mathcal{D}^2}\sum_{a,x}d_ad_x(\theta_z)^3(\theta_x)^3(\theta_a)^3
\begin{tikzpicture}[baseline={(current bounding box.center)}]
\draw (0pt,0pt) circle (10pt);
\draw[line width=4pt, draw=white] (-5pt,-18pt)--(-5pt,6pt);
\draw [>=stealth,->](-5pt,-18pt)--(-5pt,0pt);\draw(-5pt,0pt)--(-5pt,6pt);
\draw (-5pt,12pt)--(-5pt,18pt);
\draw [>=stealth,->](15pt, 8.66 pt)--(20pt,18pt);
\draw (10pt,0pt) circle (10pt);
\draw [line width=4pt, draw=white](10pt,10pt) arc (90:180:10pt);
\draw (10pt,10pt) arc (90:180:10pt);
\draw [line width=4pt, draw=white](10pt,0pt) arc (0:-90:10pt);
\draw (10pt,0pt) arc (0:-90:10pt);
\node at (-5pt,-23pt){$x$};
\node at (-5pt,23pt){$x$};
\node at (25pt,0pt){$b$};
\node at (20pt,23pt){$z$};
\node at (23pt,10pt){\small$\mu$};
\node at (15pt,0pt){$a$};
\draw [>=stealth,->] (20pt,0.5pt)--(20pt,0.51pt);
\draw [>=stealth,->] (10pt,0.5pt)--(10pt,0.51pt);
\end{tikzpicture}.
\end{split}
\ee
Including the normalization of basis vectors, one can obtain
\be
\small
\begin{split}
&\text{Tr}(w_2^{(z)})
=
\frac{1}{\mathcal{D}^2}\sum_{a,b}
\frac{d_a}{\sqrt{d_z}} (\theta_z)^3 (\theta_a)^3(\theta_b)^3\sum_{\mu}
\begin{tikzpicture}[baseline={(current bounding box.center)}]
\small
\draw (20/1.5pt,0pt) circle (20/1.5pt);
\draw[line width=6/1.5pt, draw=white] (-5/1.5pt,0pt) circle (20/1.5pt);
\draw (-5/1.5pt,0pt) circle (20/1.5pt);
\draw[line width=6/1.5pt, draw=white]  (0pt,0pt) arc (-180:-270:20/1.5pt);
\draw (0pt,0pt) arc (-180:-270:20/1.5pt);
\draw[line width=6/1.5pt, draw=white] (-30/1.5pt,0pt) circle (20/1.5pt);
\draw (-30/1.5pt,0pt) circle (20/1.5pt);
\draw[line width=6/1.5pt, draw=white]  (-25/1.5pt,0pt) arc (-180:-270:20/1.5pt);
\draw (-25/1.5pt,0pt) arc (-180:-270:20/1.5pt);
\node at (15pt,18pt){$b$};
\node at (-2/1.5pt,25/1.5pt){$a$};
\node at (-17pt,18pt){$b$};
\node at (25/1.5pt,-25/1.5pt){$\mu$};
\node at (-35/1.5pt,-25/1.5pt){$\mu$};
\draw[>=stealth]  (-30/1.5pt,-20/1.5pt)..controls (-20/1.5pt,-40/1.5pt) and (10/1.5pt,-40/1.5pt)..(20/1.5pt,-20/1.5pt);
\node at (-5/1.5pt,-42/1.5pt){$z$};
\draw [>=stealth,->] (-10/1.5pt,0.5pt)--(-10/1.5pt,0.51pt);
\draw [>=stealth,->] (15/1.5pt,0.5pt)--(15/1.5pt,0.51pt);
\draw [>=stealth,->] (40/1.5pt,0.5pt)--(40/1.5pt,0.51pt);
\draw [>=stealth,->] (-5pt,-23.3pt)--(-5-0.1pt,-23.3pt);
\end{tikzpicture}
\end{split}
\ee
which, based on Eq.\eqref{SimplifySz_maintext}, can be further simplified as
\be
\begin{split}
\text{Tr} \left(w_2^{(z)}\right)
=&
\frac{1}{\mathcal{D}^2}\sum_{a,b}
\frac{d_a d_b}{d_z} (\theta_z)^3 (\theta_a)^3(\theta_b)^3\\
&\times \sum_{x}S_{0\bar{z}}\cdot S_{\bar{z}x}^{\ast}
\begin{small}
\begin{tikzpicture}[baseline={(current bounding box.center)}]
\draw (20/1.5 pt,0pt) circle (20/1.5pt);
\draw[line width=6/1.5pt, draw=white] (-5/1.5pt,0pt) circle (20/1.5pt);
\draw (-5/1.5pt,0pt) circle (20/1.5pt);
\draw[line width=6/1.5pt, draw=white]  (0pt,0pt) arc (-180:-270:20/1.5pt);
\draw (0pt,0pt) arc (-180:-270:20/1.5pt);
\draw[line width=6/1.5pt, draw=white] (-30/1.5pt,0pt) circle (20/1.5pt);
\draw (-30/1.5pt,0pt) circle (20/1.5pt);
\draw[line width=6/1.5pt, draw=white]  (-25/1.5pt,0pt) arc (-180:-270:20/1.5pt);
\draw (-25/1.5pt,0pt) arc (-180:-270:20/1.5pt);
\node at (15pt,18pt){$b$};
\node at (-1pt,25/1.5pt){$a$};
\node at (-17pt,18pt){$b$};
\draw[>=stealth]  (-30/1.5pt,-20/1.5pt)..controls (-20/1.5pt,-40/1.5pt) and (10/1.5pt,-40/1.5pt)..(20/1.5pt,-20/1.5pt);
\draw[>=stealth]  (-37.5/1.5pt,-18/1.5pt)..controls (-20/1.5pt,-47/1.5pt) and (10/1.5pt,-47/1.5pt)..(28/1.5pt,-18/1.5pt);
\draw[line width=1/1.5pt, draw=white]  (26.6/1.5pt,-19/1.5pt) arc (-72:-91:20/1.5pt);
\draw[line width=1/1.5pt, draw=white]  (-36.2/1.5pt,-19/1.5pt) arc (-108:-89:20/1.5pt);
\draw[>=stealth, line width=3.5/1.5pt,draw=white]   (-5/1.5pt,-33/1.5pt)..controls (1/1.5pt,-18/1.5pt) and (1/1.5pt,-57/1.5pt)..(-5/1.5pt,-42/1.5pt);
\draw[>=stealth]  (-5/1.5pt,-33/1.5pt)..controls (1/1.5pt,-18/1.5pt) and (1/1.5pt,-57/1.5pt)..(-5/1.5pt,-42/1.5pt);
\node at (-5/1.5pt,-16.5pt){$x$};
\draw [>=stealth,->] (-10/1.5pt,0.5pt)--(-10/1.5pt,0.51pt);
\draw [>=stealth,->] (15/1.5pt,0.5pt)--(15/1.5pt,0.51pt);
\draw [>=stealth,->] (40/1.5pt,0.5pt)--(40/1.5pt,0.51pt);
\draw [>=stealth,->] (-0.45pt,-26.19pt)--(-0.45pt,-26.2pt);
\end{tikzpicture}.
\end{small}
\end{split}
\ee
Apparently, $W_{\Sigma_{1,1}}^{z}:=\text{Tr}(w_2^{(z)})$ 
can be expressed in terms of link invariants, and therefore it is a topological invariant.
Similarly, one can check that the trace of other words are all topological invariants. 
One basic reason is that the only possible gauge-dependent term comes from the tri-junction structure in \eqref{vertex_basis}.
By performing trace over the words with the rule in Eq.\eqref{SimplifySz_maintext}, it is found that this tri-junction structure 
is removed.

In the following, we present results of trace of the words for those in Eq.\eqref{Words}, as listed in tables \eqref{W1compare} $\sim$ \eqref{W7compare}.

\begin{table}[H]
\footnotesize
\centering
\begin{tabular}{cccccccccc}
$W^{(z)}_{\Sigma_{1,1}}$  &\vline     &$W_{\Sigma_{1,1}}^{I_0}$ &\vline     &$\sum_{i=1}^4 W_{\Sigma_{1,1}}^{I_i}$&\vline     &$W_{\Sigma_{1,1}}^{I_5,I_6}$&\vline     &$W_{\Sigma_{1,1}}^{A_1,A_2}$\\ \hline
$u=0$  &\vline     &9               &\vline          & $\color{gray} -4$               &\vline  &$4$               &\vline   &$22$              \\ \hline
$u=1$  &\vline       &$4\cos\frac{2\pi}{5}+5$ &\vline     &$\color{gray} -4$&\vline     &$4\cos\frac{2\pi}{5}$&\vline     &$22e^{\frac{2i\pi }{5}\times 1}$  \\ \hline
$u=2$  &\vline         &$4\cos\frac{4\pi}{5}+5$ &\vline     &$\color{gray} -4$&\vline     &$4\cos\frac{4\pi}{5}$&\vline     &$22e^{\frac{2i\pi }{5}\times 2}$ \\  \hline
$u=3$  &\vline       &$4\cos\frac{4\pi}{5}+5$  &\vline     &$\color{gray} -4$&\vline     &$4\cos\frac{4\pi}{5}$&\vline     &$22e^{\frac{2i\pi }{5}\times 3}$ \\ \hline
$u=4$  &\vline          &$4\cos\frac{2\pi}{5}+5$   &\vline     &$\color{gray} -4$&\vline     &$4\cos\frac{2\pi}{5}$&\vline     &$22e^{\frac{2i\pi }{5}\times 4}$ \\ \hline
\end{tabular}
\caption{$W_{\Sigma_{1,1}}^{(z)}$ with $w_1^{(z)}$ in \eqref{Words}.
}
\label{W1compare}
\end{table}

\begin{table}[H]
\footnotesize
\centering
\begin{tabular}{cccccccccc}
$W_{\Sigma_{1,1}}^{(z)}$  &\vline     &$W_{\Sigma_{1,1}}^{I_0}$ &\vline     &$\sum_{i=1}^4 W_{\Sigma_{1,1}}^{I_i}$&\vline     &$W_{\Sigma_{1,1}}^{I_5,I_6}$&\vline     &$W_{\Sigma_{1,1}}^{A_1,A_2}$\\ \hline
$u=0$  &\vline     &9               &\vline          &$\color{gray} -4$               &\vline  &$4$               &\vline   &$22$             \\ \hline
$u=1$  &\vline       &$4\cos\frac{4\pi}{5}+5$ &\vline     &$\color{gray} -4$&\vline     &$4\cos\frac{4\pi}{5}$&\vline     &$22e^{\frac{2i\pi }{5}\times 3}$  \\ \hline
$u=2$  &\vline         &$4\cos\frac{2\pi}{5}+5$ &\vline     &$\color{gray} -4$&\vline     &$4\cos\frac{2\pi}{5}$&\vline     &$22e^{\frac{2i\pi }{5}\times 1}$ \\  \hline
$u=3$  &\vline       &$4\cos\frac{2\pi}{5}+5$  &\vline     &$\color{gray} -4$&\vline     &$4\cos\frac{2\pi}{5}$&\vline     &$22e^{\frac{2i\pi }{5}\times 4}$ \\ \hline
$u=4$  &\vline          &$4\cos\frac{4\pi}{5}+5$   &\vline     &$\color{gray} -4$&\vline     &$4\cos\frac{4\pi}{5}$&\vline     &$22e^{\frac{2i\pi }{5}\times 2}$ \\ \hline
\end{tabular}
\caption{$W_{\Sigma_{1,1}}^{(z)}$ with  $w_2^{(z)}$ in \eqref{Words}.
}
\label{W2compare}
\end{table}

\begin{table}[H]
\footnotesize
\centering
\begin{tabular}{cccccccccc}
$W_{\Sigma_{1,1}}^{(z)}$  &\vline     &$W_{\Sigma_{1,1}}^{I_0}$ &\vline     &$\sum_{i=1}^4 W_{\Sigma_{1,1}}^{I_i}$&\vline     &$W_{\Sigma_{1,1}}^{I_5,I_6}$&\vline     &$W_{\Sigma_{1,1}}^{A_1,A_2}$\\ \hline
$u=0$  &\vline     &25               &\vline          &\color{gray} $0$               &\vline  &$24$              &\vline   &$66$               \\ \hline
$u=1$  &\vline       &$-5$ &\vline     &\color{gray} $0$&\vline     &$-6$ &\vline     &$66\, e^{\frac{2i\pi }{5}\times 2}$  \\ \hline
$u=2$  &\vline         &$-5$ &\vline     &\color{gray} $0$&\vline     &$-6$&\vline     &$66\, e^{\frac{2i\pi }{5}\times 4}$ \\  \hline
$u=3$  &\vline       &$-5$  &\vline     &\color{gray} $0$&\vline     &$-6$&\vline     &$66\, e^{\frac{2i\pi }{5}\times 1}$ \\ \hline
$u=4$  &\vline          &$-5$   &\vline     &\color{gray} $0$&\vline     &$-6$&\vline     &$66\, e^{\frac{2i\pi }{5}\times 3}$ \\ \hline
\end{tabular}
\caption{$W_{\Sigma_{1,1}}^{(z)}$ with $w_3^{(z)}$ in \eqref{Words}.
}
\label{W3compare}
\end{table}

\begin{table}[H]
\footnotesize
\centering
\begin{tabular}{cccccccccc}
$W_{\Sigma_{1,1}}^{(z)}$  &\vline     &$W_{\Sigma_{1,1}}^{I_0}$ &\vline     &$\sum_{i=1}^4 W_{\Sigma_{1,1}}^{I_i}$&\vline     &$W_{\Sigma_{1,1}}^{I_5,I_6}$&\vline     &$W_{\Sigma_{1,1}}^{A_1,A_2}$\\ \hline
$u=0$  &\vline     &25               &\vline          &\color{gray} $0$               &\vline  &$24$              &\vline   &$44$               \\ \hline
$u=1$  &\vline       &$5$ &\vline     &\color{gray} $0$&\vline     &$4$ &\vline     &$44\, e^{\frac{2i\pi }{5}\times 3}$  \\ \hline
$u=2$  &\vline         &$5$ &\vline     &\color{gray} $0$&\vline     &$4$&\vline     &$44\, e^{\frac{2i\pi }{5}\times 1}$ \\  \hline
$u=3$  &\vline       &$5$  &\vline     &\color{gray} $0$&\vline     &$4$&\vline     &$44\, e^{\frac{2i\pi }{5}\times 4}$ \\ \hline
$u=4$  &\vline          &$5$   &\vline     &\color{gray} $0$&\vline     &$4$&\vline     &$44\, e^{\frac{2i\pi }{5}\times 2}$ \\ \hline
\end{tabular}
\caption{$W_{\Sigma_{1,1}}^{(z)}$ with $w_4^{(z)}$ in \eqref{Words}.
}
\label{W4compare}
\end{table}

\begin{table}[H]
\footnotesize
\centering
\begin{tabular}{cccccccccc}
$W_{\Sigma_{1,1}}^{(z)}$  &\vline     &$W_{\Sigma_{1,1}}^{I_0}$ &\vline     &$\sum_{i=1}^4 W_{\Sigma_{1,1}}^{I_i}$&\vline     &$W_{\Sigma_{1,1}}^{I_5,I_6}$&\vline     &$W_{\Sigma_{1,1}}^{A_1,A_2}$\\ \hline
$u=0$  &\vline     &5               &\vline          &\color{gray} $0$              &\vline  &$4$               &\vline   &$22$            \\ \hline
$u=1$  &\vline       &$4\cos\frac{4\pi}{5}+1$  &\vline     &\color{gray} $0$  &\vline     &$4\cos\frac{4\pi}{5}$ &\vline     &$22\, e^{\frac{2i\pi }{5}\times 2}$  \\ \hline
$u=2$  &\vline         &$4\cos\frac{2\pi}{5}+1$  &\vline     &\color{gray} $0$  &\vline     &$4\cos\frac{2\pi}{5}$&\vline     &$22\, e^{\frac{2i\pi }{5}\times 4}$ \\  \hline
$u=3$  &\vline       &$4\cos\frac{2\pi}{5}+1$   &\vline     &\color{gray} $0$  &\vline     &$4\cos\frac{2\pi}{5}$&\vline     &$22\, e^{\frac{2i\pi }{5}\times 1}$ \\ \hline
$u=4$  &\vline          &$4\cos\frac{4\pi}{5}+1$   &\vline     &\color{gray} $0$  &\vline     &$4\cos\frac{4\pi}{5}$&\vline     &$22\, e^{\frac{2i\pi }{5}\times 3}$ \\ \hline
\end{tabular}
\caption{
$W_{\Sigma_{1,1}}^{(z)}$ with $w_5^{(z)}$ in \eqref{Words}.
}
\label{W5compare}
\end{table}

\begin{table}[H]
\footnotesize
\centering
\begin{tabular}{cccccccccc}
$W^{(z)}_{\Sigma_{1,1}}$  &\vline     &$W_{\Sigma_{1,1}}^{I_0}$ &\vline     &$\sum_{i=1}^4 W_{\Sigma_{1,1}}^{I_i}$&\vline     &$W_{\Sigma_{1,1}}^{I_5,I_6}$&\vline     &$W_{\Sigma_{1,1}}^{A_1,A_2}$\\ \hline
$u=0$  &\vline     &9               &\vline          &\color{gray}$-4$              &\vline  &$4$              &\vline   &$22$               \\ \hline
$u=1$  &\vline       &$4\cos\frac{2\pi}{5}+5$  &\vline     &\color{gray}$-4$&\vline     &$4\cos\frac{2\pi}{5}$ &\vline     &$22\, e^{\frac{2i\pi }{5}\times 4}$  \\ \hline
$u=2$  &\vline         &$4\cos\frac{4\pi}{5}+5$  &\vline     &\color{gray}$-4$&\vline     &$4\cos\frac{4\pi}{5}$&\vline     &$22\, e^{\frac{2i\pi }{5}\times 3}$ \\  \hline
$u=3$  &\vline       &$4\cos\frac{4\pi}{5}+5$   &\vline     &\color{gray}$-4$&\vline     &$4\cos\frac{4\pi}{5}$&\vline     &$22\, e^{\frac{2i\pi }{5}\times 2}$ \\ \hline
$u=4$  &\vline          &$4\cos\frac{2\pi}{5}+5$   &\vline     &\color{gray}$-4$&\vline     &$4\cos\frac{2\pi}{5}$&\vline     &$22\, e^{\frac{2i\pi }{5}\times 1}$ \\ \hline
\end{tabular}
\caption{
$W_{\Sigma_{1,1}}^{(z)}$ with $w_6{(z)}$ in \eqref{Words}.
}
\label{W6compare}
\end{table}

\begin{table}[H]
\footnotesize
\centering
\begin{tabular}{cccccccccc}
$W^{(z)}_{\Sigma_{1,1}}$  &\vline     &$W_{\Sigma_{1,1}}^{I_0}$ &\vline     &$\sum_{i=1}^4 W_{\Sigma_{1,1}}^{I_i}$&\vline     &$W_{\Sigma_{1,1}}^{I_5,I_6}$&\vline     &$W_{\Sigma_{1,1}}^{A_1,A_2}$\\ \hline
$u=0$  &\vline     &9               &\vline          & \color{gray}$-4$             &\vline  &$4$               &\vline   &$22$               \\ \hline
$u=1$  &\vline       &$4\cos\frac{4\pi}{5}+5$  &\vline     &\color{gray}$-4$&\vline     &$4\cos\frac{4\pi}{5}$ &\vline     &$22\, e^{\frac{2i\pi }{5}\times 2}$  \\ \hline
$u=2$  &\vline         &$4\cos\frac{2\pi}{5}+5$  &\vline     &\color{gray}$-4$&\vline     &$4\cos\frac{2\pi}{5}$&\vline     &$22\, e^{\frac{2i\pi }{5}\times 4}$ \\  \hline
$u=3$  &\vline       &$4\cos\frac{2\pi}{5}+5$   &\vline     &\color{gray}$-4$&\vline     &$4\cos\frac{2\pi}{5}$&\vline     &$22\, e^{\frac{2i\pi }{5}\times 1}$ \\ \hline
$u=4$  &\vline          &$4\cos\frac{4\pi}{5}+5$   &\vline     &\color{gray}$-4$ &\vline     &$4\cos\frac{4\pi}{5}$&\vline     &$22\, e^{\frac{2i\pi }{5}\times 3}$ \\ \hline
\end{tabular}
\caption{
$W_{\Sigma_{1,1}}^{(z)}$ with $w_7^{(z)}$ in \eqref{Words}.
}\label{W7compare}
\end{table}

Let us explain the meaning of data in Table \ref{W1compare}. 
Taking $u=1$ for example, $W_{\Sigma_{1,1}}^{(z)}$ is defined in Eq.\eqref{Wz_PuncturedTorus}.
Then we have $W^{I_0}_{\Sigma_{1,1}}=4\cos\frac{2\pi}{5}+5$, $\sum_{i=1}^4 W^{I_i}_{\Sigma_{1,1}}=W^{I_1}_{\Sigma_{1,1}}+W^{I_2}_{\Sigma_{1,1}}+W^{I_3}_{\Sigma_{1,1}}+W^{I_4}_{\Sigma_{1,1}}=-4$,
$W^{I_5}_{\Sigma_{1,1}}=W^{I_6}_{\Sigma_{1,1}}=4\cos\frac{2\pi}{5}$, $W^{A_1}_{\Sigma_{1,1}}=\sum_{i=0}^{10}W^{A_{1,i}}_{\Sigma_{1,1}}=22e^{\frac{2i\pi }{5}\times 1}$, and
$W^{A_2}_{\Sigma_{1,1}}=\sum_{i=0}^{10}W^{A_{2,i}}_{\Sigma_{1,1}}=22e^{\frac{2i\pi }{5}\times 1}$. The same definition is used in the other tables in this subsection.

The data in gray indicate they are 
not obtained in the quasi-particle basis calculation, but are obtained by
comparing with the lattice theory calculation in Sec.\ref{Sec: Lattice}.
One can find that they are independent of $u$. 
The reason is that for $z=I_1,I_2,I_3,I_4$, 
only type-$I$ and type-$A$ anyons participate in the definition of punctured $S$ matrix,
and therefore the 
results are independent of the 3-cocyles $\omega^u$ and $u$.
(Recall that only type-$B$ anyons carry the information of the 3-cocycle $\omega$.
See also the discussion at the end of Sec.\ref{Sec: punctured S and T}).
For this reason, we did not solve for $S^{z}$ with $z=I_1,I_2,I_3,I_4$ explicitly in the quasiparticle 
basis. Nevertheless, they are calculated in the lattice gauge theory approach in Sec.\ref{Sec: Lattice}.

From the results in tables \eqref{W1compare} $\sim$ \eqref{W7compare}, one can find that
for each choice of word, there are at most three sets of $W_{\Sigma_{1,1}}^{I_0}$ for
(i) $\mathcal{C}_{u=0}$, (ii) $\mathcal{C}_{u=1}$ and $\mathcal{C}_{u=4}$, 
and (iii) $\mathcal{C}_{u=2}$ and $\mathcal{C}_{u=3}$, respecively. 
Recall that $W_{\Sigma_{1,1}}^{I_0}$ are actually obtained from the modular $S$ and $T$ matrices,
this agrees with the conclusion that there are only three sets of modular data.

The pieces that can distinguish those five different categories are $W_{\Sigma_{1,1}}^{A_1}$ and $W_{\Sigma_{1,1}}^{A_2}$. 
Remarkably, the difference for the five inequivalent categories 
lies simply in the phase factor $e^{\frac{2i\pi}{5}n}$, with $n=0,1,2,3,4$.

There are some detailed feature on the phase factor $e^{\frac{2i\pi}{5}n}$
for different categories. For $u=0$ where the 3-cocycle is trivial, \textit{i.e.}, $\omega=1$,
the phase factor is always $1$. The phase factor for $u=1$ ($u=2$) are always conjugate to 
that for $u=4$ ($u=3$). An interesting observation is that by replacing $e^{\frac{2i\pi}{5}n}$ with 
$\cos\frac{2n \pi}{5}$, the information becomes degenerate. There are 3 distinct data 
instead of 5. Note that $\cos\frac{2n \pi}{5}$ appear in $W_{\Sigma_{1,1}}^{I_0}$, 
which is contributed by the modular data. From this point of view, 
the reason why the modular data cannot distinguish different categories can be
intuitively understood as the degeneracy of information.

A detailed comparison of the topological invariants with the lattice 
gauge theory approach can be found in Appendix \ref{Lattice_puncturedTorus_appendix}.

\subsubsection{Genus two}

Now we study the topological invariants defined in the Hilbert space $\mathcal{H}(\Sigma_{2,0})$ 
of degenerate ground states on genus-two surface. 
A genus-2 manifold can be obtained by gluing two punctured tori along their punctures, 
and so does the genus-2 groundstate basis.
As shown in the following, we glue two punctured-torus basis vectors by 
identifying the anyon charge at the puncture, 
\textit{i.e.}, $z'=\bar{z}$ (Note that for an anyon $z$, there is a unique $z'$ such that $z'=\bar{z}$):

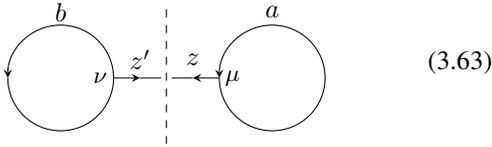
\begin{eqnarray}\label{Glue_BasisI}
\begin{tikzpicture}[baseline={(current bounding box.center)}]
\draw[>=stealth,<-] (0-10pt,0pt) arc (180:-180:20pt) ;
\draw[>=stealth,<-] (60+10pt,0pt) arc (180:-180:20pt) ;
\draw [>=stealth,->] (70pt,0pt)--(60pt,0pt); \draw (60pt,0pt)--(52pt,0pt);
\draw [>=stealth,->] (30pt,0pt)--(40pt,0pt); \draw (40pt,0pt)--(48pt, 0pt);
\node at (10pt, 25pt){$b$};
\node at (90pt, 25pt){$a$};
\node at (60pt, 7pt){$z$};
\node at (40pt, 7pt){$z'$};
\node at (25pt, 0pt){\small $\nu$};
\node at (75pt, 0pt){\small $\mu$};


\draw [dashed] (50pt, -25pt)--(50pt, 28pt);

\end{tikzpicture}
\end{eqnarray}
Then we obtain the genus-2 basis in \eqref{BasisI}.
Now we define the words 
\be
w_{\Sigma_{2,0}}^{(z)}:= I_L^{(\bar{z})}\otimes w_R^{(z)},
\ee
where $I_L^{(\bar{z})}$ is the identity matrix acting on the basis of the \textit{left} 
punctured torus in \eqref{Glue_BasisI} with fixed $z'=\bar{z}$, 
and $w_R^{(z)}$ has the same expression as \eqref{Wz} and acts on the basis of the \textit{right} punctured torus.
Then we define the following topological invariants
\be\label{W_g2_1}
W_{\Sigma_{2,0}}=\text{Tr} \left(\oplus_z w_{\Sigma_{2,0}}^{(z)}\right)=\text{Tr} \left[ \oplus_z ( I_L^{(\bar{z})}\otimes W_R^{(z)})\right],
\ee
where the `Tr' is over the Hilbert space of the degenerate ground states, \textit{i.e.}, $\mathcal{H}(\Sigma_{2,0})$.
Eq.\eqref{W_g2_1} can be further rewritten as
\be\label{W_g2}
W_{\Sigma_{2,0}}=\sum_z \text{dim}\, \mathcal{H}(\Sigma_{1,1},  \bar{z})\times W^{(z)}_{\Sigma_{1,1}},
\ee
where $W^{(z)}_{\Sigma_{1,1}}$ is the same as that in Eq.\eqref{Wz_PuncturedTorus},
and $\text{dim}\, \mathcal{H}(\Sigma_{1,1}, \bar{z})$ is the dimension of Hilbert space for a punctured torus
with an anyon charge $\bar{z}$ at the puncture. Based on Eq.\eqref{HibertSpace_Punctures_introduction}, we have
\be
\text{dim}\, \mathcal{H}(\Sigma_{1,1}, \bar{z})=\sum_{p}\frac{S_{\bar{z} p} }{S_{0p}}=\sum_p N_{p\bar{p}}^z.
\ee
For the simplest counterexamples with $G=\mathbb{Z}_{11}\rtimes \mathbb{Z}_5$ in MS MTCs, it is found that
$\text{dim}\, \mathcal{H}(\Sigma_{1,1}, \bar{z})$ is 49 for $z=I_0$, and $24$ for $z=I_i$ with $i=1\sim 6$ and all type-$A$
anyons. To track the fine structures in $W_{\Sigma_{2,0}}$, it is helpful to split $W_{\Sigma_{2,0}}$ into three pieces as follows
\be
W_{\Sigma_{2,0}}=W_{\Sigma_{2,0}}^I+W_{\Sigma_{2,0}}^{A_1}+W_{\Sigma_{2,0}}^{A_2},
\ee
where the upper indices $I$, $A_1$, and $A_2$ indicate that the summation over $z$ in Eq.\eqref{W_g2}
is performed within type-$I$, -$A_1$, and -$A_2$ anyons.

The results are summarized in tables \eqref{W1_Genus2_QP}$\sim$\eqref{W7_Genus2_QP}.
The data in gray are again contributed by $W_{\Sigma_{2,0}}^{I_i}$, with $i=1,\cdots, 4$.
They are obtained not by the quasi-particle basis calculation, but by comparing
with the lattice gauge theory calculation in Sec.\eqref{Sec: Lattice}. One can find they 
are independent of $u$. The reason is the same as that for a punctured torus, \textit{i.e.},
for $z=I_i$ with $i=1,\cdots,4$, only type-$I$ and type-$A$ anyons contribute to the punctures $S$ matrix, 
and the information of 3-cocyle $\omega^u$ will not come in.

There are at most three different sets of $W_{\Sigma_{2,0}}^{I}$ for each $W_{\Sigma_{2,0}}$, 
and cannot distinguish different categories.
The pieces of data that can distinguish the five inequivalent
categories are $W_{\Sigma_{2,0}}^{A_1}$ and $W_{\Sigma_{2,0}}^{A_2}$, with the difference in the phase factor 
$e^{\frac{2i\pi}{5}n}$, where $n=0,\, 1,\, 2,\, 3,\,4$. Similar to the case of $W_{\Sigma_{1,1}}^{(z)}$, the phase
factor for $u=1$ ($u=2$) is conjugate to that for $u=4$ ($u=3$). 

A comparison of the topological invariants with the lattice 
gauge theory approach can be found in Sec.\ref{Sec: Genus2_lattice_TopoInv}.

\begin{table}[H]
\centering
\footnotesize
\begin{tabular}{cccccccccc}
$W_{\Sigma_{2,0}}$    &\vline &$W_{\Sigma_{2,0}}^{I}$&\vline    &$W_{\Sigma_{2,0}}^{A_1}$&\vline   &$W_{\Sigma_{2,0}}^{A_2}$    \\ \hline
$u=0$  &\vline     &$633-24\times \textcolor{gray}{4}$&\vline &$528$&\vline & $528$                 \\ \hline
$u=1$  &\vline     &$245-24\times \textcolor{gray}{4}+388\cdot \cos\frac{2\pi}{5}$&\vline &$528\cdot e^{\frac{2\pi i}{5}\times 1} $&\vline   &$528\cdot e^{\frac{2\pi i}{5}\times 1} $  \\ \hline
$u=2$  &\vline     &$245-24\times \textcolor{gray}{4}+388\cdot \cos\frac{4\pi}{5}$&\vline  &$528\cdot e^{\frac{2\pi i}{5}\times 2} $&\vline    & $528\cdot e^{\frac{2\pi i}{5}\times 2} $  \\  \hline
$u=3$  &\vline      &$245-24\times \textcolor{gray}{4}+388\cdot \cos\frac{4\pi}{5}$&\vline &$528\cdot e^{\frac{2\pi i}{5}\times 3} $&\vline  &$528 \cdot e^{\frac{2i\pi}{5}\times 3}$ \\ \hline
$u=4$  &\vline      &$245-24\times \textcolor{gray}{4}+388\cdot \cos\frac{2\pi}{5}$ &\vline  &$528\cdot e^{\frac{2\pi i}{5}\times 4} $&\vline    &$528\cdot e^{\frac{2\pi i}{5}\times 4} $      \\ \hline
\end{tabular}
\caption{$W_{\Sigma_{2,0} }$, with $w_R^{(z)}=w_1^{(z)}$ in Eq.\eqref{Words}. 
}
\label{W1_Genus2_QP}
\end{table}

\begin{table}[H]
\centering
\footnotesize
\begin{tabular}{cccccccccc}
$W_{\Sigma_{2,0}}$     &\vline &$W_{\Sigma_{2,0}}^{I}$&\vline    &$W_{\Sigma_{2,0}}^{A_1}$&\vline   &$W_{\Sigma_{2,0}}^{A_2}$    \\ \hline
$u=0$  &\vline     &$633-24\times \textcolor{gray}{4}$ &\vline &$528$&\vline &    $528$    \\ \hline
$u=1$  &\vline     &$245-24\times \textcolor{gray}{4}+388\cdot \cos\frac{4\pi}{5}$&\vline &$528 \cdot e^{\frac{2i\pi}{5}\times 3}$&\vline   &$528 \cdot e^{\frac{2i\pi}{5}\times 3}$  \\ \hline
$u=2$  &\vline     &$245-24\times \textcolor{gray}{4}+388\cdot \cos\frac{2\pi}{5}$&\vline  &$528 \cdot e^{\frac{2i\pi}{5}\times 1}$&\vline    &$528 \cdot e^{\frac{2i\pi}{5}\times 1}$ \\  \hline
$u=3$  &\vline      &$245-24\times \textcolor{gray}{4}+388\cdot \cos\frac{2\pi}{5}$&\vline &$528 \cdot e^{\frac{2i\pi}{5}\times 4}$&\vline  &$528 \cdot e^{\frac{2i\pi}{5}\times 4}$ \\ \hline
$u=4$  &\vline      &$245-24\times \textcolor{gray}{4}+388\cdot \cos\frac{4\pi}{5}$ &\vline  &$528 \cdot e^{\frac{2i\pi}{5}\times 2}$&\vline    &$528 \cdot e^{\frac{2i\pi}{5}\times 2}$   \\ \hline
\end{tabular}
\caption{$W_{\Sigma_{2,0} }$, with $w_R^{(z)}=w_2^{(z)}$ in Eq.\eqref{Words}. 
}
\label{W2_Genus2_QP}
\end{table}

\begin{table}[H]
\centering
\footnotesize
\begin{tabular}{cccccccccc}
$W_{\Sigma_{2,0}}$    &\vline &$W_{\Sigma_{2,0}}^{I}$&\vline    &$W_{\Sigma_{2,0}}^{A_1}$&\vline   &$W_{\Sigma_{2,0}}^{A_2}$    \\ \hline
$u=0$  &\vline     &$2377+\textcolor{gray}{0}$ &\vline &$1584$&\vline &        $1584$            \\ \hline
$u=1$  &\vline     &$-533+\textcolor{gray}{0}$&\vline &$1584 \cdot e^{\frac{2i\pi}{5}\times 2}$&\vline   & $1584 \cdot e^{\frac{2i\pi}{5}\times 2}$  \\ \hline
$u=2$  &\vline     &$-533+\textcolor{gray}{0}$&\vline  &$1584 \cdot e^{\frac{2i\pi}{5}\times 4}$&\vline    & $1584 \cdot e^{\frac{2i\pi}{5}\times 4}$   \\  \hline
$u=3$  &\vline      &$-533+\textcolor{gray}{0}$&\vline &$1584 \cdot e^{\frac{2i\pi}{5}\times 1}$ &\vline  &$1584 \cdot e^{\frac{2i\pi}{5}\times 1}$   \\ \hline
$u=4$  &\vline      &$-533+\textcolor{gray}{0}$ &\vline  &$1584 \cdot e^{\frac{2i\pi}{5}\times 3}$&\vline    &$1584 \cdot e^{\frac{2i\pi}{5}\times 3}$      \\ \hline
\end{tabular}
\caption{$W_{\Sigma_{2,0} }$, with $w_R^{(z)}=w_3^{(z)}$ in Eq.\eqref{Words}. 
}
\label{W3_Genus2_QP}
\end{table}

\begin{table}[H]
\centering
\footnotesize
\begin{tabular}{cccccccccc}
$W_{\Sigma_{2,0}}$    &\vline &$W_{\Sigma_{2,0}}^{I}$&\vline    &$W_{\Sigma_{2,0}}^{A_1}$&\vline   &$W_{\Sigma_{2,0}}^{A_2}$    \\ \hline
$u=0$  &\vline     &$2377+\textcolor{gray}{0}$&\vline &$1056$&\vline &  $1056$                  \\ \hline
$u=1$  &\vline     &$437+\textcolor{gray}{0}$ &\vline &$1056 \cdot e^{\frac{2i\pi}{5}\times 3}$&\vline   &$1056 \cdot e^{\frac{2i\pi}{5}\times 3}$   \\ \hline
$u=2$  &\vline     &$437+\textcolor{gray}{0}$ &\vline  &$1056 \cdot e^{\frac{2i\pi}{5}\times 1}$&\vline    &$1056 \cdot e^{\frac{2i\pi}{5}\times 1}$   \\  \hline
$u=3$  &\vline      &$437+\textcolor{gray}{0}$&\vline &$1056 \cdot e^{\frac{2i\pi}{5}\times 4}$&\vline  &$1056 \cdot e^{\frac{2i\pi}{5}\times 4}$  \\ \hline
$u=4$  &\vline      &$437+\textcolor{gray}{0}$ &\vline  &$1056 \cdot e^{\frac{2i\pi}{5}\times 2}$&\vline    &$1056 \cdot e^{\frac{2i\pi}{5}\times 2}$      \\ \hline
\end{tabular}
\caption{$W_{\Sigma_{2,0} }$, with $w_R^{(z)}=w_4^{(z)}$ in Eq.\eqref{Words}. 
}
\label{W4_Genus2_QP}
\end{table}

\begin{table}[H]
\centering
\footnotesize
\begin{tabular}{cccccccccc}
$W_{\Sigma_{2,0}}$    &\vline &$W_{\Sigma_{2,0}}^{I}$&\vline    &$W_{\Sigma_{2,0}}^{A_1}$&\vline   &$W_{\Sigma_{2,0}}^{A_2}$    \\ \hline
$u=0$  &\vline     &$437+\textcolor{gray}{0}$&\vline & $528$&\vline &       $528$             \\ \hline
$u=1$  &\vline     &$49+\textcolor{gray}{0}+388\cdot \cos\frac{4\pi}{5}$&\vline &$528 \cdot e^{\frac{2i\pi}{5}\times 2}$&\vline   &$528 \cdot e^{\frac{2i\pi}{5}\times 2}$   \\ \hline
$u=2$  &\vline     &$49+\textcolor{gray}{0}+388\cdot \cos\frac{2\pi}{5}$&\vline  &$528 \cdot e^{\frac{2i\pi}{5}\times 4}$&\vline    &$528 \cdot e^{\frac{2i\pi}{5}\times 4}$   \\  \hline
$u=3$  &\vline      &$49+\textcolor{gray}{0}+388\cdot \cos\frac{2\pi}{5}$&\vline &$528 \cdot e^{\frac{2i\pi}{5}\times 1}$&\vline  &$528 \cdot e^{\frac{2i\pi}{5}\times 1}$  \\ \hline
$u=4$  &\vline      &$49+\textcolor{gray}{0}+388\cdot \cos\frac{4\pi}{5}$&\vline  &$528 \cdot e^{\frac{2i\pi}{5}\times 3}$&\vline    &$528 \cdot e^{\frac{2i\pi}{5}\times 3}$      \\ \hline
\end{tabular}
\caption{$W_{\Sigma_{2,0} }$, with $w_R^{(z)}=w_5^{(z)}$ in Eq.\eqref{Words}. 
}
\label{W5_Genus2_QP}
\end{table}

\begin{table}[H]
\centering
\footnotesize
\begin{tabular}{cccccccccc}
$W_{\Sigma_{2,0}}$    &\vline &$W_{\Sigma_{2,0}}^{I}$&\vline    &$W_{\Sigma_{2,0}}^{A_1}$&\vline   &$W_{\Sigma_{2,0}}^{A_2}$    \\ \hline
$u=0$  &\vline     &$633-24\times \textcolor{gray}{4}$&\vline & $528$&\vline &       $528$             \\ \hline
$u=1$  &\vline     &$245-24\times \textcolor{gray}{4}+388\cdot \cos\frac{2\pi}{5}$ &\vline &$528 \cdot e^{\frac{2i\pi}{5}\times 4}$&\vline   & $528 \cdot e^{\frac{2i\pi}{5}\times 4}$  \\ \hline
$u=2$  &\vline     &$245-24\times \textcolor{gray}{4}+388\cdot \cos\frac{4\pi}{5}$&\vline  &$528 \cdot e^{\frac{2i\pi}{5}\times 3}$&\vline    &$528 \cdot e^{\frac{2i\pi}{5}\times 3}$   \\  \hline
$u=3$  &\vline      &$245-24\times \textcolor{gray}{4}+388\cdot \cos\frac{4\pi}{5}$&\vline &$528 \cdot e^{\frac{2i\pi}{5}\times 2}$&\vline  &$528 \cdot e^{\frac{2i\pi}{5}\times 2}$  \\ \hline
$u=4$  &\vline      &$245-24\times \textcolor{gray}{4}+388\cdot \cos\frac{2\pi}{5}$&\vline  &$528 \cdot e^{\frac{2i\pi}{5}\times 1}$&\vline    &$528 \cdot e^{\frac{2i\pi}{5}\times 1}$      \\ \hline
\end{tabular}
\caption{$W_{\Sigma_{2,0} }$, with $w_R^{(z)}=w_6^{(z)}$ in Eq.\eqref{Words}. 
}
\label{W6_Genus2_QP}
\end{table}

\begin{table}[H]
\centering
\footnotesize
\begin{tabular}{cccccccccc}
$W_{\Sigma_{2,0}}$    &\vline &$W_{\Sigma_{2,0}}^{I}$&\vline    &$W_{\Sigma_{2,0}}^{A_1}$&\vline   &$W_{\Sigma_{2,0}}^{A_2}$    \\ \hline
$u=0$  &\vline     &$633-24\times \textcolor{gray}{4}$&\vline & $528$&\vline &        $528$            \\ \hline
$u=1$  &\vline     &$245-24\times \textcolor{gray}{4}+388\cdot \cos\frac{4\pi}{5}$ &\vline &$528 \cdot e^{\frac{2i\pi}{5}\times 2}$&\vline   &$528 \cdot e^{\frac{2i\pi}{5}\times 2}$   \\ \hline
$u=2$  &\vline     &$245-24\times \textcolor{gray}{4}+388\cdot \cos\frac{2\pi}{5}$ &\vline  &$528 \cdot e^{\frac{2i\pi}{5}\times 4}$&\vline    &$528 \cdot e^{\frac{2i\pi}{5}\times 4}$   \\  \hline
$u=3$  &\vline      &$245-24\times \textcolor{gray}{4}+388\cdot \cos\frac{2\pi}{5}$&\vline &$528 \cdot e^{\frac{2i\pi}{5}\times 1}$&\vline  &$528 \cdot e^{\frac{2i\pi}{5}\times 1}$  \\ \hline
$u=4$  &\vline      &$245-24\times \textcolor{gray}{4}+388\cdot \cos\frac{4\pi}{5}$ &\vline  &$528 \cdot e^{\frac{2i\pi}{5}\times 3}$&\vline    &$528 \cdot e^{\frac{2i\pi}{5}\times 3}$      \\ \hline
\end{tabular}
\caption{$W_{\Sigma_{2,0} }$, with $w_R^{(z)}=w_7^{(z)}$ in Eq.\eqref{Words}. 
}
\label{W7_Genus2_QP}
\end{table}

\section{Topological lattice gauge theory}
\label{Sec: Lattice}

In this section, we will use the lattice gauge theory approach to study the  representations of $\text{MCG}(\Sigma_{g,0})$ or $\text{MCG}(\Sigma_{g,n})$
for an arbitrary twisted quantum double of finite groups $G$ with 3-cocycle $\omega\in H^3(G,U(1))$, 
which is also called Dijkgraaf-Witten (DW) theory.\cite{DW1990}
Compared to the method in Sec.\ref{Sec: Quasi-particle}, the lattice gauge theory approach is a bottom-up approach 
by simply inputing the data of $G$ and $\omega$.

In general, one may consider the 
string-net model,  which is a generalization of the quantum double model for unitary tensor categories,\cite{LevinWen} 
and provides us with a general approach to translate mathematical objects to physical concepts and vice versa.
Here, for our interest of distinguishing the MS MTCs, which are twisted quantum double of finite groups, we
will focus on the DW theory.

The general idea is as follows.
DW theory in (2+1)-d is defined on a triangulation of the spacetime 3-manifold together with
a $G$-coloring, \textit{i.e.}, we assign group elements in $G$ to every $1$-simplex of the triangulation,
and a $3$-cocycle $\omega$ to each tetrahedron.
Within this approach, there is no need to solve the modular equations in Eqs.\eqref{ModularRelationSzTz} which are nonlinear.
We simply need to choose the bases of the degenerate ground states, and do modular transformations.  
The representations of MCG can be obtained by studying the overlap between the bases before
and after the transformations. The MCG representations will look different from those obtained 
in the quasiparticle basis, since we are now using a different set of bases. But the topological 
invariants, which are basis independent, constructed from these MCG representations will be the same.

For the introduction of the group cohomology on a lattice theory, 
one can refer to, \textit{e.g.}, Refs.\onlinecite{ChenGuLiuWenSPT,Ran2012} for further details.

\subsection{Exactly solvable model}

We may work either in the path-integral version or in the hamiltonian 
version of Dijkgraaf-Witten theory.\cite{Dijkgraaf1991,HuWanWu1211,Ran2012,HungWen2013}
Here we will consider the hamiltonian version.
Previous works on the mapping class group of $T^2$ and $T^3$ for a DW theory can be found, \textit{e.g.}, 
in Refs.\onlinecite{HuWanWu1211,HungWen2013, Ran2012,WangWen2014,JiangRan2014}.
Now we generalize the method to the mapping class group of higher-genus manifolds.
In particular, we will focus on the genus-2 manifold.

Our convention on the group-element action in the lattice gague theory is that
\textit{elements of a group are applied from right to left.}
For example, $g\cdot h$ means $g$ acts on $h$.

The hamiltonian version of Dijkgraaf-Witten theory is an exactly solvable lattice model 
with the following hamiltonian \cite{HuWanWu1211,Ran2012,WangWen2014} 
\be\label{Hamiltonian}
H=-\sum_v A_v-\sum_f B_f,
\ee
where $A_v=\frac{1}{|G|}\sum_{t\in G}A_v^t$ is the vertex operator defined on each vertex, and it plays the role of implementing gauge transformation
on the group elements living on the links that meet at the vertex $v$. 
$B_f$ is the face operator defined on each triangulated face $f$, and its effect is to impose zero flux (or flat connection constrain)
through the face $f$. The reason we impose flat connection is that we are interested in a TQFT, which has no local observables. 
Then with flat connection one can find that the holonomy along a closed curve only depends on its homotopy group.

Next, let us consider the Hilbert space of the (possibly) degenerate ground states.
First, the moduli space of $G$ bundles on a closed oriented surface $\Sigma_{g,0}$ of genus $g$
can be identified with $\mathcal{V}_g:=\text{Hom}(\pi_1(\Sigma_{g,0},x),G)/G$, 
where $x\in \Sigma_{g,0}$ is a basepoint, 
$\pi_1$ denotes the fundamental group of the surface, and $G$ acts by conjugation.
A representation of the fundamental group of $\Sigma_{g,0}$ consists of
elements $(g_i,\,h_i)$ ($i=1,\cdots, g$) that satisfy:
\be
\prod_{i=1}^g[g_i,h_i]=1.
\ee
where the commutation is defined as 
\be\label{Def_commutation}
[g,h]:= h^{-1}g^{-1}hg.
\ee
Then the gauge equivalent classes of $G$ bundles are determined by this representation up to conjugation.
If the 3-cocycle $\omega$ is trivial, one can find that the dimension of Hilbert space of degenerate ground states 
is $\text{dim} \mathcal{H}(\Sigma_{g,0})=|\mathcal{V}_g|$.
For a general $\omega$, one has $\text{dim} \mathcal{H}(\Sigma_{g,0})\le |\mathcal{V}_g|$.
This is because a generic 3-cocyle $\omega$ will introduce phase factors in the basis vectors, and therefore
some basis vectors may be canceled and `disappear' in the Hilbert space\cite{DW1990,HuWanWu1211}. 

Now if we consider a manifold with punctures, the case becomes more subtle.
First, the representation of the fundamental group of $\Sigma_{g,n}$ now reads\cite{massey1991basic}
\be\label{fundamental_group_puncture}
\prod_{i=1}^{g}[g_i,h_i]\, \prod_{s=1}^n c_s =1
\ee
where $c_s$ denote the homotopy class of the non-contractable loops around the punctures. 
In terms of $G$-bundle, $c_s$ correspond to the holonomy around the punctures.
Furthermore, in contrast to the closed manifold where the equivalent classes of gauge bundles are
determined by the holonomy up to conjugacy, now we have to rigid the holonomy at the puncture
without conjugacy, which will affect the dimension of the Hilbert space.\cite{freed1993chern,freed1993lectures}

Since we are considering a $(2+1)$-d theory,
we will focus on the two-dimensional (spatial) triangulated lattice, and 
assign a group element $g_{ij}$ to each oriented link $i\to j$.
An arbitrary quantum state in the Hilbert space $\mathcal{H}$ is then labeled by $|\{g_{ij}\}\rangle$.
By performing a modular transformation, we evolve the wavefunction along the `time' direction. Then the path
integral associated to the modular transformation is defined on a three dimensional `spacetime' manifold.
It is known that tetrahedra can be viewed as the building block of a 3-manifold.
To evaluate the path integral over a generic 3-manifold, let us start by assigning a complex 
number to a tetrahedra, given a three-cocycle $\omega(g,h,k)\in H^3(G,U(1))$.

There are mainly two steps: ordering and coloring. For a tetrahedron depicted as follows, 
\begin{eqnarray}\label{Tetrahedron}
\small
\begin{tikzpicture}[baseline={(current bounding box.center)}]
\draw [>=stealth,->] (-20pt,-20pt)--(0pt,-20pt);  \draw (0pt, -20pt) -- (20pt,-20pt);
\draw [>=stealth,->] (20pt,-20pt)--(30pt,-5pt); \draw (30pt,-5pt)--(40pt,10pt);
\draw [>=stealth,->][dashed] (-20pt,-20pt)--(10pt,-5pt);\draw[dashed](10pt,-5pt)--(40pt,10pt);
\draw [>=stealth,->] (-20pt,-20pt)--(-5pt,5pt); \draw (-5pt,5pt)--(10pt,30pt);
\draw [>=stealth,->] (20pt,-20pt)--(15pt,5pt); \draw (15pt,5pt)--(10pt,30pt);
\draw [>=stealth,->] (40pt,10pt)--(25pt,20pt); \draw (25pt,20pt)--(10pt,30pt);

\node at (-20pt,-25pt){${1}$};
\node at (22pt,-25pt){${2}$};
\node at (45pt,10pt){${3}$};
\node at (10pt,35pt){${4}$};

\end{tikzpicture}
\end{eqnarray}
we label the four vertices by 
$1$, $2$, $3$, and $4$ with a chosen ordering, say, $1<2<3<4$. 
For each link, the direction of the link points from $i$ to $j$ with $i<j$. 
That is, we use the convention that lower vertex labels point towards the higher vertex labels.
Then we color the tetrahedron by assigning a group element $g_{ij}\in G$
to each link pointing from $i$ to $j$, and automatically we have $g_{ji}=g_{ij}^{-1}$.
These group elements are subjected to the flat-connection constraints. For a triangle with 
link variables $[ij]$, $[jk]$, and $[ki]$, we have $g_{ij}\cdot g_{jk}\cdot g_{ki}=1$, where $1$ is the 
identity group element of $G$. 
This ensures that the holonomy along a curve only depends on the homotopy class of this curve.
With the ordering and coloring introduced above, now we can assign a $U(1)$ phase 
$\omega^{\epsilon}(g_{ij}, g_{jk}, g_{kl})$ to each tetrahedron, with $\epsilon=\pm 1$.
The sign $\epsilon$ is determined by the `chirality'. For example, for the tetrahedron 
in \eqref{Tetrahedron}, looking from the lowest labeled vertex $1$, a (counter-) clockwise 
$2\to 3\to 4$ loop means $\epsilon=-1$ ($+1$). Then one can find the phase assigned to 
the tetrahedron in \eqref{Tetrahedron} is $\omega(g_{12}, g_{23}, g_{34})$.
For convenience, we may use the definition $[ij]:=g_{ij}$ interchangeably. 

As a remark, in the topological lattice gauge theory, the 3-cocycle condition in \eqref{3cocycle_condition} has a intuitive 
geometric meaning. For example, for the tetrahedron in \eqref{Tetrahedron}, one can add one more vertex inside the tetrahedron.
Then the original tetrahedron can be triangulated into four pieces of smaller tetrahedra, each of which is associated with 
a 3-cocycle. Then one can find that the product of the four new phases is the same as the original $U(1)$ phase.\cite{Ran2012} 
This process will result in the 3-cocycle condition in \eqref{3cocycle_condition}.

The action of $A^t_v$ on vertex $v$ can be considered as creating a new vertex $v'$, with $[vv']=t$ and $t\in G$.
We use the convention of ordering such that $v'<v$, and $v'>v_i$ for arbitrary $v_i<v$.\cite{HuWanWu1211}
For those vertex $v_i$ that are linked to $v$, now we generate new links between $v_i$ and $v'$, with the group element $[v_i v']$
satisfying $[v_i v']\cdot [v'v]\cdot [vv_i]=1$.
These new generated vertex and links will introduce extra tetrahedra, and result in certain $U(1)$ phases,
as will be discussed in the following examples on manifold of genus $g=1$ and $g=2$.

For later convenience, we also introduce a useful manifold $Y\times I$, where $Y$ represents 
a 2-sphere with three punctures and $I$ is an interval, as follows:
\begin{eqnarray}\label{YtimesI}
\small
\begin{tikzpicture}[baseline={(current bounding box.center)}]

\draw [>=stealth,->] (-20pt,-20pt)--(0pt,-20pt);  \draw (0pt, -20pt) -- (20pt,-20pt);
\draw [>=stealth,->] (20pt,-20pt)--(40pt,-5pt); \draw (40pt,-5pt)--(60pt,10pt);

\node at (-27pt,30pt){${1'}$};
\node at (-25pt,-25pt){${1}$};
\node at (22pt,-25pt){${2}$};

\node at (64pt,10pt){${3}$};

\draw [>=stealth,->] (-20pt,-20+50pt)--(0pt,-20+50pt);  \draw (0pt, -20+50pt) -- (20pt,-20+50pt);
\draw [>=stealth,->] (20pt,-20+50pt)--(40pt,-5+50pt); \draw (40pt,-5+50pt)--(60pt,10+50pt);

\draw [>=stealth,->] (-20pt,-20+50pt)--(-20pt,5pt); \draw(-20pt,5pt)--(-20pt,-20pt);
\draw [>=stealth,->] (20pt,-20+50pt)--(20pt,5pt); \draw(20pt,5pt)--(20pt,-20pt);
\draw [>=stealth,->] (60pt,10+50pt)--(60pt,35pt); \draw(60pt,35pt)--(60pt,10pt);

\draw [>=stealth,->][dashed] (-20pt,-20pt)--(20pt,-5pt);\draw[dashed](20pt,-5pt)--(60pt,10pt);
\draw [>=stealth,->] (-20pt,-20+50pt)--(20pt,-5+50pt);\draw(20pt,-5+50pt)--(60pt,10+50pt);

\node at (20pt,36pt){${2'}$};
\node at (65pt,60pt){${3'}$};

\draw [dashed]  (-20pt,-20pt)--(20pt,20pt); \draw[>=stealth,->][dashed] (60pt,10+50pt)--(20pt,20pt);
\draw (20pt,-20pt)--(40pt,20pt);\draw[>=stealth,->] (60pt,10+50pt)--(40pt,20pt);
\draw  (-20pt,-20pt)--(0pt,5pt); \draw[>=stealth,->](20pt,30pt)--(0pt,5pt); 
\end{tikzpicture}
\end{eqnarray}
where we take the ordering $1'<2'<3'<1<2<3$. One can find the 
weight associated to this triangular prism can be obtained by dividing it into three tetrahedra. Then one has
$\omega([3'1], [12], [23])\cdot \omega([2'3'], [3'1], [12])\cdot \omega([1'2'],[2'3'],[3'1])
=:Y_{[123],[1'2'3']}$, which is a $U(1)$ phase.
Hereafter, for brevity we will use $Y_{[abc],[a'b'c']}$ in certain cases to represent the $U(1)$ phase
associated to the triangular prism. It is reminded that the concrete expression of $Y_{[abc],[a'b'c']}$
may depend on the ordering of vertices as well as how we triangulate the prism, but the topological invariants 
constructed from these building blocks are independent of such choices.\cite{HuWanWu1211,Ran2012}

\subsection{Modular transformation}
\label{Sec: Lattice_modularTransform}

Before we study the modular transformation on a genus-2 manifold, it is helpful to
briefly review how to perform the modular transformation for the case of a genus-1 torus.\cite{HuWanWu1211} 
Then one can generalize
the procedure to a punctured torus, and then study the genus-2 case by gluing two
punctured tori by identifying the punctures.

\subsubsection{Genus one}

Let us first check the Hilbert space of ground state on the torus $\Sigma_{1,0}$.
The simplest triangulation of a torus is as follows
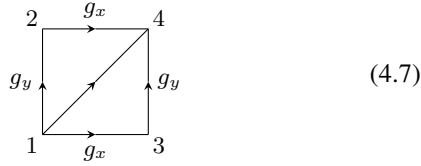
\begin{eqnarray}\label{Genus1Triangle}
\small
\begin{tikzpicture}[baseline={(current bounding box.center)}]
\draw (-20pt,20pt)--(-20pt,0pt); \draw [>=stealth,<-] (-20pt, 0pt) -- (-20pt,-20pt);
\draw (20pt,20pt)--(20pt,0pt); \draw [>=stealth,<-] (20pt, 0pt) -- (20pt,-20pt);
\draw [>=stealth,->] (-20pt,20pt)--(0pt,20pt); \draw (0pt,20pt)--(20pt,20pt);
\draw [>=stealth,->] (-20pt,-20pt)--(0pt,-20pt);  \draw (0pt, -20pt) -- (20pt,-20pt);
\node at (28pt,0pt){${g_y}$};
\node at (0pt,-27pt){${g_x}$};
\node at (0pt,27pt){${g_x}$};
\node at (-28pt,0pt){${g_y}$};
\node at (-24pt,-24pt){$1$};
\node at (-24pt,24pt){$2$};
\node at (24pt,-24pt){$3$};
\node at (24pt,24pt){$4$};
\draw [>=stealth,->](-20pt,-20pt)--(0pt,0pt); \draw(0pt,0pt)--(20pt,20pt);
\end{tikzpicture}
\end{eqnarray}
where we identity the edges $[13]$ with $[24]$, and $[12]$ with $[34]$.
We take the ordering $1<2<3<4$, and the coloring $[13]=[24]=g_x$ and $[12]=[34]=g_y$.
One can find that there is essentially only one vertex in this triangulation.
The flat connection (imposed by the term $B_f$ in the Hamiltonian) on a torus satisfies the constraint 
$[g_x,g_y]=1$, where the commutation is defined in Eq.\eqref{Def_commutation}.
The flat connection on a torus simply means that $g_xg_y=g_yg_x$. Denoting the basis in \eqref{Genus1Triangle}  
as $|g_x,g_y\rangle$ with $g_x,\,g_y \in G$, then the action of $A^t$ in \eqref{Hamiltonian} on $|g_x,g_y\rangle$ can be written as (here we simplify
$A_v^t$ as $A^t$ since there is only one vertex now): 
\be\label{At_genus1}
A^t |g_x,g_y\rangle=\eta^t(g_x,g_y)|tg_xt^{-1}, tg_yt^{-1}\rangle,
\ee
where the $U(1)$ phase is associated with the path integral defined on the following 3-manifold:
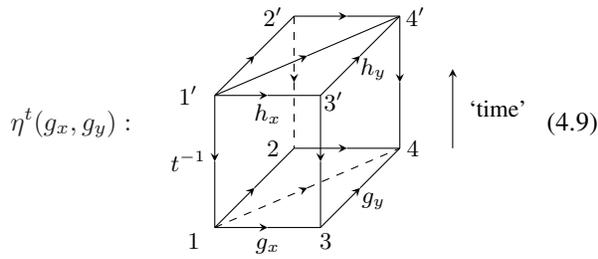
\begin{eqnarray}\label{eta_t_genus1}
\small
\eta^t (g_x,g_y):\quad
\begin{tikzpicture}[baseline={(current bounding box.center)}]
\draw [>=stealth,->] (-20pt,-20pt)--(0pt,-20pt);  \draw (0pt, -20pt) -- (20pt,-20pt);
\draw [>=stealth,->] (-20pt,-20pt)--(-5pt,-5pt); \draw(-5pt,-5pt)--(10pt,10pt);
\draw [>=stealth,->] (10pt,10pt)--(30pt,10pt); \draw (30pt,10pt)--(50pt,10pt);
\draw [>=stealth,->] (20pt,-20pt)--(35pt,-5pt); \draw (35pt,-5pt)--(50pt,10pt);
\node at (40pt,-10pt){${g_y}$};
\node at (0pt,-27pt){${g_x}$};
\node at (-28pt,-25pt){${1}$};
\node at (22pt,-25pt){${3}$};

\node at (2pt,10pt){${2}$};
\node at (55pt,10pt){${4}$};

\draw [>=stealth,->] (-20pt,-20+50pt)--(0pt,-20+50pt);  \draw (0pt, -20+50pt) -- (20pt,-20+50pt);
\draw [>=stealth,->] (-20pt,-20+50pt)--(-5pt,-5+50pt); \draw(-5pt,-5+50pt)--(10pt,10+50pt);
\draw [>=stealth,->] (10pt,10+50pt)--(30pt,10+50pt); \draw (30pt,10+50pt)--(50pt,10+50pt);
\draw [>=stealth,->] (20pt,-20+50pt)--(35pt,-5+50pt); \draw (35pt,-5+50pt)--(50pt,10+50pt);

\draw [>=stealth,->] (-20pt,-20+50pt)--(-20pt,5pt); \draw(-20pt,5pt)--(-20pt,-20pt);
\draw [>=stealth,->] (20pt,-20+50pt)--(20pt,5pt); \draw(20pt,5pt)--(20pt,-20pt);
\draw [>=stealth,->] (50pt,10+50pt)--(50pt,35pt); \draw(50pt,35pt)--(50pt,10pt);
\draw [>=stealth,->] [dashed](10pt,10+50pt)--(10pt,35pt); \draw[dashed](10pt,35pt)--(10pt,10pt);

\draw [>=stealth,->][dashed] (-20pt,-20pt)--(15pt,-5pt);\draw[dashed](15pt,-5pt)--(50pt,10pt);
\draw [>=stealth,->] (-20pt,-20+50pt)--(15pt,45pt); \draw(15pt,45pt)--(50pt,60pt); 

\node at (-30pt,30pt){${1'}$};
\node at (25pt,28pt){${3'}$};

\node at (2pt,60pt){${2'}$};
\node at (56pt,60pt){$4'$};
\node at (0pt,23pt){${h_x}$};
\node at (40pt,40pt){${h_y}$};

\node at (-30pt,5pt){${t^{-1}}$};

\draw [>=stealth,->] (70pt,10pt)--(70pt,40pt);
\node at (87pt,25pt){${\text{`time'}}$};

\end{tikzpicture}
\end{eqnarray}
The group elements on the links along the `time' direction are $[11']=[22']=[33']=[44']=t$.
The new basis $|h_x,h_y\rangle:=|[1'3'], [3'4']\rangle$ are given by $h_x=t g_x t^{-1}$ and $h_y=t g_y t^{-1}$
due to the flat connection constraint.
Then one can evaluate 
$\eta^t(g_x,g_y)$ explicitly by dividing the cube into two triangular prisms, with each 
prism containing three tetrahedra as shown in \eqref{YtimesI}. From the convention of defining $A^t$ in the previous section,  we have the ordering 
$1'<1<2'<2<3'<3<4'<4$ in the triangulation in \eqref{eta_t_genus1}. 
The $U(1)$ phase $\eta^t (g_x,g_y)$ is then expressed as:
\be
\small
\begin{split}
&Y_{[124],[1'2'4']}\cdot Y_{[134],[1'3'4']},\\
=&\big[\omega([12],[24'],[4'4])^{-1}\cdot \omega([12'],[2'2],[24']) \\
&\cdot \omega([1'1],[12'],[2'4'])^{-1}\big]\cdot\big[\omega([13],[34'],[4'4])\\
&\cdot \omega([13'],[3'3],[34'])^{-1}\cdot \omega([1'1],[13'],[3'4']) \big]\\
=&\big[\omega(g_y,t g_x,t^{-1})^{-1}\cdot \omega(t g_y,t,tg_x)\\
&\cdot \omega(t^{-1},tg_y,tg_xt^{-1})^{-1}\big]\cdot \big[\omega(g_x,tg_y,t^{-1})\\
&\cdot \omega(tg_x,t^{-1},tg_y)^{-1}\cdot \omega(t^{-1}, tg_x, tg_yt^{-1})\big]
=:\eta^t(g_x,g_y).
\end{split}
\ee
One can check that $A^t A^{t'}=A^{t\cdot t'}$ by using the 3-cocyle condition in Eq.\eqref{3cocycle_condition}.
Based on Eq.\eqref{At_genus1}, the ground states are spanned by the vectors
 \be\label{GSD_torus1}
\Big\{ \frac{1}{|G|}\sum_{x\in G} \eta^t(g_x,g_y)|t g_x t^{-1}, t g_y t^{-1}\rangle \,\big| [g_x,g_y]=1
\Big\}.
 \ee

Now let us consider the mapping class group for the torus $\Sigma_{1,0}$.
There are two generators for $\text{MCG}(\Sigma_{1,0})$, which can be chosen as the two Dehn twists $\mathfrak{t}_x$ and $\mathfrak{t}_y$
along $x$ and $y$ directions, 
or alternatively their combination such as $\mathfrak{s}$ and $\mathfrak{t}$ transformations (See introduction). 
They are related by 
\be
\mathfrak{t}:=\mathfrak{t}_x, \quad \mathfrak{s}:=\mathfrak{t}_y\cdot \mathfrak{t}_x^{-1}\cdot \mathfrak{t}_y.
\ee
One can check that the modular relation $\mathfrak{t}_y\cdot \mathfrak{t}_x^{-1}\cdot \mathfrak{t}_y
=\mathfrak{t}_x^{-1}\cdot \mathfrak{t}_y\cdot \mathfrak{t}_x^{-1}$ is equivalent to $(\mathfrak{s}\mathfrak{t})^3=\mathfrak{s}^2$.
For a lattice gauge theory, it is more convenient to perform Dehn twists directly, based on which we can get the  representations 
$T_x$ and $T_y$ for $\mathfrak{t}_x$ and $\mathfrak{t}_y$, respectively. Then $S$ and $T$ matrices can be defined as
\be\label{T_S_def}
T:=T_x, \quad S:=T_y\cdot T_x^{-1}\cdot T_y.
\ee
In the following, let us check how the representations of Dehn twists $T_x$ and $T_y$ act on the basis $|g_x,g_y\rangle$ in \eqref{Genus1Triangle}.
For the effect of $T_x$ on the basis vector $|g_x,\,g_y\rangle$, we have\cite{HuWanWu1211}
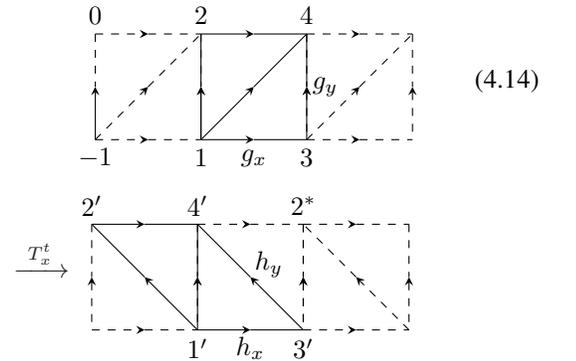
\begin{eqnarray}\label{Da_Inverse}
\small
&\begin{tikzpicture}[baseline={(current bounding box.center)}]
\draw [dashed](-60pt,20pt)--(-60pt,0pt); \draw [>=stealth,<-] (-60pt, 0pt) -- (-60pt,-20pt);
\draw [dashed](-20pt,20pt)--(-20pt,0pt); \draw [>=stealth,<-] (-20pt, 0pt) -- (-20pt,-20pt);
\draw [dashed][>=stealth,->] (-60pt,20pt)--(-40pt,20pt); \draw [dashed](-40pt,20pt)--(-20pt,20pt);
\draw [dashed][>=stealth,->] (-60pt,-20pt)--(-40pt,-20pt);  \draw [dashed](-40pt, -20pt) -- (-20pt,-20pt);
\draw [dashed][>=stealth,->] (-60pt,-20pt)--(-40pt,0pt);  \draw [dashed](-40pt, 0pt) -- (-20pt,20pt);

\draw (-20pt,20pt)--(-20pt,0pt); \draw [>=stealth,<-] (-20pt, 0pt) -- (-20pt,-20pt);
\draw (20pt,20pt)--(20pt,0pt); \draw [>=stealth,<-] (20pt, 0pt) -- (20pt,-20pt);
\draw [>=stealth,->] (-20pt,20pt)--(0pt,20pt); \draw (0pt,20pt)--(20pt,20pt);
\draw [>=stealth,->] (-20pt,-20pt)--(0pt,-20pt);  \draw (0pt, -20pt) -- (20pt,-20pt);
\draw [>=stealth,->] (-20pt,-20pt)--(0pt,0pt);  \draw (0pt, 0pt) -- (20pt,20pt);

\node at (-60pt,-27pt){${-1}$};
\node at (-20pt,-27pt){${1}$};
\node at (20pt,-27pt){${3}$};

\node at (-60pt,27pt){${0}$};
\node at (-20pt,27pt){${2}$};
\node at (20pt,27pt){${4}$};

\draw [dashed](20pt,20pt)--(20pt,0pt); \draw [dashed][>=stealth,<-] (20pt, 0pt) -- (20pt,-20pt);
\draw [dashed](60pt,20pt)--(60pt,0pt); \draw [dashed][>=stealth,<-] (60pt, 0pt) -- (60pt,-20pt);
\draw [dashed][>=stealth,->] (20pt,20pt)--(40pt,20pt); \draw [dashed](40pt,20pt)--(60pt,20pt);
\draw [dashed][>=stealth,->] (20pt,-20pt)--(40pt,-20pt);  \draw [dashed](40pt, -20pt) -- (60pt,-20pt);
\draw [dashed][>=stealth,->] (20pt,-20pt)--(40pt,0pt);  \draw [dashed](40pt, 0pt) -- (60pt,20pt);
\node at (27pt,0pt){${g_y}$};
\node at (0pt,-27pt){${g_x}$};
\end{tikzpicture}
\\
\xrightarrow{\,\,T_x^t\,\,}
&\begin{tikzpicture}[baseline={(current bounding box.center)}]

\draw [dashed](60pt,20pt)--(60pt,0pt); \draw [>=stealth,<-] [dashed](60pt, 0pt) -- (60pt,-20pt);
\draw [dashed](100pt,20pt)--(100pt,0pt); \draw [>=stealth,<-] [dashed](100pt, 0pt) -- (100pt,-20pt);
\draw [dashed][>=stealth,->] (60pt,20pt)--(80pt,20pt); \draw [dashed](80pt,20pt)--(100pt,20pt);
\draw [dashed][>=stealth,->] (60pt,-20pt)--(80pt,-20pt);  \draw [dashed](80pt, -20pt) -- (100pt,-20pt);
\draw [dashed][>=stealth,->] (100pt,-20pt)--(80pt,0pt);  \draw [dashed](80pt, 0pt) -- (60pt,20pt);

\draw [dashed](-20pt,20pt)--(-20pt,0pt); \draw [dashed][>=stealth,<-] (-20pt, 0pt) -- (-20pt,-20pt);
\draw [dashed](20pt,20pt)--(20pt,0pt); \draw [dashed][>=stealth,<-] (20pt, 0pt) -- (20pt,-20pt);
\draw [>=stealth,->] (-20pt,20pt)--(0pt,20pt); \draw(0pt,20pt)--(20pt,20pt);
\draw [dashed][>=stealth,->] (-20pt,-20pt)--(0pt,-20pt);  \draw [dashed](0pt, -20pt) -- (20pt,-20pt);
\draw [>=stealth,->] (20pt,-20pt)--(0pt,0pt);  \draw (0pt, 0pt) -- (-20pt,20pt);

\draw (20pt,20pt)--(20pt,0pt); \draw [>=stealth,<-] (20pt, 0pt) -- (20pt,-20pt);
\draw [dashed](60pt,20pt)--(60pt,0pt); \draw [dashed][>=stealth,<-] (60pt, 0pt) -- (60pt,-20pt);
\draw [dashed][>=stealth,->] (20pt,20pt)--(40pt,20pt); \draw [dashed](40pt,20pt)--(60pt,20pt);
\draw [>=stealth,->] (20pt,-20pt)--(40pt,-20pt);  \draw (40pt, -20pt) -- (60pt,-20pt);
\draw [>=stealth,->] (60pt,-20pt)--(40pt,0pt);  \draw (40pt, 0pt) -- (20pt,20pt);

\node at (47pt,5pt){${h_y}$};
\node at (40pt,-27pt){${h_x}$};
\node at (20pt,-27pt){${1'}$};
\node at (60pt,-27pt){${3'}$};
\node at (20pt,27pt){${4'}$};
\node at (60pt,27pt){${{2^{\ast}}}$};
\node at (-20pt,27pt){$2'$};
\end{tikzpicture}
\nonumber
\end{eqnarray}
The basis $|g_x,g_y\rangle$ is transformed to 
$|h_x,h_y\rangle:=|[1'3'],[3'4']\rangle=|tg_x t^{-1}, t(g_x^{-1}\cdot g_y)t^{-1}\rangle$, where $t\in G$,
up to a $U(1)$ phase $u_x^t(\Sigma_{1,0})$.
By restricting to the unit squares spanned by $1,2,3,4$ and $1',3',4',2^{\ast}$, one can find that the $U(1)$ phase $u_x^t(\Sigma_{1,0})$ is associated with the path integral as follows:
\begin{eqnarray}\label{TxPhase}
\small
u^t_x(\Sigma_{1,0}):\quad
\begin{tikzpicture}[baseline={(current bounding box.center)}]

\draw [>=stealth,->] (-20pt,-20pt)--(0pt,-20pt);  \draw (0pt, -20pt) -- (20pt,-20pt);
\draw [>=stealth,->] (-20pt,-20pt)--(-5pt,-5pt); \draw(-5pt,-5pt)--(10pt,10pt);
\draw [>=stealth,->] (10pt,10pt)--(30pt,10pt); \draw (30pt,10pt)--(50pt,10pt);
\draw [>=stealth,->] (20pt,-20pt)--(35pt,-5pt); \draw (35pt,-5pt)--(50pt,10pt);
\node at (40pt,-10pt){${g_y}$};
\node at (0pt,-27pt){${g_x}$};
\node at (-28pt,-25pt){${1}$};
\node at (22pt,-25pt){${3}$};

\node at (2pt,10pt){${2}$};
\node at (55pt,10pt){${4}$};

\draw [>=stealth,->] (-20pt,-20+50pt)--(0pt,-20+50pt);  \draw (0pt, -20+50pt) -- (20pt,-20+50pt);
\draw [>=stealth,->] (-20pt,-20+50pt)--(-5pt,-5+50pt); \draw(-5pt,-5+50pt)--(10pt,10+50pt);
\draw [>=stealth,->] (10pt,10+50pt)--(30pt,10+50pt); \draw (30pt,10+50pt)--(50pt,10+50pt);
\draw [>=stealth,->] (20pt,-20+50pt)--(35pt,-5+50pt); \draw (35pt,-5+50pt)--(50pt,10+50pt);

\draw [>=stealth,->] (-20pt,-20+50pt)--(-20pt,5pt); \draw(-20pt,5pt)--(-20pt,-20pt);
\draw [>=stealth,->] (20pt,-20+50pt)--(20pt,5pt); \draw(20pt,5pt)--(20pt,-20pt);
\draw [>=stealth,->] (50pt,10+50pt)--(50pt,35pt); \draw(50pt,35pt)--(50pt,10pt);
\draw [>=stealth,->] [dashed](10pt,10+50pt)--(10pt,35pt); \draw[dashed](10pt,35pt)--(10pt,10pt);

\draw [>=stealth,->][dashed] (-20pt,-20pt)--(15pt,-5pt);\draw[dashed](15pt,-5pt)--(50pt,10pt);
\draw [>=stealth,->] (20pt,30pt)--(15pt,45pt); \draw(15pt,45pt)--(10pt,60pt);

\node at (-30pt,30pt){${1'}$};
\node at (25pt,28pt){${3'}$};

\node at (2pt,60pt){${4'}$};
\node at (58pt,60pt){${2^{\ast}}$};
\node at (0pt,23pt){${h_x}$};
\node at (22pt,45pt){${h_y}$};

\node at (-30pt,5pt){${t^{-1}}$};

\end{tikzpicture}
\end{eqnarray}
with the ordering: $1'<3'<4'<2^{\ast}<1<2<3<4$.
The group elements on the links along the `time' direction are
$[11']=[33']=[24']=[42^{\ast}]=t$, with $t\in G$.
With the flat connection condition, $h_x=t g_x t^{-1}$ and $h_y=t (g_x^{-1}\cdot g_y)t^{-1}$ is automatically satisfied.

The phase in \eqref{TxPhase} can be decomposed into the product of three
pieces. The first two pieces correspond to the two triangular prisms as follows
\begin{eqnarray}\label{DaPhaseDecomposition}
\begin{tikzpicture}[baseline={(current bounding box.center)}]
\draw [>=stealth,->] (-20*0.8pt,-20*0.8pt)--(-5*0.8pt,-5*0.8pt); \draw(-5*0.8pt,-5*0.8pt)--(10*0.8pt,10*0.8pt);
\draw [>=stealth,->] (10*0.8pt,10*0.8pt)--(30*0.8pt,10*0.8pt); \draw (30*0.8pt,10*0.8pt)--(50*0.8pt,10*0.8pt);

\node at (-25*0.8pt,-25*0.8pt){${1}$};
\node at (2*0.8pt,10*0.8pt){${2}$};
\node at (55*0.8pt,10*0.8pt){${4}$};
\node at (-28*0.8pt,30*0.8pt){${1'}$};

\draw [>=stealth,->] (-20*0.8pt,-20*0.8+50*0.8pt)--(-5*0.8pt,-5*0.8+50*0.8pt); \draw(-5*0.8pt,-5*0.8+50*0.8pt)--(10*0.8pt,10*0.8+50*0.8pt);
\draw [>=stealth,->] (10*0.8pt,10*0.8+50*0.8pt)--(30*0.8pt,10*0.8+50*0.8pt); \draw (30*0.8pt,10*0.8+50*0.8pt)--(50*0.8pt,10*0.8+50*0.8pt);

\draw [>=stealth,->] (-20*0.8pt,-20*0.8+50*0.8pt)--(-20*0.8pt,5*0.8pt); \draw(-20*0.8pt,5*0.8pt)--(-20*0.8pt,-20*0.8pt);
\draw [>=stealth,->] [dashed](10*0.8pt,10*0.8+50*0.8pt)--(10*0.8pt,35*0.8pt); \draw[dashed](10*0.8pt,35*0.8pt)--(10*0.8pt,10*0.8pt);

\draw [>=stealth,->] (-20*0.8pt,-20*0.8pt)--(15*0.8pt,-5*0.8pt);\draw(15*0.8pt,-5*0.8pt)--(50*0.8pt,10*0.8pt);
\draw [>=stealth,->] (-20*0.8pt,-20*0.8+50*0.8pt)--(15*0.8pt,-5*0.8+50*0.8pt);\draw(15*0.8pt,-5*0.8+50*0.8pt)--(50*0.8pt,10*0.8+50*0.8pt);
\draw [>=stealth,->] (50*0.8pt,10*0.8+50*0.8pt)--(50*0.8pt,35*0.8pt);\draw(50*0.8pt,35*0.8pt)--(50*0.8pt,10*0.8pt);

\node at (2*0.8pt,60*0.8pt){${4'}$};
\node at (58*0.8pt,60*0.8pt){${2^{\ast}}$};
\end{tikzpicture}
\small
\begin{tikzpicture}[baseline={(current bounding box.center)}]

\draw [>=stealth,->] (-20*0.8pt,-20*0.8pt)--(0pt,-20*0.8pt);  \draw (0pt, -20*0.8pt) -- (20*0.8pt,-20*0.8pt);
\draw [>=stealth,->] (20*0.8pt,-20*0.8pt)--(35*0.8pt,-5*0.8pt); \draw (35*0.8pt,-5*0.8pt)--(50*0.8pt,10*0.8pt);

\node at (-27*0.8pt,30*0.8pt){${1'}$};
\node at (-25*0.8pt,-25*0.8pt){${1}$};
\node at (22*0.8pt,-25*0.8pt){${3}$};

\node at (55*0.8pt,10*0.8pt){${4}$};

\draw [>=stealth,->] (-20*0.8pt,-20*0.8+50*0.8pt)--(0pt,-20*0.8+50*0.8pt);  \draw (0pt, -20*0.8+50*0.8pt) -- (20*0.8pt,-20*0.8+50*0.8pt);
\draw [>=stealth,->] (20*0.8pt,-20*0.8+50*0.8pt)--(35*0.8pt,-5*0.8+50*0.8pt); \draw (35*0.8pt,-5*0.8+50*0.8pt)--(50*0.8pt,10*0.8+50*0.8pt);

\draw [>=stealth,->] (-20*0.8pt,-20*0.8+50*0.8pt)--(-20*0.8pt,5*0.8pt); \draw(-20*0.8pt,5*0.8pt)--(-20*0.8pt,-20*0.8pt);
\draw [>=stealth,->] (20*0.8pt,-20*0.8+50*0.8pt)--(20*0.8pt,5*0.8pt); \draw(20*0.8pt,5*0.8pt)--(20*0.8pt,-20*0.8pt);
\draw [>=stealth,->] (50*0.8pt,10*0.8+50*0.8pt)--(50*0.8pt,35*0.8pt); \draw(50*0.8pt,35*0.8pt)--(50*0.8pt,10*0.8pt);

\draw [>=stealth,->][dashed] (-20*0.8pt,-20*0.8pt)--(15*0.8pt,-5*0.8pt);\draw[dashed](15*0.8pt,-5*0.8pt)--(50*0.8pt,10*0.8pt);
\draw [>=stealth,->] (-20*0.8pt,-20*0.8+50*0.8pt)--(15*0.8pt,-5*0.8+50*0.8pt);\draw(15*0.8pt,-5*0.8+50*0.8pt)--(50*0.8pt,10*0.8+50*0.8pt);

\node at (25*0.8pt,28*0.8pt){${3'}$};
\node at (58*0.8pt,60*0.8pt){${2^{\ast}}$};

\end{tikzpicture}
\end{eqnarray}
with each prism containing three tetrahedra and expressed in 
terms of the product of three 3-cocycles. The third piece is

\begin{eqnarray}\label{DaPhase3rd}
\begin{tikzpicture}[baseline={(current bounding box.center)}]
\draw [>=stealth,->] (-20pt,-20+50pt)--(0pt,-20+50pt);  \draw (0pt, -20+50pt) -- (20pt,-20+50pt);
\draw [>=stealth,->] (-20pt,-20+50pt)--(-5pt,-5+50pt); \draw(-5pt,-5+50pt)--(10pt,10+50pt);
\draw [>=stealth,->] (10pt,10+50pt)--(30pt,10+50pt); \draw (30pt,10+50pt)--(50pt,10+50pt);
\draw [>=stealth,->] (20pt,-20+50pt)--(35pt,-5+50pt); \draw (35pt,-5+50pt)--(50pt,10+50pt);

\draw [>=stealth,->][dashed] (-20pt,-20+50pt)--(32.5pt,2.5+50pt);\draw[dashed](32.5pt,2.5+50pt)--(50pt,10+50pt);
\draw [>=stealth,->] (20pt,30pt)--(12.5pt,52.5pt); \draw(12.5pt,52.5pt)--(10pt,60pt);

\node at (-30pt,30pt){${1'}$};
\node at (25pt,28pt){${3'}$};

\node at (2pt,60pt){${4'}$};
\node at (58pt,60pt){${2^{\ast}}$};
\end{tikzpicture}
\end{eqnarray}
which contributes a single 3-cocycle $\omega([1'3'],[3'4'],[4'2^{\ast}])$.
Combining all the contributions above, the phase in \eqref{TxPhase} is the product of seven
$3$-cocycles as follows
\be\label{u_x^t}
\small
\begin{split}
&Y_{[124],[1'4'2^{\ast}]}\cdot Y_{[134],[1'3'2^{\ast}]}\cdot \omega ([1'3'],[3'4'],[4'2^{\ast}])\\
=&\big[\omega ^{-1}([2^{\ast}1],[12],[24])\cdot \omega ^{-1}([4'2^{\ast}],[2^{\ast}1],[12])\\
&\cdot\omega ^{-1}([1'4'],[4'2^{\ast}],[2^{\ast}1])\big]\cdot
\big[
\omega ([2^{\ast}1],[13],[34])\\
&\cdot\omega ([3'2^{\ast}],[2^{\ast}1],[13])\cdot\omega ([1'3'],[3'2^{\ast}],[2^{\ast}1])\big]\\
&\cdot \omega ([1'3'],[3'4'],[4'2^{\ast}])\\
=&\omega ^{-1}(g_x^{-1}g_y^{-1}\,t^{-1}, g_y, g_x)\cdot \omega ^{-1} (t\, g_x\, t^{-1}, g_x^{-1}g_y^{-1} t^{-1}, g_y)\\
&\cdot \omega ^{-1}(tg_yt^{-1}, tg_x t^{-1}, g_x^{-1}g_y^{-1}t^{-1})\cdot \omega (g_x^{-1}g_y^{-1}t^{-1},g_x,g_y)\\
&\cdot \omega (t g_y t^{-1}, g_x^{-1}g_y^{-1}t^{-1}, g_x)\cdot \omega (t g_x t^{-1}, t g_y t^{-1}, g_x^{-1}g_y^{-1} t^{-1})\\
&\cdot \omega (t  g_x t^{-1}, t g_x^{-1}g_y t^{-1}, t g_x t^{-1})
=:u_x^t(\Sigma_{1,0}).
\end{split}
\ee
Then the effect of $T_x^t$ on the basis $|g_x,g_y\rangle$ is
\be
T_x^t \,|g_x,g_y \rangle=u_x^t(\Sigma_{1,0})\, |t\,g_x\,t^{-1}, t\,(g_x^{-1}\cdot g_y)\,t^{-1}\rangle,
\ee
 with the phase $u_x^t(\Sigma_{1,0})$ expressed in Eq.\eqref{u_x^t}.

Next we consider the effect of $T_y$ on the basis $|g_x, g_y\rangle$ as follows:
\begin{eqnarray}\label{Db_Inverse}
\small
\begin{tikzpicture}[baseline={(current bounding box.center)}]
\draw [dashed](-20pt,60pt)--(-20pt,40pt); \draw [dashed][>=stealth,<-] (-20pt, 40pt) -- (-20pt,20pt);
\draw [dashed](20pt,60pt)--(20pt,40pt); \draw [dashed][>=stealth,<-] (20pt, 40pt) -- (20pt,20pt);
\draw [dashed][>=stealth,->] (-20pt,60pt)--(0pt,60pt); \draw [dashed](0pt,60pt)--(20pt,60pt);
\draw [dashed][>=stealth,->] (-20pt,20pt)--(0pt,20pt);  \draw [dashed](0pt, 20pt) -- (20pt,20pt);
\draw [dashed][>=stealth,->] (-20pt,20pt)--(0pt,40pt);  \draw [dashed](0pt, 40pt) -- (20pt,60pt);
\node at (-28pt,-60pt){${-3}$};
\node at (-25pt,-20pt){${1}$};
\node at (25pt,-20pt){${3}$};
\node at (27pt,-58pt){${-1}$};
\node at (-25pt,20pt){${2}$};
\node at (25pt,20pt){${4}$};

\draw (-20pt,20pt)--(-20pt,0pt); \draw [>=stealth,<-] (-20pt, 0pt) -- (-20pt,-20pt);
\draw (20pt,20pt)--(20pt,0pt); \draw [>=stealth,<-] (20pt, 0pt) -- (20pt,-20pt);
\draw [>=stealth,->] (-20pt,20pt)--(0pt,20pt); \draw (0pt,20pt)--(20pt,20pt);
\draw [>=stealth,->] (-20pt,-20pt)--(0pt,-20pt);  \draw (0pt, -20pt) -- (20pt,-20pt);
\draw [>=stealth,->] (-20pt,-20pt)--(0pt,0pt);  \draw (0pt, 0pt) -- (20pt,20pt);

\draw [dashed](-20pt,-20pt)--(-20pt,-40pt); \draw [dashed][>=stealth,<-] (-20pt, -40pt) -- (-20pt,-60pt);
\draw [dashed](20pt,-20pt)--(20pt,-40pt); \draw [dashed][>=stealth,<-] (20pt, -40pt) -- (20pt,-60pt);
\draw [dashed][>=stealth,->] (-20pt,-20pt)--(0pt,-20pt); \draw [dashed](0pt,-20pt)--(20pt,-20pt);
\draw [dashed][>=stealth,->] (-20pt,-60pt)--(0pt,-60pt);  \draw [dashed](0pt, -60pt) -- (20pt,-60pt);
\draw [dashed][>=stealth,->] (-20pt,-60pt)--(0pt,-40pt);  \draw [dashed](0pt, -40pt) -- (20pt,-20pt);

\node at (27pt,0pt){${g_y}$};
\node at (0pt,-27pt){${g_x}$};
\end{tikzpicture}
\quad
\xrightarrow{\,\, T_y^t\,\,}
\quad
\begin{tikzpicture}[baseline={(current bounding box.center)}]

\draw [dashed](-20pt,60pt)--(-20pt,40pt); \draw [dashed][>=stealth,<-] (-20pt, 40pt) -- (-20pt,20pt);
\draw [dashed](20pt,60pt)--(20pt,40pt); \draw [dashed][>=stealth,<-] (20pt, 40pt) -- (20pt,20pt);
\draw [dashed][>=stealth,->] (-20pt,60pt)--(0pt,60pt); \draw [dashed](0pt,60pt)--(20pt,60pt);
\draw [dashed][>=stealth,->] (-20pt,20pt)--(0pt,20pt);  \draw [dashed](0pt, 20pt) -- (20pt,20pt);
\draw [dashed][>=stealth,->] (-20pt,60pt)--(0pt,40pt);  \draw [dashed](0pt, 40pt) -- (20pt,20pt);

\draw (-20pt,20pt)--(-20pt,0pt); \draw [>=stealth,<-] (-20pt, 0pt) -- (-20pt,-20pt);
\draw [dashed](20pt,20pt)--(20pt,0pt); \draw [dashed][>=stealth,<-] (20pt, 0pt) -- (20pt,-20pt);
\draw [dashed][>=stealth,->] (-20pt,20pt)--(0pt,20pt); \draw [dashed](0pt,20pt)--(20pt,20pt);
\draw [>=stealth,->] (-20pt,20pt)--(0pt,0pt);  \draw (0pt, 0pt) -- (20pt,-20pt);
\node at (-25pt,20pt){${2'}$};
\node at (25pt,-60pt){$$};
\node at (-25pt,-20pt){${1'}$};
\node at (27pt,-20pt){${4'}$};
\node at (27pt,20pt){${3^{\ast}}$};
\node at (27pt,-60pt){${3'}$};

\draw [dashed](-20pt,-20pt)--(-20pt,-40pt); \draw [dashed][>=stealth,<-] (-20pt, -40pt) -- (-20pt,-60pt);
\draw (20pt,-20pt)--(20pt,-40pt); \draw [>=stealth,<-] (20pt, -40pt) -- (20pt,-60pt);
\draw [>=stealth,->] (-20pt,-20pt)--(0pt,-20pt); \draw (0pt,-20pt)--(20pt,-20pt);
\draw [dashed][>=stealth,->] (-20pt,-60pt)--(0pt,-60pt);  \draw [dashed](0pt, -60pt) -- (20pt,-60pt);
\draw [>=stealth,->] (-20pt,-20pt)--(0pt,-40pt);  \draw(0pt, -40pt) -- (20pt,-60pt);

\node at (7pt,0pt){${h_x}$};
\node at (27pt,-37pt){${h_y}$};
\end{tikzpicture}
\nonumber
\end{eqnarray}
Similar to the $T_x$ transformation, here the basis vector $|g_x,g_y\rangle$ is transformed to 
$|h_x,h_y\rangle:=|[1'3'],[3'4']\rangle=
|t(g_y^{-1} g_x)t^{-1}, tg_yt^{-1}\rangle$ up to a $U(1)$ phase $u_y^t(\Sigma_{1,0})$.
By restricting to the unit squares spanned by $1,2,3,4$ and $1',2',4',3^{\ast}$, one can find that $u_y^t(\Sigma_{1,0})$ is 
associated with the path integral defined on the following 3-manifold:
\begin{eqnarray}\label{TyPhase}
\small
u_y^t(\Sigma_{1,0}):\quad
\begin{tikzpicture}[baseline={(current bounding box.center)}]

\draw [>=stealth,->] (-20pt,-20pt)--(0pt,-20pt);  \draw (0pt, -20pt) -- (20pt,-20pt);
\draw [>=stealth,->] (-20pt,-20pt)--(-5pt,-5pt); \draw(-5pt,-5pt)--(10pt,10pt);
\draw [>=stealth,->] (10pt,10pt)--(30pt,10pt); \draw (30pt,10pt)--(50pt,10pt);
\draw [>=stealth,->] (20pt,-20pt)--(35pt,-5pt); \draw (35pt,-5pt)--(50pt,10pt);
\node at (40pt,-10pt){${g_y}$};
\node at (0pt,-27pt){${g_x}$};
\node at (-28pt,-25pt){${1}$};
\node at (22pt,-25pt){${3}$};

\node at (2pt,10pt){${2}$};
\node at (55pt,10pt){${4}$};

\draw [>=stealth,->] (-20pt,-20+50pt)--(0pt,-20+50pt);  \draw (0pt, -20+50pt) -- (20pt,-20+50pt);
\draw [>=stealth,->] (-20pt,-20+50pt)--(-5pt,-5+50pt); \draw(-5pt,-5+50pt)--(10pt,10+50pt);
\draw [>=stealth,->] (10pt,10+50pt)--(30pt,10+50pt); \draw (30pt,10+50pt)--(50pt,10+50pt);
\draw [>=stealth,->] (20pt,-20+50pt)--(35pt,-5+50pt); \draw (35pt,-5+50pt)--(50pt,10+50pt);

\draw [>=stealth,->] (-20pt,-20+50pt)--(-20pt,5pt); \draw(-20pt,5pt)--(-20pt,-20pt);
\draw [>=stealth,->] (20pt,-20+50pt)--(20pt,5pt); \draw(20pt,5pt)--(20pt,-20pt);
\draw [>=stealth,->] (50pt,10+50pt)--(50pt,35pt); \draw(50pt,35pt)--(50pt,10pt);
\draw [>=stealth,->] [dashed](10pt,10+50pt)--(10pt,35pt); \draw[dashed](10pt,35pt)--(10pt,10pt);

\draw [>=stealth,->][dashed] (-20pt,-20pt)--(15pt,-5pt);\draw[dashed](15pt,-5pt)--(50pt,10pt);
\draw [>=stealth,->] (10pt,60pt)--(15pt,45pt); \draw (15pt,45pt)--(20pt,30pt);

\node at (-30pt,30pt){${1'}$};
\node at (25pt,28pt){${4'}$};

\node at (2pt,60pt){${2'}$};
\node at (58pt,60pt){${3^{\ast}}$};
\node at (-15pt,45pt){${h_y}$};
\node at (22pt,45pt){${h_x}$};

\node at (-30pt,5pt){${t^{-1}}$};

\end{tikzpicture}
\end{eqnarray}
with the ordering: $1'<2'<4'<3^{\ast}<1<2<3<4$.
Again, the group elements on the links in the `time' direction are $[11']=[22']=[34']=[43^{\ast}]=t$. With the flat connection condition one can find
$h_x=t(g_y^{-1}g_x)t^{-1}$ and $h_y=tg_yt^{-1}$.
Similar to \eqref{TxPhase}, the phase $u_y^t(\Sigma_{1,0})$ corresponding to \eqref{TyPhase} 
is the product of seven $3$-cocycles as
\be
\small
\begin{split}
&Y_{[124],[1'2'3^{\ast}]}\cdot Y_{[134],[1'4'3^{\ast}]}\cdot \omega^{-1}([1'2'],[2'4'],[4'3^{\ast}])\\
=&\big[\omega^{-1}([3^{\ast}1],[12],[24])\cdot \omega^{-1}([2'3^{\ast}],[3^{\ast}1],[12])\\
&\cdot \omega^{-1}([1'2'],[2'3^{\ast}],[3^{\ast}1])\big]\cdot
\big[\omega([3^{\ast}1],[13],[34])\\
&\cdot \omega([4'3^{\ast}],[3^{\ast}1],[13])\cdot\omega([1'4'],[4'3^{\ast}],[3^{\ast}1])\big]\\
&\cdot \omega^{-1} ([1'2'],[2'4'],[4'3^{\ast}])\\
=&\omega^{-1}(g_x^{-1}g_y^{-1}\,t^{-1}, g_y, g_x)\cdot \omega^{-1} (t\, g_x\, t^{-1}, g_x^{-1}g_y^{-1} t^{-1}, g_y)\\
&\cdot \omega^{-1}(t\,g_y\,t^{-1}, t\,g_x\,t^{-1}, g_x^{-1}g_y^{-1}\,t^{-1})\cdot \omega(g_x^{-1}g_y^{-1}\,t^{-1},g_x,g_y)\\
&\cdot \omega(t\,g_y\,t^{-1}, g_x^{-1}g_y^{-1}t^{-1}, g_x)\cdot \omega(t\,g_x\,t^{-1}, t\,g_y\,t^{-1}, g_x^{-1}g_y^{-1}t^{-1})\\
&\cdot \omega^{-1}(t\, g_y\, t^{-1}, t\,g_y^{-1}g_x\,t^{-1}, t\,g_y\,t^{-1})
=:u_y^t(\Sigma_{1,0}).
\end{split}
\ee
Then the effect of $T_y^t$ on the basis $|g_x,g_y\rangle$ is
\be
T_y^t \,|g_x,g_y \rangle=u_y^t(\Sigma_{1,0})\, |t(g_y^{-1} g_x)t^{-1}, tg_yt^{-1}\rangle.
\ee
Having known how $T_{x,y}^t$ act on the basis $|g_x,g_y\rangle$, 
now we can define $T_{x,y}:=\frac{1}{|G|}\sum_{t\in G}T_{x,y}^t$.
It can be checked that 
$T_{x,y}^t\cdot A^{t'}=A^{t}\cdot T_{x,y}^{t'}=T_{x,y}^{t\cdot t'}$.
The moduar $S$ and $T$ matrices are defined as
\be\label{SfromT}
S:=\frac{1}{|G|}\sum_{t\in G} S^t, \quad T:=\frac{1}{|G|}\sum_{t\in G} T^t, 
\ee
where $T^t=T_x^t$, and $S^t=T_y^{t'=1}\cdot (T_x^{t'=1})^{-1}\cdot T_y^{t'=1}\cdot A^t
=A^t\cdot T_y^{t'=1}\cdot (T_x^{t'=1})^{-1}\cdot T_y^{t'=1}$. 
Then one can calculate the 
modular $S$ and $T$ matrices with the groundstate bases in \eqref{GSD_torus1}.\cite{HuWanWu1211}
 
Before we end this part, we remark that for the genus-1 torus $\Sigma_{1,0}$, one can construct the 
quasi-particle basis $|g,\chi\rangle$ corresponding to \eqref{AnyonData} in terms of the group-element basis $|g_x,g_y\rangle$, 
so that the $T$ matrix is diagonal with the diagonal elements being topological spins of anyons (see appendix \ref{SubSec: Zp}
for an illustration).
With the quasi-particle basis $|g,\chi\rangle$, one can obtain the modular $S$ and $T$ matrices exactly of the form in
Eqs.\eqref{TopologicalSpin0} and \eqref{Sab1}.\cite{HuWanWu1211}

\subsubsection{Punctured torus}
\label{Sec: Lattice_PuncturedTorus}

Now we generalize the previous procedure to the punctured torus $\Sigma_{1,1}$.
Here we are interested in $\Sigma_{1,1}$ because the genus-2 manifold $\Sigma_{2,0}$ can be obtained by gluing two copies of $\Sigma_{1,1}$
by identifying their punctures.

The simplest configuration for a punctured torus is as follows:
\begin{eqnarray}\label{BasisPuncturedTorus}
\small
\begin{tikzpicture}[baseline={(current bounding box.center)}]
\draw (-20pt,20pt)--(-20pt,0pt); \draw [>=stealth,<-] (-20pt, 0pt) -- (-20pt,-20pt);
\draw (20pt,20pt)--(20pt,0pt); \draw [>=stealth,<-] (20pt, 0pt) -- (20pt,-20pt);
\draw [>=stealth,->] (-20pt,20pt)--(0pt,20pt); \draw (0pt,20pt)--(20pt,20pt);
\draw [>=stealth,->] (-20pt,-20pt)--(0pt,-20pt);  \draw (0pt, -20pt) -- (20pt,-20pt);
\node at (28pt,0pt){${g_y}$};
\node at (0pt,-27pt){${g_x}$};
\node at (0pt,27pt){${g_x}$};
\node at (-28pt,0pt){${g_y}$};
\draw [>=stealth,->](-20pt,-20pt)..controls (0pt,15pt) and (0pt,-10pt)..(-20pt,-20pt);
\node at (0pt,0pt){${k}$};
\end{tikzpicture}
\quad
\leftrightarrow
\quad
\begin{tikzpicture}[baseline={(current bounding box.center)}]
\draw (-20pt,20pt)--(-20pt,5pt); \draw [>=stealth,<-] (-20pt, 5pt) -- (-20pt,-10pt);
\draw (20pt,20pt)--(20pt,0pt); \draw [>=stealth,<-] (20pt, 0pt) -- (20pt,-20pt);
\draw [>=stealth,->] (-20pt,20pt)--(0pt,20pt); \draw (0pt,20pt)--(20pt,20pt);
\draw [>=stealth,->] (-10pt,-20pt)--(5pt,-20pt);  \draw (5pt, -20pt) -- (20pt,-20pt);
\draw [>=stealth,->] (-20pt,-10pt)--(-15pt,-15pt); \draw (-15pt,-15pt)--(-10pt,-20pt);
\node at (27pt,0pt){${g_y}$};
\node at (5pt,-27pt){${g_x}$};
\node at (-19pt, -19pt) {$k$};
\node at (0pt,27pt){${g_x}$};
\node at (-28pt,0pt){${g_y}$};
\node at (-10pt, -26pt){$1$};
\node at (-23pt, -10pt){$\bar{1}$};
\node at (22pt, -26pt){$3$};
\node at (-23pt, 20pt){$2$};
\node at (24pt, 20pt){$4$};
\end{tikzpicture}
\end{eqnarray}
where the left and right configurations are equivalent to each other. 
For the configuration on the right, we identify the edges $[13]$ with $[24]$, $[\bar{1}2]$ with $[34]$, and therefore
there is essentially only one vertex.
The triangulation of the punctured torus is:
\begin{eqnarray}\label{PuncturedTorusBasis1}
\begin{tikzpicture}[baseline={(current bounding box.center)}]
\draw (-20pt,20pt)--(-20pt,5pt); \draw [>=stealth,<-] (-20pt, 5pt) -- (-20pt,-10pt);
\draw (20pt,20pt)--(20pt,0pt); \draw [>=stealth,<-] (20pt, 0pt) -- (20pt,-20pt);
\draw [>=stealth,->] (-20pt,20pt)--(0pt,20pt); \draw (0pt,20pt)--(20pt,20pt);
\draw [>=stealth,->] (-10pt,-20pt)--(5pt,-20pt);  \draw (5pt, -20pt) -- (20pt,-20pt);
\draw [>=stealth,->] (-20pt,-10pt)--(-15pt,-15pt); \draw (-15pt,-15pt)--(-10pt,-20pt);
\node at (27pt,0pt){${g_y}$};
\node at (5pt,-27pt){${g_x}$};
\node at (-19pt, -19pt) {$k$};
\node at (-10pt, -26pt){$1$};
\node at (-23pt, -10pt){$\bar{1}$};
\node at (22pt, -26pt){$3$};
\node at (-23pt, 20pt){$2$};
\node at (24pt, 20pt){$4$};
\node at (0pt,27pt){${g_x}$};
\node at (-28pt,5pt){${g_y}$};
\draw [>=stealth,->] (-20pt,-10pt)--(0pt, 5pt);\draw (0pt,5pt)--(20pt,20pt);
\draw [>=stealth,->] (-10pt,-20pt)--(5pt,0pt);\draw(5pt,0pt)--(20pt, 20pt);
\end{tikzpicture}
\end{eqnarray}
where we take the ordering: $\bar{1}<1<2<3<4$, and the coloring $[13]=[24]=g_x$, $[\bar{1}2]=[34]=g_y$, and $[\bar{1}1]=k$.
The flat connection condition is [see Eq.\eqref{fundamental_group_puncture}]
\be\label{flat_connection_pucturedtorus}
k\cdot [g_x, g_y]=1,
\ee
\textit{i.e.}, $k\cdot g_y^{-1}g_x^{-1}g_yg_x=1$, where $1$ is the identity group element.
Here the holonomy $k$ measures the `flux' of anyon $z$ in  
the single-puncture torus basis in \eqref{PuncturedTorusBasis}.

Denoting the punctured-tours basis in \eqref{PuncturedTorusBasis1} as $|[13], [34]; [\bar{1}1]\rangle=|g_x, g_y; k\rangle$, we will study how modular transformations act on this basis vector.
It is known that MCG$(\Sigma_{1,1})$ is generated by the same $\mathfrak{s}$ and $\mathfrak{t}$ (or equivalently the two Dehn twists $\mathfrak{t}_x$ and 
$\mathfrak{t}_y$) as those for MCG$(\Sigma_{1,0})$. The difference is that for a punctured torus one has $\mathfrak{s}^4=\mathfrak{r}^{-1}$, where $\mathfrak{r}$
represents the Dehn twist along the closed curve that encloses the puncture, and for the torus without puncture one simply has $\mathfrak{s}^4=1$.
Following the procedure in the previous section, we denote $T_{x,y}:=\frac{1}{|G|}\sum_{t\in G}T_{x,y}^t$ as the representation of Dehn twists $\mathfrak{t}_{x,y}$.
Then one can find the effect of $T_x^t$ on $|g_x, g_y; k\rangle$ as
\be\label{Tx_punctured_torus}
T_x^t \, |g_x,g_y;k\rangle=u_x^t(\Sigma_{1,1}) |t g_x t^{-1}, t (g_x^{-1}\cdot g_y )t^{-1};tkt^{-1}\rangle,
\ee
where the $U(1)$ phase $u_x^t(\Sigma_{1,1})$ is associated to the path integral on the following 3-manifold:
(It is helpful to compare this with the phase $u_x^t(\Sigma_{1,0})$ in \eqref{TxPhase}.)
\begin{eqnarray}\label{TxPhasePuncture}
\footnotesize
u_x^t(\Sigma_{1,1}):\quad
\begin{tikzpicture}[baseline={(current bounding box.center)}]

\draw [>=stealth,->] (-5pt,-20pt)--(10pt,-20pt);  \draw (10pt, -20pt) -- (20pt,-20pt);

\draw [>=stealth,->] (-10pt,-10pt)--(0pt,0pt); \draw(0pt,0pt)--(10pt,10pt);

\draw [>=stealth,->] (-10pt,-10pt)--(-7.5pt,-15pt); \draw(-7.5pt,-15pt)--(-5pt,-20pt);

\draw [>=stealth,->][dashed] (-5pt,-20pt)--(22.5pt,-5pt); \draw[dashed](22.5pt,-5pt)--(50pt,10pt);

\draw [>=stealth,->][dashed] (-10pt,-10pt)--(20pt,0pt); \draw[dashed](20pt,0pt)--(50pt,10pt);

\draw [>=stealth,->] (10pt,10pt)--(30pt,10pt); \draw (30pt,10pt)--(50pt,10pt);
\draw [>=stealth,->] (20pt,-20pt)--(35pt,-5pt); \draw (35pt,-5pt)--(50pt,10pt);
\node at (40pt,-10pt){${g_y}$};
\node at (10pt,-27pt){${g_x}$};
\node at (-5pt,-25pt){${1}$};
\node at (-15pt,-12pt){${\bar{1}}$};
\node at (22pt,-25pt){${3}$};

\node at (5pt,12pt){${2}$};
\node at (55pt,10pt){${4}$};

\draw [>=stealth,->] (-5pt,-20+50pt)--(10pt,-20+50pt);  \draw (10pt, -20+50pt) -- (20pt,-20+50pt);

\draw [>=stealth,->] (-10pt,-10+50pt)--(0pt,0+50pt); \draw(0pt,0+50pt)--(10pt,10+50pt);

\draw [>=stealth,->] (-10pt,-10+50pt)--(-7.5pt,-15+50pt); \draw(-7.5pt,-15+50pt)--(-5pt,-20+50pt);

\draw [>=stealth,->] (-5pt,30pt)--(2.5pt,45pt); \draw(2.5pt,45pt)--(10pt,60pt);

\draw [>=stealth,->] (10pt,10+50pt)--(30pt,10+50pt); \draw (30pt,10+50pt)--(50pt,10+50pt);
\draw [>=stealth,->] (20pt,-20+50pt)--(35pt,-5+50pt); \draw (35pt,-5+50pt)--(50pt,10+50pt);

\draw [>=stealth,->] (-5pt,-20+50pt)--(-5pt,5pt); \draw(-5pt,5pt)--(-5pt,-20pt);
\draw [>=stealth,->] (-10pt,-10+50pt)--(-10pt,15pt); \draw(-10pt,15pt)--(-10pt,-10pt);

\draw [>=stealth,->] (20pt,-20+50pt)--(20pt,5pt); \draw(20pt,5pt)--(20pt,-20pt);

\draw [>=stealth,->] (50pt,10+50pt)--(50pt,35pt); \draw(50pt,35pt)--(50pt,10pt);
\draw [>=stealth,->] [dashed](10pt,10+50pt)--(10pt,35pt); \draw[dashed](10pt,35pt)--(10pt,10pt);

\draw [>=stealth,->] (20pt,30pt)--(15pt,45pt); \draw(15pt,45pt)--(10pt,60pt);

\node at (-5pt,28pt){${1'}$};
\node at (25pt,28pt){${3'}$};
\node at (-15pt,-12+50pt){${\bar{1}'}$};

\node at (5pt,62pt){${4'}$};
\node at (58pt,60pt){${2^{\ast}}$};
\node at (8pt,25pt){${h_x}$};
\node at (22pt,45pt){${h_y}$};

\node at (60pt,40pt){${t^{-1}}$};

\end{tikzpicture}
\end{eqnarray}
The ordering is taken as: $\bar{1}'<\bar{1}<1'<3'<4'<2^{\ast}<1<2<3<4$.
The group elements living on the vertical links are $[\bar{1}\bar{1}']=[11']=[24']=[33']=[42^{\ast}]=t$.
With the flat connection condition, one can find that $|h_x,h_y;k'\rangle:=|[1'3'],[3'4'];[\bar{1}'1']\rangle=|tg_xt^{-1}, t(g_x^{-1}g_y)t^{-1};tkt^{-1}\rangle$.
The phase $u_x^t(\Sigma_{1,1})$ in \eqref{TxPhasePuncture} is composed of $3\times 3+2=11$ 3-cocycles as follows:
\be\label{u_x^t_PuncturedTorus}
\small
\begin{split}
u_x^t(\Sigma_{1,1})=&Y_{[\bar{1}24],[\bar{1}'4'2^{\ast}]}\cdot Y_{[\bar{1}14],[\bar{1}'1'2^{\ast}]}\cdot Y_{[134],[1'3'2^{\ast}]}\\
&\cdot \omega([\bar{1}'1'],[1'4'],[4'2^{\ast}])\cdot \omega([1'3'],[3'4'],[4'2^{\ast}]),
\end{split}
\ee
where each $Y_{[abc],[a'b'c']}$ is composed of three 3-cocyles, corresponding to the triangular prism spanned by $[abc]$ and $[a'b'c']$.
The explicit expression of $u_x^t(\Sigma_{1,1})$ can be found in Eqs.\eqref{u_x^t_PuncturedTorus_appendix}-\eqref{LastTwoCocyle_appendix} 
in Appendix \ref{Sec: appendix_PI}. The last two 3-cocyles in Eq.\eqref{u_x^t_PuncturedTorus} correspond to the following two tetrahedra:
\begin{eqnarray}\label{TxPhasePuncture_last2_plot}
\footnotesize
\begin{tikzpicture}[baseline={(current bounding box.center)}]

\draw [dashed][gray][>=stealth,->] (-5pt,-20pt)--(10pt,-20pt);  \draw [dashed][gray](10pt, -20pt) -- (20pt,-20pt);

\draw [>=stealth,->] (-10pt,-10pt)--(0pt,0pt); \draw(0pt,0pt)--(10pt,10pt);

\draw [>=stealth,->] (-10pt,-10pt)--(-7.5pt,-15pt); \draw(-7.5pt,-15pt)--(-5pt,-20pt);

\draw [>=stealth,->] (-5pt,-20pt)--(22.5pt,-5pt); \draw(22.5pt,-5pt)--(50pt,10pt);

\draw [>=stealth,->][dashed] (-10pt,-10pt)--(20pt,0pt); \draw[dashed](20pt,0pt)--(50pt,10pt);

\draw [>=stealth,->] (10pt,10pt)--(30pt,10pt); \draw (30pt,10pt)--(50pt,10pt);
\draw [dashed][gray][>=stealth,->] (20pt,-20pt)--(35pt,-5pt); \draw [dashed][gray](35pt,-5pt)--(50pt,10pt);

\draw [>=stealth,->] (-5pt,-20pt)--(2.5pt,-5pt); \draw(2.5pt,-5pt)--(10pt,10pt);

\node at (-5pt,-26pt){${1'}$};
\node at (-15pt,-12pt){${\bar{1}'}$};
\node at (22pt,-26pt){$\color{gray}{3'}$};

\node at (5pt,12pt){${4'}$};
\node at (55pt,10pt){${2^{\ast}}$};

\end{tikzpicture}
\quad
\begin{tikzpicture}[baseline={(current bounding box.center)}]

\draw [>=stealth,->] (-5pt,-20pt)--(10pt,-20pt);  \draw (10pt, -20pt) -- (20pt,-20pt);

\draw [gray][dashed][>=stealth,->] (-10pt,-10pt)--(0pt,0pt); \draw[gray][dashed](0pt,0pt)--(10pt,10pt);

\draw [dashed][gray][>=stealth,->] (-10pt,-10pt)--(-7.5pt,-15pt); \draw [dashed][gray](-7.5pt,-15pt)--(-5pt,-20pt);

\draw [dashed][>=stealth,->] (-5pt,-20pt)--(22.5pt,-5pt); \draw[dashed](22.5pt,-5pt)--(50pt,10pt);


\draw [>=stealth,->] (10pt,10pt)--(30pt,10pt); \draw (30pt,10pt)--(50pt,10pt);
\draw [>=stealth,->] (20pt,-20pt)--(35pt,-5pt); \draw (35pt,-5pt)--(50pt,10pt);

\draw [>=stealth,->] (-5pt,-20pt)--(2.5pt,-5pt); \draw(2.5pt,-5pt)--(10pt,10pt);

\draw [>=stealth,->] (20pt,-20pt)--(15pt,-5pt); \draw(15pt,-5pt)--(10pt,10pt);

\node at (-5pt,-26pt){${1'}$};
\node at (-15pt,-12pt){$\color{gray}{\bar{1}'}$};
\node at (22pt,-26pt){${3'}$};

\node at (5pt,12pt){${4'}$};
\node at (55pt,10pt){${2^{\ast}}$};

\end{tikzpicture}
\end{eqnarray}
Similarly, the Dehn twist along $y$ direction acts on the basis $|g_x,g_y;k\rangle$ in the following way
\be\label{Ty_punctured_torus}
T_y^t|g_x,g_y;k\rangle=u_y^t(\Sigma_{1,1})|tg_y^{-1}g_xt^{-1}, tg_yt_{-1}; tkt^{-1}\rangle.
\ee
The $U(1)$ phase $u_y^t(\Sigma_{1,1})$ is associated to the 3-manifold 
(It is helpful to compare this with the phase $u_y^t(\Sigma_{1,0})$ in \eqref{TyPhase}.)
\begin{eqnarray}\label{TyPhasePuncture}
\footnotesize
u_y^t(\Sigma_{1,1}):\quad
\begin{tikzpicture}[baseline={(current bounding box.center)}]
\draw [>=stealth,->] (-5pt,-20pt)--(10pt,-20pt);  \draw (10pt, -20pt) -- (20pt,-20pt);
\draw [>=stealth,->] (-10pt,-10pt)--(0pt,0pt); \draw(0pt,0pt)--(10pt,10pt);
\draw [>=stealth,->] (-10pt,-10pt)--(-7.5pt,-15pt); \draw(-7.5pt,-15pt)--(-5pt,-20pt);
\draw [>=stealth,->][dashed] (-5pt,-20pt)--(22.5pt,-5pt); \draw[dashed](22.5pt,-5pt)--(50pt,10pt);
\draw [>=stealth,->][dashed] (-10pt,-10pt)--(20pt,0pt); \draw[dashed](20pt,0pt)--(50pt,10pt);

\draw [>=stealth,->] (10pt,10pt)--(30pt,10pt); \draw (30pt,10pt)--(50pt,10pt);
\draw [>=stealth,->] (20pt,-20pt)--(35pt,-5pt); \draw (35pt,-5pt)--(50pt,10pt);
\node at (40pt,-10pt){${g_y}$};
\node at (10pt,-27pt){${g_x}$};
\node at (-5pt,-25pt){${1}$};
\node at (-15pt,-12pt){${\bar{1}}$};
\node at (22pt,-25pt){${3}$};

\node at (5pt,12pt){${2}$};
\node at (55pt,10pt){${4}$};

\draw [>=stealth,->] (-5pt,-20+50pt)--(10pt,-20+50pt);  \draw (10pt, -20+50pt) -- (20pt,-20+50pt);
\draw [>=stealth,->] (-10pt,-10+50pt)--(0pt,0+50pt); \draw(0pt,0+50pt)--(10pt,10+50pt);
\draw [>=stealth,->] (-10pt,-10+50pt)--(-7.5pt,-15+50pt); \draw(-7.5pt,-15+50pt)--(-5pt,-20+50pt);
\draw [>=stealth,->] (-5pt,30pt)--(2.5pt,45pt); \draw(2.5pt,45pt)--(10pt,60pt);
\draw [>=stealth,->] (10pt,10+50pt)--(30pt,10+50pt); \draw (30pt,10+50pt)--(50pt,10+50pt);
\draw [>=stealth,->] (20pt,-20+50pt)--(35pt,-5+50pt); \draw (35pt,-5+50pt)--(50pt,10+50pt);
\draw [>=stealth,->] (-5pt,-20+50pt)--(-5pt,5pt); \draw(-5pt,5pt)--(-5pt,-20pt);
\draw [>=stealth,->] (-10pt,-10+50pt)--(-10pt,15pt); \draw(-10pt,15pt)--(-10pt,-10pt);

\draw [>=stealth,->] (20pt,-20+50pt)--(20pt,5pt); \draw(20pt,5pt)--(20pt,-20pt);
\draw [>=stealth,->] (50pt,10+50pt)--(50pt,35pt); \draw(50pt,35pt)--(50pt,10pt);
\draw [>=stealth,->] [dashed](10pt,10+50pt)--(10pt,35pt); \draw[dashed](10pt,35pt)--(10pt,10pt);
\draw [>=stealth,->](10pt,60pt)--(15pt,45pt); \draw (15pt,45pt)--(20pt,30pt);

\node at (-5pt,28pt){${1'}$};
\node at (25pt,28pt){${4'}$};
\node at (-15pt,-12+50pt){${\bar{1}'}$};

\node at (5pt,62pt){${2'}$};
\node at (58pt,60pt){${3^{\ast}}$};
\node at (-10pt,49pt){${h_y}$};
\node at (22pt,45pt){${h_x}$};

\node at (60pt,40pt){${t^{-1}}$};
\end{tikzpicture}
\end{eqnarray}
The ordering is taken as: $\bar{1}'<\bar{1}<1'<2'<4'<3^{\ast}<1<2<3<4$.
Again, the group elements living on the vertical links are $[\bar{1}\bar{1}']=[11']=[22']=[34']=[43^{\ast}]=t$.
With the flat connection condition, one can find the old basis is transformed to 
$|h_x,h_y;k'\rangle:=|[2'4'],[\bar{1}'2'];[\bar{1}'1']\rangle=|t(g_y^{-1}g_x)t^{-1},tg_yt^{-1};tkt^{-1}\rangle$.
The phase $u_y^t(\Sigma_{1,1})$ in \eqref{TyPhasePuncture} is composed of $3\times 3+2=11$ three-cocycles:
\be\label{u_y^t_PuncturedTorus}
\small
\begin{split}
u_y^t(\Sigma_{1,1})=&Y_{[\bar{1}24],[\bar{1}'2'3^{\ast}]}\cdot Y_{[\bar{1}14],[\bar{1}'1'3^{\ast}]}\cdot Y_{[134],[1'4'3^{\ast}]}\\
&\cdot \omega([\bar{1}'1'],[1'2'],[2'3^{\ast}])\cdot \omega([1'2'],[2'4'],[4'3^{\ast}])^{-1},
\end{split}
\ee
where each $Y_{[abc],[a'b'c']}$ corresponding to a triangular prism is the product of three 3-cocycles. 
An explicit expression of $u_y^t(\Sigma_{1,1})$ can be found in 
Eqs.\eqref{u_y^t_PuncturedTorus_appendix}-\eqref{LastTwoCocyle_Ty_appendix} in Appendix \ref{Sec: appendix_PI}.
The last two terms in Eq.\eqref{u_y^t_PuncturedTorus} are contributed by the following two tetrahedra:
\begin{eqnarray}\label{TyPhasePuncture_last2_plot}
\footnotesize
\begin{tikzpicture}[baseline={(current bounding box.center)}]

\draw [dashed][gray][>=stealth,->] (-5pt,-20pt)--(10pt,-20pt);  \draw [dashed][gray](10pt, -20pt) -- (20pt,-20pt);

\draw [>=stealth,->] (-10pt,-10pt)--(0pt,0pt); \draw(0pt,0pt)--(10pt,10pt);

\draw [>=stealth,->] (-10pt,-10pt)--(-7.5pt,-15pt); \draw(-7.5pt,-15pt)--(-5pt,-20pt);

\draw [>=stealth,->] (-5pt,-20pt)--(22.5pt,-5pt); \draw(22.5pt,-5pt)--(50pt,10pt);

\draw [>=stealth,->][dashed] (-10pt,-10pt)--(20pt,0pt); \draw[dashed](20pt,0pt)--(50pt,10pt);

\draw [>=stealth,->] (10pt,10pt)--(30pt,10pt); \draw (30pt,10pt)--(50pt,10pt);
\draw [dashed][gray][>=stealth,->] (20pt,-20pt)--(35pt,-5pt); \draw [dashed][gray](35pt,-5pt)--(50pt,10pt);

\draw [>=stealth,->] (-5pt,-20pt)--(2.5pt,-5pt); \draw(2.5pt,-5pt)--(10pt,10pt);

\node at (-5pt,-26pt){${1'}$};
\node at (-15pt,-12pt){${\bar{1}'}$};
\node at (22pt,-26pt){$\color{gray}{4'}$};

\node at (5pt,12pt){${2'}$};
\node at (55pt,10pt){${3^{\ast}}$};

\end{tikzpicture}
\quad
\begin{tikzpicture}[baseline={(current bounding box.center)}]

\draw [>=stealth,->] (-5pt,-20pt)--(10pt,-20pt);  \draw (10pt, -20pt) -- (20pt,-20pt);

\draw [gray][dashed][>=stealth,->] (-10pt,-10pt)--(0pt,0pt); \draw[gray][dashed](0pt,0pt)--(10pt,10pt);

\draw [dashed][gray][>=stealth,->] (-10pt,-10pt)--(-7.5pt,-15pt); \draw [dashed][gray](-7.5pt,-15pt)--(-5pt,-20pt);

\draw [dashed][>=stealth,->] (-5pt,-20pt)--(22.5pt,-5pt); \draw[dashed](22.5pt,-5pt)--(50pt,10pt);


\draw [>=stealth,->] (10pt,10pt)--(30pt,10pt); \draw (30pt,10pt)--(50pt,10pt);
\draw [>=stealth,->] (20pt,-20pt)--(35pt,-5pt); \draw (35pt,-5pt)--(50pt,10pt);

\draw [>=stealth,->] (-5pt,-20pt)--(2.5pt,-5pt); \draw(2.5pt,-5pt)--(10pt,10pt);

\draw [>=stealth,->]  (10pt,10pt)--(15pt,-5pt); \draw (15pt,-5pt)--(20pt,-20pt); 

\node at (-5pt,-26pt){${1'}$};
\node at (-15pt,-12pt){$\color{gray}{\bar{1}'}$};
\node at (22pt,-26pt){${4'}$};

\node at (5pt,12pt){${2'}$};
\node at (55pt,10pt){${3^{\ast}}$};

\end{tikzpicture}
\end{eqnarray}
and the concrete expression can be found in Eq.\eqref{LastTwoCocyle_Ty_appendix}.

Till now, we have studied how to perform Denh twists on the basis $|g_x,g_y;k\rangle$ in \eqref{BasisPuncturedTorus}
for a punctured torus. In the next section, we will use these basic operations to study the modular transformations
of the degenerate ground states on a genus-2 manifold.

\subsubsection{Genus two}

The simplest triangulation of a genus-2 manifold is an octagon as follows:
\begin{eqnarray}\label{Genus2Triangle0}
\begin{tikzpicture}[baseline={(current bounding box.center)}]
\draw [>=stealth,->](-15.3073pt,36.9552pt)--(0pt,36.9552pt); \draw (0pt,36.9552pt)--(15.3073pt,36.9552pt);
\draw [>=stealth,->](15.3073pt,36.9552pt)--(26.1312pt,26.1312pt); \draw (26.1312pt,26.1312pt)--(36.9552pt,15.3073pt);
\draw [>=stealth,->](15.3073pt,-36.9552pt)--(0pt,-36.9552pt); \draw (0pt,-36.9552pt)--(-15.3073pt,-36.9552pt);
\draw [>=stealth,->](-15.3073pt,-36.9552pt)--(-26.1312pt,-26.1312pt); \draw (-26.1312pt,-26.1312pt)--(-36.9552pt,-15.3073pt);
\draw [>=stealth,->](15.3073pt,-36.9552pt)--(26.1312pt,-26.1312pt); \draw (26.1312pt,-26.1312pt)--(36.9552pt,-15.3073pt);
\draw [>=stealth,->](-15.3073pt,36.9552pt)--(-26.1312pt,26.1312pt); \draw (-26.1312pt,26.1312pt)--(-36.9552pt,15.3073pt);
\draw [>=stealth,->](36.9552pt,-15.3073pt)--(36.9552pt,0pt); \draw (36.9552pt,0pt)--(36.9552pt,15.3073pt);
\draw [>=stealth,->](-36.9552pt,15.3073pt)--(-36.9552pt,0pt); \draw (-36.9552pt,0pt)--(-36.9552pt,-15.3073pt);
\node at (-15pt,42pt){$\bar{1}$};
\node at (15pt,42pt){$2$};

\node at (-15pt,-42pt){$6$};
\node at (15pt,-42pt){$1$};

\node at (41pt,16pt){$4$};
\node at (41pt,-16pt){$3$};

\node at (-41pt,16pt){$5$};
\node at (-41pt,-16pt){$7$};

\draw [>=stealth,->](-15.3073pt,36.9552pt)--(0pt,0pt); \draw (0pt,0pt)--(15.3073pt,-36.9552pt);
\draw [>=stealth,->](-15.3073pt,36.9552pt)--(10.8239pt,26.1312pt); \draw (10.8239pt,26.1312pt)--(36.9552pt,15.3073pt); 
\draw [>=stealth,->](15.3073pt,-36.9552pt)--(-10.8239pt,-26.1312pt); \draw (-10.8239pt,-26.1312pt)--(-36.9552pt,-15.3073pt); 
\draw [>=stealth,->](15.3073pt,-36.9552pt)--(26.1312pt,-10.8239pt); \draw (26.1312pt,-10.8239pt)--(36.9552pt,15.3073pt);
\draw [>=stealth,->](-15.3073pt,36.9552pt)--(-26.1312pt,10.8239pt); \draw (-26.1312pt,10.8239pt)--(-36.9552pt,-15.3073pt);

\end{tikzpicture}
\end{eqnarray}
where we identify the edges $[13]$ with $[24]$, $[34]$ with $[\bar{1}2]$, $[16]$ with $[57]$, and $[67]$ with $[\bar{1}5]$, 
and take the ordering $\bar{1}<1<2<3<4<5<6<7$. This triangulation has six triangle faces and only one vertex.

To compare with the configuration of a punctured torus, it is convenient to replot the octagon in \eqref{Genus2Triangle0} in the following way:
\begin{eqnarray}\label{Genus2Triangle}
\begin{small}
\begin{tikzpicture}[baseline={(current bounding box.center)}]
\draw (-20pt,20pt)--(-20pt,0pt); \draw [>=stealth,<-] (-20pt, 0pt) -- (-20pt,-10pt);
\draw (20pt,20pt)--(20pt,0pt); \draw [>=stealth,<-] (20pt, 0pt) -- (20pt,-20pt);
\draw [>=stealth,->] (-20pt,20pt)--(0pt,20pt); \draw (0pt,20pt)--(20pt,20pt);
\draw [>=stealth,->] (-10pt,-20pt)--(0pt,-20pt);  \draw (0pt, -20pt) -- (20pt,-20pt);
\draw [>=stealth,->] (-20pt,-10pt)--(-15pt,-15pt); \draw (-15pt,-15pt)--(-10pt,-20pt);
\node at (30pt,0pt){${g_{1,y}}$};
\node at (10pt,-26pt){${g_{1,x}}$};
\node at (-19pt, -19pt) {$k$};
\node at (-6pt, -26pt){$1$};
\node at (-25pt, -5pt){$\bar{1}$};
\node at (24pt, -20pt){$3$};
\node at (-23pt, 20pt){$2$};
\node at (24pt, 20pt){$4$};
\draw [>=stealth,->] (-20pt,-10pt)--(0pt, 5pt);\draw (0pt,5pt)--(20pt,20pt);
\draw [>=stealth,->] (-10pt,-20pt)--(5pt,0pt);\draw(5pt,0pt)--(20pt, 20pt);

\draw [>=stealth,->] (-20pt,-10pt)--(-35pt,-10pt); \draw(-35pt,-10pt)--(-50pt,-10pt);
\draw [>=stealth,->] (-10pt,-20pt)--(-10pt,-35pt);\draw(-10pt,-35pt)--(-10pt,-50pt);
\draw (-50pt,-50pt)--(-30pt,-50pt); \draw [>=stealth,<-] (-30pt,-50pt)--(-10pt,-50pt);
\draw (-50pt,-50pt)--(-50pt,-30pt); \draw[>=stealth,<-] (-50pt,-30pt)--(-50pt,-10pt);
\node at (-53pt, -8pt){$5$};
\node at (-53pt, -50pt){$7$};
\node at (-6pt, -50pt){$6$};

\draw [>=stealth,->] (-10pt,-20pt)--(-30pt,-35pt); \draw (-30pt,-35pt)--(-50pt,-50pt);
\draw [>=stealth,->] (-20pt,-10pt)--(-35pt,-30pt); \draw (-35pt,-30pt)--(-50pt,-50pt);

\node at (-30pt,-55pt){${g_{2,x}}$};
\node at (-58pt,-30pt){${g_{2,y}}$};
\end{tikzpicture}
\end{small}
\end{eqnarray}
where we have colored the triangulation with $[13]=[24]=g_{1,x}$, $[\bar{1}2]=[34]=g_{1,y}$,
$[\bar{1}5]=[67]=g_{2,x}$, $[16]=[57]=g_{2,y}$, and $[\bar{1}1]=k$, with $g_{i,x}$, $g_{i,y}$, $k\in G$.
We denote the basis vector corresponding to the configuration in \eqref{Genus2Triangle} as 
\be
|g_{1,x},g_{1,y}; g_{2,x}, g_{2,y}\rangle.
\ee
By comparing with the punctured-torus basis in \eqref{PuncturedTorusBasis1}, one can find that the genus-2
basis in \eqref{Genus2Triangle}
can be obtained by gluing two copies of punctured-torus bases along the puncture as follows:
\begin{eqnarray}\label{Genus2_gluing}
\begin{small}
\begin{tikzpicture}[baseline={(current bounding box.center)}]
\draw (-20+5pt,20+5pt)--(-20+5pt,0+5pt); \draw [>=stealth,<-] (-20+5pt, 0+5pt) -- (-20+5pt,-10+5pt);
\draw (20+5pt,20+5pt)--(20+5pt,0+5pt); \draw [>=stealth,<-] (20+5pt, 0+5pt) -- (20+5pt,-20+5pt);
\draw [>=stealth,->] (-20+5pt,20+5pt)--(0+5pt,20+5pt); \draw (0+5pt,20+5pt)--(20+5pt,20+5pt);
\draw [>=stealth,->] (-10+5pt,-20+5pt)--(0+5pt,-20+5pt);  \draw (0+5pt, -20+5pt) -- (20+5pt,-20+5pt);
\draw [>=stealth,->] (-20+5pt,-10+5pt)--(-15+5pt,-15+5pt); \draw (-15+5pt,-15+5pt)--(-10+5pt,-20+5pt);
\node at (30+5pt,0+5pt){${g_{1,y}}$};
\node at (10+5pt,-26+5pt){${g_{1,x}}$};

\node at (-6+5pt, -26+5pt){$1$};
\node at (-25+5pt, -5+5pt){$\bar{1}$};
\node at (24+5pt, -20+5pt){$3$};
\node at (-23+5pt, 20+5pt){$2$};
\node at (24+5pt, 20+5pt){$4$};
\draw [>=stealth,->] (-20+5pt,-10+5pt)--(0+5pt, 5+5pt);\draw (0+5pt,5+5pt)--(20+5pt,20+5pt);
\draw [>=stealth,->] (-10+5pt,-20+5pt)--(5+5pt,0+5pt);\draw(5+5pt,0+5pt)--(20+5pt, 20+5pt);
\node at (-6pt, -6pt) {$k$};


\node at (-6pt, -26pt){$1'$};
\node at (-25pt, -5pt){$\bar{1}'$};

\draw [>=stealth,->] (-20pt,-10pt)--(-15pt,-15pt); \draw (-15pt,-15pt)--(-10pt,-20pt);
\draw [>=stealth,->] (-20pt,-10pt)--(-35pt,-10pt); \draw(-35pt,-10pt)--(-50pt,-10pt);
\draw [>=stealth,->] (-10pt,-20pt)--(-10pt,-35pt);\draw(-10pt,-35pt)--(-10pt,-50pt);
\draw (-50pt,-50pt)--(-30pt,-50pt); \draw [>=stealth,<-] (-30pt,-50pt)--(-10pt,-50pt);
\draw (-50pt,-50pt)--(-50pt,-30pt); \draw[>=stealth,<-] (-50pt,-30pt)--(-50pt,-10pt);
\node at (-53pt, -8pt){$5$};
\node at (-53pt, -50pt){$7$};
\node at (-6pt, -50pt){$6$};
\node at (-19pt, -19pt) {$k'$};
\draw [>=stealth,->] (-10pt,-20pt)--(-30pt,-35pt); \draw (-30pt,-35pt)--(-50pt,-50pt);
\draw [>=stealth,->] (-20pt,-10pt)--(-35pt,-30pt); \draw (-35pt,-30pt)--(-50pt,-50pt);

\node at (-30pt,-55pt){${g_{2,x}}$};
\node at (-58pt,-30pt){${g_{2,y}}$};
\end{tikzpicture}
\end{small}
\end{eqnarray}
with the holonomy around the puncture identified, \textit{i.e.}, $k=k'$.

For the ground state of the Hamiltonian in Eq.\eqref{Hamiltonian}, the terms $B_f$ enforce the flat 
connection condition for each face (triangle), and then we have
\be\label{Genus2_FlatConnection}
[g_{2,x}, g_{2,y}]\cdot [g_{1,x}, g_{1,y}]=1,
\ee
where the commutation relation $[g,h]$ is defined in Eq.\eqref{Def_commutation}.
The holonomy $k$ is determined by $k\cdot [g_{1,x}, g_{1,y}]=k^{-1}\cdot  [g_{2,x}, g_{2,y}]=1$.
For the basis $|g_{1,x},g_{1,y}; g_{2,x}, g_{2,y}\rangle$ in \eqref{Genus2Triangle},
one can find that $A^t$ in Eq.\eqref{Hamiltonian} acts on it as 
\be\label{At_genus2}
\small
\begin{split}
&A^t\, |g_{1,x},g_{1,y}; g_{2,x}, g_{2,y}\rangle\\
=&\eta^t(g_{1,x}, g_{1,y}; g_{2,x}, g_{2,y} ) \, |t g_{1,x}t^{-1}, tg_{1,y}t^{-1}; t g_{2,x}t^{-1}, t g_{2,y}t^{-1}\rangle
\end{split}
\ee
The $U(1)$ phase 
$\eta^t(g_{1,x}, g_{1,y}; g_{2,x}, g_{2,y} )$, also written as $\eta^t(\Sigma_{2,0})$ for brevity,
 corresponds to the path integral over the following 3-manifold:

\begin{eqnarray}\label{GaugeTransformGenus2}
\footnotesize
\eta^t(\Sigma_{2,0}):\quad
\begin{tikzpicture}[baseline={(current bounding box.center)}]

\draw [>=stealth,->] (-5pt,-20pt)--(10pt,-20pt);  \draw (10pt, -20pt) -- (20pt,-20pt);

\draw [>=stealth,->] (-10pt,-10pt)--(0pt,0pt); \draw(0pt,0pt)--(10pt,10pt);

\draw [dashed][>=stealth,->] (-10pt,-10pt)--(-7.5pt,-15pt); \draw[dashed](-7.5pt,-15pt)--(-5pt,-20pt);

\draw [>=stealth,->][dashed] (-5pt,-20pt)--(22.5pt,-5pt); \draw[dashed](22.5pt,-5pt)--(50pt,10pt);

\draw [>=stealth,->][dashed] (-10pt,-10pt)--(20pt,0pt); \draw[dashed](20pt,0pt)--(50pt,10pt);

\draw [>=stealth,->] (10pt,10pt)--(30pt,10pt); \draw (30pt,10pt)--(50pt,10pt);
\draw [>=stealth,->] (20pt,-20pt)--(35pt,-5pt); \draw (35pt,-5pt)--(50pt,10pt);
\node at (40pt,-10pt){\textcolor{teal}{${g_{1,y}}$}};
\node at (10pt,-27pt){\textcolor{teal}{${g_{1,x}}$}};
\node at (-5pt,-25pt){${1}$};
\node at (-15pt,-5pt){${\bar{1}}$};
\node at (22pt,-25pt){${3}$};

\node at (2pt,10pt){${2}$};
\node at (55pt,10pt){${4}$};

\draw [>=stealth,->] (-5pt,-20+50pt)--(10pt,-20+50pt);  \draw (10pt, -20+50pt) -- (20pt,-20+50pt);

\draw [>=stealth,->] (-10pt,-10+50pt)--(0pt,0+50pt); \draw(0pt,0+50pt)--(10pt,10+50pt);

\draw [>=stealth,->] (-10pt,-10+50pt)--(-7.5pt,-15+50pt); \draw(-7.5pt,-15+50pt)--(-5pt,-20+50pt);

\draw [>=stealth,->] (-5pt,30pt)--(22.5pt,45pt);\draw(22.5pt,45pt)--(10+40pt,10+50pt);

\draw [>=stealth,->] (10pt,10+50pt)--(30pt,10+50pt); \draw (30pt,10+50pt)--(50pt,10+50pt);
\draw [>=stealth,->] (20pt,-20+50pt)--(35pt,-5+50pt); \draw (35pt,-5+50pt)--(50pt,10+50pt);

\draw [>=stealth,->] (-5pt,-20+50pt)--(-5pt,5pt); \draw(-5pt,5pt)--(-5pt,-20pt);
\draw [dashed][>=stealth,->] (-10pt,-10+50pt)--(-10pt,15pt); \draw[dashed](-10pt,15pt)--(-10pt,-10pt);

\draw [>=stealth,->] (20pt,-20+50pt)--(20pt,5pt); \draw(20pt,5pt)--(20pt,-20pt);

\draw [>=stealth,->] (50pt,10+50pt)--(50pt,35pt); \draw(50pt,35pt)--(50pt,10pt);
\draw [>=stealth,->] [dashed](10pt,10+50pt)--(10pt,35pt); \draw[dashed](10pt,35pt)--(10pt,10pt);

\draw [>=stealth,->] (-10pt,-10+50pt)--(20pt,50pt);\draw(20pt,50pt)--(10+40pt,10+50pt);

\node at (-5pt,25pt){${1'}$};
\node at (25pt,28pt){${3'}$};
\node at (-13pt,-12+57pt){${\bar{1}'}$};

\node at (2pt,60pt){${2'}$};
\node at (58pt,60pt){${4'}$};
\node at (8pt,25pt){\textcolor{teal}{${h_{1,x}}$}};

\node at (60pt,35pt){\textcolor{teal}{${t^{-1}}$}};


\draw [>=stealth,->] (-10pt,-10+50pt)--(-22.5pt,-10+50pt);\draw(-22.5pt,-10+50pt)--(-35pt,-10+50pt);

\draw [>=stealth,->] (-5pt,-20+50pt)--(-15pt,-30+50pt); \draw(-15pt,-30+50pt)--(-25pt,-40+50pt);

\draw [>=stealth,->] (-25pt,-40+50pt)--(-25pt,-15pt); \draw(-25pt,-15pt)--(-25pt,-40pt);

\draw [>=stealth,->] (-25pt,-40+50pt)--(-45pt,-40+50pt);\draw(-45pt,-40+50pt)--(-65pt,-40+50pt);

\draw [>=stealth,->] (-65pt,-40+50pt)--(-65pt,-15pt);\draw(-65pt,-15pt)--(-65pt,-40pt);

\draw [>=stealth,->] (-35pt,-10+50pt)--(-50pt,-25+50pt);\draw(-50pt,-25+50pt)--(-65pt,-40+50pt);

\draw [dashed][>=stealth,->] (-35pt,-10+50pt)--(-35pt,15pt);\draw[dashed](-35pt,15pt)--(-35pt,-10pt);

\draw [>=stealth,->] (-5pt,-20+50pt)--(-35pt,-30+50pt); \draw(-35pt,-30+50pt)--(-65pt,-40+50pt);

\draw [>=stealth,->] (-10pt,-10+50pt)--(-37.5pt,-25+50pt); \draw(-37.5pt,-25+50pt)--(-65pt,-40+50pt);

\node at (-38pt,-6pt){${5}$};
\node at (-25pt,-45pt){${6}$};
\node at (-65pt,-44pt){${7}$};

\node at (-38pt,-6+50pt){${5'}$};
\node at (-25pt,-45+59pt){${6'}$};
\node at (-65pt,-44+59pt){${7'}$};


\draw [>=stealth,->] (-10pt,-10pt)--(-22.5pt,-10pt);\draw(-22.5pt,-10pt)--(-35pt,-10pt);

\draw [>=stealth,->] (-5pt,-20pt)--(-15pt,-30pt); \draw(-15pt,-30pt)--(-25pt,-40pt);

\draw [>=stealth,->] (-25pt,-40pt)--(-45pt,-40pt);\draw(-45pt,-40pt)--(-65pt,-40pt);

\draw [>=stealth,->] (-35pt,-10pt)--(-50pt,-25pt);\draw(-50pt,-25pt)--(-65pt,-40pt);

\draw [dashed][>=stealth,->] (-5pt,-20pt)--(-35pt,-30pt); \draw[dashed](-35pt,-30pt)--(-65pt,-40pt);

\draw [>=stealth,->] [dashed](-10pt,-10pt)--(-37.5pt,-25pt); \draw[dashed](-37.5pt,-25pt)--(-65pt,-40pt);

\node at (38pt,38pt){\textcolor{teal}{${h_{1,y}}$}};

\node at (-55pt,-20pt){\textcolor{teal}{${g_{2,y}}$}};
\node at (-40pt,-45pt){\textcolor{teal}{${g_{2,x}}$}};

\node at (-55pt,-20+50pt){\textcolor{teal}{${h_{2,y}}$}};
\node at (-40pt,-45+50pt){\textcolor{teal}{${h_{2,x}}$}};
\end{tikzpicture}
\end{eqnarray}
with the ordering $i'<i<(i+1)'$, and $\bar{1}<1$. 
The group elements on all the vertical links are $[ii']=t$.
The new basis $|h_{1,x}, h_{1,y}; h_{2,x}, h_{2,y}\rangle:=|[1'3'], [3'4']; [1'6'], [\bar{1}'5']\rangle
=|tg_{1,x}t^{-1}, tg_{1,y}t^{-1}; tg_{2,x}t^{-1}, tg_{2,y}t^{-1}\rangle$ automatically satisfies the 
flat connection condition. 
One can find that $\eta^t(\Sigma_{2,0})$ in \eqref{GaugeTransformGenus2}
is the product of $6\times 3=18$ three-cocyles:
\be\label{PhaseGaugeTransf_genus2}
\begin{split}
\eta^t(\Sigma_{2,0})=&Y_{[167],[1'6'7']}\cdot Y_{[\bar{1}17],[\bar{1}'1'7']}\cdot Y_{[\bar{1}57],[\bar{1}'5'7']}\\
&\cdot Y_{[134],[1'3'4']}\cdot Y_{[\bar{1}14],[\bar{1}'1'4']}\cdot Y_{[\bar{1}24],[\bar{1}'2'4']},
\end{split}
\ee
where each $Y_{[abc],[a'b'c']}$ corresponding to a triangular prism is the product of three 3-cocyles. 
The explicit expression of $\eta^t(\Sigma_{2,0})$ can be found in Eqs.\eqref{PhaseGaugeTransf_genus2_appendix}-\eqref{GaugeTransf_genus2_B} 
in the appendix.
Similar to the case of $\Sigma_{1,0}$, one can check that $A^t\cdot A^{t'}=A^{t\cdot t'}$ by using 
the 3-cocyle condition in Eq.\eqref{3cocycle_condition}. 

It follows from Eq.\eqref{At_genus2} that the degenerate ground states on $\Sigma_{2,0}$ are spanned by the vectors:
\be\label{GS_genus2}
\small
\Big\{
\frac{1}{|G|}\sum_{t \in G} \eta^t(\Sigma_{2,0}) |t g_{1,x}t^{-1}, t g_{1,y}t^{-1}, t g_{2,x}t^{-1}, t g_{2,y}t^{-1}\rangle
\Big\},
\ee
where $g_{1,x}, \, g_{1,y}, \, g_{2,x}$, and $g_{2,y}$ satisfy the flat connection constraint in Eq.\eqref{Genus2_FlatConnection}.

One remark here:
The expression in \eqref{GS_genus2} may suggest that the ground state degeneracy (GSD) on $\Sigma_{2,0}$ is
counted by Hom$(\pi_1(\Sigma_{2,0}),G)/G$, where $G$ acts by conjugation. As emphasized in Ref.\onlinecite{DW1990},
this is true for the case with trivial three-cocyle $\omega=1$. For general $\omega\in H^3(G,U(1))$, we have 
$\small \text{GSD}\le |\text{Hom}(\pi_1(\Sigma_{2,0}),G)/G|$. This is because for certain $g_{i,x}$ and $g_{i,y}$, the summation in \eqref{GS_genus2}
may vanish because of the phase factor $\eta^t(\Sigma_{2,0})$.\cite{HuWanWu1211} Therefore, one may 
overcount the states with Hom$(\pi_1(\Sigma_{2,0}),G)/G$. For the MS MTCs as studied in this work, it is found that 
$\small \text{GSD}= |\text{Hom}(\pi_1(\Sigma_{2,0}),G)/G|$.

Now let us look at how the modular transformations act on the basis $|g_{1,x},g_{1,y}; g_{2,x}, g_{2,y}\rangle$.
In particular, we will focus on the Dehn twists along the closed curves $a_1$, $b_1$, $a_2$, and $b_2$ in \eqref{Genus2_5Dehntwists}.
Considering that the genus-2 manifold is obtained by gluing two punctured tori along the punctures, 
we can act the Denh twists on the two punctured tori separately, and then the results on the punctured torus in the 
previous section can be applied here. 
Let us denote the punctured torus spanned by $[1\bar{1}567]$ in \eqref{Genus2Triangle} as the left
punctured torus, and that spanned by $[1\bar{1}234]$ as the right punctured torus.  
Now we consider how the Dehn twists act on, without loss of generality, the right punctured torus of the basis in 
\eqref{Genus2Triangle}. That is, we will consider the operation $I^t_L\otimes T_{R,\alpha}^t$, where $I$ means no 
modular transformation, and $\alpha=x,y$. 
For $\alpha=x$, we have
\be
\small
\begin{split}
&I^t_L\otimes T_{R,x}^t\, |g_{1,x}, g_{1,y}; g_{2,x}, g_{2,y}\rangle\\
=&u^t_{R,x}(\Sigma_{2,0}) \, |tg_{1,x}t^{-1}, t (g^{-1}_{1,x}\cdot g_{1,y}) t^{-1}; t g_{2,x} t^{-1}, t g_{2,y} t^{-1}\rangle\\
\end{split}
\ee
Here the $U(1)$ phase $u^t_{R,x}(\Sigma_{2,0}) $ is associated with the path integral defined on the following 3-manifold:

\begin{eqnarray}\label{TxPhasePuncture_Genus2}
\footnotesize
u_{R,x}^t(\Sigma_{2,0}):\quad
\begin{tikzpicture}[baseline={(current bounding box.center)}]

\draw [>=stealth,->] (-5pt,-20pt)--(10pt,-20pt);  \draw (10pt, -20pt) -- (20pt,-20pt);

\draw [>=stealth,->] (-10pt,-10pt)--(0pt,0pt); \draw(0pt,0pt)--(10pt,10pt);

\draw [dashed][>=stealth,->] (-10pt,-10pt)--(-7.5pt,-15pt); \draw[dashed](-7.5pt,-15pt)--(-5pt,-20pt);

\draw [>=stealth,->][dashed] (-5pt,-20pt)--(22.5pt,-5pt); \draw[dashed](22.5pt,-5pt)--(50pt,10pt);

\draw [>=stealth,->][dashed] (-10pt,-10pt)--(20pt,0pt); \draw[dashed](20pt,0pt)--(50pt,10pt);

\draw [>=stealth,->] (10pt,10pt)--(30pt,10pt); \draw (30pt,10pt)--(50pt,10pt);
\draw [>=stealth,->] (20pt,-20pt)--(35pt,-5pt); \draw (35pt,-5pt)--(50pt,10pt);
\node at (40pt,-10pt){\textcolor{teal}{${g_{1,y}}$}};
\node at (10pt,-27pt){\textcolor{teal}{${g_{1,x}}$}};
\node at (-5pt,-25pt){${1}$};
\node at (-15pt,-5pt){${\bar{1}}$};
\node at (22pt,-25pt){${3}$};

\node at (2pt,10pt){${2}$};
\node at (55pt,10pt){${4}$};

\draw [>=stealth,->] (-5pt,-20+50pt)--(10pt,-20+50pt);  \draw (10pt, -20+50pt) -- (20pt,-20+50pt);

\draw [>=stealth,->] (-10pt,-10+50pt)--(0pt,0+50pt); \draw(0pt,0+50pt)--(10pt,10+50pt);

\draw [>=stealth,->] (-10pt,-10+50pt)--(-7.5pt,-15+50pt); \draw(-7.5pt,-15+50pt)--(-5pt,-20+50pt);

\draw [>=stealth,->] (-5pt,30pt)--(2.5pt,45pt); \draw(2.5pt,45pt)--(10pt,60pt);

\draw [>=stealth,->] (10pt,10+50pt)--(30pt,10+50pt); \draw (30pt,10+50pt)--(50pt,10+50pt);
\draw [>=stealth,->] (20pt,-20+50pt)--(35pt,-5+50pt); \draw (35pt,-5+50pt)--(50pt,10+50pt);

\draw [>=stealth,->] (-5pt,-20+50pt)--(-5pt,5pt); \draw(-5pt,5pt)--(-5pt,-20pt);
\draw [dashed][>=stealth,->] (-10pt,-10+50pt)--(-10pt,15pt); \draw[dashed](-10pt,15pt)--(-10pt,-10pt);

\draw [>=stealth,->] (20pt,-20+50pt)--(20pt,5pt); \draw(20pt,5pt)--(20pt,-20pt);

\draw [>=stealth,->] (50pt,10+50pt)--(50pt,35pt); \draw(50pt,35pt)--(50pt,10pt);
\draw [>=stealth,->] [dashed](10pt,10+50pt)--(10pt,35pt); \draw[dashed](10pt,35pt)--(10pt,10pt);

\draw [>=stealth,->] (20pt,30pt)--(15pt,45pt); \draw(15pt,45pt)--(10pt,60pt);

\node at (-5pt,25pt){${1'}$};
\node at (25pt,28pt){${3'}$};
\node at (-13pt,-12+57pt){${\bar{1}'}$};

\node at (2pt,60pt){${4'}$};
\node at (58pt,60pt){${2^{\ast}}$};
\node at (8pt,25pt){\textcolor{teal}{${h_{1,x}}$}};



\draw [>=stealth,->] (-10pt,-10+50pt)--(-22.5pt,-10+50pt);\draw(-22.5pt,-10+50pt)--(-35pt,-10+50pt);

\draw [>=stealth,->] (-5pt,-20+50pt)--(-15pt,-30+50pt); \draw(-15pt,-30+50pt)--(-25pt,-40+50pt);

\draw [>=stealth,->] (-25pt,-40+50pt)--(-25pt,-15pt); \draw(-25pt,-15pt)--(-25pt,-40pt);

\draw [>=stealth,->] (-25pt,-40+50pt)--(-45pt,-40+50pt);\draw(-45pt,-40+50pt)--(-65pt,-40+50pt);

\draw [>=stealth,->] (-65pt,-40+50pt)--(-65pt,-15pt);\draw(-65pt,-15pt)--(-65pt,-40pt);

\draw [>=stealth,->] (-35pt,-10+50pt)--(-50pt,-25+50pt);\draw(-50pt,-25+50pt)--(-65pt,-40+50pt);

\draw [dashed][>=stealth,->] (-35pt,-10+50pt)--(-35pt,15pt);\draw[dashed](-35pt,15pt)--(-35pt,-10pt);

\draw [>=stealth,->] (-5pt,-20+50pt)--(-35pt,-30+50pt); \draw(-35pt,-30+50pt)--(-65pt,-40+50pt);

\draw [>=stealth,->] (-10pt,-10+50pt)--(-37.5pt,-25+50pt); \draw(-37.5pt,-25+50pt)--(-65pt,-40+50pt);

\node at (-38pt,-6pt){${5}$};
\node at (-25pt,-45pt){${6}$};
\node at (-65pt,-44pt){${7}$};

\node at (-38pt,-6+50pt){${5'}$};
\node at (-25pt,-45+59pt){${6'}$};
\node at (-65pt,-44+59pt){${7'}$};


\draw [>=stealth,->] (-10pt,-10pt)--(-22.5pt,-10pt);\draw(-22.5pt,-10pt)--(-35pt,-10pt);

\draw [>=stealth,->] (-5pt,-20pt)--(-15pt,-30pt); \draw(-15pt,-30pt)--(-25pt,-40pt);

\draw [>=stealth,->] (-25pt,-40pt)--(-45pt,-40pt);\draw(-45pt,-40pt)--(-65pt,-40pt);

\draw [>=stealth,->] (-35pt,-10pt)--(-50pt,-25pt);\draw(-50pt,-25pt)--(-65pt,-40pt);

\draw [dashed][>=stealth,->] (-5pt,-20pt)--(-35pt,-30pt); \draw[dashed](-35pt,-30pt)--(-65pt,-40pt);

\draw [>=stealth,->] [dashed](-10pt,-10pt)--(-37.5pt,-25pt); \draw[dashed](-37.5pt,-25pt)--(-65pt,-40pt);

\node at (22pt,48pt){\textcolor{teal}{${h_{1,y}}$}};

\node at (-55pt,-20pt){\textcolor{teal}{${g_{2,y}}$}};
\node at (-40pt,-45pt){\textcolor{teal}{${g_{2,x}}$}};

\node at (-55pt,-20+50pt){\textcolor{teal}{${h_{2,y}}$}};
\node at (-40pt,-45+50pt){\textcolor{teal}{${h_{2,x}}$}};

\node at (60pt,35pt){\textcolor{teal}{${t^{-1}}$}};
\end{tikzpicture}
\end{eqnarray}
The ordering is taken as $\bar{1}'<1'<3'<4'<2^{\ast}<5'<6'<7'<\bar{1}<1<2<3<4<5<6<7$.
All the group elements assigned on the directed vertical links are $t^{-1}$.
The new basis after Dehn twist is $|h_{1,x}, h_{1,y}; h_{2,x}, h_{2,y}\rangle:=|[1'3'], [3'4']; [6'7'],[1'6']\rangle=
|t g_{1,x}t^{-1}, t (g^{-1}_{1,x}\cdot g_{1,y} )t^{-1}; t g_{2,x} t^{-1}, t g_{2,y} t^{-1}\rangle$ which satisfies the
flat connection condition. The $U(1)$ phase $u_{R,x}^t(\Sigma_{2,0})$ in \eqref{TxPhasePuncture_Genus2}
can be expressed as
\be\label{u_Rx_genus2}
u_{R,x}^t(\Sigma_{2,0})=I_L^t(\Sigma_{1,1}) \cdot u_x^t(\Sigma_{1,1})
\ee
where the second term, which is the same as that in \eqref{u_x^t_PuncturedTorus}, is contributed by the Dehn twist on the right punctured torus. 
The first term $I_L^t(\Sigma_{1,1})$ is contributed by the left punctured torus, with the expression given in Eq.\eqref{It_L_appendix}.

For the Dehn twist $T_{R,y}$ acting on the right punctured torus, we have
\be
\small
\begin{split}
&I^t_L\otimes T_{R,y}^t\, |g_{1,x}, g_{1,y}; g_{2,x}, g_{2,y}\rangle\\
=&u^t_{R,y}(\Sigma_{2,0}) \, |t(g_{1,y}^{-1}\cdot g_{1,x})t^{-1}, t g_{1,y} t^{-1}; t g_{2,x} t^{-1}, t g_{2,y} t^{-1}\rangle\\
\end{split}
\ee
where the $U(1)$ phase $u_{R,y}^t(\Sigma_{2,0})$ is associated with the path integral defined on the following 3-manifold:
\begin{eqnarray}\label{TyPhasePuncture_Genus2}
\footnotesize
u_{R,y}^t(\Sigma_{2,0}):\quad
\begin{tikzpicture}[baseline={(current bounding box.center)}]

\draw [>=stealth,->] (-5pt,-20pt)--(10pt,-20pt);  \draw (10pt, -20pt) -- (20pt,-20pt);

\draw [>=stealth,->] (-10pt,-10pt)--(0pt,0pt); \draw(0pt,0pt)--(10pt,10pt);

\draw [dashed][>=stealth,->] (-10pt,-10pt)--(-7.5pt,-15pt); \draw[dashed](-7.5pt,-15pt)--(-5pt,-20pt);

\draw [>=stealth,->][dashed] (-5pt,-20pt)--(22.5pt,-5pt); \draw[dashed](22.5pt,-5pt)--(50pt,10pt);

\draw [>=stealth,->][dashed] (-10pt,-10pt)--(20pt,0pt); \draw[dashed](20pt,0pt)--(50pt,10pt);

\draw  [>=stealth,->] (10pt,10pt)--(30pt,10pt); \draw (30pt,10pt)--(50pt,10pt);
\draw [>=stealth,->] (20pt,-20pt)--(35pt,-5pt); \draw (35pt,-5pt)--(50pt,10pt);
\node at (40pt,-10pt){\textcolor{teal}{${g_{1,y}}$}};
\node at (10pt,-27pt){\textcolor{teal}{${g_{1,x}}$}};
\node at (-5pt,-25pt){${1}$};
\node at (-15pt,-5pt){${\bar{1}}$};
\node at (22pt,-25pt){${3}$};

\node at (2pt,10pt){${2}$};
\node at (55pt,10pt){${4}$};

\draw [>=stealth,->] (-5pt,-20+50pt)--(10pt,-20+50pt);  \draw (10pt, -20+50pt) -- (20pt,-20+50pt);

\draw [>=stealth,->] (-10pt,-10+50pt)--(0pt,0+50pt); \draw(0pt,0+50pt)--(10pt,10+50pt);

\draw [>=stealth,->] (-10pt,-10+50pt)--(-7.5pt,-15+50pt); \draw(-7.5pt,-15+50pt)--(-5pt,-20+50pt);

\draw [>=stealth,->] (-5pt,30pt)--(2.5pt,45pt); \draw(2.5pt,45pt)--(10pt,60pt);

\draw [>=stealth,->] (10pt,10+50pt)--(30pt,10+50pt); \draw (30pt,10+50pt)--(50pt,10+50pt);
\draw [>=stealth,->] (20pt,-20+50pt)--(35pt,-5+50pt); \draw (35pt,-5+50pt)--(50pt,10+50pt);

\draw [>=stealth,->] (-5pt,-20+50pt)--(-5pt,5pt); \draw(-5pt,5pt)--(-5pt,-20pt);
\draw [dashed][>=stealth,->] (-10pt,-10+50pt)--(-10pt,15pt); \draw[dashed](-10pt,15pt)--(-10pt,-10pt);

\draw [>=stealth,->] (20pt,-20+50pt)--(20pt,5pt); \draw(20pt,5pt)--(20pt,-20pt);

\draw [>=stealth,->] (50pt,10+50pt)--(50pt,35pt); \draw(50pt,35pt)--(50pt,10pt);
\draw [>=stealth,->] [dashed](10pt,10+50pt)--(10pt,35pt); \draw[dashed](10pt,35pt)--(10pt,10pt);

\draw  [>=stealth,->] (10pt,60pt)--(15pt,45pt); \draw(15pt,45pt)--(20pt,30pt);

\node at (-5pt,25pt){${1'}$};
\node at (25pt,28pt){${4'}$};
\node at (-13pt,-12+57pt){${\bar{1}'}$};

\node at (2pt,60pt){${2'}$};
\node at (58pt,60pt){${3^{\ast}}$};
\node at (-10pt,55pt){\textcolor{teal}{${h_{1,y}}$}};



\draw [>=stealth,->] (-10pt,-10+50pt)--(-22.5pt,-10+50pt);\draw(-22.5pt,-10+50pt)--(-35pt,-10+50pt);

\draw [>=stealth,->] (-5pt,-20+50pt)--(-15pt,-30+50pt); \draw(-15pt,-30+50pt)--(-25pt,-40+50pt);

\draw [>=stealth,->] (-25pt,-40+50pt)--(-25pt,-15pt); \draw(-25pt,-15pt)--(-25pt,-40pt);

\draw [>=stealth,->] (-25pt,-40+50pt)--(-45pt,-40+50pt);\draw(-45pt,-40+50pt)--(-65pt,-40+50pt);

\draw [>=stealth,->] (-65pt,-40+50pt)--(-65pt,-15pt);\draw(-65pt,-15pt)--(-65pt,-40pt);

\draw [>=stealth,->] (-35pt,-10+50pt)--(-50pt,-25+50pt);\draw(-50pt,-25+50pt)--(-65pt,-40+50pt);

\draw [dashed][>=stealth,->] (-35pt,-10+50pt)--(-35pt,15pt);\draw[dashed](-35pt,15pt)--(-35pt,-10pt);

\draw [>=stealth,->] (-5pt,-20+50pt)--(-35pt,-30+50pt); \draw(-35pt,-30+50pt)--(-65pt,-40+50pt);

\draw [>=stealth,->] (-10pt,-10+50pt)--(-37.5pt,-25+50pt); \draw(-37.5pt,-25+50pt)--(-65pt,-40+50pt);

\node at (-38pt,-6pt){${5}$};
\node at (-25pt,-45pt){${6}$};
\node at (-65pt,-44pt){${7}$};

\node at (-38pt,-6+50pt){${5'}$};
\node at (-25pt,-45+59pt){${6'}$};
\node at (-65pt,-44+59pt){${7'}$};


\draw [>=stealth,->] (-10pt,-10pt)--(-22.5pt,-10pt);\draw(-22.5pt,-10pt)--(-35pt,-10pt);

\draw [>=stealth,->] (-5pt,-20pt)--(-15pt,-30pt); \draw(-15pt,-30pt)--(-25pt,-40pt);

\draw [>=stealth,->] (-25pt,-40pt)--(-45pt,-40pt);\draw(-45pt,-40pt)--(-65pt,-40pt);

\draw [>=stealth,->] (-35pt,-10pt)--(-50pt,-25pt);\draw(-50pt,-25pt)--(-65pt,-40pt);

\draw [dashed][>=stealth,->] (-5pt,-20pt)--(-35pt,-30pt); \draw[dashed](-35pt,-30pt)--(-65pt,-40pt);

\draw [>=stealth,->] [dashed](-10pt,-10pt)--(-37.5pt,-25pt); \draw[dashed](-37.5pt,-25pt)--(-65pt,-40pt);

\node at (22pt,48pt){\textcolor{teal}{${h_{1,x}}$}};

\node at (-55pt,-20pt){\textcolor{teal}{${g_{2,y}}$}};
\node at (-40pt,-45pt){\textcolor{teal}{${g_{2,x}}$}};

\node at (-55pt,-20+50pt){\textcolor{teal}{${h_{2,y}}$}};
\node at (-40pt,-45+50pt){\textcolor{teal}{${h_{2,x}}$}};
\node at (60pt,35pt){\textcolor{teal}{${t^{-1}}$}};
\end{tikzpicture}
\end{eqnarray}
with the ordering 
$\bar{1}'<1'<2'<4'<3^{\ast}<5'<6'<7'<\bar{1}<1<2<3<4<5<6<7$.
Again the group elements on the directed vertical links are $t^{-1}$.
The new basis
$|h_{1,x}, h_{1,y}; h_{2,x}, h_{2,y}\rangle:=|[2'4'], [\bar{1}'2']; [6'7'],[1'6']\rangle=
|t(g_{1,y}^{-1}\cdot g_{1,x})t^{-1}, t g_{1,y} t^{-1}; t g_{2,x} t^{-1}, t g_{2,y} t^{-1}\rangle$ satisfies the 
flat connection condition. 
The $U(1)$ phase $u_{R,y}^t(\Sigma_{2,0})$ in \eqref{TyPhasePuncture_Genus2} can be written as:
\be\label{u_Ry_genus2}
u_{R,y}^t(\Sigma_{2,0})=I_L^t(\Sigma_{1,1}) \cdot u_y^t(\Sigma_{1,1})
\ee
where the first term $I_L^t(\Sigma_{1,1})$ is the same as that in Eq.\eqref{u_Rx_genus2}, with
the expression given in Eq.\eqref{It_L_appendix}. The second term $ u_y^t(\Sigma_{1,1})$,
which is the same as \eqref{TyPhasePuncture},
is contributed by the Denh twist acting on the right punctured torus.

Similar to the case of $\Sigma_{1,0}$, 
it can checked that 
$\small (I^t_L\otimes T_{R,x(y)}^t )\cdot A^{t'}=A^{t}\cdot (I^{t'}_L\otimes T_{R, x(y)}^{t'})=I^{t\cdot t'}_L\otimes T_{R,x(y)}^{t\cdot t'}$,
with $A^t$ defined in Eq.\eqref{At_genus2}.
We can define the following quantities
\be\label{TR_SR}
T_{R}:=\frac{1}{|G|}\sum_{t\in G} I^t\otimes T^t_{R,x}, \quad S_R:=\frac{1}{|G|}\sum_{t\in G} I^t\otimes S^t_R,
\ee
where $\small I^t\otimes S^t_R=(I^{t'=1}\otimes T_{R,y}^{t'=1})\cdot (I^{t'=1}\otimes(T_{R,x}^{t'=1})^{-1})\cdot (I^{t'=1}\otimes T_{R,y}^{t'=1})\cdot A^t
=A^t\cdot (I^{t'=1}\otimes T_{R,y}^{t'=1})\cdot (I^{t'=1}\otimes(T_{R,x}^{t'=1})^{-1})\cdot (I^{t'=1}\otimes T_{R,y}^{t'=1})$.
It is noted that here we act the modular transformations only on the right punctured torus in the basis \eqref{Genus2Triangle}. One can 
certainly consider the same procedure by performing modular transformations on the left punctured torus.

Given $T_R$ and $S_R$ in Eq.\eqref{TR_SR}, we can calculate the matrix representations 
$(T_{R})_{ij}=\langle \Phi_i|T_R|\Phi_j\rangle$ and$(S_{R})_{ij}=\langle \Phi_i|S_R|\Phi_j\rangle$, 
with the ground state bases $|\Phi_i\rangle$ in \eqref{GS_genus2}.

\subsection{Topological invariants}
\label{Sec: Genus2_lattice_TopoInv}

The lattice gauge theory approach discussed in the previous subsections work for general Dijkgraaf-Witten theories, and here we apply it to the MS MTCs.
In particular, we focus on the simplest counterexamples in MS MTCs with $G=\mathbb{Z}_{11}\rtimes \mathbb{Z}_5$,
twisted by 3-cocyles $\omega\in H^3(G, U(1))\cong \mathbb{Z}_5$.
One can find that even for these simplest examples the GSD on a genus-2 manifold is $18529$.

As introduced in Sec.\ref{Sec: Quasi-particle}, we construct the topological invariants 
based on the `words' composed of $S_R$ and $T_R$ in Eq.\eqref{TR_SR} as:
\be\label{Topological_Invariant_genus2_lattice}
W_{\Sigma_{2,0}}:=\text{Tr}\Big[\big(
S_R\big)^{n_1}\big(T_R\big)^{n_2}\big(S_R\big)^{n_3}\big(T_R\big)^{n_4}\cdots 
\Big],
\ee
where the trace is over the Hilbert space of degenerate ground states on a genus-2 manifold in Eq.\eqref{GS_genus2}.

To compare with the results in Sec.\ref{Sec: Quasi-particle} in more details, 
let us first point out some fine structures in the degenerate ground states in Eq.\eqref{GS_genus2}.
For each ground state $\frac{1}{|G|}\sum_{t \in G} \eta^t(\Sigma_{2,0}) |t g_{1,x}t^{-1}, t g_{1,y}t^{-1}, t g_{2,x}t^{-1}, t g_{2,y}t^{-1}\rangle$, 
one can find that the conjugacy class of `$k$' in \eqref{Genus2Triangle} is well defined, with $k=t\cdot [g_{1,x}, g_{1,y}]^{-1}\cdot t^{-1}
=t\cdot [g_{2,x}, g_{2,y}]\cdot t^{-1}$, where $t\in G$.
For the MS MTCs we consider here, 
it is found that the conjugacy class $[k]$ can only be $[1]$, $[a^1]$, and $[a^2]$ (see Table \ref{ConjugacyClass}).
Considering that $[k]$ measures the `magnetic flux' of anyon $z$ in \eqref{BasisI}, this means the anyon type for $z$ can only be type-$I$ or type-$A$, which agrees with the 
fusion rules as studied in Sec.\ref{Sec: Quasi-particle}. 

According to the conjugacy class of $k$, now we divide the ground states in Eq.\eqref{GS_genus2} into three groups with $[k]=[1]$, $[a^1]$, and $[a^2]$, respectively.
One can find that the number of degenerate ground states is $5857$ for $[k]=[1]$, $6336$ for $[k]=[a^1]$, and $6336$ for $[k]=[a^2]$
(This agrees with the fact that the GSD on a genus-2 surface is 18529).
Then the topological invariant defined in Eq.\eqref{Topological_Invariant_genus2_lattice} can be written as
\be
\begin{split}
W_{\Sigma_{2,0}}=&W_{\Sigma_{2,0}}^{[1]}+W_{\Sigma_{2,0}}^{[a^1]}+W_{\Sigma_{2,0}}^{[a^2]}\\
=:&W_{\Sigma_{2,0}}^{I}+W_{\Sigma_{2,0}}^{A_1}+W_{\Sigma_{2,0}}^{A_2},
\end{split}
\ee
where $W_{\Sigma_{2,0}}^{[k]}$ is defined in \eqref{Topological_Invariant_genus2_lattice}, except that now the trace is 
taken over the degenerate ground states with a fixed $[k]$. 

In the following, we study the trace of word samples in \eqref{Words}. 
More explicitly, the word $w^{(z)}_1=(\theta^{(z)})^2\cdot (T^{(z)})^7S^{(z)}=(S^{(z)})^{-7}(T^{(z)})^7S^{(z)}$ in 
Eq.\eqref{Words} corresponds to $w_1=S_R^{-8} \cdot T_R^7\cdot S_R$ in the lattice gauge theory approach here, and similarly for other words.
 The results for the trace over different words are summarized in Tables \ref{W1_Genus2} - \ref{W7_Genus2}.

For convenience, we compare the results of quasi-particle basis calculation and the lattice gauge theory calculation for the 
trace of word $w_1^{(z)}$ (see Eq.\ref{Words}) in Table \ref{W1_Genus2}.
The comparisons for other words are similar and straightforward.
It is remarkable that although we work with two independent methods and the ground state bases are chosen 
in different ways, the topological invariants we obtained are the same. 

It is reminded that for the quasi-particle basis result in Table \ref{W1_Genus2}, the data in gray are contributed by 
 $W_{\Sigma_{2,0}}^{I_i}$, with $i=1,\cdots, 4$.
As explained in Sec.\ref{Sec: Topological_invariants}, for $z=I_i$ with $i=1,\cdots,4$, only type-$I$ and type-$A$ anyons
appear in the punctured $S$ matrix, and therefore the information of 3-cocyle $\omega^u$ will not come in (Recall that only
type-$B$ anyons carry the information of 3-cocycles).
This is more transparent by considering the lattice gauge theory approach:
For the basis vector in \eqref{Genus2Triangle}, $z=I_i$ ($i=1,\cdots,4$) correspond to the case that the group element $k=1$,
where $1$ denotes the identity group element.
Then with the fusion rules in Sec.\ref{Sec: Z11Z5_data}, one can find for $a\times \bar{a}=\sum_z N_{a\bar{a}}^z\, z$ where $z=I_i$
($i=1,\cdots,4$), $N_{a\bar{a}}^z$ is nonzero only when $a$ are type-$I$ or type-$A$ anyons. This indicates that the
group elements $g_{1,x}$ and $g_{1,y}$ in \eqref{Genus2Triangle} can only be of the form $(a^l, b^0)$ with $l=0,\cdots,10$.
By performing modular transformation on the right half torus in \eqref{Genus2Triangle}, one can find that the 3-cocycles $\omega$ 
involved in this process are always trivial with $\omega=1$.  This explains why  $W_{\Sigma_{2,0}}^{I_i}$ ($i=1,\cdots, 4$)
are independent of the three-cocycle $\omega^u$ and $u$.

The invariants that can distinguish different categories are $W_{\Sigma_{2,0}}^{A_1}$ and $W_{\Sigma_{2,0}}^{A_2}$, with the difference
in the phase factor $e^{\frac{2i\pi}{5}n}$, where $n=0,\, 1,\, 2,\, 3,\,4$. In addition, the phase
factor for $u=1$ ($u=2$) is conjugate to that for $u=4$ ($u=3$).

It is emphasized that the topological invariants $W_{\Sigma_{2,0}}$ we defined 
is independent of the permutation of anyons. 
This is different from the method we used in Sec.\ref{PuncturedS_distinguish_MS_MC}, 
where we need to track how the topological invariants (diagonal elements of $S^{(z)}$ and $T$ matrix) transform
with the bijection of anyons between two categories. Here, a single number is enough to 
distinguish different categories.

One remark here:  As mentioned in Sec.\ref{Sec: Topological_invariants},
corresponding to the five different phase factors $e^{\frac{2i\pi}{5}n}$ with $n=0,\, 1,\, 2,\, 3,\,4$, 
if we consider $\cos\left(\frac{2\pi}{5}n\right)$,
there are only three different values left. The distinct information becomes degenerate now. 
In fact, as seen from Sec.\ref{Sec: Topological_invariants}, the invariants that contain $\cos\left(\frac{2\pi}{5}n\right)$ are contributed by 
the modular data with $z=I_0$. This agrees with the fact that there are only three distinct sets of modular data 
in MS MTCs. From this point of view, the reason that the modular data is not enough to distinguish 
different categories is because the information becomes degenerate. 
With the MCG representations of higher genus manifold, we can split this information degeneracy and 
distinguish different categories.

\begin{table}[H]
\centering
\footnotesize
\underline
{Quasi-particle basis result}\\
\begin{tabular}{cccccccccc}
$W_{\Sigma_{2,0}}$    &\vline &$W_{\Sigma_{2,0}}^{I}$&\vline    &$W_{\Sigma_{2,0}}^{A_1}$&\vline   &$W_{\Sigma_{2,0}}^{A_2}$    \\ \hline
$u=0$  &\vline     &$633-24\times \textcolor{gray}{4}$&\vline &$528$&\vline & $528$                 \\ \hline
$u=1$  &\vline     &$245-24\times \textcolor{gray}{4}+388\cdot \cos\frac{2\pi}{5}$&\vline &$528\cdot e^{\frac{2\pi i}{5}\times 1} $&\vline   &$528\cdot e^{\frac{2\pi i}{5}\times 1} $  \\ \hline
$u=2$  &\vline     &$245-24\times \textcolor{gray}{4}+388\cdot \cos\frac{4\pi}{5}$&\vline  &$528\cdot e^{\frac{2\pi i}{5}\times 2} $&\vline    & $528\cdot e^{\frac{2\pi i}{5}\times 2} $  \\  \hline
$u=3$  &\vline      &$245-24\times \textcolor{gray}{4}+388\cdot \cos\frac{4\pi}{5}$&\vline &$528\cdot e^{\frac{2\pi i}{5}\times 3} $&\vline  &$528 \cdot e^{\frac{2i\pi}{5}\times 3}$ \\ \hline
$u=4$  &\vline      &$245-24\times \textcolor{gray}{4}+388\cdot \cos\frac{2\pi}{5}$ &\vline  &$528\cdot e^{\frac{2\pi i}{5}\times 4} $&\vline    &$528\cdot e^{\frac{2\pi i}{5}\times 4} $      \\ \hline
\end{tabular}
\\
\underline
{Lattice gauge theory result}\\
\begin{tabular}{cccccccccc}
$W_{\Sigma_{2,0}}$    &\vline &$W_{\Sigma_{2,0}}^{I}$&\vline    &$W_{\Sigma_{2,0}}^{A_1}$&\vline   &$W_{\Sigma_{2,0}}^{A_2}$    \\ \hline
$u=0$  &\vline     &$537$&\vline &$528$&\vline & $528$                 \\ \hline
$u=1$  &\vline     &$149+388\cdot \cos\frac{2\pi}{5}$&\vline &$528\cdot e^{\frac{2\pi i}{5}\times 1} $&\vline   &$528\cdot e^{\frac{2\pi i}{5}\times 1} $  \\ \hline
$u=2$  &\vline     &$149+388\cdot \cos\frac{4\pi}{5}$&\vline  &$528\cdot e^{\frac{2\pi i}{5}\times 2} $&\vline    & $528\cdot e^{\frac{2\pi i}{5}\times 2} $  \\  \hline
$u=3$  &\vline      &$149+388\cdot \cos\frac{4\pi}{5}$&\vline &$528\cdot e^{\frac{2\pi i}{5}\times 3} $&\vline  &$528 \cdot e^{\frac{2i\pi}{5}\times 3}$ \\ \hline
$u=4$  &\vline      &$149+388\cdot \cos\frac{2\pi}{5}$ &\vline  &$528\cdot e^{\frac{2\pi i}{5}\times 4} $&\vline    &$528\cdot e^{\frac{2\pi i}{5}\times 4} $      \\ \hline
\end{tabular}
\caption{
Comparison of the quasi-particle basis results in Table \eqref{W1_Genus2_QP}
and the lattice gauge theory results, for $W_{\Sigma_{2,0}}$ with the word $w_1^{(z)}$ in Eq.\eqref{Words}.
}
\label{W1_Genus2}
\end{table}

\begin{table}[H]
\centering
\footnotesize
\begin{tabular}{cccccccccc}
$W_{\Sigma_{2,0}}$     &\vline &$W_{\Sigma_{2,0}}^{I}$&\vline    &$W_{\Sigma_{2,0}}^{A_1}$&\vline   &$W_{\Sigma_{2,0}}^{A_2}$    \\ \hline
$u=0$  &\vline     &$537$ &\vline &$528$&\vline &    $528$    \\ \hline
$u=1$  &\vline     &$149+388\cdot \cos\frac{4\pi}{5}$&\vline &$528 \cdot e^{\frac{2i\pi}{5}\times 3}$&\vline   &$528 \cdot e^{\frac{2i\pi}{5}\times 3}$  \\ \hline
$u=2$  &\vline     &$149+388\cdot \cos\frac{2\pi}{5}$&\vline  &$528 \cdot e^{\frac{2i\pi}{5}\times 1}$&\vline    &$528 \cdot e^{\frac{2i\pi}{5}\times 1}$ \\  \hline
$u=3$  &\vline      &$149+388\cdot \cos\frac{2\pi}{5}$&\vline &$528 \cdot e^{\frac{2i\pi}{5}\times 4}$&\vline  &$528 \cdot e^{\frac{2i\pi}{5}\times 4}$ \\ \hline
$u=4$  &\vline      &$149+388\cdot \cos\frac{4\pi}{5}$ &\vline  &$528 \cdot e^{\frac{2i\pi}{5}\times 2}$&\vline    &$528 \cdot e^{\frac{2i\pi}{5}\times 2}$   \\ \hline
\end{tabular}
\caption{$W_{\Sigma_{2,0}}$ with the word $w_2^{(z)}$ in Eq.\eqref{Words}, based on the lattice gauge theory calculation.
}
\label{W2_Genus2}
\end{table}

\begin{table}[H]
\centering
\footnotesize
\begin{tabular}{cccccccccc}
$W_{\Sigma_{2,0}}$    &\vline &$W_{\Sigma_{2,0}}^{I}$&\vline    &$W_{\Sigma_{2,0}}^{A_1}$&\vline   &$W_{\Sigma_{2,0}}^{A_2}$    \\ \hline
$u=0$  &\vline     &$2377$ &\vline &$1584$&\vline &        $1584$            \\ \hline
$u=1$  &\vline     &$-533$&\vline &$1584 \cdot e^{\frac{2i\pi}{5}\times 2}$&\vline   & $1584 \cdot e^{\frac{2i\pi}{5}\times 2}$  \\ \hline
$u=2$  &\vline     &$-533$&\vline  &$1584 \cdot e^{\frac{2i\pi}{5}\times 4}$&\vline    & $1584 \cdot e^{\frac{2i\pi}{5}\times 4}$   \\  \hline
$u=3$  &\vline      &$-533$&\vline &$1584 \cdot e^{\frac{2i\pi}{5}\times 1}$ &\vline  &$1584 \cdot e^{\frac{2i\pi}{5}\times 1}$   \\ \hline
$u=4$  &\vline      &$-533$ &\vline  &$1584 \cdot e^{\frac{2i\pi}{5}\times 3}$&\vline    &$1584 \cdot e^{\frac{2i\pi}{5}\times 3}$      \\ \hline
\end{tabular}
\caption{$W_{\Sigma_{2,0}}$ with the word $w_3^{(z)}$ in Eq.\eqref{Words}, based on the lattice gauge theory calculation.
}
\label{W3_Genus2}
\end{table}

\begin{table}[H]
\centering
\footnotesize
\begin{tabular}{cccccccccc}
$W_{\Sigma_{2,0}}$    &\vline &$W_{\Sigma_{2,0}}^{I}$&\vline    &$W_{\Sigma_{2,0}}^{A_1}$&\vline   &$W_{\Sigma_{2,0}}^{A_2}$    \\ \hline
$u=0$  &\vline     &$2377$&\vline &$1056$&\vline &  $1056$                  \\ \hline
$u=1$  &\vline     &$437$ &\vline &$1056 \cdot e^{\frac{2i\pi}{5}\times 3}$&\vline   &$1056 \cdot e^{\frac{2i\pi}{5}\times 3}$   \\ \hline
$u=2$  &\vline     &$437$ &\vline  &$1056 \cdot e^{\frac{2i\pi}{5}\times 1}$&\vline    &$1056 \cdot e^{\frac{2i\pi}{5}\times 1}$   \\  \hline
$u=3$  &\vline      &$437$&\vline &$1056 \cdot e^{\frac{2i\pi}{5}\times 4}$&\vline  &$1056 \cdot e^{\frac{2i\pi}{5}\times 4}$  \\ \hline
$u=4$  &\vline      &$437$ &\vline  &$1056 \cdot e^{\frac{2i\pi}{5}\times 2}$&\vline    &$1056 \cdot e^{\frac{2i\pi}{5}\times 2}$      \\ \hline
\end{tabular}
\caption{$W_{\Sigma_{2,0}}$ with the word $w_4^{(z)}$ in Eq.\eqref{Words}, based on the lattice gauge theory calculation.
}
\label{W4_Genus2}
\end{table}

\begin{table}[H]
\centering
\footnotesize
\begin{tabular}{cccccccccc}
$W_{\Sigma_{2,0}}$    &\vline &$W_{\Sigma_{2,0}}^{I}$&\vline    &$W_{\Sigma_{2,0}}^{A_1}$&\vline   &$W_{\Sigma_{2,0}}^{A_2}$    \\ \hline
$u=0$  &\vline     &$437$&\vline & $528$&\vline &       $528$             \\ \hline
$u=1$  &\vline     &$49+388\cdot \cos\frac{4\pi}{5}$&\vline &$528 \cdot e^{\frac{2i\pi}{5}\times 2}$&\vline   &$528 \cdot e^{\frac{2i\pi}{5}\times 2}$   \\ \hline
$u=2$  &\vline     &$49+388\cdot \cos\frac{2\pi}{5}$&\vline  &$528 \cdot e^{\frac{2i\pi}{5}\times 4}$&\vline    &$528 \cdot e^{\frac{2i\pi}{5}\times 4}$   \\  \hline
$u=3$  &\vline      &$49+388\cdot \cos\frac{2\pi}{5}$&\vline &$528 \cdot e^{\frac{2i\pi}{5}\times 1}$&\vline  &$528 \cdot e^{\frac{2i\pi}{5}\times 1}$  \\ \hline
$u=4$  &\vline      &$49+388\cdot \cos\frac{4\pi}{5}$&\vline  &$528 \cdot e^{\frac{2i\pi}{5}\times 3}$&\vline    &$528 \cdot e^{\frac{2i\pi}{5}\times 3}$      \\ \hline
\end{tabular}
\caption{$W_{\Sigma_{2,0}}$ with the word $w_5^{(z)}$ in Eq.\eqref{Words}, based on the lattice gauge theory calculation.
}
\label{W5_Genus2}
\end{table}

\begin{table}[H]
\centering
\footnotesize
\begin{tabular}{cccccccccc}
$W_{\Sigma_{2,0}}$    &\vline &$W_{\Sigma_{2,0}}^{I}$&\vline    &$W_{\Sigma_{2,0}}^{A_1}$&\vline   &$W_{\Sigma_{2,0}}^{A_2}$    \\ \hline
$u=0$  &\vline     &$537$&\vline & $528$&\vline &       $528$             \\ \hline
$u=1$  &\vline     &$149+388\cdot \cos\frac{2\pi}{5}$ &\vline &$528 \cdot e^{\frac{2i\pi}{5}\times 4}$&\vline   & $528 \cdot e^{\frac{2i\pi}{5}\times 4}$  \\ \hline
$u=2$  &\vline     &$149+388\cdot \cos\frac{4\pi}{5}$&\vline  &$528 \cdot e^{\frac{2i\pi}{5}\times 3}$&\vline    &$528 \cdot e^{\frac{2i\pi}{5}\times 3}$   \\  \hline
$u=3$  &\vline      &$149+388\cdot \cos\frac{4\pi}{5}$&\vline &$528 \cdot e^{\frac{2i\pi}{5}\times 2}$&\vline  &$528 \cdot e^{\frac{2i\pi}{5}\times 2}$  \\ \hline
$u=4$  &\vline      &$149+388\cdot \cos\frac{2\pi}{5}$&\vline  &$528 \cdot e^{\frac{2i\pi}{5}\times 1}$&\vline    &$528 \cdot e^{\frac{2i\pi}{5}\times 1}$      \\ \hline
\end{tabular}
\caption{$W_{\Sigma_{2,0}}$ with the word $w_6^{(z)}$ in Eq.\eqref{Words}, based on the lattice gauge theory calculation.
}
\label{W6_Genus2}
\end{table}

\begin{table}[H]
\centering
\footnotesize
\begin{tabular}{cccccccccc}
$W_{\Sigma_{2,0}}$    &\vline &$W_{\Sigma_{2,0}}^{I}$&\vline    &$W_{\Sigma_{2,0}}^{A_1}$&\vline   &$W_{\Sigma_{2,0}}^{A_2}$    \\ \hline
$u=0$  &\vline     &$537$&\vline & $528$&\vline &        $528$            \\ \hline
$u=1$  &\vline     &$149+388\cdot \cos\frac{4\pi}{5}$ &\vline &$528 \cdot e^{\frac{2i\pi}{5}\times 2}$&\vline   &$528 \cdot e^{\frac{2i\pi}{5}\times 2}$   \\ \hline
$u=2$  &\vline     &$149+388\cdot \cos\frac{2\pi}{5}$ &\vline  &$528 \cdot e^{\frac{2i\pi}{5}\times 4}$&\vline    &$528 \cdot e^{\frac{2i\pi}{5}\times 4}$   \\  \hline
$u=3$  &\vline      &$149+388\cdot \cos\frac{2\pi}{5}$&\vline &$528 \cdot e^{\frac{2i\pi}{5}\times 1}$&\vline  &$528 \cdot e^{\frac{2i\pi}{5}\times 1}$  \\ \hline
$u=4$  &\vline      &$149+388\cdot \cos\frac{4\pi}{5}$ &\vline  &$528 \cdot e^{\frac{2i\pi}{5}\times 3}$&\vline    &$528 \cdot e^{\frac{2i\pi}{5}\times 3}$      \\ \hline
\end{tabular}
\caption{$W_{\Sigma_{2,0}}$ with the word $w_7^{(z)}$ in Eq.\eqref{Words}, based on the lattice gauge theory calculation.
}
\label{W7_Genus2}
\end{table}

\section{Discussion and conclusion}
\label{Sec: DiscussAndConclusion}

In this work, we study in detail the  representations of mapping class group of a closed genus-2 manifold 
$\Sigma_{2,0}$ and a punctured torus $\Sigma_{1,1}$.
These data provide more information than the modular data obtained on a torus $\Sigma_{1,0}$.
We use these data to construct topological invariants, which are used to distinguish the simplest 
counterexamples in MS MTCs that are indistinguishable by the modular data.

The mapping class group representation itself can be used to distinguish the counterexamples as well. 
For example, for the punctured $S$ matrix, although containing 
gauge redundancy in the entries, the diagonal elements $\sum_{\mu}(S^{(z)}_{a,\mu;a,\mu})$ are gauge invariant. 
It is found that the diagonal elements of $S^{(z)}$ together with the modular $T$ matrix can be used to 
distinguish different categories, in the sense that $\sum_{\mu}(S^{(z)}_{a,\mu;a,\mu})$ and $T$ matrix do 
not transform in a consistent way as we consider the bijection of anyons between two different categories.

Furthermore, we propose a more convenient and efficient way to distinguish different MS MTCs by constructing 
topological invariants. These topological invariants are obtained by tracing over the `words' where the `letters' are
 representations of $\text{MCG}(\Sigma_{2,0})$ or $\text{MCG}(\Sigma_{1,1})$.
With these topological invariants, there is no need to permute anyons among different categories. That is, a single
number is enough to distinguish different MS MTCs.
 
In addition, we use the lattice gauge theory approach to study the mapping class group representations 
of $\Sigma_{2,0}$ and $\Sigma_{1,1}$ for a general Dijkgraaf-Witten theory.
This approach is practical since one only needs to input the group data and the 3-cocyle $\omega$. 
This approach can in principle be used to study any counterexamples in the MS MTCs with
$G=\mathbb{Z}_q\rtimes \mathbb{Z}_p$, and the only constraint is the large size of Hilbert space for large $q$ and $p$.
In this work, we apply this approach to the simplest counterexamples in MS MTCs with $q=11$ and $p=5$. 
The topological invariants obtained from the lattice gauge theory approach agree with those obtained from
the quasi-particle basis calculation.

Comparing the two different approaches, the MCG representations may look totally different, since two different 
sets of basis vectors are considered, one with quasi-particle basis colored by anyons, and the other with group-element basis.
However, the topological invariants (which are basis independent) constructed from these MCG representations are the same.

There are some open questions to be studied in the future. We mention a few as follows.

-- There are five generators for the  representations of $\text{MCG}(\Sigma_{2,0})$ in \eqref{Genus2_5Dehntwists},
and we only study four of them, \textit{i.e.}, $S_i$ and $T_i$ ($i=1,2$).
The fifth generator $T_c$ corresponds to the Dehn twist along the closed curve $c$ in \eqref{Genus2_5Dehntwists}.
By choosing the basis vectors in \eqref{BasisI_introduction},
it can be found that to obtain $T_c$ one needs to evaluate the twice-punctured $S$ matrix as:
\be
\begin{tikzpicture}[baseline={(current bounding box.center)}]
\draw (20pt,0pt) circle (20pt);
\draw[line width=6pt, draw=white] (0pt,0pt) circle (20pt);
\draw (0pt,0pt) circle (20pt);
\draw[line width=6pt, draw=white]  (0pt,0pt) arc (-180:-270:20pt);
\draw (0pt,0pt) arc (-180:-270:20pt);
\node at (25pt,0pt){$a$};
\node at (45pt,0pt){$b$};
\draw[>=stealth,->]  (20pt,-20pt)..controls (15pt,-30pt) and (5pt,-30pt)..(0pt,-20pt);
\draw[>=stealth,->] (0pt,20pt)..controls (5pt,30pt) and (15pt,30pt)..(20pt,20pt);
\node at (10pt,32pt){$z$};
\node at (10pt,-32pt){$z'$};
\draw [>=stealth,->] (20pt,0.1pt)--(20pt,0.11pt);
\draw [>=stealth,->] (40pt,0.1pt)--(40pt,0.11pt);
\end{tikzpicture}
\ee
It is straightforward to check the twice-punctured $S$ matrix is different from the punctured $S$ matrix in \eqref{PuncturedSmatrix}
by two $F$ transformations. From this point of view, the twice-punctured $S$ matrix $S^{(z,z')}$ may contain more information
than the punctured $S$ matrix $S^{(z)}$.
Practically, once the fifth generator $T_c$ is introduced, the modular relations for the five generators of $\text{MCG}(\Sigma_{2,0})$ 
become quite complicate.\cite{Wajnryb1983,Moore1989} Solving the modular relations in the quasi-particle basis 
would become challenging.  
Nevertheless, it is still possible to calculate $T_c$ with the lattice gauge theory approach as introduced in Sec.\ref{Sec: Lattice}.

-- We only focus on the simplest counterexamples in MS MTCs with $G=\mathbb{Z}_{11}\rtimes \mathbb{Z}_5$.
It is noted there are a family of counterexamples with $G=\mathbb{Z}_q\rtimes_n \mathbb{Z}_p$
($p$ and $q$ are primes with $p|(q-1)$) that are indistinguishable by the modular data.\cite{MS1708}
It is interesting to check those counterexamples with our methods. In particular, the application of 
the lattice gauge theory approach to other counterexamples is straightforward, 
at least for the next few examples with small $p$ and $q$. 

Moreover, one possible way to show that the  representations of $\text{MCG}(\Sigma_{g,0})$ 
may distinguish \textit{all} the MS MTCs can be considered as follows.
In Ref.\onlinecite{MS1806}, it is found that the link invariant of a Borromean ring together with $T$ 
matrix are enough to distinguish all the counterexamples in MS-MTCs.
If we can express the Borromean ring in terms of MCG representations of $\Sigma_{g,0}$,
this will indicate that the representations of $\text{MCG}(\Sigma_{g,0})$ can distinguish all the counterexamples
in MS MTCs. A nice illustration of this method for $\text{MCG}(\Sigma_{2,0})$ can be found in Ref.\onlinecite{Shakirov1504}, 
as also briefly introduced in appendix. \ref{Appendix: MCG and Link invariants}
It is found that the invariant of a figure-eight knot can be obtained by evaluating the expectation value of
combinations of $\text{MCG}(\Sigma_{2,0})$ representations within a certain ground state on 
$\Sigma_{2,0}$.\cite{Shakirov1504}
On the other hand, based on a computer search, it is found that the figure-eight knot invariant together with 
the modular $T$ matrix can be used to distinguish the simplest counterexamples in MS MTCs.
Therefore, based on the connection of the figure-eight knot and $\text{MCG}(\Sigma_{2,0})$, one can 
immediately conclude that the representations of $\text{MCG}(\Sigma_{2,0})$ can be used to distinguish 
the simplest counterexamples in MS MTCs. Now, to distinguish \textit{all} the MS MTCs,
it is interesting to generalize this construction to the Borromean ring.

-- We mainly focus on the representations of $\text{MCG}(\Sigma_{g,0})$ with $g=2$ in this work. It is interesting to ask if 
there is more information contained in the representations of $\text{MCG}(\Sigma_{g,0})$ with $g>2$.
By choosing the canonical basis in the quantum Hilbert space $\mathcal{H}(\Sigma_{g,0})$ in \eqref{Genus_g_basis},
one can find that as $g$ grows, there is no new structure beyond the 
following building block:
\be\label{BasicStructure}
\begin{tikzpicture}[baseline={(current bounding box.center)}]
\draw (35pt,0pt) arc (180:-180:12pt) ;
\draw (24pt,0pt)--(35pt,0pt);
\draw (59pt,0pt)--(70pt,0pt);
\end{tikzpicture}
\ee
Considering the generators of $\text{MCG}(\Sigma_{g,0})$in \eqref{MCG_g}, this indicates that 
there will be no more information beyond the representations of $\text{MCG}(\Sigma_{3,0})$.
As a remark, it is noted that the  representations of MCG$(\Sigma_{g,0})$ has been studied in terms of 
$F$ and $R$ symbols in Ref.\onlinecite{bloomquist2018topological}.
One can check explicitly there is no more new information for the cases of $g>3$. 
In the lattice gauge theory approach, the structure in \eqref{BasicStructure} can be studied by considering the following triangulation 
(with appropriate ordering of vertices)
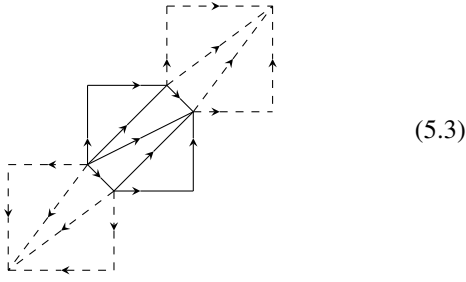
\begin{eqnarray}\label{Genus3Triangle}
\begin{small}
\begin{tikzpicture}[baseline={(current bounding box.center)}]
\draw (-20pt,20pt)--(-20pt,0pt); \draw [>=stealth,<-] (-20pt, 0pt) -- (-20pt,-10pt);
\draw (20pt,10pt)--(20pt,0pt); \draw [>=stealth,<-] (20pt, 0pt) -- (20pt,-20pt);
\draw [>=stealth,->] (-20pt,20pt)--(0pt,20pt); \draw (0pt,20pt)--(10pt,20pt);
\draw [>=stealth,->] (-10pt,-20pt)--(0pt,-20pt);  \draw (0pt, -20pt) -- (20pt,-20pt);
\draw [>=stealth,->] (-20pt,-10pt)--(-15pt,-15pt); \draw (-15pt,-15pt)--(-10pt,-20pt);

\draw [>=stealth,->] (-20pt,-10pt)--(-5pt, 5pt);\draw (-5pt,5pt)--(10pt,20pt);
\draw [>=stealth,->] (-10pt,-20pt)--(5pt,-5pt);\draw(5pt,-5pt)--(20pt, 10pt);
\draw [>=stealth,->] (-20pt,-10pt)--(0pt, 0pt);\draw (0pt,0pt)--(20pt,10pt);

\draw [dashed][>=stealth,->] (-20pt,-10pt)--(-35pt,-10pt); \draw[dashed](-35pt,-10pt)--(-50pt,-10pt);
\draw [dashed][>=stealth,->] (-10pt,-20pt)--(-10pt,-35pt);\draw[dashed](-10pt,-35pt)--(-10pt,-50pt);
\draw [dashed](-50pt,-50pt)--(-30pt,-50pt); \draw [dashed][>=stealth,<-] (-30pt,-50pt)--(-10pt,-50pt);
\draw [dashed](-50pt,-50pt)--(-50pt,-30pt); \draw[dashed][>=stealth,<-] (-50pt,-30pt)--(-50pt,-10pt);

\draw [dashed][>=stealth,->] (-10pt,-20pt)--(-30pt,-35pt); \draw [dashed](-30pt,-35pt)--(-50pt,-50pt);
\draw [dashed][>=stealth,->] (-20pt,-10pt)--(-35pt,-30pt); \draw [dashed](-35pt,-30pt)--(-50pt,-50pt);


\draw [dashed](-20+30pt,20+30pt)--(-20+30pt,0+30pt); \draw [dashed][>=stealth,<-] (-20+30pt, 0+30pt) -- (-20+30pt,-10+30pt);
\draw [dashed](20+30pt,20+30pt)--(20+30pt,0+30pt); \draw [dashed][>=stealth,<-] (20+30pt, 0+30pt) -- (20+30pt,-20+30pt);
\draw [dashed][>=stealth,->] (-20+30pt,20+30pt)--(0+30pt,20+30pt); \draw [dashed](0+30pt,20+30pt)--(20+30pt,20+30pt);
\draw [dashed][>=stealth,->] (-10+30pt,-20+30pt)--(0+30pt,-20+30pt);  \draw [dashed](0+30pt, -20+30pt) -- (20+30pt,-20+30pt);
\draw [>=stealth,->] (-20+30pt,-10+30pt)--(-15+30pt,-15+30pt); \draw (-15+30pt,-15+30pt)--(-10+30pt,-20+30pt);

\draw [dashed][>=stealth,->] (-20+30pt,-10+30pt)--(0+30pt, 5+30pt);\draw [dashed](0+30pt,5+30pt)--(20+30pt,20+30pt);
\draw [dashed][>=stealth,->] (-10+30pt,-20+30pt)--(5+30pt,0+30pt);\draw [dashed](5+30pt,0+30pt)--(20+30pt, 20+30pt);

\end{tikzpicture}
\end{small}
\end{eqnarray}
where we take a genus-3 manifold for example.
The action of Dehn twists on this basis vectors can be performed straightforwardly following the procedure in
Sec.\ref{Sec: Lattice_modularTransform}, as briefly described in appendix \ref{appendix: more punctures}.
It will be interesting to work out the details of the MCG$(\Sigma_{g,0})$ representations for MS MTCs.

-- For a (twisted) quantum double of a finite group $G$ with $\omega\in H^3(G,U(1))$, the modular $S$ and $T$ matrices can be constructed 
in terms of the group data and 3-cocyle $\omega$ (see Eqs.\eqref{Sab1} and \eqref{TopologicalSpin0}).\cite{CGR00}
It is interesting to find an explicit expression for the punctured $S$ matrix in terms of the group data and $\omega$, at least for the 
diagonal elements $\sum_{\mu}(S^{(z)}_{a,\mu;a,\mu})$  which are gauge invariant.
This question is also related to how to make a direct connection of the quasi-particle basis approach in Sec.\ref{Sec: Quasi-particle} and 
the lattice gauge theory approach in Sec.\ref{Sec: Lattice}. More explicitly, in the lattice gauge theory, it is known how to express the quasi-particle
basis of the ground state on a torus $\Sigma_{1,0}$ in terms of the group-element basis.\cite{HuWanWu1211} 
Then one can obtain the modular $S$ and $T$ matrices in terms of 
group data and 3-cocyle $\omega$. For the punctured torus $\Sigma_{1,1}$ or a genus-2 manifold $\Sigma_{2,0}$, it is open to us how to 
write the quasi-particle basis (see, \textit{e.g.}, \ref{BasisI_introduction}) in terms of the group-element basis  (see \ref{Genus2Triangle}).
Understanding this correspondence will help to express the MCG$(\Sigma_{g,n})$ representations in terms of the group data and 3-cocycle $\omega$
for a general twisted quantum double of a finite group $G$.

\section*{Acknowledgement} 

We thank for helpful conversations and communications with Xie Chen, Liang Kong, Tian Lan, Shinsei Ryu, Peter Schauenburg, and Chenjie Wang.
This research is partially supported by the Gordon and Betty Moore Foundation’s EPiQS initiative through Grant No. GBMF4303 at MIT (X.W.), 
NSF Grant No. DMR-1506475 and DMS-1664412 (X-.G.W.).

\appendix

\section{More on twisted quantum double of $G=\mathbb{Z}_q \rtimes_n \mathbb{Z}_p$}
\label{Appendix_ZqZp}

\subsection{Basic property of $G=\mathbb{Z}_q \rtimes_n \mathbb{Z}_p$}

In this subsection, we introduce the basic property of $G=\mathbb{Z}_q \rtimes_n \mathbb{Z}_p$, 
including the conjugacy class and the centralizer subgroup.
Elements of $G$ are denoted as $(a^l, b^m)$, with $a^q=b^p=1$, $l\in \{0, 1, \cdots, q-1\}$, $m\in \{0, 1, \cdots, p-1\}$,
and the multiplication is
\be
\begin{split}
(a^l,b^m)\cdot (a^{l'}, b^{m'})=&(a^l (b^m a^{l'} b^{-m}),b^{m+m'})\\
=&(a^{l+n^m l'},b^{m+m'}),
\end{split}
\ee
where we have used $bab^{-1}=a^n$. The inverse of $(a^l, b^m)$ is 
\be
\begin{split}
(a^l, b^m)^{-1}=&(b^{-m} a^{-l} b^m, b^{-m})\\
=&(b^{p-m} a^{q-l} b^{-(p-m)}, b^{p-m})\\
=&(a^{(q-l)\cdot n^{p-m}}, b^{p-m}).
\end{split}
\ee
Then one can find the conjugate of $(a^{l_0}, b^{m_0})$ by $(a^l, b^m)$ as:
\be\label{Conjugate_general}
\begin{split}
&(a^l,b^m)\cdot (a^{l_0},b^{m_0})\cdot (a^l,b^{m})^{-1}\\
=&(a^{x}, b^{m_0}),
\end{split}
\ee
where $x:=l+(q-l)\cdot n^{p+m_0}+l_0 \cdot n^m$.

For the simplest case of $l_0=m_0=0$, one has $x=0$ mod $q$. 
That is, the conjugacy class of $1:=(a^0, b^0)$ is $1$ itself.

Now let us look at the case of $l_0\neq 0$ and $m_0=0$, which correspond to the conjugacy classes for type-$A$ anyons in
the main text. One has 
\be\label{ConjugateM0}
(a^l,b^m)\cdot (a^{l_0},b^{0})\cdot (a^l,b^{m})^{-1}=(a^x,b^0),
\ee
where $x=l+(q-l)\cdot n^p+l_0\cdot n^m$. Since $n^p=1$ mod $q$, then $x$ can be simplified as
\be
x=l_0\cdot n^m\quad \text{mod} \,\, q.
\ee
Then it is convenient to choose $(a^l,b^0)$ as the presentative in the corresponding conjugacy class, with
$l\in \mathbb{Z}_q^{\times}/\langle n\rangle$, which is Eq.\eqref{ConjugacyA}. 
Then the conjugacy class $[(a^l,b^0)]$ is $(a^x,b^0)$ with $x=l\cdot n^m$, where $m=0, 1, \cdots, p-1$.
Apparently, the size of conjugacy class $[(a^l,b^0)]$ is $p$, and the number of conjugacy classes of this type
are $\frac{q-1}{p}$.

Then let us look at the case of $m_0\neq 0$, \textit{i.e.}, $m\in \{1,2,\cdots, p-1\}$ 
in Eq.\eqref{Conjugate_general}. This corresponds to the conjugacy class
for the type-$B$ anyon. Let us check $l_0=0$ for simplicity.
Then $x$ in Eq.\eqref{Conjugate_general} can be further simplified as $x:=l(1- n^{m_0})$ mod $q$.
It is found that by choosing all possible $l$ with $l\in\{0, 1, \cdots, q-1\}$, $x$ can be any value in $\{0, 1, \cdots, q-1\}$.
That is, the conjugacy class $[(a^l, b^m)]=\{(a^0,b^m), (a^1,b^m),\cdots, (a^{q-1},b^m)\}$.
The size of conjugacy class $[(a^l, b^m)]$ is $q$, and the number of conjugacy classes $[(a^l, b^m)]$ with $m\neq 0$ is $p-1$,
corresponding to $m=1,2,\cdots, p-1$.

Next, let us check the centralizer subgroup of $(a^{l_0}, b^{m_0})$.
First, it is straightforward to see that for $l_0=m_0=0$, the corresponding centralizer subgroup is $G$.
Then for $l_0\neq 0$ and $m_0=0$, the group elements in the centralizer subgroup $\{(a^l, b^m)\}$ of $(a^{l_0}, b^0)$ should satisfy
Eq.\eqref{ConjugateM0} with $x=l_0$. That is, $l_0=l+(q-l)\cdot n^p+l_0\cdot n^m$ mod $q$. Note that $n^p=1$ mod $q$, then one has
$l_0=l_0\cdot n^m$ mod $q$ for arbitrary $l$. Here $m$ can only be chosen as $m=0$. Therefore, the centralizer subgroup of 
$(a^{l_0},b^0)$ is $\mathbb{Z}_q=\{(a^l,b^0)| l=0, 1,\cdots, q-1\}$. 

Then let us check the centralizer subgroup of $(a^{l_0},b^{m_0})$ with $m_0\neq 0$.
The group elements $(a^l,b^m)$ in the centralizer subgroup satisfy Eq.\eqref{Conjugate_general}
with $x=l_0$, \textit{i.e.}, $l_0=l+(q-l)\cdot n^{p+m_0}+l_0\cdot n^m$ mod $q$. This can be further simplified as
$l_0(1-n^m)=l(1-n^{m_0})$. Note that for the case of $l_0=0$, one has $l=0$. That is, the centralizer of
$(a^0, b^{m_0})$ is  $\mathbb{Z}_p=\{(a^0, b^m)| m=0, 1, \cdots, p-1\}$. For general $l_0\neq 0$, the centralizer of
$(a^{l_0}, b^{m_0})$ is also an abelian group $\mathbb{Z}_p=\{(a^l,b^m)| l_0(1-n^m)=l(1-n^{m_0}), m=0, 1, 2, \cdots, p-1\}$.
For arbitrary two elements $(a^l, b^m)$ and $(a^{l'},b^{m'})$ in the centralizer, one can check
that $l(1-n^{m'})=l'(1-n^m)$, and thus the two group elements commute with each other.

\subsection{More on modular $S$ matrix}
\label{Appendix: Smatrix}

In this part, we discuss more properties of the modular $S$-matrix.
We show that for MS MTCs with $G=\mathbb{Z}_q\rtimes \mathbb{Z}_p$, only $S_{B_{k,n}, B_{k',n'}}$
depend on the 3-cocycle $\omega\in H^3(G, U(1))$. In addition, we give explicit expressions for the modular $S$ matrix
of MS MTCs with $G=\mathbb{Z}_{11}\rtimes \mathbb{Z}_5$.

In Sec.\ref{MS MTCs: anyons}, we show there are three types of anyons in MS MTCs, \textit{i.e.}, type-$I$ anyons $(1,\chi_i)$, 
type-$A$ anyons $(a^l, \chi_m)$, and type-$B$ anyons $(b^k, \tilde{\chi}_n )$. Only the characters $\tilde{\chi}$ in type-$B$ anyons are
. For the modular $S$ matrix in Eq.\eqref{Sab1}, it is expressed as a combination of different characters corresponding
to the types of anyons. One can find that if we require $S_{ab}$ depends on the 3-cocycle $\omega\in H^3(G, U(1))$, at least one of $a$ and $b$
should be type-$B$ anyons. With this observation, now let us check the following three cases: (1) $a=I_i$ and $b=B_{k,n}$, or $a=B_{k,n}$ and $b=I_i$;
(2) $a=A_{l,m}$ and $b=B_{k,n}$, or $a=B_{k,n}$ and $b=A_{l,m}$;  (3) $a=B_{k,m}$ and $b=B_{k',n}$.

For case (1), since $S_{ab}=S_{ba}$, we just need to consider $a=I_i$ and $b=B_{k,n}$.
Based on Eq.\eqref{Sab1}, one can find that
\be
S_{(1,\chi_i), (b^k, \tilde{\chi}_n)}=\frac{1}{|G|}\sum_{h\in [b^k]} \big(\chi_i(h)\big)^{\ast} \cdot \big(\tilde{\chi}^{h}_n (1) \big)^{\ast}.
\ee
From Eq.\eqref{Character_B}, the  irreducible representation is related to the linear one as 
$\tilde{\chi}^{h}_n(x)=\chi_n (x)\cdot \epsilon_g(x)$. Here $\chi_n$ is the linear representation of $\mathbb{Z}_p$, and $\epsilon_g(x)$
is expressed in Eq.\eqref{epsilon_bk}. One can find that $\tilde{\chi}^{h}_n (1)=1$ for arbitrary $h\in [b^k]$.
Therefore, we have $S_{(1,\chi_i), (b^k, \tilde{\chi}_n)}=\frac{1}{|G|}\sum_{h\in [b^k]} \big(\chi_i(h)\big)^{\ast}$, which is independent of
the 3-cocycle $\omega$.

Simlarly, for case (2), let us consider $a=A_{l,m}$ and $b=B_{k,n}$.
One can check that for arbitrary $g\in [a^l]$ and $h\in [b^k]$, $g$ and $h$ will not commute with each other.
This can be straightforwardly seen as follows. For $h=(a^{r}, b^k)\in [b^k]$, from the analysis in the previous section, if $g=(a^{l'}, b^0)\in [a^l]$
commutes with $h$, then one has $r(1-n^0)=l' (1-n^{k})$. Since $n^k\neq 1$ mod $q$, and $l'\neq 0$ mod $q$, the above equation cannot hold.
Therefore, $g$ and $h$ cannot commute with each other. Then from Eq.\eqref{Sab1}, one can find that $S_{A_{l,m}, B_{k,n}}=0$.

For case (3), when both indexes $a$ and $b$ belong to type-$B$ anyons, $S_{ab}$ has a very concise expression.
We denote the two anyons as $(b^k, \tilde{\chi}_m)$ and $(b^{k'},\tilde{\chi}_n)$.
Recall that the conjugacy class of $b^k$ is $[b^k]=\{(a^0,b^k), (a^1,b^k),\cdots, (a^{p-1}, b^k)\}$, then for $g\in [b^k]$, there is a unique
$h\in [b^{k'}]$ so that $[g,h]=1$. Denoting $g=(a^l, b^k)$, then $h=(a^{l'}, b^{k'})$ is uniquely determined by
$l(1-n^{k'})=l'(1-n^k)$. One can find that
\be
\tilde{\chi}^{g}_m(h)=e^{\frac{2\pi i}{p}\cdot mk'}\cdot e^{\frac{2\pi i}{p^2} k k' u},
\ee
and 
\be
\tilde{\chi}^{h}_n(g)=e^{\frac{2\pi i}{p}\cdot nk}\cdot e^{\frac{2\pi i}{p^2} k k' u}.
\ee
Then, the modular $S$matrix in Eq.\eqref{Sab1} can be expressed as
\be\label{SBB_appendix}
\small
\begin{split}
S^{(u)}_{(b^k,\tilde{\chi}_m),(b^{k'},\tilde{\chi}_n)}=&\frac{1}{|G|}\cdot q \cdot \exp\Big( -\frac{2\pi i}{p^2} \big[2u k k'+p(kn + k'm)\big]\Big),\\
=&\frac{1}{p}\exp\Big( -\frac{2\pi i}{p^2} \big[2u k k'+p(kn + k'm)\big]\Big),
\end{split}
\ee
where we have considered $|G|=pq$. Then we obtain Eq.\eqref{S_BB} in the main text.

In the following, we will focus on the simplest exmples  of MS MTCs with $G=\mathbb{Z}_{11}\rtimes \mathbb{Z}_5$, and give the expression of $S$ matrix explicitly. 
The simple objects are labeled 
as [see Eq.\eqref{Z11Z5Simples}]: 
$I_i:=\left(1,\chi_i\right)$, $A_{l,m}=\left(a^l,\omega_{11}^m\right)$, and $B_{k,n}=\left(b^k,\tilde{\omega}_5^n\right)$.
Here $\chi_i$, $\omega_{11}^m$, and $\tilde{\omega}_{5}^n$ represent the () irreducible representations of $G$, $\mathbb{Z}_{11}$, and $\mathbb{Z}_5$,
respectively. We will check $S_{ab}$ for different cases explicitly:

(1) $a,b\in \{I_i\}$.

(2) $a,b\in \{A_{l,m}\}$.

(3) $a,b \in \{B_{k,n}\}$.

(4) $a \in \{I_i\}$ and $b\in \{A_{l,m}\}$, or $a\in \{A_{l,m}\}$ and $b \in \{I_i\}$.

(5) $a \in \{I_i\}$ and $b\in \{B_{k,n}\}$, or $a\in \{B_{k,n}\}$ and $b \in \{I_i\}$

(6) $a\in \{A_{l,m}\}$ and $b\in \{B_{k,n}\}$, or $a\in \{B_{k,n}\}$ and $b\in \{A_{l,m}\}$

The details are as follows.

(1) $a,b\in \{I_i\}$.
That is, we check $S_{I_i,I_j}$ first. Based on Eq.\eqref{Sab1}, one can obtain
\be
\small
S_{I_i, I_j}=\frac{1}{|G|} \chi^{\ast}_i (1) \chi^{\ast}_j(1),
\ee
where $1$ is the identity group element in $G$. For $G=\mathbb{Z}_{11}\rtimes \mathbb{Z}_5$, the characters are shown in Eq.\eqref{Gcharacter}.
One can find that for $I_i, I_j\in \{I_{1}, \cdots, I_5\}$, $S_{I_i, I_j}=\frac{1}{55}$; for $I_i, I_j\in \{I_6, I_7\}$, then $S_{I_i,I_j}=\frac{5\times 5}{55}=\frac{5}{11}$.
If $I_i, I_j$ belong to $\{I_{1}, \cdots, I_5\}$ and $I_6, I_7$ separately, then one has $S_{I_i,I_j}=\frac{1\times 5}{55}=\frac{1}{11}$.

(2) $a,b\in \{A_{l,m}\}$.
Now we check $S_{A_{l_1,m_1},A_{l_2,m_2}}$, with $l_1, l_2=1, 2$, and $m_1,\, m_2=0, 1,\cdots, 10$. 
The conjugacy class of $a^l:=(a^l,b^0)$ is shown in Table \ref{ConjugacyClass}, and the centralizer of 
$a^l$ is $\mathbb{Z}_{11}=\{a^0, a^1, \cdots, a^{10}\}$. Then, based on Eq.\eqref{Sab1}, we have
\be
\small
S_{A_{l_1,m_1}, A_{l_2,m_2}}=\frac{1}{|G|}\sum_{\substack{a^{l_1'}\in [a^{l_1}], \\a^{l_2'}\in [a^{l_2}]}} \big(\omega_{11}^{m_1}(a^{l_2'})\big)^{\ast} \cdot \big(\omega_{11}^{m_2}(a^{l_1'})\big)^{\ast}
\ee
where $\omega_{11}^m(a^l)=e^{\frac{2\pi i}{11} ml}$. Then $S_{A_{l_1,m_1}, A_{l_2,m_2}}$ can be further written as
\be
S_{A_{l_1,m_1}, A_{l_2,m_2}}=\frac{1}{|G|}\sum_{a^{l_1'}\in [a^{l_1}], a^{l_2'}\in [a^{l_2}]}e^{-
\frac{2\pi i}{11}(m_1\cdot l_2'+m_2\cdot l_1')
}
\ee
There are in total $25$ terms in the summation. This can be simplified in certain cases.
In particular, for $m_1=m_2=0$, the above formula can be simplified as
$S_{A_{l_1,0}, A_{l_2,0}}=\frac{25}{55}=\frac{5}{11}$.
If one of $m_1$ and $m_2$ is zero, say $m_1=0$, then one has
$S_{A_{l_1,0}, A_{l_2,m_2}}=\frac{|[a^{l_2}]|}{|G|}\sum_{a^{l_1'}\in[a^{l_1}]} \big(\omega^{m_2}_{11}(a^{l_1'})\big)^{\ast}$.

(3) $a,b \in \{B_{k,n}\}$.
The results are obtained in Eq.\eqref{SBB_appendix}. For $G=\mathbb{Z}_{11}\rtimes \mathbb{Z}_5$, one has
\be\label{SBBZ11Z5_appendix}
\small
S^{(u)}_{(b^k,\tilde{\chi}_m),(b^{k'},\tilde{\chi}_n)}
=\frac{1}{5}\exp\Big( -\frac{2\pi i}{25} \big[2u k k'+5(kn + k'm)\big]\Big),
\ee
where $u=0,1,2,3,4$.

(4) $a \in \{I_i\}$ and $b\in \{A_{l,m}\}$, or $a\in \{A_{l,m}\}$ and $b \in \{I_i\}$.

We consider the case of $a \in \{I_i\}$ and $b\in \{A_{l,m}\}$, and the later case can be obtained based on the symmetric 
property of modular $S$ matrix, \textit{i.e.}, $S_{ab}=S_{ba}$. One can find
\be
S_{(1, \chi_i), (a^l, \omega_{11}^m)}=\frac{1}{|G|}\sum_{h\in [a^l]} \chi^{\ast}_i(h),
\ee
where we have considered $\omega_{11}^m(1)=1$.
If $I_i=I_0, I_1, I_2, I_3$ or $I_4$, then $\chi_i([a^l])=1$, and the $S$ matrix can be simplified as 
$S_{I_i, A_{l,m}}=\frac{1}{11}$.
If $I_i=I_5,I_6$, then from the character table in Eq.\eqref{Gcharacter}, one can obtain
$S_{I_5, A_{1,m}}=S_{I_6, A_{2,m}}=\frac{1}{11}\sigma^{\ast}$, and 
$S_{I_5, A_{2,m}}=S_{I_6, A_{1,m}}=\frac{1}{11}\sigma$, where 
$\sigma=e^{\frac{2\pi i}{11}1}+e^{\frac{2\pi i}{11}3}+e^{\frac{2\pi i}{11}4}+e^{\frac{2\pi i}{11}5}+e^{\frac{2\pi i}{11}9}$.

(5) $a \in \{I_i\}$ and $b\in \{B_{k,n}\}$, or $a\in \{B_{k,n}\}$ and $b \in \{I_i\}$.
Let us consider $a \in \{I_i\}$ and $b\in \{B_{k,n}\}$.
Then based on Eq.\eqref{Sab1}, one can find that
\be
\small
S_{(1,\chi_i), (b^k, \tilde{\omega}_5^n)}=\frac{1}{|G|}\sum_{h\in [b^k]} \chi_i(h) \cdot \tilde{\omega}_5^n(1)=\frac{1}{|G|}\sum_{h\in [b^k]} \chi_i(h),
\ee
where we have considered the fact that $\tilde{\omega}_5(1)=1$.
In particular, for $I_i=I_5, I_6$, from the character table in Eq.\eqref{Gcharacter}, one can find that $\chi_i([b^k])=0$, and then $S_{I_i, B_{k,n}}=0$.
For $I_i=I_0$, one has $\chi_0([b^k])=1$, and then $S_{I_0, B_{k,n}}=\frac{1}{5}$. For $I_j=I_1, I_2, I_3, I_4$, one has
$\chi_j([b^k])=e^{\frac{2\pi i}{5}\cdot j\cdot k}$. Then the $S$ matrix has the form
$S_{I_j, B_{k,n}}=S_{B_{k,n},I_j}=\frac{1}{5}e^{-\frac{2\pi i}{5}\cdot j\cdot k}$.

(6) $a\in \{A_{l,m}\}$ and $b\in \{B_{k,n}\}$, or $a\in \{B_{k,n}\}$ and $b\in \{A_{l,m}\}$. Let us consider the case of 
$a\in \{A_{l,m}\}$ and $b\in \{B_{k,n}\}$. One can find that for arbitrary $g\in [a^l]$ and $h\in [b^k]$ in Eq.\eqref{Sab1}, 
they do not communicate with each other. Then we have
$S_{A_{l,m}, B_{k,n}}=0$. In fact, this result applies to an arbitrary $G=\mathbb{Z}_q\rtimes \mathbb{Z}_p$ in MS MTCs.

As a short summary, only $S_{(b^k,\tilde{\chi}_m),(b^{k'},\tilde{\chi}_n)}$ depend on the equivalent cohomology class $u$.
Other components of the modular $S$ matrix are determined by the finite group $G$ itself.

\section{On algebraic theory of anyons and others}
\label{Sec: Appendix_anyon}

In this appendix, we introduce some notions and conventions on algebraic theory of anyons that are necessary 
for studying the punctured $S$ and $T$ matrices, as well as the topological invariants in the main text.
A more complete description of the algebraic theory of anyons can be found, \textit{e.g.}, 
in Refs.\onlinecite{Moore1989,Kitaev2006,bonderson2007}.
We also introduce other choices of quasi-particle basis in $\mathcal{H}(\Sigma_{g,0})$, and 
express the  representation of MCG$(\Sigma_{g,0})$ in terms of $F$ and $R$ matrices.

We assign a fusion vector space $V_{ab}^c$ to each fusion product of two anyons $a\times b=\sum_c N_{ab}^c c$.
The dual space, also called `splitting space', is denoted as $V^{ab}_c$.
The numbers $N_{ab}^c=\text{dim}(V_{ab}^c)=\text{dim}(V^{ab}_c)$ are called fusion multiplicities.
If $N_{ab}^c$ are equal to 0 or 1, we will call such fusion rules multiplicity free.
The orthonormal basis vectors $|a,b;c,\mu\rangle\in V_c^{ab}$ and $\langle a,b;c,\mu| \in V_{ab}^c$ can be
diagrammatically expressed as
\be\label{BasisVectors}
\begin{split}
(d_c/d_ad_b)^{1/4}
\begin{tikzpicture}[baseline={(current bounding box.center)}]
\draw [>=stealth,->] (0pt,-15pt)--(0pt,-6pt);\draw (0pt,-6pt)--(0pt,0pt);
\draw [>=stealth,->] (0pt,0pt)--(-6pt,8pt); \draw(-6pt,8pt)--(-7.5pt,10pt);
\draw [>=stealth,->] (0pt,0pt)--(6pt,8pt); \draw(6pt,8pt)--(7.5pt,10pt);
\node at (-11pt,12pt){$a$};
\node at (11pt,12pt){$b$};
\node at (0pt,-20pt){$c$};
\node at (6pt,-2pt){\small$\mu$};
\end{tikzpicture}
&=: |a,b;c,\mu\rangle \in V_c^{ab},\\
(d_c/d_ad_b)^{1/4}
\begin{tikzpicture}[baseline={(current bounding box.center)}]
\draw [>=stealth,->] (0pt,-15+15pt)--(0pt,-6+15pt);\draw (0pt,-6+15pt)--(0pt,0+13pt);
\draw [>=stealth,->] (-8pt,-10pt)--(-4pt,-5pt); 
\draw(-4pt,-5pt)--(0pt,0pt);
\draw [>=stealth,->](8pt,-10pt)--(4pt,-5pt); 
\draw(4pt,-5pt)--(0pt,0pt);
\node at (4pt,2pt){\small$\mu$};
\node at (-11pt,-12pt){$a$};
\node at (11pt,-12pt){$b$};
\node at (0pt,17pt){$c$};
\end{tikzpicture}
&=: \langle a,b;c,\mu| \in V^c_{ab},
\end{split}
\ee
where  the normalization factor $(d_c/d_ad_b)^{1/4}$ is introduced so that diagrams are in the
isotopy invariant convention.\cite{Kitaev2006,bonderson2007}

In Eq.\eqref{CzMatrixElement} in the main text, 
we use the operation of $R$ move, which describes the braiding of two anyons.
The braiding operations of pairs of anyons can be expressed as:\cite{Moore1989,Kitaev2006,bonderson2007}
\be\label{R_move}
\small
R_{ab}=
\begin{tikzpicture}[baseline={(current bounding box.center)}]
\draw [>=stealth,->] (10pt,-10pt)--(-5pt,5pt);\draw(-5pt,5pt)--(-10pt,10pt);
\draw [line width=4pt, draw=white] (-10pt,-10pt)--(10pt,10pt);
\draw [>=stealth,->]  (-10pt,-10pt)--(5pt,5pt); \draw(5pt,5pt)--(10pt,10pt);
\node at (-12pt,-12pt){$a$};
\node at (12pt,-12pt){$b$};
\end{tikzpicture},
\quad
R_{ab}^{-1}=R_{ab}^{\dag}=
\begin{tikzpicture}[baseline={(current bounding box.center)}]
\draw [>=stealth,->]  (-10pt,-10pt)--(5pt,5pt); \draw(5pt,5pt)--(10pt,10pt);
\draw [line width=4pt, draw=white] (10pt,-10pt)--(-10pt,10pt);
\draw [>=stealth,->] (10pt,-10pt)--(-5pt,5pt);\draw(-5pt,5pt)--(-10pt,10pt);
\node at (-12pt,-12pt){$b$};
\node at (12pt,-12pt){$a$};
\end{tikzpicture}.
\ee
By acting on the basis vectors in Eq.\eqref{BasisVectors}, we have
\be
\small
\begin{split}
&R_{ab}|a,b;c,\mu\rangle=\sum_{\nu}[R^{ab}_c]_{\mu\nu} |b,a;c,\nu\rangle,\\
& R_{ab}^{-1}|b,a;c,\mu\rangle=\sum_{\nu}[R^{ab}_c]^{-1}_{\mu\nu} |a,b;c,\nu\rangle,
\end{split}
\ee
which are represented diagrammatically as 
\be\label{Rab_c}
\begin{tikzpicture}[baseline={(current bounding box.center)}]
\draw [>=stealth,->] (0pt,-15pt)--(0pt,-6pt);\draw (0pt,-6pt)--(0pt,0pt);
\draw [>=stealth,->] (0pt,0pt)..controls (7pt,2pt) and (7pt,4pt)..(-8pt,12pt);
\draw [line width=4pt, draw=white](0pt,0pt)..controls (-7pt,2pt) and (-7pt,4pt)..(8pt,12pt);
\draw [>=stealth,->] (0pt,0pt)..controls (-7pt,2pt) and (-7pt,4pt)..(8pt,12pt);
\node at (-12pt,15pt){$b$};
\node at (12pt,15pt){$a$};
\node at (0pt,-20pt){$c$};
\node at (6pt,-2pt){\small$\mu$};
\end{tikzpicture}
=R_{ab}
\begin{tikzpicture}[baseline={(current bounding box.center)}]
\draw [>=stealth,->] (0pt,-15pt)--(0pt,-6pt);\draw (0pt,-6pt)--(0pt,0pt);
\draw [>=stealth,->] (0pt,0pt)--(-6pt,8pt); \draw(-6pt,8pt)--(-7.5pt,10pt);
\draw [>=stealth,->] (0pt,0pt)--(6pt,8pt); \draw(6pt,8pt)--(7.5pt,10pt);
\node at (-11pt,12pt){$a$};
\node at (11pt,12pt){$b$};
\node at (0pt,-20pt){$c$};
\node at (6pt,-2pt){\small$\mu$};
\end{tikzpicture}
=\sum_{\nu} (R^{ab}_c)_{\mu\nu}
\begin{tikzpicture}[baseline={(current bounding box.center)}]
\draw [>=stealth,->] (0pt,-15pt)--(0pt,-6pt);\draw (0pt,-6pt)--(0pt,0pt);
\draw [>=stealth,->] (0pt,0pt)--(-6pt,8pt); \draw(-6pt,8pt)--(-7.5pt,10pt);
\draw [>=stealth,->] (0pt,0pt)--(6pt,8pt); \draw(6pt,8pt)--(7.5pt,10pt);
\node at (-11pt,11pt){$b$};
\node at (11pt,12pt){$a$};
\node at (0pt,-20pt){$c$};
\node at (6pt,-2pt){\small$\nu$};
\end{tikzpicture},
\ee
and
\be\label{Rba_c}
\begin{tikzpicture}[baseline={(current bounding box.center)}]
\draw [>=stealth,->] (0pt,-15pt)--(0pt,-6pt);\draw (0pt,-6pt)--(0pt,0pt);
\draw [>=stealth,->] (0pt,0pt)..controls (-7pt,2pt) and (-7pt,4pt)..(8pt,12pt);

\draw [line width=4pt, draw=white](0pt,0pt)..controls (7pt,2pt) and (7pt,4pt)..(-8pt,12pt);
\draw [>=stealth,->] (0pt,0pt)..controls (7pt,2pt) and (7pt,4pt)..(-8pt,12pt);
\node at (-12pt,15pt){$a$};
\node at (12pt,15pt){$b$};
\node at (0pt,-20pt){$c$};
\node at (6pt,-2pt){\small$\mu$};
\end{tikzpicture}
=R^{-1}_{ab}
\begin{tikzpicture}[baseline={(current bounding box.center)}]
\draw [>=stealth,->] (0pt,-15pt)--(0pt,-6pt);\draw (0pt,-6pt)--(0pt,0pt);
\draw [>=stealth,->] (0pt,0pt)--(-6pt,8pt); \draw(-6pt,8pt)--(-7.5pt,10pt);
\draw [>=stealth,->] (0pt,0pt)--(6pt,8pt); \draw(6pt,8pt)--(7.5pt,10pt);
\node at (-11pt,12pt){$b$};
\node at (11pt,12pt){$a$};
\node at (0pt,-20pt){$c$};
\node at (6pt,-2pt){\small$\mu$};
\end{tikzpicture}
=\sum_{\nu} (R^{ab}_c)^{-1}_{\mu\nu}
\begin{tikzpicture}[baseline={(current bounding box.center)}]
\draw [>=stealth,->] (0pt,-15pt)--(0pt,-6pt);\draw (0pt,-6pt)--(0pt,0pt);
\draw [>=stealth,->] (0pt,0pt)--(-6pt,8pt); \draw(-6pt,8pt)--(-7.5pt,10pt);
\draw [>=stealth,->] (0pt,0pt)--(6pt,8pt); \draw(6pt,8pt)--(7.5pt,10pt);
\node at (-11pt,12pt){$a$};
\node at (11pt,12pt){$b$};
\node at (0pt,-20pt){$c$};
\node at (6pt,-2pt){\small$\nu$};
\end{tikzpicture}.
\ee
The $R$ matrices satisfy the so-called ribbon property:
\be\label{Ribbon}
\small
\sum_{\lambda}[R^{ab}_c]_{\mu\lambda}[R^{ba}_c]_{\lambda,\nu}=\frac{\theta_c}{\theta_a \theta_b}\delta_{\mu,\nu}.
\ee
Another useful quantity is the so-called $F$ matrix, or $F$-symbol, defined as
\be\label{F_matrix}
\small
\begin{tikzpicture}[baseline={(current bounding box.center)}]
\draw [>=stealth,->] (0pt,0pt)--(-7pt,7pt);\draw (-7pt,7pt)--(-10pt,10pt);
\draw [>=stealth,->] (-10pt,10pt)--(-17pt,17pt);\draw (-17pt,17pt)--(-20pt,20pt);
\draw [>=stealth,->] (-20pt,20pt)--(-27pt,27pt);\draw (-27pt,27pt)--(-30pt,30pt);

\draw [>=stealth,->] (-20pt,20pt)--(-20+7pt,20+7pt);\draw (-20+7pt,20+7pt)--(-20+10pt,20+10pt);
\draw [>=stealth,->] (-10pt,10pt)--(-10+12pt,10+12pt);\draw (-10+12pt,10+12pt)--(-10+20pt,10+20pt);
\node at (-30pt,35pt){$a$};
\node at (-10pt,35pt){$b$};
\node at (10pt,35pt){$c$};
\node at (4pt,-2pt){$d$};

\node at (-12pt,17pt){$e$};
\node at (-22pt,16pt){$\alpha$};
\node at (-12pt, 5pt){$\beta$};
\end{tikzpicture}
=\sum_{f,\mu,\nu} [F^{abc}_d]_{(e,\alpha,\beta),(f,\mu,\nu)}
\begin{tikzpicture}[baseline={(current bounding box.center)}]
\draw [>=stealth,->] (0pt,0pt)--(-7pt,7pt);\draw (-7pt,7pt)--(-10pt,10pt);
\draw [>=stealth,->] (-10pt,10pt)--(-10-12pt,10+12pt);\draw (-10-12pt,10+12pt)--(-30pt,30pt);

\draw [>=stealth,->] (-10pt,10pt)--(-10+7pt,10+7pt);\draw (-10+7pt,10+7pt)--(-10+10pt,10+10pt);
\draw [>=stealth,->] (0pt,20pt)--(0+7pt,20+7pt);\draw (0+7pt,20+7pt)--(10pt,20+10pt);
\draw [>=stealth,->] (0pt,20pt)--(0-7pt,20+7pt); \draw (0-7pt,20+7pt)--(-10pt,30pt);
\node at (4pt,-2pt){$d$};
\node at (-30pt,35pt){$a$};
\node at (-10pt,35pt){$b$};
\node at (10pt,35pt){$c$};

\node at (-12pt, 6pt){$\nu$};
\node at (2pt, 16pt){$\mu$};
\node at (-8pt, 17pt){$f$};
\end{tikzpicture}
\ee
where $\alpha, \beta, \mu$ and $\nu$ denote the fusion channels.
As will be seen later, $F$-symbol will be used in expressing the  representation for 
MCG$(\Sigma_{g,0})$ in Sec.\ref{Appendix: Other basis}.
We may use $F$, $R$ symbols and $F$, $R$ matrices interchangeably.
The $F$ and $R$ symbols satisfy consistency conditions called Pentagon and Hexagon equations.\cite{Kitaev2006}
In certain cases, it is useful to consider the following $F$ transformation:
\be\label{F_move_II}
\small
\begin{tikzpicture}[baseline={(current bounding box.center)}]
\draw [>=stealth,->](-8pt,-16pt)--(-8pt,-6pt);\draw (-8pt,-6pt)--(-8pt,0pt);
\draw [>=stealth,->](-8pt,0pt)--(-8pt,10pt); \draw(-8pt,10pt)--(-8pt,16pt);

\draw [>=stealth,->](8pt,-16pt)--(8pt,-6pt);\draw (8pt,-6pt)--(8pt,0pt);
\draw [>=stealth,->](8pt,0pt)--(8pt,10pt); \draw(8pt,10pt)--(8pt,16pt);

\draw [>=stealth,->](8pt,-4pt)--(0pt,0pt); \draw(0pt,0pt)--(-8pt,4pt);

\node at (-8pt,-20pt){$c$};
\node at (-8pt,20pt){$a$};
\node at (8pt,-20pt){$d$};
\node at (8pt,21pt){$b$};
\node at (0pt,4pt){$e$};
\node at (-12pt,5pt){$\alpha$};
\node at ( 12pt,-4pt){$\beta$};
\end{tikzpicture}
=
\sum_{f,\mu,\nu}
\left[
F^{ab}_{cd}
\right]_{(e,\alpha,\beta),(f,\mu,\nu)}
\begin{tikzpicture}[baseline={(current bounding box.center)}]
\draw [>=stealth,->](-10pt,-18pt)--(-5pt,-13pt);\draw(-5pt,-13pt)--(0pt,-8pt);
\draw [>=stealth,->](10pt,-18pt)--(5pt,-13pt);\draw(5pt,-13pt)--(0pt,-8pt);
\draw [>=stealth,->](0pt,-8pt)--(0pt,2pt); \draw(0pt,2pt)--(0pt,8pt);
\draw [>=stealth,->](0pt,8pt)--(-7pt,15pt);\draw(-7pt,15pt)--(-10pt,18pt);
\draw [>=stealth,->](0pt,8pt)--(7pt,15pt);\draw(7pt,15pt)--(10pt,18pt);
\node at (-13pt,-18pt){$c$};
\node at (-13pt,18pt){$a$};
\node at (13pt,-18pt){$d$};
\node at (13pt,18pt){$b$};
\node[right] at (0pt,6pt){$\mu$};
\node[right] at (-10pt,0pt){$f$};
\node[right] at (0pt,-6pt){$\nu$};
\end{tikzpicture}
\ee
where 
\be
\small
\left[
F^{ab}_{cd}
\right]_{(e,\alpha,\beta),(f,\mu,\nu)}
=
\sqrt{\frac{d_e d_f}{d_a d_d}} \left[
F^{ceb}_f
\right]^{\ast}_{(a,\alpha,\mu),(d,\beta,\nu)}.
\ee

In the following are some relations we will use in Sec.\ref{Sec: Quasi-particle} in the main text.
\begin{eqnarray}\label{Fuse_ab}
\small
\begin{tikzpicture}[baseline={(current bounding box.center)}]
\draw [>=stealth,->](-8pt,-16pt)--(-8pt,2pt);\draw(-8pt,2pt)--(-8pt,16pt);
\draw [>=stealth,->](8pt,-16pt)--(8pt,2pt); \draw(8pt,2pt)--(8pt,16pt);
\node at (-8pt,-20pt){$a$};
\node at (-8pt,20pt){$a$};
\node at (8pt,-20pt){$b$};
\node at (8pt,21pt){$b$};
\end{tikzpicture}
=
\sum_{c,\mu}
\sqrt{\frac{d_c}{d_ad_b}}
\begin{tikzpicture}[baseline={(current bounding box.center)}]
\draw [>=stealth,->](-10pt,-18pt)--(-5pt,-13pt);\draw(-5pt,-13pt)--(0pt,-8pt);
\draw [>=stealth,->](10pt,-18pt)--(5pt,-13pt);\draw(5pt,-13pt)--(0pt,-8pt);
\draw [>=stealth,->](0pt,-8pt)--(0pt,2pt); \draw(0pt,2pt)--(0pt,8pt);
\draw [>=stealth,->](0pt,8pt)--(-7pt,15pt);\draw(-7pt,15pt)--(-10pt,18pt);
\draw [>=stealth,->](0pt,8pt)--(7pt,15pt);\draw(7pt,15pt)--(10pt,18pt);
\node at (-13pt,-18pt){$a$};
\node at (-13pt,18pt){$a$};
\node at (13pt,-18pt){$b$};
\node at (13pt,18pt){$b$};
\node[right] at (0pt,6pt){$\mu$};
\node[right] at (-10pt,0pt){$c$};
\node[right] at (0pt,-6pt){$\mu$};
\end{tikzpicture}
\end{eqnarray}
Based on Eqs.\eqref{Fuse_ab} and \eqref{Rab_c}, one may express $R_{ab}$ in Eq.\eqref{R_move} as
\begin{eqnarray}\label{R_ab}
\small
R_{ab}
=
\sum_{c,\mu,\nu}
\sqrt{\frac{d_c}{d_ad_b}} [R^{ab}_c]_{\mu\nu}
\begin{tikzpicture}[baseline={(current bounding box.center)}]
\draw [>=stealth,->](-10pt,-18pt)--(-5pt,-13pt);\draw(-5pt,-13pt)--(0pt,-8pt);
\draw [>=stealth,->](10pt,-18pt)--(5pt,-13pt);\draw(5pt,-13pt)--(0pt,-8pt);
\draw [>=stealth,->](0pt,-8pt)--(0pt,2pt); \draw(0pt,2pt)--(0pt,8pt);
\draw [>=stealth,->](0pt,8pt)--(-7pt,15pt);\draw(-7pt,15pt)--(-10pt,18pt);
\draw [>=stealth,->](0pt,8pt)--(7pt,15pt);\draw(7pt,15pt)--(10pt,18pt);
\node at (-13pt,-18pt){$a$};
\node at (-13pt,18pt){$b$};
\node at (13pt,-18pt){$b$};
\node at (13pt,18pt){$a$};
\node[right] at (0pt,6pt){$\nu$};
\node[right] at (-10pt,0pt){$c$};
\node[right] at (0pt,-6pt){$\mu$};
\end{tikzpicture}
\end{eqnarray}
In addition, we have
\begin{eqnarray}
\small
\begin{tikzpicture}[baseline={(current bounding box.center)}]
\draw [>=stealth,->](0pt,-20pt)--(0pt,-13pt);\draw(0pt,-13pt)--(0pt, -9pt);
\draw [>=stealth,->](0pt,9pt)--(0pt,16pt);\draw(0pt,16pt)--(0pt, 20pt);
\draw (0pt,-9pt)..controls (-7pt,0pt) and (-7pt,0pt)..(0pt, 9pt);
\draw (0pt,-9pt)..controls (7pt,0pt) and (7pt,0pt)..(0pt, 9pt);
\draw [>=stealth,->](-5.2pt,1.45pt)--(-5.2pt,1.46pt);
\draw [>=stealth,->](5.2pt,1.45pt)--(5.2pt,1.46pt);
\node at (0pt,-24pt){$c$};
\node at (0pt,24pt){$c'$};
\node at (-9pt,0pt){$a$};
\node at (9pt,0pt){$b$};
\node [right] at (0pt, 10pt){$\mu$};
\node [right] at (0pt, -10pt){$\mu'$}; 
\end{tikzpicture}= \delta_{c,c'}\delta_{\mu,\mu'}\sqrt{\frac{d_ad_b}{d_c} }\quad
\begin{tikzpicture}[baseline={(current bounding box.center)}]
\draw [>=stealth,->] (0pt, -18pt)--(0pt,2pt);
\draw (0pt,2pt)--(0pt,18pt);
\node at (0pt,-22pt){$c$};
\node at (0pt,22pt){$c$};
\end{tikzpicture} 
\end{eqnarray}
\begin{eqnarray}\label{Shrinking}
\small
\frac{1}{\mathcal{D}^2} \sum_a d_a \
\begin{tikzpicture}[baseline={(current bounding box.center)}]
\draw (-10pt,0pt) arc (180:0:10pt);
\draw [>=stealth,->](-10pt,0pt) arc (-180:0:10pt);
\draw[line width=4pt, draw=white] (-4pt,0pt)--(-4pt,20pt);
\draw [>=stealth,->](-4pt,-7pt)--(-4pt,5pt);\draw(-4pt,5pt)--(-4pt,18pt);
\draw (-4pt,-13pt)--(-4pt,-20pt);
\draw[line width=4pt, draw=white] (4pt,0pt)--(4pt,20pt);
\draw [>=stealth,->](4pt,-7pt)--(4pt,5pt);\draw(4pt,5pt)--(4pt,18pt);
\draw (4pt,-13pt)--(4pt,-20pt);
\node at (15pt,0pt){$a$};
\node at (-4pt,-25pt){${x}$};
\node at (-4pt,23pt){${x}$};
\node at (4pt,-25pt){${y}$};
\node at (4pt,23pt){${y}$};
\end{tikzpicture}=\frac{1}{d_x}
\delta_{x,\bar{y}}
\begin{tikzpicture}[baseline={(current bounding box.center)}]
\draw [>=stealth,->](-10pt,-18pt)--(-5pt,-13pt);\draw(-5pt,-13pt)--(0pt,-8pt);
\draw [>=stealth,->](10pt,-18pt)--(5pt,-13pt);\draw(5pt,-13pt)--(0pt,-8pt);
\draw [dashed][>=stealth,->](0pt,-8pt)--(0pt,2pt); \draw[dashed](0pt,2pt)--(0pt,8pt);
\draw [>=stealth,->](0pt,8pt)--(-7pt,15pt);\draw(-7pt,15pt)--(-10pt,18pt);
\draw [>=stealth,->](0pt,8pt)--(7pt,15pt);\draw(7pt,15pt)--(10pt,18pt);
\node at (-13pt,-18pt){$x$};
\node at (-13pt,18pt){$x$};
\node at (13pt,-18pt){$y$};
\node at (13pt,18pt){$y$};
\node[right] at (-10pt,0pt){$1$};
\end{tikzpicture}
\end{eqnarray}
where the dashed line indicates the world line of the identity anyon $1$.
One may not be confused with the simple currents $I_k$ in Eq.\eqref{split}
in the main text.

It is also useful to introduce the $\omega$ loop, which is defined as\cite{bonderson2013twisted}
\be
\small
\begin{tikzpicture}[baseline={(current bounding box.center)}]
\draw (10pt,-10pt) arc (-90:90:10pt) ;
\draw [>=stealth,->](10pt,10pt) arc (90:270:10pt) ;
\node at (10pt,-15pt){$\omega_z$};
\end{tikzpicture}
\,\,
=
\,\,
\begin{tikzpicture}[baseline={(current bounding box.center)}]
\draw (10pt,-10pt) arc (270:90:10pt) ;
\draw [>=stealth,->](10pt,10pt) arc (90:-90:10pt) ;
\node at (10pt,-15pt){$\omega_{\bar{z}}$};
\end{tikzpicture}
=\,\,
\sum_x S_{0z}S_{zx}^{\ast}
\begin{tikzpicture}[baseline={(current bounding box.center)}]
\draw (10pt,-10pt) arc (-90:90:10pt) ;
\draw [>=stealth,->](10pt,10pt) arc (90:270:10pt) ;
\node at (10pt,-15pt){$x$};
\end{tikzpicture}
\ee 
For a modular tensor category, in which the modular $S$ matrix is unitary, one can find that 
the $\omega$ loop acts as a projector on the total charge of anyons that go through the $\omega$ loop.
For example, we have
\be\label{omega_loop_1}
\small
\begin{tikzpicture}[baseline={(current bounding box.center)}]
\draw (-10pt,0pt) arc (180:0:10pt);
\draw [>=stealth,->](-10pt,0pt) arc (-180:0:10pt);
\draw[line width=4pt, draw=white] (0pt,0pt)--(0pt,18pt);
\draw [>=stealth,->](0pt,-7pt)--(0pt,5pt);\draw(0pt,5pt)--(0pt,18pt);
\draw (0pt,-13pt)--(0pt,-20pt);
\node at (17pt,0pt){$\omega_z$};
\node at (0pt,23pt){${a}$};
\node at (0pt,-25pt){${a}$};
\end{tikzpicture}
\,\,
=
\,\,
\sum_x S_{0z}S^{\ast}_{zx}\,
\begin{tikzpicture}[baseline={(current bounding box.center)}]
\draw (-10pt,0pt) arc (180:0:10pt);
\draw [>=stealth,->](-10pt,0pt) arc (-180:0:10pt);
\draw[line width=4pt, draw=white] (0pt,0pt)--(0pt,18pt);
\draw [>=stealth,->](0pt,-7pt)--(0pt,5pt);\draw(0pt,5pt)--(0pt,18pt);
\draw (0pt,-13pt)--(0pt,-20pt);
\node at (17pt,0pt){$x$};
\node at (0pt,23pt){${a}$};
\node at (0pt,-25pt){${a}$};
\end{tikzpicture}
\,\,
=
\,\,
\delta_{za} \,\,
\begin{tikzpicture}[baseline={(current bounding box.center)}]
\draw [>=stealth,->](0pt,-18pt)--(0pt,2pt); \draw (0pt,2pt)--(0pt,18pt);
\node at (0pt,21pt){${a}$};
\node at (0pt,-23pt){${a}$};
\end{tikzpicture}
\ee
where in the last step we have used Eq.\eqref{local_shrink} and the unitarity property
of modular $S$ matrix. Similarly, one can find that
\be\label{omega_loop_2}
\small
\begin{tikzpicture}[baseline={(current bounding box.center)}]
\draw (-10pt,0pt) arc (180:0:10pt);
\draw [>=stealth,->](-10pt,0pt) arc (-180:0:10pt);
\draw[line width=4pt, draw=white] (-4pt,0pt)--(-4pt,18pt);
\draw [>=stealth,->](-4pt,-7pt)--(-4pt,5pt);\draw(-4pt,5pt)--(-4pt,18pt);
\draw (-4pt,-13pt)--(-4pt,-20pt);
\draw[line width=4pt, draw=white] (4pt,0pt)--(4pt,18pt);
\draw [>=stealth,->](4pt,-7pt)--(4pt,5pt); \draw(4pt,5pt)--(4pt,18pt);
\draw (4pt,-13pt)--(4pt,-20pt);
\node at (17pt,0pt){$\omega_z$};
\node at (-4pt,-25pt){${a}$};
\node at (-4pt,23pt){${a}$};
\node at (4pt,-25pt){${b}$};
\node at (4pt,23pt){${b}$};
\end{tikzpicture}
\,
=
\,
\sum_x S_{0z}S^{\ast}_{zx}\,
\begin{tikzpicture}[baseline={(current bounding box.center)}]
\draw (-10pt,0pt) arc (180:0:10pt);
\draw [>=stealth,->](-10pt,0pt) arc (-180:0:10pt);
\draw[line width=4pt, draw=white] (-4pt,0pt)--(-4pt,18pt);
\draw [>=stealth,->](-4pt,-7pt)--(-4pt,5pt);\draw(-4pt,5pt)--(-4pt,18pt);
\draw (-4pt,-13pt)--(-4pt,-20pt);
\draw[line width=4pt, draw=white] (4pt,0pt)--(4pt,18pt);
\draw [>=stealth,->](4pt,-7pt)--(4pt,5pt); \draw(4pt,5pt)--(4pt,18pt);
\draw (4pt,-13pt)--(4pt,-20pt);
\node at (15pt,0pt){$x$};
\node at (-4pt,-25pt){${a}$};
\node at (-4pt,23pt){${a}$};
\node at (4pt,-25pt){${b}$};
\node at (4pt,23pt){${b}$};
\end{tikzpicture}
\,
=
\,
\sum_{\mu}\sqrt{\frac{d_z}{d_a d_b}}
\begin{tikzpicture}[baseline={(current bounding box.center)}]
\draw [>=stealth,->](-10pt,-18pt)--(-5pt,-13pt);\draw(-5pt,-13pt)--(0pt,-8pt);
\draw [>=stealth,->](10pt,-18pt)--(5pt,-13pt);\draw(5pt,-13pt)--(0pt,-8pt);
\draw [>=stealth,->](0pt,-8pt)--(0pt,2pt); \draw(0pt,2pt)--(0pt,8pt);
\draw [>=stealth,->](0pt,8pt)--(-7pt,15pt);\draw(-7pt,15pt)--(-10pt,18pt);
\draw [>=stealth,->](0pt,8pt)--(7pt,15pt);\draw(7pt,15pt)--(10pt,18pt);
\node at (-13pt,-18pt){$a$};
\node at (-13pt,18pt){$a$};
\node at (13pt,-18pt){$b$};
\node at (13pt,18pt){$b$};
\node[right] at (0pt,6pt){$\mu$};
\node[right] at (-10pt,0pt){$z$};
\node[right] at (0pt,-6pt){$\mu$};
\end{tikzpicture}
\ee
where we have used Eqs.\eqref{Fuse_ab} and \eqref{omega_loop_1}.
Eq.\eqref{omega_loop_2} will be used to remove the vertex structures in $\sum_{\mu} S^{(z)}_{a,\mu; a,\mu}$ and
the words introduced in Eq.\eqref{Words}.

One useful formula in proving the modular relations 
is $\sum_a d_a\theta_a S_{a\bar{x}}=\frac{1}{\mathcal{D}} \sum_{a,y} d_a \theta_a N_{ax}^y \frac{\theta_y}{\theta_a \theta_x}d_y=\frac{1}{\mathcal{D}}\sum_y d_y^2 \theta_y\cdot d_x \theta_x^{\ast}=\Theta d_x\theta_x^{\ast}$, 
where we used the fact that $S_{ab}=\frac{1}{\mathcal{D}}\sum_c N_{a\bar{b}}^c\frac{\theta_c}{\theta_a \theta_b}d_c$, 
and $\Theta:=\mathcal{D}^{-1}\sum_a d_a^2 \theta_a$ is a phase factor. 
For the MS MTCs we are interested in here, one can check that $\Theta=1$.
Nevertheless, we will keep the phase factor $\Theta$ in the following discussion.
Diagrammatically, one has
\be
\small
\frac{1}{\mathcal{D}}\sum_a d_a \theta_a\,
\begin{tikzpicture}[baseline={(current bounding box.center)}]
\draw (-10pt,0pt) arc (180:0:10pt);
\draw [>=stealth,->](-10pt,0pt) arc (-180:0:10pt);
\draw[line width=4pt, draw=white] (0pt,0pt)--(0pt,18pt);
\draw [>=stealth,->](0pt,-7pt)--(0pt,5pt);\draw(0pt,5pt)--(0pt,18pt);
\draw (0pt,-13pt)--(0pt,-20pt);
\node at (17pt,0pt){$a$};
\node at (0pt,23pt){${x}$};
\node at (0pt,-25pt){${x}$};
\end{tikzpicture}
=\Theta \,\,
\begin{tikzpicture}[baseline={(current bounding box.center)}]
\draw [>=stealth,->](2pt,-20pt)--(2pt,-10pt); \draw (2pt,-10pt)--(2pt,-2pt);
\draw [>=stealth,->](4pt,8pt)..controls (10pt,15pt) and (18pt,0pt)..(10pt,-4pt)..controls (3pt,-6pt) and (2pt,4pt)..(2pt,20pt);
\node at (2pt,-23pt){$x$};
\node at (2pt,23pt){$x$};
\end{tikzpicture}
\ee
It can be generalized to the following case:
\be\label{Twist_2}
\small
\frac{1}{\mathcal{D}}\sum_a d_a \theta_a\,
\begin{tikzpicture}[baseline={(current bounding box.center)}]
\draw (-10pt,0pt) arc (180:0:10pt);
\draw [>=stealth,->](-10pt,0pt) arc (-180:0:10pt);
\draw[line width=4pt, draw=white] (-4pt,0pt)--(-4pt,20pt);
\draw [>=stealth,->](-4pt,-7pt)--(-4pt,5pt);\draw(-4pt,5pt)--(-4pt,18pt);
\draw (-4pt,-13pt)--(-4pt,-20pt);
\draw[line width=4pt, draw=white] (4pt,0pt)--(4pt,20pt);
\draw [>=stealth,->](4pt,-7pt)--(4pt,5pt);\draw(4pt,5pt)--(4pt,18pt);
\draw (4pt,-13pt)--(4pt,-20pt);
\node at (15pt,0pt){$a$};
\node at (-4pt,-25pt){${x}$};
\node at (-4pt,23pt){${x}$};
\node at (4pt,-25pt){${y}$};
\node at (4pt,23pt){${y}$};
\end{tikzpicture}
=\Theta \,\,
\begin{tikzpicture}[baseline={(current bounding box.center)}]
\draw [>=stealth,->](-4pt,-20pt)..controls (-3.5pt,-12pt) and (-2.5pt,-10pt)..(-2pt,-4pt);
\draw [>=stealth,->](-1pt,1pt)..controls (3pt,10pt) and (5pt,12pt)..(9pt,12pt)..controls (19pt,13pt) and (25pt,-6pt)..(12pt,-10pt)..controls (6pt,-12pt) and (2pt,-7pt)..(-1pt,-2pt)..
controls (-2.5pt,1pt) and (-3.5pt,10pt)..(-4pt,20pt);
\draw [>=stealth,->](0pt,-20pt)..controls (0.3pt,-16pt) and (0.7pt,-14pt)..(1.5pt,-10pt);
\draw (2pt,-4pt)..controls (3pt,0pt) and (4pt,1.5pt)..(5pt,3.2pt);
\draw [line width=3pt, draw=white] (5pt,3.2pt)..controls (9pt,8pt) and (15.5pt,8pt)..(16pt,0pt)..controls (15pt,-6pt) and (10pt,-8pt)..(6pt,-3pt)..controls (2.5pt,2pt) and (1.5pt,10pt)..(0pt,20pt);
\draw [>=stealth,->](5pt,3.2pt)..controls (9pt,8pt) and (15.5pt,8pt)..(16pt,0pt)..controls (15pt,-6pt) and (10pt,-8pt)..(6pt,-3pt)..
controls (2.5pt,2pt) and (1.5pt,10pt)..(0pt,20pt);
\node at (-5pt,-23pt){$x$};
\node at (-5pt,23pt){$x$};
\node at (2pt,-23pt){$y$};
\node at (2pt,23pt){$y$};
\end{tikzpicture}
\,=\,
\Theta\cdot \theta_x^{\ast}\theta_y^{\ast}
\begin{tikzpicture}[baseline={(current bounding box.center)}]
\draw [>=stealth,->] (-15pt,-20pt)..controls (-2pt,-15pt) and (-0.5pt,-4pt)..(0pt,0pt);
\draw [line width=4pt, draw=white] (15-8pt,-20pt)..controls (2-8pt,-15pt) and (0.5-8pt,-4pt)..(0-8pt,0pt);
\draw [>=stealth,->] (15-8pt,-20pt)..controls (2-8pt,-15pt) and (0.5-8pt,-4pt)..(0-8pt,0pt);

\draw (0-8pt,0pt)..controls (0.5-8pt,4pt) and (2-8pt,15pt)..(15-8pt,20pt);

\draw [line width=4pt, draw=white](0pt,0pt)..controls (-0.5pt,4pt) and (-2pt,15pt)..(-15pt,20pt);
\draw (0pt,0pt)..controls (-0.5pt,4pt) and (-2pt,15pt)..(-15pt,20pt);
\node at (-18pt,-21pt){$x$};
\node at (-18pt,21pt){$x$};
\node at (10pt,-21pt){$y$};
\node at (10pt,21pt){$y$};
\end{tikzpicture}
\ee

\subsection{Properties of punctured $S$ matrix}
Based on the fusion and braiding of anyons introduced above, 
now we are ready to discuss the properties of punctured $S$ matrix.
The punctured $S$ matrix is defined through the following action:
\begin{eqnarray}\label{Sz_appendix}
S^{(z)}
\begin{tikzpicture}[baseline={(current bounding box.center)}]
\draw [>=stealth,->] (0pt,-15pt)--(0pt,-6pt);\draw (0pt,-6pt)--(0pt,0pt);
\draw [>=stealth,->] (0pt,0pt)--(-6pt,8pt); \draw(-6pt,8pt)--(-7.5pt,10pt);
\draw [>=stealth,->] (0pt,0pt)--(6pt,8pt); \draw(6pt,8pt)--(7.5pt,10pt);
\node at (-8pt,15pt){$b$};
\node at (8pt,15pt){$z$};
\node at (0pt,-20pt){$b$};
\node at (6pt,-2pt){\small$\mu$};
\end{tikzpicture}
:=\frac{1}{\mathcal{D}} \sum_a d_a \,\,
\begin{tikzpicture}[baseline={(current bounding box.center)}]
\draw (10pt,0pt) arc (0:180:10pt);
\draw [>=stealth,->](-10pt,0pt) arc (-180:0:10pt);
\draw[line width=4pt, draw=white] (-5pt,-20pt)--(-5pt,5pt);
\draw [>=stealth,->](-5pt,-18pt)--(-5pt,0pt); \draw(-5pt,0pt)--(-5pt,5pt);
\draw (-5pt, 12pt)--(-5pt,18pt);
\draw [>=stealth,->](5pt, 8.66 pt)--(8pt,14.264pt);\draw(8pt,14.264pt)--(10pt,18pt);
\node at (-5pt,-23pt){$a$};
\node at (-5pt,23pt){$a$};
\node at (15pt,0pt){$b$};
\node at (10pt,23pt){$z$};
\node at (11pt,10pt){\small$\mu$};
\end{tikzpicture}.
\end{eqnarray}
Similarly, for $(S^{(z)})^{\dag}$, we have
\begin{eqnarray}
\small
(S^{(z)})^{\dag}
\begin{tikzpicture}[baseline={(current bounding box.center)}]
\draw [>=stealth,->] (0pt,-15pt)--(0pt,-6pt);\draw (0pt,-6pt)--(0pt,0pt);
\draw [>=stealth,->] (0pt,0pt)--(-6pt,8pt); \draw(-6pt,8pt)--(-7.5pt,10pt);
\draw [>=stealth,->] (0pt,0pt)--(6pt,8pt); \draw(6pt,8pt)--(7.5pt,10pt);
\node at (-8pt,15pt){$b$};
\node at (8pt,15pt){$z$};
\node at (0pt,-20pt){$b$};
\node at (6pt,-2pt){\small$\mu$};
\end{tikzpicture}
:=\frac{1}{\mathcal{D}} \sum_a d_a \,\,
\begin{tikzpicture}[baseline={(current bounding box.center)}]
\draw (10pt,0pt) arc (0:180:10pt);
\draw [>=stealth,->](-10pt,0pt) arc (-180:0:10pt);
\draw[line width=4pt, draw=white] (-5pt,5pt)--(-5pt,18pt);
\draw (-5pt,-18pt)--(-5pt,-12pt);
\draw [>=stealth,->](-5pt, -5pt)--(-5pt,6pt);\draw(-5pt,6pt)--(-5pt,18pt);

\draw [>=stealth,->](5pt, 8.66 pt)--(8pt,14.264pt);\draw(8pt,14.264pt)--(10pt,18pt);

\node at (-5pt,-23pt){$a$};
\node at (-5pt,23pt){$a$};
\node at (15pt,0pt){$b$};
\node at (10pt,23pt){$z$};
\node at (13pt,10pt){$\mu$};
\end{tikzpicture}.
\end{eqnarray}
Then the the unitarity property of $S^{(z)}$ can be shown as follows.
\be
\begin{split}
&(S^{(z)})^\dagger S^{(z)}
\begin{tikzpicture}[baseline={(current bounding box.center)}]
\draw [>=stealth,->] (0pt,-15pt)--(0pt,-6pt);\draw (0pt,-6pt)--(0pt,0pt);
\draw [>=stealth,->] (0pt,0pt)--(-6pt,8pt); \draw(-6pt,8pt)--(-7.5pt,10pt);
\draw [>=stealth,->] (0pt,0pt)--(6pt,8pt); \draw(6pt,8pt)--(7.5pt,10pt);
\node at (-8pt,15pt){$b$};
\node at (8pt,15pt){$z$};
\node at (0pt,-20pt){$b$};
\node at (6pt,-2pt){\small$\mu$};
\end{tikzpicture}
=
\frac{1}{\mathcal{D}^2} \sum_{x,a} d_x d_a  \
\begin{tikzpicture}[baseline={(current bounding box.center)}]
\draw (0pt,0pt) circle (10pt);
\draw[line width=4pt, draw=white] (-5pt,20pt)--(-5pt,-5pt);
\draw [>=stealth,->](-5pt,-5pt)--(-5pt,5pt); \draw(-5pt,5pt)--(-5pt,18pt);
\draw (-5pt, -12pt)--(-5pt,-18pt);
\draw [>=stealth,->](15pt, 8.66 pt)--(18pt,14.264pt); \draw(18pt,14.264pt)--(20pt,18pt);
\draw (10pt,0pt) circle (10pt);
\draw [line width=4pt, draw=white](10pt,10pt) arc (90:180:10pt);
\draw (10pt,10pt) arc (90:180:10pt);
\draw [line width=4pt, draw=white](10pt,0pt) arc (0:-90:10pt);
\draw (10pt,0pt) arc (0:-90:10pt);
\node at (-5pt,-23pt){$x$};
\node at (-5pt,23pt){$x$};
\node at (25pt,0pt){$b$};
\node at (20pt,23pt){$z$};
\begin{small}
\node at (23pt,10pt){$\mu$};
\end{small}
\node at (15pt,0pt){$a$};
\draw [>=stealth,->] (10pt,1.0pt)--(10pt,1.01pt);
\draw [>=stealth,->] (20pt,1.0pt)--(20pt,1.01pt);
\end{tikzpicture}\\
=&
\sum_{x} \delta_{xb}  \
\begin{tikzpicture}[baseline={(current bounding box.center)}]
\draw [>=stealth,->] (0pt,-15pt)--(0pt,-6pt);\draw (0pt,-6pt)--(0pt,0pt);
\draw [>=stealth,->] (0pt,0pt)--(-6pt,8pt); \draw(-6pt,8pt)--(-7.5pt,10pt);
\draw [>=stealth,->] (0pt,0pt)--(6pt,8pt); \draw(6pt,8pt)--(7.5pt,10pt);
\node at (-8pt,15pt){$x$};
\node at (8pt,15pt){$z$};
\node at (0pt,-20pt){$x$};
\begin{small}
\node at (8pt,0pt){$\mu$};
\end{small}
\end{tikzpicture}
=
\begin{tikzpicture}[baseline={(current bounding box.center)}]
\draw [>=stealth,->] (0pt,-15pt)--(0pt,-6pt);\draw (0pt,-6pt)--(0pt,0pt);
\draw [>=stealth,->] (0pt,0pt)--(-6pt,8pt); \draw(-6pt,8pt)--(-7.5pt,10pt);
\draw [>=stealth,->] (0pt,0pt)--(6pt,8pt); \draw(6pt,8pt)--(7.5pt,10pt);
\node at (-8pt,15pt){$b$};
\node at (8pt,15pt){$z$};
\node at (0pt,-20pt){$b$};
\node at (6pt,-2pt){\small$\mu$};
\end{tikzpicture},
\end{split}
\ee
where we used Eq.\eqref{Shrinking} in the last second step.
Similarly, for $S^{(z)}(S^{(z)})^\dagger$, we have
\be
\begin{split}
&S^{(z)}(S^{(z)})^\dagger 
\begin{tikzpicture}[baseline={(current bounding box.center)}]
\draw [>=stealth,->] (0pt,-15pt)--(0pt,-6pt);\draw (0pt,-6pt)--(0pt,0pt);
\draw [>=stealth,->] (0pt,0pt)--(-6pt,8pt); \draw(-6pt,8pt)--(-7.5pt,10pt);
\draw [>=stealth,->] (0pt,0pt)--(6pt,8pt); \draw(6pt,8pt)--(7.5pt,10pt);
\node at (-8pt,15pt){$b$};
\node at (8pt,15pt){$z$};
\node at (0pt,-20pt){$b$};
\node at (6pt,-2pt){\small$\mu$};
\end{tikzpicture}
=
\frac{1}{\mathcal{D}^2} \sum_{x,a} d_x d_a  \
\begin{tikzpicture}[baseline={(current bounding box.center)}]
\draw (0pt,0pt) circle (10pt);
\draw [>=stealth,->](15pt, 8.66 pt)--(18pt,14.264pt);\draw(18pt,14.264pt)--(20pt,18pt);
\draw (10pt,0pt) circle (10pt);
\draw [line width=4pt, draw=white](0pt,10pt) arc (90:0:10pt);
\draw (0pt,10pt) arc (90:0:10pt);
\draw [line width=4pt, draw=white](10pt,-10pt) arc (-90:-180:10pt);
\draw (10pt,-10pt) arc (-90:-180:10pt);
\draw[line width=4pt, draw=white] (-5pt,-18pt)--(-5pt,5pt);
\draw (-5pt,18pt)--(-5pt,12pt);
\draw [>=stealth,->] (-5pt, -18pt)--(-5pt,0pt);\draw(-5pt,0pt)--(-5pt,5pt);
%
\node at (-5pt,-23pt){$x$};
\node at (-5pt,23pt){$x$};
\node at (25pt,0pt){$b$};
\node at (20pt,23pt){$z$};
\begin{small}
\node at (23pt,10pt){$\mu$};
\end{small}
\node at (15pt,0pt){$a$};
\draw [>=stealth,->] (10pt,1.0pt)--(10pt,1.01pt);
\draw [>=stealth,->] (20pt,1.0pt)--(20pt,1.01pt);
\end{tikzpicture}\\
=&
\sum_{x} \delta_{xb}  \
\begin{tikzpicture}[baseline={(current bounding box.center)}]
\draw [>=stealth,->] (0pt,-15pt)--(0pt,-6pt);\draw (0pt,-6pt)--(0pt,0pt);
\draw [>=stealth,->] (0pt,0pt)--(-6pt,8pt); \draw(-6pt,8pt)--(-7.5pt,10pt);
\draw [>=stealth,->] (0pt,0pt)--(6pt,8pt); \draw(6pt,8pt)--(7.5pt,10pt);
\node at (-8pt,15pt){$x$};
\node at (8pt,15pt){$z$};
\node at (0pt,-20pt){$x$};
\begin{small}
\node at (8pt,0pt){$\mu$};
\end{small}
\end{tikzpicture}
=
\begin{tikzpicture}[baseline={(current bounding box.center)}]
\draw [>=stealth,->] (0pt,-15pt)--(0pt,-6pt);\draw (0pt,-6pt)--(0pt,0pt);
\draw [>=stealth,->] (0pt,0pt)--(-6pt,8pt); \draw(-6pt,8pt)--(-7.5pt,10pt);
\draw [>=stealth,->] (0pt,0pt)--(6pt,8pt); \draw(6pt,8pt)--(7.5pt,10pt);
\node at (-8pt,15pt){$b$};
\node at (8pt,15pt){$z$};
\node at (0pt,-20pt){$b$};
\node at (6pt,-2pt){\small$\mu$};
\end{tikzpicture}.
\end{split}
\ee
Therefore, we have $(S^{(z)})^\dagger S^{(z)}=S^{(z)}(S^{(z)})^\dagger=1$.

Next, we prove the modular relations in Eq.\eqref{ModularRelationSzTz}.
Let us prove $\left(S^{(z)}\right)^2=C^{(z)}$ first. By acting $\left(S^{(z)}\right)^2$ on the basis vector,
one has
\be\label{SzSquare_appendix}
\begin{split}
&\big(S^{(z)}\big)^2
\begin{tikzpicture}[baseline={(current bounding box.center)}]
\draw [>=stealth,->] (0pt,-15pt)--(0pt,-6pt);\draw (0pt,-6pt)--(0pt,0pt);
\draw [>=stealth,->] (0pt,0pt)--(-6pt,8pt); \draw(-6pt,8pt)--(-7.5pt,10pt);
\draw [>=stealth,->] (0pt,0pt)--(6pt,8pt); \draw(6pt,8pt)--(7.5pt,10pt);
\node at (-8pt,15pt){$b$};
\node at (8pt,15pt){$z$};
\node at (0pt,-20pt){$b$};
\node at (6pt,-2pt){\small$\mu$};
\end{tikzpicture}
=
\frac{1}{\mathcal{D}^2} \sum_{x,a} d_x d_a  \
\begin{tikzpicture}[baseline={(current bounding box.center)}]
\draw (0pt,0pt) circle (10pt);
\draw [>=stealth,->](15pt, 8.66 pt)--(18pt,14.264pt);\draw(18pt,14.264pt)--(20pt,18pt);
\draw (10pt,0pt) circle (10pt);
\draw [line width=4pt, draw=white](10pt,10pt) arc (90:180:10pt);
\draw (10pt,10pt) arc (90:180:10pt);
\draw [line width=4pt, draw=white](10pt,0pt) arc (0:-90:10pt);
\draw (10pt,0pt) arc (0:-90:10pt);
\draw[line width=4pt, draw=white] (-5pt,-18pt)--(-5pt,-5pt);
\draw (-5pt,18pt)--(-5pt,10pt);
\draw [>=stealth,->] (-5pt,-18pt)--(-5pt, 0pt); \draw (-5pt,0pt)--(-5pt,5pt);
\node at (-5pt,-23pt){$x$};
\node at (-5pt,23pt){$x$};
\node at (25pt,0pt){$b$};
\node at (20pt,23pt){$z$};
\begin{small}
\node at (23pt,10pt){$\mu$};
\end{small}
\node at (15pt,0pt){$a$};
\draw [>=stealth,->] (10pt,1.0pt)--(10pt,1.01pt);
\draw [>=stealth,->] (20pt,1.0pt)--(20pt,1.01pt);
\end{tikzpicture}\\
=&
\frac{1}{\mathcal{D}^2} \sum_{x,a} d_x d_a  \theta_x^{\ast}\
\begin{tikzpicture}[baseline={(current bounding box.center)}]
\draw (0pt,0pt) circle (10pt);
\draw [>=stealth,->](15pt, 8.66 pt)--(18pt,14.264pt);\draw(18pt,14.264pt)--(20pt,18pt);
\draw (10pt,0pt) circle (10pt);
\draw [line width=4pt, draw=white](10pt,10pt) arc (90:180:10pt);
\draw (10pt,10pt) arc (90:180:10pt);
\draw [line width=4pt, draw=white](10pt,0pt) arc (0:-90:10pt);
\draw (10pt,0pt) arc (0:-90:10pt);
\draw (-5pt,0pt)..controls (-6pt,-15pt) and (-15pt,-15pt)..(-25pt,15pt);
\draw [line width=4pt, draw=white](-5pt,0pt)..controls (-6pt,15pt) and (-15pt,15pt)..(-25pt,-15pt);
\draw (-5pt,0pt)..controls (-6pt,15pt) and (-15pt,15pt)..(-25pt,-15pt);
\draw [line width=4pt, draw=white](-10pt,0pt) arc (-180:-90:10pt);
\draw [>=stealth,->] (-5pt,0.1pt)--(-5pt,-0.1pt);
\draw [>=stealth,->] (10pt,1.0pt)--(10pt,1.01pt);
\draw [>=stealth,->] (20pt,1.0pt)--(20pt,1.01pt);
\draw (-10pt,0pt) arc (-180:-90:10pt);
\node at (-25pt,-20pt){$x$};
\node at (-25pt,20pt){$x$};
\node at (25pt,0pt){$b$};
\node at (20pt,23pt){$z$};
\begin{small}
\node at (23pt,10pt){$\mu$};
\end{small}
\node at (15pt,0pt){$a$};
\end{tikzpicture}\\
=&\sum_x \delta_{x,\bar{b}}\,\theta_x^{\ast}
\begin{tikzpicture}[baseline={(current bounding box.center)}]
\draw [>=stealth,->](0pt,0pt)..controls (-4pt,-15pt) and (-10pt,-15pt)..(-20pt,10pt);
\draw[line width=6pt, draw=white] (0pt,0pt)..controls (-4pt,15pt) and (-10pt,15pt)..(-20pt,-14pt);
\draw (0pt,0pt)..controls (-4pt,15pt) and (-10pt,15pt)..(-20pt,-14pt);
\draw [>=stealth,->](-16pt,-3.5pt)--(-16+0.2pt,-3.5+0.5pt);
\draw [>=stealth,->] (0pt,0pt)--(6pt,8pt); \draw(6pt,8pt)--(7.5pt,10pt);
\node at (-20pt,16pt){$x$};
\node at (8pt,15pt){$z$};
\node at (-18pt,-20pt){$x$};
\node at (6pt,0pt){\small$\mu$};
\end{tikzpicture}
=
\theta_b^{\ast}
\begin{tikzpicture}[baseline={(current bounding box.center)}]
\draw [>=stealth,->](0pt,0pt)..controls (-4pt,-15pt) and (-10pt,-15pt)..(-20pt,10pt);
\draw[line width=6pt, draw=white] (0pt,0pt)..controls (-4pt,15pt) and (-10pt,15pt)..(-20pt,-14pt);
\draw (0pt,0pt)..controls (-4pt,15pt) and (-10pt,15pt)..(-20pt,-14pt);
\draw [>=stealth,->](-16pt,-3.5pt)--(-16+0.2pt,-3.5+0.5pt);
\draw [>=stealth,->] (0pt,0pt)--(6pt,8pt); \draw(6pt,8pt)--(7.5pt,10pt);
\node at (-20pt,16pt){$\bar{b}$};
\node at (8pt,15pt){$z$};
\node at (-18pt,-20pt){$\bar{b}$};
\node at (6pt,0pt){\small$\mu$};
\end{tikzpicture},
\end{split}
\ee
where we have considered the fact $\theta_b=\theta_{\bar{b}}$. Comparing \eqref{SzSquare_appendix} 
with the definition of $C^{(z)}$ in Eq.\eqref{Cz}, one can find that $\big(S^{(z)}\big)^2=C^{(z)}$.

Then, to prove $\big(S^{(z)}\big)^4=\big(C^{(z)}\big)^2=\theta_z^{\ast}$,
since we have already proved that $\big(S^{(z)}\big)^2=C^{(z)}$, we only need to show $\big(C^{(z)}\big)^2=\theta_z^{\ast}$.
Based on the definition of $C^{(z)}$ in Eq.\eqref{Cz}, one can find that
\be\label{CzSquare}
\big(C^{(z)}\big)^2
\begin{tikzpicture}[baseline={(current bounding box.center)}]
\draw [>=stealth,->] (0pt,-15pt)--(0pt,-6pt);\draw (0pt,-6pt)--(0pt,0pt);
\draw [>=stealth,->] (0pt,0pt)--(-6pt,8pt); \draw(-6pt,8pt)--(-7.5pt,10pt);
\draw [>=stealth,->] (0pt,0pt)--(6pt,8pt); \draw(6pt,8pt)--(7.5pt,10pt);
\node at (-8pt,15pt){$b$};
\node at (8pt,15pt){$z$};
\node at (0pt,-20pt){$b$};
\node at (6pt,-2pt){\small$\mu$};
\end{tikzpicture}
=
(\theta_b^{\ast})^2
\begin{tikzpicture}[baseline={(current bounding box.center)}]
\draw (0pt,0pt)..controls (-5pt,-12pt) and (-10pt,-12pt)..(-15pt,0pt)..controls (-17.5pt,8pt) and (-21pt,9pt)..(-22.5pt,9pt);
\draw [line width=4pt, draw=white] (-10pt,12pt)--(-20pt,-12pt);
\draw (0pt,0pt)..controls (-5pt,12pt) and (-10pt,12pt)..(-15pt,0pt)..controls (-17.5pt,-8pt) and (-21pt,-9pt)..(-22.5pt,-9pt);
\draw (-22.5pt,-9pt)..controls (-23pt,-9pt) and (-27.5pt,-8pt)..(-30pt,0pt)..controls (-32.5pt,8pt) and (-37.5pt,10pt)..(-40pt,10pt);
\draw [line width=4pt, draw=white] (-27.5pt,6pt)--(-32.5pt,-6pt);
\draw (-22.5pt,9pt)..controls (-23pt,9pt) and (-27.5pt,8pt)..(-30pt,0pt)..controls (-32.5pt,-8pt) and (-37.5pt,-10pt)..(-40pt,-10pt);
\draw [>=stealth,->] (-33.5pt,-10+3.61pt)--(-33.4pt,-9.91+3.61pt);
\draw [>=stealth,->] (-36.4pt, 9.91-1.01pt)--(-36.5pt, 9.99-1.01pt);

\draw [>=stealth,->] (0pt,0pt)--(6pt,8pt); \draw(6pt,8pt)--(7.5pt,10pt);
\node at (-38pt,16pt){$b$};
\node at (8pt,15pt){$z$};
\node at (-38pt,-16pt){$b$};
\node at (6pt,0pt){\small$\mu$};
\end{tikzpicture}
=
\theta_z^{\ast}
\begin{tikzpicture}[baseline={(current bounding box.center)}]
\draw [>=stealth,->] (0pt,-15pt)--(0pt,-6pt);\draw (0pt,-6pt)--(0pt,0pt);
\draw [>=stealth,->] (0pt,0pt)--(-6pt,8pt); \draw(-6pt,8pt)--(-7.5pt,10pt);
\draw [>=stealth,->] (0pt,0pt)--(6pt,8pt); \draw(6pt,8pt)--(7.5pt,10pt);
\node at (-8pt,15pt){$b$};
\node at (8pt,15pt){$z$};
\node at (0pt,-20pt){$b$};
\begin{small}
\node at (8pt,0pt){$\mu$};
\end{small}
\end{tikzpicture}
\ee
where in the last step we have used the ribbon property in Eq.\eqref{Ribbon}.

Next, we give the proof of  $\big(S^{(z)}T^{(z)}\big)^3=\Theta \big(S^{(z)}\big)^2$ in Eq.\eqref{ModularRelationSzTz}:
\be
\begin{split}
&\big(S^{(z)}T^{(z)}\big)^3
\begin{tikzpicture}[baseline={(current bounding box.center)}]
\draw [>=stealth,->] (0pt,-15pt)--(0pt,-6pt);\draw (0pt,-6pt)--(0pt,0pt);
\draw [>=stealth,->] (0pt,0pt)--(-6pt,8pt); \draw(-6pt,8pt)--(-7.5pt,10pt);
\draw [>=stealth,->] (0pt,0pt)--(6pt,8pt); \draw(6pt,8pt)--(7.5pt,10pt);
\node at (-8pt,15pt){$b$};
\node at (8pt,15pt){$z$};
\node at (0pt,-20pt){$b$};
\node at (6pt,-2pt){\small$\mu$};
\end{tikzpicture}
=
\frac{1}{\mathcal{D}^3} \sum_{x,a} d_y d_x d_a \theta_x \theta_a \theta_b \
\begin{tikzpicture}[baseline={(current bounding box.center)}]
\draw (0pt,0pt) circle (10pt);
\draw [>=stealth,->](20pt, 8.66 pt)--(18+5pt,14.264pt);\draw(18+5pt,14.264pt)--(25pt,18pt);
\draw (15pt,0pt) circle (10pt);
\draw [line width=4pt, draw=white](15pt,10pt) arc (90:180:10pt);
\draw (15pt,10pt) arc (90:180:10pt);
\draw [line width=4pt, draw=white](10pt,0pt) arc (0:-90:10pt);
\draw (10pt,0pt) arc (0:-90:10pt);
\draw (-15pt,0pt) circle (10pt);
\draw [line width=4pt, draw=white](0pt,10pt) arc (90:180:10pt);
\draw (0pt,10pt) arc (90:180:10pt);
\draw [line width=4pt, draw=white](-5pt,0pt) arc (0:-90:10pt);
\draw (-5pt,0pt) arc (0:-90:10pt);
\draw[line width=4pt, draw=white] (-20pt,-18pt)--(-20pt,-5pt);
\draw (-20pt,18pt)--(-20pt,10pt);
\draw [>=stealth,->] (-20pt,-18pt)--(-20pt, 0pt); \draw (-20pt,0pt)--(-20pt,5pt);
\node at (-20pt,-23pt){$y$};
\node at (-20pt,23pt){$y$};
\node at (29pt,1pt){$b$};
\node at (25pt,23pt){$z$};
\begin{small}
\node at (28pt,10pt){$\mu$};
\end{small}
\node at (15pt,0pt){$a$};
\node at (-1pt,0pt){$x$};
\draw [>=stealth,->] (-5pt,1.0pt)--(-5pt,1.01pt);
\draw [>=stealth,->] (10pt,1.0pt)--(10pt,1.01pt);
\draw [>=stealth,->] (25pt,1.0pt)--(25pt,1.01pt);
\end{tikzpicture}\\
=&
\frac{1}{\mathcal{D}^3} \sum_{x,a} \theta_y^{\ast} d_y d_x d_a \theta_x \theta_a \theta_b \
\begin{tikzpicture}[baseline={(current bounding box.center)}]
\draw (0pt,0pt) circle (10pt);
\draw [>=stealth,->](20pt, 8.66 pt)--(18+5pt,14.264pt);\draw(18+5pt,14.264pt)--(25pt,18pt);
\draw (15pt,0pt) circle (10pt);
\draw [line width=4pt, draw=white](15pt,10pt) arc (90:180:10pt);
\draw (15pt,10pt) arc (90:180:10pt);
\draw [line width=4pt, draw=white](10pt,0pt) arc (0:-90:10pt);
\draw (10pt,0pt) arc (0:-90:10pt);
\draw (-15pt,0pt) circle (10pt);
\draw [line width=4pt, draw=white](0pt,10pt) arc (90:180:10pt);
\draw (0pt,10pt) arc (90:180:10pt);
\draw [line width=4pt, draw=white](-5pt,0pt) arc (0:-90:10pt);
\draw (-5pt,0pt) arc (0:-90:10pt);
\draw (-5-15pt,0pt)..controls (-6-15pt,-15pt) and (-15-15pt,-15pt)..(-25-15pt,15pt);
\draw [line width=4pt, draw=white](-5-15pt,0pt)..controls (-6-15pt,15pt) and (-15-15pt,15pt)..(-25-15pt,-15pt);
\draw (-5-15pt,0pt)..controls (-6-15pt,15pt) and (-15-15pt,15pt)..(-25-15pt,-15pt);
\draw [line width=4pt, draw=white](-10-15pt,0pt) arc (-180:-90:10pt);
\draw (-10-15pt,0pt) arc (-180:-90:10pt);
\node at (29pt,1pt){$b$};
\node at (25pt,23pt){$z$};
\begin{small}
\node at (28pt,10pt){$\mu$};
\end{small}
\node at (15pt,0pt){$a$};
\node at (-1pt,0pt){$x$};
\node at (-25-15pt,-20pt){$y$};
\node at (-25-15pt,20pt){$y$};
\draw [>=stealth,->] (-5pt,1.0pt)--(-5pt,1.01pt);
\draw [>=stealth,->] (10pt,1.0pt)--(10pt,1.01pt);
\draw [>=stealth,->] (25pt,1.0pt)--(25pt,1.01pt);
\draw [>=stealth,->] (-20pt,-0.5pt)--(-20pt,-0.51pt);
\end{tikzpicture}\\
\end{split}
\ee
Now we sum over $x$ by using Eq.\eqref{Twist_2}. Then the above equation can be simplified as
\be
\begin{split}
&\big(S^{(z)}T^{(z)}\big)^3
\begin{tikzpicture}[baseline={(current bounding box.center)}]
\draw [>=stealth,->] (0pt,-15pt)--(0pt,-6pt);\draw (0pt,-6pt)--(0pt,0pt);
\draw [>=stealth,->] (0pt,0pt)--(-6pt,8pt); \draw(-6pt,8pt)--(-7.5pt,10pt);
\draw [>=stealth,->] (0pt,0pt)--(6pt,8pt); \draw(6pt,8pt)--(7.5pt,10pt);
\node at (-8pt,15pt){$b$};
\node at (8pt,15pt){$z$};
\node at (0pt,-20pt){$b$};
\node at (6pt,-2pt){\small$\mu$};
\end{tikzpicture}\\
=&
\frac{1}{\mathcal{D}^2} \sum_{a,y}\Theta\cdot \theta_y^{\ast} d_y  d_a  \theta_a \theta_b (\theta_{\bar{y}}^{\ast} \theta_{\bar{a}}^{\ast}) \
\begin{tikzpicture}[baseline={(current bounding box.center)}]
\draw (0pt,0pt) circle (10pt);
\draw [>=stealth,->](20pt, 8.66 pt)--(18+5pt,14.264pt);\draw(18+5pt,14.264pt)--(25pt,18pt);
\draw (15pt,0pt) circle (10pt);
\draw [line width=4pt, draw=white](15pt,10pt) arc (90:180:10pt);
\draw (15pt,10pt) arc (90:180:10pt);
\draw [line width=4pt, draw=white](10pt,0pt) arc (0:-90:10pt);
\draw (10pt,0pt) arc (0:-90:10pt);
\draw (-5pt,0pt)..controls (-6pt,-15pt) and (-15pt,-15pt)..(-25pt,15pt);
\draw [line width=4pt, draw=white](-5pt,0pt)..controls (-6pt,15pt) and (-15pt,15pt)..(-25pt,-15pt);
\draw (-5pt,0pt)..controls (-6pt,15pt) and (-15pt,15pt)..(-25pt,-15pt);
\draw [line width=4pt, draw=white](-10pt,0pt) arc (-180:-90:10pt);
\draw (-10pt,0pt) arc (-180:-90:10pt);
\node at (29pt,1pt){$b$};
\node at (25pt,23pt){$z$};
\begin{small}
\node at (28pt,10pt){$\mu$};
\end{small}
\node at (15pt,0pt){$a$};
\node at (-25pt,-20pt){$y$};
\node at (-25pt,20pt){$y$};
\draw [>=stealth,->] (-5pt,0.0pt)--(-5pt,-0.01pt);
\draw [>=stealth,->] (10pt,1.0pt)--(10pt,1.01pt);
\draw [>=stealth,->] (25pt,1.0pt)--(25pt,1.01pt);
\end{tikzpicture}\\
=&
\Theta\cdot
\sum_y \delta_{y,\bar{b}}\,\theta_y^{\ast} \theta_b \theta_y^{\ast}
\begin{tikzpicture}[baseline={(current bounding box.center)}]
\draw [>=stealth,->](0pt,0pt)..controls (-4pt,-15pt) and (-10pt,-15pt)..(-20pt,10pt);
\draw[line width=6pt, draw=white] (0pt,0pt)..controls (-4pt,15pt) and (-10pt,15pt)..(-20pt,-14pt);
\draw (0pt,0pt)..controls (-4pt,15pt) and (-10pt,15pt)..(-20pt,-14pt);
\draw [>=stealth,->](-16pt,-3.5pt)--(-16+0.2pt,-3.5+0.5pt);
\draw [>=stealth,->] (0pt,0pt)--(6pt,8pt); \draw(6pt,8pt)--(7.5pt,10pt);
\node at (-20pt,16pt){$y$};
\node at (8pt,15pt){$z$};
\node at (-18pt,-20pt){$y$};
\node at (6pt,0pt){\small$\mu$};
\end{tikzpicture}
=
\Theta\cdot
\theta_b^{\ast}
\begin{tikzpicture}[baseline={(current bounding box.center)}]
\draw [>=stealth,->](0pt,0pt)..controls (-4pt,-15pt) and (-10pt,-15pt)..(-20pt,10pt);
\draw[line width=6pt, draw=white] (0pt,0pt)..controls (-4pt,15pt) and (-10pt,15pt)..(-20pt,-14pt);
\draw (0pt,0pt)..controls (-4pt,15pt) and (-10pt,15pt)..(-20pt,-14pt);
\draw [>=stealth,->](-16pt,-3.5pt)--(-16+0.2pt,-3.5+0.5pt);
\draw [>=stealth,->] (0pt,0pt)--(6pt,8pt); \draw(6pt,8pt)--(7.5pt,10pt);
\node at (-20pt,16pt){$\bar{b}$};
\node at (8pt,15pt){$z$};
\node at (-18pt,-20pt){$\bar{b}$};
\node at (6pt,0pt){\small$\mu$};
\end{tikzpicture},
\end{split}
\ee
where in the last second step we have summed over $a$ by using Eq.\eqref{Shrinking}.
By comparing with Eq.\eqref{SzSquare_appendix}, one can find that $\big(S^{(z)}T^{(z)}\big)^3=\Theta \big(S^{(z)}\big)^2$.

Similar to $S^{(z)}$, which corresponds to the $S$ transformation of a punctured torus, one can generalize to the 
case of a torus with more than one punctures. For example, for a torus with two punctures, 
the corresponding  twice-punctured $S$ matrix can be defined as
\be\label{Sz1z2}
\small
S^{(z_1,z_2)}
\begin{tikzpicture}[baseline={(current bounding box.center)}]
\draw [>=stealth,->](0pt,-15pt)--(0pt,-9.5pt); \draw(0pt,-9.5pt)--(0pt,-4pt);
\draw [>=stealth,->](0pt,-4pt)--(7pt,1pt);\draw(7pt,1pt)--(14pt,6pt);
\draw [>=stealth,->](0pt,-4pt)--(-2pt,1pt);\draw(-2pt,1pt)--(-4pt,6pt);
\draw [>=stealth,->](-4pt,6pt)--(-9.5pt,10.5pt);\draw(-9.5pt,10.5pt)--(-15pt,15pt);
\draw [>=stealth,->](-4pt,6pt)--(1pt,10.5pt);\draw(1pt,10.5pt)--(6pt,15pt);
\node at (-15pt,20pt){$b$};
\node at (12pt,17pt){$z_1$};
\node at (20pt,7pt){$z_2$};
\node at (0pt,-20pt){$b$};
\node at (2pt,4pt){$p$};
\end{tikzpicture}
:=
\frac{1}{\mathcal{D}} \sum_a d_a \,\,
\begin{tikzpicture}[baseline={(current bounding box.center)}]
\draw [>=stealth,->](0pt,-4pt)--(7pt,1pt);\draw(7pt,1pt)--(14pt,6pt);
\draw [>=stealth,->](0pt,-4pt)--(-2pt,1pt);\draw(-2pt,1pt)--(-4pt,6pt);
\draw [>=stealth,->](-4pt,6pt)--(1pt,10.5pt);\draw(1pt,10.5pt)--(6pt,15pt);
\node at (-19pt,18pt){$a$};
\node at (-15pt,-25pt){$a$};
\node at (12pt,17pt){$z_1$};
\node at (20pt,7pt){$z_2$};
\node at (-5pt,-17pt){$b$};
\node at (2pt,4pt){$q$};
\draw (-4pt,6pt) arc (50:352:10.6pt);
\draw [line width=4pt,draw=white](-15pt,5pt)--(-15pt,-20pt);
\draw [>=stealth,->](-15pt,-20pt)--(-15pt,0pt);\draw(-15pt,0pt)--(-15pt,5pt);
\draw (-15pt,9pt)--(-15pt,18pt);
\draw [>=stealth,->](-5pt,-12+1.2pt)--(-4.9pt,-11.94+1.2pt);
\end{tikzpicture}
\ee
On the other hand, by using $F$-move and once-punctured $S$ transformation, one has
\be\label{F_PunctureS}
\begin{split}
&\begin{tikzpicture}[baseline={(current bounding box.center)}]
\draw [>=stealth,->](0pt,-15pt)--(0pt,-9.5pt); \draw(0pt,-9.5pt)--(0pt,-4pt);
\draw [>=stealth,->](0pt,-4pt)--(7pt,1pt);\draw(7pt,1pt)--(14pt,6pt);
\draw [>=stealth,->](0pt,-4pt)--(-2pt,1pt);\draw(-2pt,1pt)--(-4pt,6pt);
\draw [>=stealth,->](-4pt,6pt)--(-9.5pt,10.5pt);\draw(-9.5pt,10.5pt)--(-15pt,15pt);
\draw [>=stealth,->](-4pt,6pt)--(1pt,10.5pt);\draw(1pt,10.5pt)--(6pt,15pt);
\node at (-15pt,20pt){$b$};
\node at (12pt,17pt){$z_1$};
\node at (20pt,7pt){$z_2$};
\node at (0pt,-20pt){$b$};
\node at (2pt,4pt){$p$};
\end{tikzpicture}
\xrightarrow{\,\,F\,\,}
\begin{tikzpicture}[baseline={(current bounding box.center)}]
\draw (0pt,2pt)--(0pt,-15pt);
\draw (0pt,2pt)--(-15pt,15pt);
\draw [>=stealth,->](0pt,2pt)--(4pt,6pt);\draw(4pt,6pt)--(6pt,8pt);
\node at (-15pt,20pt){$b$};
\draw [>=stealth,->] (6pt,8pt)--(10pt,15pt);
\draw [>=stealth,->] (6pt,8pt)--(15pt,10pt);
\node at (14pt,19pt){$z_1$};
\node at (20pt,7pt){$z_2$};
\node at (0pt,-20pt){$b$};
\node at (0pt,9pt){\small$w$};
\end{tikzpicture}
\xrightarrow{S^{(w)}}\\
&
\begin{tikzpicture}[baseline={(current bounding box.center)}]
\draw [>=stealth,->] (0pt,0pt) circle (10pt);
\draw [>=stealth,->] (10pt,1.0pt)--(10pt,1.01pt);
\draw[line width=4pt, draw=white] (-5pt,-20pt)--(-5pt,5pt);
\draw [>=stealth,->](-5pt,-18pt)--(-5pt,0pt); \draw(-5pt,0pt)--(-5pt,5pt);
\draw (-5pt, 12pt)--(-5pt,18pt);
\draw [>=stealth,->] (8.5pt, 5pt)--(12.9pt,8.2pt); \draw(12.9pt,8.2pt)--(14pt,9pt);
\draw [>=stealth,->] (14pt,9pt)--(20pt,18pt);
\draw [>=stealth,->] (14pt,9pt)--(24pt,11pt);
\node at (-5pt,-23pt){$a$};
\node at (-5pt,23pt){$a$};
\node at (15pt,0pt){$b$};
\node at (20pt,22pt){$z_1$};
\node at (29pt,10pt){$z_2$};
\node at (10pt,12pt){\small$w$};
\end{tikzpicture}
\xrightarrow{F^{-1}}
\begin{tikzpicture}[baseline={(current bounding box.center)}]
\draw [>=stealth,->](0pt,-4pt)--(7pt,1pt);\draw(7pt,1pt)--(14pt,6pt);
\draw [>=stealth,->](0pt,-4pt)--(-2pt,1pt);\draw(-2pt,1pt)--(-4pt,6pt);
\draw [>=stealth,->](-4pt,6pt)--(1pt,10.5pt);\draw(1pt,10.5pt)--(6pt,15pt);
\node at (-19pt,18pt){$a$};
\node at (-15pt,-25pt){$a$};
\node at (12pt,17pt){$z_1$};
\node at (20pt,7pt){$z_2$};
\node at (-5pt,-17pt){$b$};
\node at (2pt,4pt){$q$};
\draw (-4pt,6pt) arc (50:352:10.6pt);
\draw [line width=4pt,draw=white](-15pt,5pt)--(-15pt,-20pt);
\draw [>=stealth,->](-15pt,-20pt)--(-15pt,0pt);\draw(-15pt,0pt)--(-15pt,5pt);
\draw (-15pt,9pt)--(-15pt,18pt);
\draw [>=stealth,->](-5pt,-12+1.2pt)--(-4.9pt,-11.94+1.2pt);
\end{tikzpicture}
\end{split}
\ee
By comparing Eqs.\eqref{Sz1z2} and \eqref{F_PunctureS}, one can find that the twice-punctured $S$-matrix 
and once-punctued $S$-matrix are related through the $F$-move as follows:
\be
S^{(z_1,z_2)}=F\cdot S^{(w)}\cdot F^{-1},
\ee
where we have neglected the indices of punctured $S$ matrix and $F$ symbols for brevity.

Based on the $F$ moves, one can find the procedure above also applies to a multi-punctured $S$ matrix.
That is, a multi-punctured $S$ matrix is related to the single-punctured $S$ matrix through $F$ moves.

\subsubsection{Other punctured $S$ matrices} 
\label{Sec: Other_Sz}

In the main text, we have seen that the (gauge invariant) diagonal elements of certain $S^{(z)}$ can be used 
to distinguish different categories. 
In this subsection, we present the following two results on the punctured $S$ matrix.
(i) We summarize how $S_{B_{i,j},B_{i,j}}^{(z)}$ are mapped between 
$\mathcal{C}_{u}$ and $\mathcal{C}_{u'}$. We will consider $u=1$ and $u'=4$ for example,
and the mapping between $\mathcal{C}_{u=2}$ and $\mathcal{C}_{u'=3}$ is similar.
(ii) Not all punctured $S$ matrices can distinguish different categories.

\textit{(i) Mapping of $S^{(z)}_{B_{i,j},B_{i,j}}$ between $\mathcal{C}_{u=1}$ and $\mathcal{C}_{u'=4}$.  }

It is interesting that all the diagonal elements $S^{(z),u=1}_{B_{i,j},B_{i,j}}$ 
can be mapped to $S^{(z'),u=4}_{B_{i',j'},B_{i',j'}}$ by permuting anyons. 
This permutation is different from that of the modular $T$ matrix in two aspects:
(i) For $S^{(z)}$, we need to permute
the anyons $z$ in $\mathcal{C}_{u=1}$ to $z'$ in $\mathcal{C}_{u'=4}$ where $\theta_z\neq \theta_{z'}$. 
For the $T$ matrix, one can only permute anyons that have the same topological spin.
(ii) Even for the permutation of type-$B$ anyons in $S^{(z)}_{B_{i,j},B_{i,j}}$, the permutations
are different from that of $T$ matrix. This is used to distinguish different categories in Sec.\ref{PuncturedS_distinguish_MS_MC}
in the main text. The mappings from $S^{(z),u=1}_{B_{i,j},B_{i,j}}$ to $S^{(z'),u=4}_{B_{i',j'},B_{i',j'}}$
are summarized in Table \ref{MappingSzDiagonal_appendix}.

\begin{table}[h]
\centering
\footnotesize
\begin{tabular}{cccccccccccc}
       &$u=1$  &\vline         &$u=4$       \\ \hline\hline
            &$S^{A_{1,1}(A_{2,6})}_{B_{1(4),i},B_{1(4),i}}$  &\vline             &$S^{A_{1,9}(A_{2,10})}_{B_{2(3),j},B_{2(3),j}}$    \\ 
            &$S^{A_{1,1}(A_{2,6})}_{B_{2(3),i},B_{2(3),i}}$  &\vline             &$S^{A_{1,5}(A_{2,8})}_{B_{1(4),j},B_{1(4),j}}$      \\ \hline
            
           &$S^{A_{1,2}(A_{2,1})}_{B_{1(4),i},B_{1(4),i}}$   &\vline             &$S^{A_{1,7}(A_{2,9})}_{B_{2(3),j},B_{2(3),j}}$   \\ 
           &$S^{A_{1,2}(A_{2,1})}_{B_{2(3),i},B_{2(3),i}}$  &\vline              &$S^{A_{1,10}(A_{2,5})}_{B_{1(4),j},B_{1(4),j}}$    \\ \hline
     
            &$S^{A_{1,3}(A_{2,7})}_{B_{1(4),i},B_{1(4),i}}$  &\vline             &$S^{A_{1,5}(A_{2,8})}_{B_{2(3),j},B_{2(3),j}}$    \\ 
            &$S^{A_{1,3}(A_{2,7})}_{B_{2(3),i},B_{2(3),i}}$  &\vline             &$S^{A_{1,4}(A_{2,2})}_{B_{1(4),j},B_{1(4),j}}$      \\ \hline           
            
            &$S^{A_{1,4}(A_{2,2})}_{B_{1(4),i},B_{1(4),i}}$  &\vline             &$S^{A_{1,3}(A_{2,7})}_{B_{2(3),j},B_{2(3),j}}$    \\ 
            &$S^{A_{1,4}(A_{2,2})}_{B_{2(3),i},B_{2(3),i}}$  &\vline             &$S^{A_{1,9}(A_{2,10})}_{B_{1(4),j},B_{1(4),j}}$      \\ \hline            
                  
            &$S^{A_{1,5}(A_{2,8})}_{B_{1(4),i},B_{1(4),i}}$  &\vline             &$S^{A_{1,1}(A_{2,6})}_{B_{2(3),j},B_{2(3),j}}$    \\ 
            &$S^{A_{1,5}(A_{2,8})}_{B_{2(3),i},B_{2(3),i}}$  &\vline             &$S^{A_{1,3}(A_{2,7})}_{B_{1(4),j},B_{1(4),j}}$      \\ \hline   
            
            &$S^{A_{1,6}(A_{2,3})}_{B_{1(4),i},B_{1(4),i}}$  &\vline             &$S^{A_{1,10}(A_{2,5})}_{B_{2(3),j},B_{2(3),j}}$    \\ 
            &$S^{A_{1,6}(A_{2,3})}_{B_{2(3),i},B_{2(3),i}}$  &\vline             &$S^{A_{1,8}(A_{2,4})}_{B_{1(4),j},B_{1(4),j}}$      \\ \hline         
            
            &$S^{A_{1,7}(A_{2,9})}_{B_{1(4),i},B_{1(4),i}}$  &\vline             &$S^{A_{1,8}(A_{2,4})}_{B_{2(3),j},B_{2(3),j}}$    \\ 
            &$S^{A_{1,7}(A_{2,9})}_{B_{2(3),i},B_{2(3),i}}$  &\vline             &$S^{A_{1,2}(A_{2,1})}_{B_{1(4),j},B_{1(4),j}}$      \\ \hline          
            
            &$S^{A_{1,8}(A_{2,4})}_{B_{1(4),i},B_{1(4),i}}$  &\vline             &$S^{A_{1,6}(A_{2,3})}_{B_{2(3),j},B_{2(3),j}}$    \\ 
            &$S^{A_{1,8}(A_{2,4})}_{B_{2(3),i},B_{2(3),i}}$  &\vline             &$S^{A_{1,7}(A_{2,9})}_{B_{1(4),j},B_{1(4),j}}$      \\ \hline

            &$S^{A_{1,9}(A_{2,10})}_{B_{1(4),i},B_{1(4),i}}$  &\vline             &$S^{A_{1,4}(A_{2,2})}_{B_{2(3),j},B_{2(3),j}}$    \\ 
            &$S^{A_{1,9}(A_{2,10})}_{B_{2(3),i},B_{2(3),i}}$  &\vline             &$S^{A_{1,1}(A_{2,6})}_{B_{1(4),j},B_{1(4),j}}$      \\ \hline          
            
            &$S^{A_{1,10}(A_{2,5})}_{B_{1(4),i},B_{1(4),i}}$  &\vline             &$S^{A_{1,2}(A_{2,1})}_{B_{2(3),j},B_{2(3),j}}$    \\ 
            &$S^{A_{1,10}(A_{2,5})}_{B_{2(3),i},B_{2(3),i}}$  &\vline             &$S^{A_{1,6}(A_{2,3})}_{B_{1(4),j},B_{1(4),j}}$      \\ \hline                                           
\end{tabular}
\caption{Mapping of the diagonal elements of 
$S^{(z)}$ between $\mathcal{C}_{u=1}$ and $\mathcal{C}_{u'=4}$.
The concrete values of $S^{(z)}_{B_{i,j},B_{i,j}}$ and the indices $i,j \in \{0,\cdots,4\}$ can be found in online materials.\cite{OnlineData} }
\label{MappingSzDiagonal_appendix}
\end{table}

More explicitly, in the first row of Table \ref{MappingSzDiagonal_appendix}, the mapping between diagonal elements 
$S^{A_{1,1}}_{B_{1,i}, B_{1,i}}$ in $\mathcal{C}_{u=1}$ and
$S^{A_{1,9}}_{B_{2,j}, B_{2,j}}$ in $\mathcal{C}_{u=4}$ means by considering the bijection of anyons 
$A^{(u=1)}_{1,1}\leftrightarrow A^{(u=4)}_{1,9}$, $B^{(u=1)}_{1,i}\leftrightarrow B^{(u=4)}_{2,j}$, $S^{A_{1,1}}_{B_{1,i}, B_{1,i}}$
in $\mathcal{C}_{u=1}$  can be sent to $S^{A_{1,9}}_{B_{2,j}, B_{2,j}}$ in $\mathcal{C}_{u=4}$, and vice versa. 
The mappings are similar for other diagonal elements in Table \ref{MappingSzDiagonal_appendix}.
The concrete values of $S^{(z)}_{B_{i,j},B_{i,j}}$ and the indices $i(j)$ can be found in online materials.\cite{OnlineData}

\textit{(ii) $S^{(z)}$ that cannot distinguish different categories.}

It is noted that not all punctured $S$ matrices (together with the modular $T$ matrix) can be used to 
distinguish different categories. This is as expected for $z=I_i$ with $i=1,\cdots, 4$. In this case, based on the fusion rules 
in Sec.\ref{Sec: Z11Z5_data}, one can find that only type-$I$ and type-$A$ anyons are involved in $S^{(z)}$.
Recalling that only type-$B$ anyons carry the information of 3-cocycle $\omega^u$, then $S^{(z)}$ with $z=I_i$ 
will be independent of $u$.

Here we want to emphasize that even for certain $z$ which are type-$A$ anyons, $S^{(z)}$ and $T$ matrices may still not 
be enough to distinguish different categories.
For example, let us consider $z=A_{1,0}$ ($A_{2,0}$) in $\mathcal{C}_{u=1}$ and $\mathcal{C}_{u=4}$ categories.

For convenience, we write down the topological spins defined through Eq.\eqref{TopologicalSpin} as follows:
\be
\small
\begin{split}
u=1,\quad s\in\{&1,6,11,16,21;\quad 4, 14, 24, 9, 19;\\
& 9, 24,14,4,19;\quad 16,11,6, 1, 21\},\\
u=4,\quad s\in\{&4,9,14,19,24;\quad 16,1,11,21,6;\\
&11,1,16,6,21;\quad 14,9,4,24,19\}.
\end{split}
\ee
It can be found that the punctured $S$ matrix in Table \ref{SaaSampleA10} 
together with the modular data cannot distinguish different categories.
More precisely, under a certain anyon bijection/permutation, the modular $T$ matrix in $\mathcal{C}_{u=1}$ category
is mapped to that in $\mathcal{C}_{u=4}$ category. With the \textit{same} anyon permutation, 
$S^{z=A_{1,0}}_{B_{k,n},B_{k,n}}$ in $\mathcal{C}_{u=1}$
is mapped to $S^{z=A_{1,0}}_{B_{k',n'},B_{k',n'}}$in $\mathcal{C}_{u=4}$.
For example, an anyon permutation that sends $\theta^{(u=1)}_{B_{1,0}}$ to $\theta^{(u=4)}_{B_{2,1}}$
will also send $S^{z=A_{1,0},(u=1)}_{B_{1,0},B_{1,0}}$ to $S^{z=A_{1,0},(u=4)}_{B_{2,1},B_{2,1}}$.
Similarly, one can check the permutations for other elements of $T$ and $S^{z=A_{1,0} (A_{2,0})}$.

One may ask that we only focus on the diagonal elements of $S^{z=A_{1,0} (A_{2,0})}$
in the discussions above, but it is possible that by considering the off-diagonal elements of $S^{z=A_{1,0} (A_{2,0})}$ 
together with the $T$ matrix 
we may be able to distinguish different categories. 
To check this, we also study the topological invariants in \eqref{Wz_PuncturedTorus}
which contain the information of both diagonal and off-diagonal elements of $S^{z=A_{1,0} (A_{2,0})}$.
It is found that these topological invariants cannot distinguish different categories. 
Therefore, $S^{(z)}$ with $z=A_{1,0} (A_{2,0})$ cannot be used to distinguish different categories.

Our observation is that the punctured $S$ matrix $S^{(z)}$ that can be used to distinguish different categories 
must have nontrivial topological spins $\theta_z\neq 1$ for the anyon $z$.

\begin{table}[h]
\centering
\footnotesize
\begin{tabular}{cccccccccccc}
$B_{k,n}$  		 &\vline        &$u=1$  &\vline         &$u=4$\\ \hline\hline
$B_{1,0}$  &\vline            &$\frac{1}{5}e^{-\frac{4}{25} i\pi}$  &\vline             &$\frac{1}{5}e^{-\frac{16}{25} i\pi}$ \\
$B_{1,1}$  &\vline             &$\frac{1}{5}e^{-\frac{24}{25} i\pi}$  &\vline             &$\frac{1}{5}e^{\frac{14}{25} i\pi}$  \\ 
$B_{1.2}$  &\vline           &$\frac{1}{5}e^{\frac{6}{25} i\pi}$  &\vline             &$\frac{1}{5}e^{-\frac{6}{25} i\pi}$  \\ 
$B_{1.3}$  &\vline            &$\frac{1}{5}e^{-\frac{14}{25} i\pi}$  &\vline              &$\frac{1}{5}e^{\frac{24}{25} i\pi}$  \\ 
$B_{1,4}$  &\vline            &$\frac{1}{5}e^{\frac{16}{25} i\pi}$  &\vline              &$\frac{1}{5}e^{\frac{4}{25} i\pi}$   \\ \hline

$B_{2,0}$  &\vline            &$\frac{1}{5}e^{-\frac{16}{25} i\pi}$  &\vline             &$\frac{1}{5}e^{-\frac{14}{25} i\pi}$   \\
$B_{2,1}$  &\vline             &$\frac{1}{5}e^{-\frac{6}{25} i\pi}$ &\vline              &$\frac{1}{5}e^{-\frac{4}{25} i\pi}$   \\ 
$B_{2.2}$  &\vline            &$\frac{1}{5}e^{\frac{4}{25} i\pi}$  &\vline              &$\frac{1}{5}e^{\frac{6}{25} i\pi}$   \\ 
$B_{2.3}$  &\vline             &$\frac{1}{5}e^{\frac{14}{25} i\pi}$  &\vline              &$\frac{1}{5}e^{\frac{16}{25} i\pi}$    \\ 
$B_{2,4}$  &\vline             &$\frac{1}{5}e^{\frac{24}{25} i\pi}$  &\vline              &$\frac{1}{5}e^{-\frac{24}{25} i\pi}$  \\ \hline

$B_{3,0}$  &\vline             &$\frac{1}{5}e^{\frac{14}{25} i\pi}$  &\vline              &$\frac{1}{5}e^{\frac{6}{25} i\pi}$   \\
$B_{3,1}$  &\vline            &$\frac{1}{5}e^{\frac{4}{25} i\pi}$  &\vline             &$\frac{1}{5}e^{-\frac{4}{25} i\pi}$    \\ 
$B_{3.2}$  &\vline             &$\frac{1}{5}e^{-\frac{6}{25} i\pi}$  &\vline              &$\frac{1}{5}e^{-\frac{14}{25} i\pi}$    \\ 
$B_{3.3}$  &\vline              &$\frac{1}{5}e^{-\frac{16}{25} i\pi}$  &\vline              &$\frac{1}{5}e^{-\frac{24}{25} i\pi}$    \\ 
$B_{3,4}$  &\vline             &$\frac{1}{5}e^{\frac{24}{25} i\pi}$  &\vline            &$\frac{1}{5}e^{\frac{16}{25} i\pi}$  \\ \hline

$B_{4,0}$  &\vline              &$\frac{1}{5}e^{-\frac{24}{25} i\pi}$   &\vline             &$\frac{1}{5}e^{-\frac{6}{25} i\pi}$   \\
$B_{4,1}$  &\vline             &$\frac{1}{5}e^{\frac{6}{25} i\pi}$   &\vline             &$\frac{1}{5}e^{\frac{14}{25} i\pi}$   \\ 
$B_{4.2}$  &\vline              &$\frac{1}{5}e^{-\frac{24}{25} i\pi}$  &\vline              &$\frac{1}{5}e^{-\frac{16}{25} i\pi}$   \\ 
$B_{4.3}$  &\vline               &$\frac{1}{5}e^{-\frac{4}{25} i\pi}$   &\vline              &$\frac{1}{5}e^{\frac{4}{25} i\pi}$   \\ 
$B_{4,4}$  &\vline             &$\frac{1}{5}e^{\frac{16}{25} i\pi}$   &\vline              &$\frac{1}{5}e^{\frac{24}{25} i\pi}$   \\ \hline
\end{tabular}
\caption{The diagonal elements $S^{(z)}_{B_{k,n}B_{k,n}}$ of the punctured $S$ matrix with $z=A_{1,0}$ and $A_{2,0}$ ($A_{1,0}$ and $A_{2,0}$ are dual anyons)
for $u=1$ and $4$, respectively. The complete data for $S^{(z)}_{a,\mu;b,\nu}$ can be found in materials online.}
\label{SaaSampleA10}
\end{table}

\subsection{More on  representations of mapping class group}
\label{Appendix: Other basis}

In the main text, we choose the canonical basis in \eqref{Genus_g_basis}
to represent the quasi-particle basis in $\mathcal{H}(\Sigma_{g,0})$.
In fact, there are many other ways to choose the bases, which
correspond to the ways of decomposing the surface into pairs of pants. 
This is related to trivalent graphs with $g$ loops in mathematics. Up to homemorphism, there are \textit{finitely}
many trivalent graphs with $g$ loops. \cite{McMullen2018}
See $g=2$ and $g=3$ for example:

\be\label{g=2TrivalentGraph}
\begin{split}
g=2:&\quad  
\begin{tikzpicture}[baseline={(current bounding box.center)}]
\draw(0pt,0pt) arc (180:-180:12pt) ;
\draw (40pt,0pt) arc (180:-180:12pt) ;
\draw (24pt,0pt)--(40pt,0pt);
\end{tikzpicture}
\quad,\quad
\begin{tikzpicture}[baseline={(current bounding box.center)}]
\draw(0pt,0pt) arc (180:-180:12pt) ;
\draw (12pt,-12pt)--(12pt,12pt);
\end{tikzpicture}\\
g=3:&\quad 
\begin{tikzpicture}[baseline={(current bounding box.center)}]
\draw(0pt,0pt) arc (180:-180:12pt) ;
\draw (35pt,0pt) arc (180:-180:12pt) ;
\draw (24pt,0pt)--(35pt,0pt);
\draw (70pt,0pt) arc (180:-180:12pt) ;
\draw (59pt,0pt)--(70pt,0pt);
\end{tikzpicture}
\,\, ,\,\,
\begin{tikzpicture}[baseline={(current bounding box.center)}]
\draw(0pt,0pt) arc (180:-180:12pt) ;
\draw (40pt,0pt) arc (180:-180:12pt) ;
\draw (24pt,0pt)--(40pt,0pt);
\draw (52pt,-12pt)--(52pt,12pt);
\end{tikzpicture}\,\,,\\
&
\begin{tikzpicture}[baseline={(current bounding box.center)}]
\draw(0pt,0pt) arc (180:-180:12pt) ;
\draw (40pt,0pt) arc (180:-180:12pt) ;
\draw (24pt,0pt)--(40pt,0pt);
\draw (32pt,15pt) arc (-90:270:12pt) ;
\draw (32pt,0pt)--(32pt,15pt);
\end{tikzpicture}
\,\, ,\,\,
\begin{tikzpicture}[baseline={(current bounding box.center)}]
\draw(0pt,0pt) arc (180:-180:12pt) ;
\draw (12pt,-12pt)--(12pt,12pt);
\draw (0pt,0pt)..controls(-10pt,5pt) and (0pt,25pt)..(12pt,25pt)..
controls (24pt,25pt) and (34pt,5pt)..(24pt,0pt);
\end{tikzpicture}
\,\, ,\,\,
\begin{tikzpicture}[baseline={(current bounding box.center)}]
\draw(0pt,0pt) arc (180:-180:12pt) ;
\draw (35pt,0pt) arc (180:-180:12pt) ;
\draw (12pt,12pt)--(47pt,12pt);
\draw (12pt,-12pt)--(47pt,-12pt);
\end{tikzpicture}
\end{split}
\ee
In the quasi-particle basis, each line above 
may be colored with an anyon type, and the tri-junctions are characterized by the 
fusion channels [see the discussions under \eqref{Genus_g_basis}].

With the basis in \eqref{Genus_g_basis} for $\mathcal{H}(\Sigma_{g,0})$, 
Ref.\onlinecite{bloomquist2018topological} studies the  representations of 
the $(2g+1)$ generators of MCG$(\Sigma_{g,0})$, corresponding to the $(2g+1)$ Dehn 
twists in Fig.\ref{MCG_g}.
These  representations are expressed in terms of $F$ and $R$ matrices.

It can be found there is another convenient choice of basis vectors for $\mathcal{H}(\Sigma_{g,0})$
as follows:
\be\label{Basis_alternative}
\begin{tikzpicture}[baseline={(current bounding box.center)}]
\draw  (-55pt,11pt) arc (90:270:12pt);
\draw  (-60pt,0pt) arc (-120:-60:20pt);
\draw (-56pt,-1.5pt) arc (130:50:10pt);
\draw  (-60+35pt,0pt) arc (-120:-60:20pt);
\draw  (-56+35pt,-1.5pt) arc (130:50:10pt);
\draw  (-60+70pt,0pt) arc (-120:-60:20pt);
\draw  (-56+70pt,-1.5pt) arc (130:50:10pt);
\draw (-55pt,11pt)--(45pt,11pt);
\draw (-55pt,-13pt)--(45pt,-13pt);
\draw (-33pt,11pt)--(-33pt,-13pt); 
\draw (-33+35pt,11pt)--(-33+35pt,-13pt); 
\draw (-33+70pt,11pt)--(-33+70pt,-13pt); 
\node at (55pt,-1pt){$\cdots$};
\draw (64pt,11pt)--(100pt,11pt);
\draw (64pt,-13pt)--(100pt,-13pt);
\draw  (100pt,11pt) arc (90:-90:12pt);
\draw  (-60+143pt,0pt) arc (-120:-60:20pt);
\draw  (-56+143pt,-1.5pt) arc (130:50:10pt);
\draw (75pt,11pt)--(75pt,-13pt); 
\end{tikzpicture}
\ee
With the basis above, one can find that by performing Dehn twists along the simple curves $b_i$ ($i=1,2$) and 
$c_j$ ($j=1,\cdots, g-1$) in Fig.\ref{MCG_g}, the corresponding  representations are simply diagonal 
matrices with the diagonal elements being the topological spins.
Then we only need to study the Dehn twists along the simple curves $a_i$ ($i=1,\cdots, g$) in Fig.\ref{MCG_g}.
This can be done as follows.
Considering the local structure around the $i$-th genus in the basis \eqref{Basis_alternative},
we perform two $F$ transformations as
\be\label{Basis_ai_F_move}
\begin{tikzpicture}[baseline={(current bounding box.center)}]
\draw  (-60+70pt,0pt) arc (-120:-60:20pt);
\draw  (-56+70pt,-1.5pt) arc (130:50:10pt);
\draw (-10pt,11pt)--(50pt,11pt);
\draw (-10pt,-13pt)--(50pt,-13pt);
\draw (5pt,-13pt)--(5pt,0pt);\draw(5pt,0pt)--(5pt,11pt); 
\draw (35pt,-13pt)--(35pt,0pt);\draw(35pt,0pt)--(35pt,11pt); 
\end{tikzpicture}
\quad
\xrightarrow{\,\,F\,\,}
\quad
\begin{tikzpicture}[baseline={(current bounding box.center)}]
\draw  (-60+70pt,0pt) arc (-120:-60:20pt);
\draw  (-56+70pt,-1.5pt) arc (130:50:10pt);

\draw (35-27pt,0pt) arc (180:-180:12pt) ;
\draw (26-27pt,0pt)--(35-27pt,0pt);
\draw (59-27pt,0pt)--(68-27pt,0pt);

\draw (-1pt,0pt)--(-6pt,8pt);\draw (-1pt,0pt)--(-6pt,-8pt);
\draw (41pt,0pt)--(46pt,8pt);\draw (41pt,0pt)--(46pt,-8pt);
\end{tikzpicture}
\ee
With the new basis, one can perform the Dehn twist along the closed curve $a_i$ (see Fig.\ref{MCG_g})
which has been studied in Ref.\onlinecite{bloomquist2018topological}. 
After the Denh twist, we do the inverse of $F$ transformations in \eqref{Basis_ai_F_move}.
Then we can obtain the representation of the Dehn twist along $a_i$ in basis \eqref{Basis_alternative}.

As an application, we apply the idea above to the Dehn twist representation $T_c$ in \eqref{Genus2_5Dehntwists}
with the basis in \eqref{BasisI_introduction}. 
Denoting the two bases in \eqref{BasisI} and \eqref{BasisII} as $|\psi^{\text{I}}_{ab;z}\rangle$
and $|\psi^{\text{II}}_{ab;z}\rangle$, where we consider the case of multiplicity free for simplicity.
Then one has
\be
|\psi^{\text{I}}_{ab;z}\rangle=\sum_{z'} [F^{b\bar{a}}_{b\bar{a}}]_{(z,z')}  |\psi^{\text{II}}_{ab;z'}\rangle,
\ee
with the $F$ matrix defined in Eq.\eqref{F_move_II}.
Then the overlap of two sets of basis vectors are $\langle \psi^{\text{II}}_{ab;z'}|\psi^{\text{I}}_{ab;z}\rangle=[F^{b\bar{a}}_{b\bar{a}}]_{(z,z')}$.
Noting that the action of $T_c$ on the basis$|\psi^{\text{II}}_{ab;z'}\rangle$ simply results in a phase, \textit{i.e.},
$T_c |\psi^{\text{II}}_{ab;z'}\rangle=\theta_{z'} |\psi^{\text{II}}_{ab;z'}\rangle$, then transforming back to basis I, 
we have
\be
\hat{T}_c\, |\psi^{\text{I}}_{ab;z}\rangle=\sum_{z',z''} [F^{b\bar{a}}_{b\bar{a}}]_{(z,z')} \cdot \theta_{z'}
\cdot [F^{b\bar{a}}_{b\bar{a}}]^{-1}_{(z',z'')} 
 |\psi^{\text{I}}_{ab;z''}\rangle.
\ee
Then the Dehn twist representation $T_c$ in the basis in \eqref{BasisI_introduction} has the form
\be
\langle \psi^{\text{I}}_{ab;z''}|\hat{T}_c|\psi^{\text{I}}_{ab;z}\rangle=
\sum_{z'} [F^{b\bar{a}}_{b\bar{a}}]_{(z,z')} \cdot \theta_{z'}
\cdot [F^{b\bar{a}}_{b\bar{a}}]^{-1}_{(z',z'')}. 
\ee
In general, $T_c$ is not a diagonal matrix. 
However, if the theory is abelian, then one has  $z=\mathbf{1}$ and $[F^{ab}_{ab}]_{\mathbf{1}z'}=N_{ab}^{z'}$.
In this case, $T_c$ is a diagonal matrix with the matrix elements
\be
\langle \psi^{\text{I}}_{ab;z''}|\hat{T}_c|\psi^{\text{I}}_{ab;z}\rangle=
\delta_{\mathbf{1}, z} \delta_{\mathbf{1},z''} \theta_{z'} N_{b\bar{a}}^{z'}.
\ee

\subsection{Mapping class group of genus-2 surface and knot/link invariants}
\label{Appendix: MCG and Link invariants}

As shown in Fig.\ref{Genus2_5Dehntwists}, MCG$(\Sigma_{2,0})$ can be generated by the five Dehn twists
around the closed curves $a_1$, $a_2$, $b_1$, $b_2$, and $c$. 
The modular relations of these generators can be found in Ref.\onlinecite{Wajnryb1983}.
In this appendix, with the surgery approach in TQFT, we give an intuitive picture why the 
representation of MCG$(\Sigma_{2,0})$ are related to the knots/links invariant 
that can distinguish MS MTCs.

In Ref.\onlinecite{DT1806}, based on a computer search, it is found that the simplest knot invariant that
can distinguish different MS MTCs with $G=\mathbb{Z}_{11}\rtimes \mathbb{Z}_5$ and the 
three-cocyle $\omega\in H^3(G,U(1))$ is the so-called figure-eight knot. 
In Ref.\onlinecite{arthamonov2015refined}, for some other motivations, the authors study how to use the representations 
of MCG$(\Sigma_{2,0})$ to construct a family of genus-2 pretzel knots including the figure-eight knot.
Here we give a short review of this construction.

We consider a solid genus-2 manifold $\mathcal{M}_{g=2}^{3d}$, which is a 3-dimentional open manifold,
with $\partial \mathcal{M}_{g=2}^{3d}=\Sigma_{2,0}$. 
Then we color $\mathcal{M}_{g=2}^{3d}$ by the trivalent graph in \eqref{BasisI_introduction}.
In TQFT, this colored open 3-manifold represents the wavefunction $|\psi_{(a,\mu),(b,\nu),z}\rangle$.
The first question is how to glue two $M_{g=2}^{3d}$ to form a $S^3$. This is fulfilled by gluing two 
solid genus-2 manifold with the operation:
\be\label{I_genus2}
I=T_{b_2} T_{a_2} T_c T_{a_1} T_{b_1^{-1}},
\ee
where $T_{\gamma}$ is the representation of Dehn twist around the closed curve $\gamma$ with the direction labeled in \eqref{Genus2_5Dehntwists}.
If we do Dehn twist along the opposite direction, then one has $T_{\gamma^{-1}}=T^{-1}_{\gamma}$.
With the operation in \eqref{I_genus2}, we have
\be
\langle \mathcal{M}_{g=2}^{3d}|I|\mathcal{M}_{g=2}^{3d}\rangle=\mathcal{Z}(S^3),
\ee
or equivalently, 
$\langle \psi_{1,1,1}| I |\psi_{1,1,1}\rangle=\mathcal{Z}(S^3)=1/\mathcal{D}$,
where $\mathcal{Z}(S^3)$ denotes the partition function on $S^3$, $\mathcal{D}$ is the total quantum dimension of the theory,
and the index `1' in $|\psi_{1,1,1}\rangle$ denotes the identity anyon.

Next, one can create a figure-eight knot in $S^3$ by starting from the following wavefunction
defined on a solid genus-2 manifold:
\be\label{WF_fighre8knot}
\small
\begin{tikzpicture}[baseline={(current bounding box.center)}]
\draw  (-55pt,11pt) arc (90:270:12pt);
\draw  (-60pt,0pt) arc (-120:-60:20pt);
\draw (-56pt,-1.5pt) arc (130:50:10pt);
\draw  (-60+35pt,0pt) arc (-120:-60:20pt);
\draw  (-56+35pt,-1.5pt) arc (130:50:10pt);
\draw  (-9pt,11pt) arc (90:-90:12pt);
\draw [>=stealth,->](-9pt,11pt)--(-34pt,11pt);\draw(-34pt,11pt)--(-55pt,11pt);
\draw [>=stealth,->](-55pt,-13pt)--(-30pt,-13pt); \draw(-30pt,-13pt)--(-9pt,-13pt);
\node at (-32pt,17pt){$a$};
\end{tikzpicture}
\ee
which we denote as $|\psi_{\text{unknot}}^a\rangle$. Based on Eq.\eqref{Fuse_ab},
one can expand $|\psi_{\text{unknot}}^a\rangle$ with the complete basis vectors in \eqref{BasisI_introduction} as
$|\psi_{\text{unknot}}^a\rangle=\sum_{z,\mu}\sqrt{\frac{d_z}{d_a^2}}|\psi_{(a,\mu),(a,\mu),z}\rangle$.
Then it is observed in Ref.\onlinecite{arthamonov2015refined} that the figure-eight knot invariant can be
created by
\be
\mathcal{Z}(S^3, \text{figure-eight knot})=\langle \psi_{1,1,1}|I\cdot U|\psi_{\text{unknot}}^a\rangle,
\ee
where $U=T_cT^{-1}_{a_1}T^{-1}_{b_1}T^{-1}_{a_2}T_{b_2}$, and $\mathcal{Z}(S^3, \text{figure-eight knot})$
denotes the partition function on $S^3$ with a figure-eight knot inserted.
That is, with the basis vectors in \eqref{BasisI_introduction} and the representations of MCG($\Sigma_{2,0}$),
one can construct certain nontrivial knot invariants that can distinguish MS MTCs. This illustrates 
why the MCG($\Sigma_{2,0}$) representations can be used to distinguish MS MTCs beyond modular data.

It will be interesting to study how to produce other nontrivial 
knot/link invariants with the generators of MCG$(\Sigma_{g,0})$.
For example, if we can construct the Borromean ring with the genus-$g$ basis \eqref{Genus_g_basis}
and the MCG$(\Sigma_{g,0})$ representations, then this will indirectly 
prove that the genus-$g$ data will be able to distinguish \textit{all} the MS MTCs.

\section{More on topological lattice gauge theory}

\subsection{Topological invariants based on a punctured torus}
\label{Lattice_puncturedTorus_appendix}

For a lattice gauge theory on a manifold with punctures, the Hilbert space is discussed carefully in Ref.\onlinecite{freed1993chern,freed1993lectures}.
Different from the case of a closed manifold, now we need to fix the 
choice of group elements on the punctures.

The reason why we need to rigid the group elements on the punctures is due to the axiom of gluing in TQFTs.
In TQFTs, the gluing of two punctured manifolds along 
the punctures is well defined only when we choose basepoints at the punctures.\cite{freed1993chern,freed1993lectures}
Let us consider gluing two punctured tori and gauge bundles along the punctures as follows:
\be\label{gluing_torus}
\small
\begin{tikzpicture}[baseline={(current bounding box.center)}]
\draw  (-12-10pt,0pt) arc (-60:-120:20pt);
\draw  (-16-10pt,-1.5pt) arc (50:130:10pt);
\draw  (12pt,0pt) arc (-120:-60:20pt);
\draw  (16pt,-1.5pt) arc (130:50:10pt);

\draw 
(25pt,12.5pt)..controls (32.5pt,12.5pt) and (41pt,7.5pt)..
(41pt,0pt).. controls (41pt,-7.5pt) and (32.5pt,-12.5pt).. 
(25pt,-12.5pt);

\draw (25pt,12.5pt)..controls (15pt,12.5pt) and (10pt,10.5pt)..(0pt,5pt);
\draw (25pt,-12.5pt)..controls (15pt,-12.5pt) and (10pt,-10.5pt)..(0pt,-5pt);

\draw 
(-25-10pt,12.5pt)..controls (-32.5-10pt,12.5pt) and (-41-10pt,7.5pt)..
(-41-10pt,0pt).. controls (-41-10pt,-7.5pt) and (-32.5-10pt,-12.5pt).. 
(-25-10pt,-12.5pt);

\draw (-25-10pt,12.5pt)..controls (-15-10pt,12.5pt) and (-10-10pt,10.5pt)..(0-10pt,5pt);
\draw (-25-10pt,-12.5pt)..controls (-15-10pt,-12.5pt) and (-10-10pt,-10.5pt)..(0-10pt,-5pt);

\draw (0,0) ellipse (0.05cm and 0.18cm);
\draw (-10pt,0) ellipse (0.05cm and 0.18cm);


\draw[black,fill=black] (1.4pt,0pt) circle (1.0pt);
\draw[black,fill=black] (-10-1.4pt,0pt) circle (1.0pt);

\node at (-12pt,13pt){$S^1$};
\node at ( 3pt,13pt){$S^1$};

\end{tikzpicture}
\xrightarrow{\,\,\text{glue}\,\,}
\begin{tikzpicture}[baseline={(current bounding box.center)}]
\draw  (-12pt,0pt) arc (-60:-120:20pt);
\draw  (-16pt,-1.5pt) arc (50:130:10pt);
\draw  (12pt,0pt) arc (-120:-60:20pt);
\draw  (16pt,-1.5pt) arc (130:50:10pt);

\draw 
(25pt,12.5pt)..controls (32.5pt,12.5pt) and (41pt,7.5pt)..
(41pt,0pt).. controls (41pt,-7.5pt) and (32.5pt,-12.5pt).. 
(25pt,-12.5pt);

\draw (25pt,12.5pt)..controls (15pt,12.5pt) and (10pt,10.5pt)..(0pt,5pt);
\draw (25pt,-12.5pt)..controls (15pt,-12.5pt) and (10pt,-10.5pt)..(0pt,-5pt);

\draw 
(-25pt,12.5pt)..controls (-32.5pt,12.5pt) and (-41pt,7.5pt)..
(-41pt,0pt).. controls (-41pt,-7.5pt) and (-32.5pt,-12.5pt).. 
(-25pt,-12.5pt);

\draw (-25pt,12.5pt)..controls (-15pt,12.5pt) and (-10pt,10.5pt)..(0pt,5pt);
\draw (-25pt,-12.5pt)..controls (-15pt,-12.5pt) and (-10pt,-10.5pt)..(0pt,-5pt);

\draw [dashed ](0,0) ellipse (0.05cm and 0.18cm);

\node at ( 3pt,13pt){$S^1$};

\end{tikzpicture}
\ee
where the punctures are represented by a circle $\partial \Sigma_{1,1}=S^1$.
It is noted that the bundle over $S^1$ with a basepoint has a definite holonomy
$h\in G$. If no basepoint is chosen, the holonomy is determined 
only up to group conjugation. The two punctures in \eqref{gluing_torus}
can be glued only when the holonomies along the two $S^1$ agree with each other.

The procedure in \eqref{gluing_torus} is more straightforwardly understood by 
considering the gluing in \eqref{Genus2_gluing}. 
In \eqref{Genus2_gluing}, the holonomies along the punctures are $k$ and $k'$, 
respectively. The two punctured tori can be glued only when $k=k'$.
Note this condition is different from $[k]=[k']$, where $[k]$ is the conjugacy class of $k$.
Once we glue the two punctured tori along the puncture $S^1$, we are supposed to forget
the basepoint, and then the holonomy is only determined up to conjugacy.
From this point of view, the gluing in \eqref{gluing_torus} corresponds to the map
\be
\text{gluing}:\quad G\to \text{Conj}(G)
\ee
for the holonomy around $S^1$ in \eqref{gluing_torus}. 
This mapping assigns to each $k\in G$ its conjugacy class $[k]$.

Based on the flat connection condition in Eq.\eqref{flat_connection_pucturedtorus}, one can find 
there are in total $|G|^2$ equivalence classes of bundles over $\Sigma_{1,1}$.
That is, in the lattice gauge theory, we have $\text{dim}\,\mathcal{H}(\Sigma_{1,1})=|G|^2$.
In addition, it is noted that for the modular transfomation in Sec.\ref{Sec: Lattice_PuncturedTorus},
since we need to rigid the holonomy around the puncture, one needs to choose the 
group element $t=1$ in Eqs.\eqref{Tx_punctured_torus} and \eqref{Ty_punctured_torus}.
That is, no conjugacy is performed during the modular transformation.

Now let us compare with the result of quasi-particle basis calculation 
in Sec.\ref{Sec: Quasi-particle}. 
The dimension of Hilbert space with a fixed anyon $z$ at the puncture on $\Sigma_{1,1}$ has the expression
\be\label{appendix_puncture_Hilbert}
\text{dim}\, \mathcal{H}(\Sigma_{1,1}, z)=\sum_{a} N^z_{a\bar{a}},
\ee
where $N^z_{a\bar{a}}$ is the fusion coefficient. 
The meaning of the expression in \eqref{appendix_puncture_Hilbert} is apparent. By fixing the anyon type $z$ at the puncture,
the dimension of the Hilbert space of degenerate ground states is the total number of fusion channels that $a$ and $\bar{a}$
fuse into $z$ for all possible $a\in \Pi_{\mathcal{C}}$.
Now by including all the possible anyon types $z$, 
one can find the following relation
\be\label{Relation_HilbertSpace_puncture_torus}
\sum_z \sum_a N^z_{a\bar{a}}\, d_z=|G|^2,
\ee
where on the right is the dimension of Hilbert space $\text{dim}\,\mathcal{H}(\Sigma_{1,1})$ obtained 
from the lattice gauge theory of a finite group $G$.
Considering that the integral quantum dimension $d_z$ describes the internal degree of freedom for the anyon $z$
at the puncture, the meaning of \eqref{Relation_HilbertSpace_puncture_torus} is also straightforward.
Eq.\eqref{Relation_HilbertSpace_puncture_torus} can be rewritten as
\be\label{Relation_HilbertSpace_puncture_torus_b}
\sum_z \text{dim}\, \mathcal{H}(\Sigma_{1,1}, z) \cdot d_z=\text{dim}\,\mathcal{H}(\Sigma_{1,1}).
\ee
This is as expected by considering that in defining $\text{dim}\,\mathcal{H}(\Sigma_{1,1})$ we have to rigid the 
holonomy around the puncture without conjugacy. That is, the internal degree of freedom for anyon $z$ at the puncture
is included in this definition.

With the discussion above, now we are ready to compare the topological invariants obtained from the lattice gauge theory
with those obtained from the quasi-particle basis calculation in Sec.\ref{Sec: Quasi-particle}.
It is noted that the $T^t_{x(y)}$ in Eqs.\eqref{Tx_punctured_torus} and \eqref{Ty_punctured_torus}
are used as building blocks for the genus-2 case, and therefore a gauge transformation is introduced.
For $\Sigma_{1,1}$ itself, to have a definite holonomy around the puncture, we should not perform gauge transformation, 
and therefore the group elements $t$ are fixed as  $t=1$. Denoting $T_{x(y)}:=T^{t=1}_{x(y)}$,  the punctured $S$ and $T$ matrices
are obtained with the definition
$T:=T_x$, and $S:=T_y\cdot T_x^{-1}\cdot T_y$.
Then we can construct the topological invariants  
with the words in \eqref{Words}.

Depending on the holonomy $k$ around the puncture [see \eqref{PuncturedTorusBasis1}], the topological invariants 
for the MS MTCs with $G=\mathbb{Z}_{11}\rtimes \mathbb{Z}_5$ are grouped into three pieces:
\be
W_{\Sigma_{1,1}}=W^{I}_{\Sigma_{1,1}}+W^{A_1}_{\Sigma_{1,1}}+W^{A_2}_{\Sigma_{1,1}},
\ee
which are obtained by tracing over the basis vectors $|g_x, g_y; k\rangle$ in  \eqref{PuncturedTorusBasis1}
with $k\in [1], \, [a^1]$ and $[a^2]$, respectively.

\begin{table}[H]
\footnotesize
\underline{Quasi-particle basis result:}
\centering
\begin{tabular}{cccccccccc}
$W^{(z)}_1$  &\vline     &$W^{I_0}_{\Sigma_{1,1}}$ &\vline     &$\sum_{i=1}^4 W_{\Sigma_{1,1}}^{I_i}$&\vline     &$W_{\Sigma_{1,1}}^{I_5,I_6}$&\vline     &$W_{\Sigma_{1,1}}^{A_1,A_2}$\\ \hline
$u=0$  &\vline     &9               &\vline          & \textcolor{gray}{$-4$}               &\vline  &$4$               &\vline   &$22$              \\ \hline
$u=1$  &\vline       &$4\cos\frac{2\pi}{5}+5$ &\vline     & \textcolor{gray}{$-4$}&\vline     &$4\cos\frac{2\pi}{5}$&\vline     &$22e^{\frac{2i\pi }{5}\times 1}$  \\ \hline
$u=2$  &\vline         &$4\cos\frac{4\pi}{5}+5$ &\vline     & \textcolor{gray}{$-4$}&\vline     &$4\cos\frac{4\pi}{5}$&\vline     &$22e^{\frac{2i\pi }{5}\times 2}$ \\  \hline
$u=3$  &\vline       &$4\cos\frac{4\pi}{5}+5$  &\vline     & \textcolor{gray}{$-4$}&\vline     &$4\cos\frac{4\pi}{5}$&\vline     &$22e^{\frac{2i\pi }{5}\times 3}$ \\ \hline
$u=4$  &\vline          &$4\cos\frac{2\pi}{5}+5$   &\vline     & \textcolor{gray}{$-4$}&\vline     &$4\cos\frac{2\pi}{5}$&\vline     &$22e^{\frac{2i\pi }{5}\times 4}$ \\ \hline
\end{tabular}
\underline{Lattice gauge theory result:}
\\
\centering
\footnotesize
\begin{tabular}{cccccccccccc}
$W^{(z)}_1$  &\vline     &   &   &$W_{\Sigma_{1,1}}^{I}$&  & &\vline     &$W_{\Sigma_{1,1}}^{A_1}$ &\vline  &$W_{\Sigma_{1,1}}^{A_2}$\\ \hline
$u=0$  &\vline     &              &          & $45$               & 	&               &\vline   &110 &\vline   &110              \\ \hline
$u=1$  &\vline       &  &     & $44\cos\frac{2\pi}{5}+1$&     & &\vline     &$110\,e^{\frac{2i\pi }{5}\times 1}$ &\vline  &$110\,e^{\frac{2i\pi }{5}\times 1}$  \\ \hline
$u=2$  &\vline         & &     &$44\cos\frac{4\pi}{5}+1$&   & &\vline     &$110\, e^{\frac{2i\pi }{5}\times 2}$ &\vline  &$110\, e^{\frac{2i\pi }{5}\times 2}$   \\  \hline
$u=3$  &\vline       &  &  &$44\cos\frac{4\pi}{5}+1$&   & &\vline     &$110\, e^{\frac{2i\pi }{5}\times 3}$ &\vline  &$110\, e^{\frac{2i\pi }{5}\times 3}$  \\ \hline
$u=4$  &\vline          &  &    &$44\cos\frac{2\pi}{5}+1$&   & &\vline     &$110\, e^{\frac{2i\pi }{5}\times 4}$ &\vline  &$110\, e^{\frac{2i\pi }{5}\times 4}$ \\ \hline
\end{tabular}
\caption{
Comparison of the quasi-particle basis results in Table \ref{W1compare}
and the lattice gauge theory results, for $W_{\Sigma_{1,1}}$ with the word $w_1^{(z)}$ in Eq.\eqref{Words}.
}
\label{W1compareAppendix}
\end{table}

In comparison with the results obtained from the quasiparticle basis in Sec.\ref{Sec: Punctured Torus_QuasiParticleBasis}, 
based on the relation in Eq.\eqref{Relation_HilbertSpace_puncture_torus_b}, we have the following correspondence:
\be
\begin{split}
W_{\Sigma_{1,1}}^{I,\text{lattice}}=&\sum_{i=0,\cdots,6} W_{\Sigma_{1,1}}^{I_i,\text{q.p.}}\cdot d_{I_i},\\
W_{\Sigma_{1,1}}^{A_1,\text{lattice}}=& W_{\Sigma_{1,1}}^{A_1,\text{q.p.}}\cdot d_{A_1},\\
W_{\Sigma_{1,1}}^{A_2,\text{lattice}}=& W_{\Sigma_{1,1}}^{A_2,\text{q.p.}}\cdot d_{A_2},\\
\end{split}
\ee
where `$\text{q.p}$' means the results are obtained from the quasi-particle basis calculation.
Let us take $u=1$ in Table.\ref{W1compareAppendix} for example. Then we have 
$\small W_{\Sigma_{1,1}}^{I,\text{lattice}}=\sum_{i=0,\cdots,6} W_{\Sigma_{1,1}}^{I_i,\text{q.p.}}\cdot d_{I_i}
=(4\cos\frac{2\pi}{5}+5)\textcolor{gray}{-4}\times 1
+4\cos\frac{2\pi}{5}\times 2\times 5=44 \cos\frac{2\pi}{5}+1$,
$\small W_{\Sigma_{1,1}}^{I,\text{lattice}}=W_{\Sigma_{1,1}}^{A_1,\text{q.p.}}\cdot d_{A_1}
=22\, e^{\frac{2i\pi }{5}\times 1}\times 5=110\, e^{\frac{2i\pi }{5}\times 1}$, 
and $\small W_{\Sigma_{1,1}}^{A_2,\text{lattice}}=W_{\Sigma_{1,1}}^{A_2,\text{q.p.}}\cdot d_{A_2}
=110\, e^{\frac{2i\pi }{5}\times 1}$, where we have used the concrete value of quantum dimensions 
in Table \ref{49Anyons}. The comparison of topological invariants for other words in Tables \ref{W2compareAppendix}
$\sim$\ref{W7compareAppendix} can be made in a similar way.

\begin{table}[H]
\footnotesize
\centering
\begin{tabular}{cccccccccccc}
$W^{(z)}_2$  &\vline     &   &   &$W_{\Sigma_{1,1}}^{I}$&  & &\vline     &$W_{\Sigma_{1,1}}^{A_1}$&\vline     &$W_{\Sigma_{1,1}}^{A_2}$\\ \hline
$u=0$  &\vline     &              &          & $45$               & 	&               &\vline   &110  &\vline   &110              \\ \hline
$u=1$  &\vline       &  &     & $44\cos\frac{4\pi}{5}+1$&     & &\vline     &$110\,e^{\frac{2i\pi }{5}\times 3}$ &\vline     &$110\,e^{\frac{2i\pi }{5}\times 3}$ \\ \hline
$u=2$  &\vline         & &     &$44\cos\frac{2\pi}{5}+1$&   & &\vline     &$110\,e^{\frac{2i\pi }{5}\times 1}$ &\vline     &$110\,e^{\frac{2i\pi }{5}\times 1}$   \\  \hline
$u=3$  &\vline       &  &  &$44\cos\frac{2\pi}{5}+1$&   & &\vline     &$110\,e^{\frac{2i\pi }{5}\times 4}$  &\vline     &$110\,e^{\frac{2i\pi }{5}\times 4}$ \\ \hline
$u=4$  &\vline          &  &    &$44\cos\frac{4\pi}{5}+1$&   & &\vline     &$110\,e^{\frac{2i\pi }{5}\times 2}$ &\vline     &$110\,e^{\frac{2i\pi }{5}\times 2}$\\ \hline
\end{tabular}
\caption{$W_{\Sigma_{1,1}}$ with the word $w_2^{(z)}$ in Eq.\eqref{Words}, based on the lattice gauge theory calculation.
}
\label{W2compareAppendix}
\end{table}

\begin{table}[H]
\footnotesize
\centering
\begin{tabular}{cccccccccccc}
$W^{(z)}_3$  &\vline     &   &   &$W_{\Sigma_{1,1}}^{I}$&  & &\vline     &$W_{\Sigma_{1,1}}^{A_1}$ &\vline     &$W_{\Sigma_{1,1}}^{A_1}$\\ \hline
$u=0$  &\vline     &              &          & $265$               & 	&               &\vline   &330         &\vline   &330        \\ \hline
$u=1$  &\vline       &  &     & $-65$&     & &\vline     &$330\,e^{\frac{2i\pi }{5}\times 2}$  &\vline     &$330\,e^{\frac{2i\pi }{5}\times 2}$  \\ \hline
$u=2$  &\vline         & &     &$-65$&   & &\vline     &$330\,e^{\frac{2i\pi }{5}\times 4}$ &\vline     &$330\,e^{\frac{2i\pi }{5}\times 4}$\\  \hline
$u=3$  &\vline       &  &  &$-65$&   & &\vline     &$330\,e^{\frac{2i\pi }{5}\times 1}$ &\vline     &$330\,e^{\frac{2i\pi }{5}\times 1}$  \\ \hline
$u=4$  &\vline          &  &    &$-65$&   & &\vline     &$330\,e^{\frac{2i\pi }{5}\times 3}$ &\vline     &$330\,e^{\frac{2i\pi }{5}\times 3}$ \\ \hline
\end{tabular}
\caption{
$W_{\Sigma_{1,1}}$ with the word $w_3^{(z)}$ in Eq.\eqref{Words}, based on the lattice gauge theory calculation.
}
\label{W3compareAppendix}
\end{table}

\begin{table}[H]
\footnotesize
\centering
\begin{tabular}{cccccccccccc}
$W^{(z)}_4$  &\vline     &   &   &$W_{\Sigma_{1,1}}^I$&  & &\vline     &$W_{\Sigma_{1,1}}^{A_1}$&\vline     &$W_{\Sigma_{1,1}}^{A_2}$\\ \hline
$u=0$  &\vline     &              &          & $265$               & 	&               &\vline   &220 &\vline   &220              \\ \hline
$u=1$  &\vline       &  &     & $45$&     & &\vline     &$220\,e^{\frac{2i\pi }{5}\times 3}$&\vline     &$220\,e^{\frac{2i\pi }{5}\times 3}$ \\ \hline
$u=2$  &\vline         & &     &$45$&   & &\vline     &$220\,e^{\frac{2i\pi }{5}\times 1}$&\vline     &$220\,e^{\frac{2i\pi }{5}\times 1}$ \\  \hline
$u=3$  &\vline       &  &  &$45$&   & &\vline     &$220\,e^{\frac{2i\pi }{5}\times 4}$ &\vline     &$220\,e^{\frac{2i\pi }{5}\times 4}$  \\ \hline
$u=4$  &\vline          &  &    &$45$&   & &\vline     &$220\,e^{\frac{2i\pi }{5}\times 2}$&\vline     &$220\,e^{\frac{2i\pi }{5}\times 2}$ \\ \hline
\end{tabular}
\caption{$W_{\Sigma_{1,1}}$ with the word $w_4^{(z)}$ in Eq.\eqref{Words}, based on lattice gauge theory calculation.
}
\label{W4compareAppendix}
\end{table}

\begin{table}[H]
\centering
\footnotesize
\begin{tabular}{cccccccccccc}
$W^{(z)}_5$  &\vline     &   &   &$W^I_{\Sigma_{1,1}}$&  & &\vline     &$W_{\Sigma_{1,1}}^{A_1}$&\vline     &$W_{\Sigma_{1,1}}^{A_2}$\\ \hline
$u=0$  &\vline     &              &          & $45$               & 	&               &\vline   &110  &\vline   &110              \\ \hline
$u=1$  &\vline       &  &     & $44\cos\frac{4\pi}{5}+1$&     & &\vline     &$110\,e^{\frac{2i\pi }{5}\times 2}$&\vline     &$110\,e^{\frac{2i\pi }{5}\times 2}$ \\ \hline
$u=2$  &\vline         & &     &$44\cos\frac{2\pi}{5}+1$&   & &\vline     &$110\,e^{\frac{2i\pi }{5}\times 4}$&\vline     &$110\,e^{\frac{2i\pi }{5}\times 4}$  \\  \hline
$u=3$  &\vline       &  &  &$44\cos\frac{2\pi}{5}+1$&   & &\vline     &$110\,e^{\frac{2i\pi }{5}\times 1}$&\vline     &$110\,e^{\frac{2i\pi }{5}\times 1}$  \\ \hline
$u=4$  &\vline          &  &    &$44\cos\frac{4\pi}{5}+1$&   & &\vline     &$110\,e^{\frac{2i\pi }{5}\times 3}$&\vline     &$110\,e^{\frac{2i\pi }{5}\times 3}$ \\ \hline
\end{tabular}
\caption{
$W_{\Sigma_{1,1}}$ with the word $w_5^{(z)}$ in Eq.\eqref{Words}, based on the lattice gauge theory calculation.
}
\label{W5compareAppendix}
\end{table}

\begin{table}[H]
\centering
\footnotesize
\begin{tabular}{cccccccccccc}
$W^{(z)}_6$  &\vline     &   &   &$W^I_{\Sigma_{1,1}}$&  & &\vline     &$W^{A_1}_{\Sigma_{1,1}}$ &\vline     &$W^{A_2}_{\Sigma_{1,1}}$\\ \hline
$u=0$  &\vline     &              &          & $45$               & 	&               &\vline   &110    &\vline   &110              \\ \hline
$u=1$  &\vline       &  &     & $44\cos\frac{2\pi}{5}+1$&     & &\vline     &$110\,e^{\frac{2i\pi }{5}\times 4}$ &\vline     &$110\,e^{\frac{2i\pi }{5}\times 4}$  \\ \hline
$u=2$  &\vline         & &     &$44\cos\frac{4\pi}{5}+1$&   & &\vline     &$110\,e^{\frac{2i\pi }{5}\times 3}$&\vline     &$110\,e^{\frac{2i\pi }{5}\times 3}$\\  \hline
$u=3$  &\vline       &  &  &$44\cos\frac{4\pi}{5}+1$&   & &\vline     &$110\,e^{\frac{2i\pi }{5}\times 2}$&\vline     &$110\,e^{\frac{2i\pi }{5}\times 2}$ \\ \hline
$u=4$  &\vline          &  &    &$44\cos\frac{2\pi}{5}+1$&   & &\vline     &$110\,e^{\frac{2i\pi }{5}\times 1}$&\vline     &$110\,e^{\frac{2i\pi }{5}\times 1}$ \\ \hline
\end{tabular}
\caption{
$W_{\Sigma_{1,1}}$ with the word $w_6^{(z)}$ in Eq.\eqref{Words}, based on lattice gauge theory calculation.
}
\label{W6compareAppendix}
\end{table}

\begin{table}[H]
\centering
\footnotesize
\begin{tabular}{cccccccccccc}
$W^{(z)}_7$  &\vline     &   &   &$W^I_{\Sigma_{1,1}}$&  & &\vline     &$W^{A_1}_{\Sigma_{1,1}}$&\vline     &$W^{A_2}_{\Sigma_{1,1}}$\\ \hline
$u=0$  &\vline     &              &          & $45$               & 	&               &\vline   &110   &\vline   &110              \\ \hline
$u=1$  &\vline       &  &     & $44\cos\frac{4\pi}{5}+1$&     & &\vline     &$110\,e^{\frac{2i\pi }{5}\times 2}$&\vline     &$110\,e^{\frac{2i\pi }{5}\times 2}$ \\ \hline
$u=2$  &\vline         & &     &$44\cos\frac{2\pi}{5}+1$&   & &\vline     &$110\,e^{\frac{2i\pi }{5}\times 4}$ &\vline     &$110\,e^{\frac{2i\pi }{5}\times 4}$ \\  \hline
$u=3$  &\vline       &  &  &$44\cos\frac{2\pi}{5}+1$&   & &\vline     &$110\,e^{\frac{2i\pi }{5}\times 1}$ &\vline     &$110\,e^{\frac{2i\pi }{5}\times 1}$ \\ \hline
$u=4$  &\vline          &  &    &$44\cos\frac{4\pi}{5}+1$&   & &\vline     &$110\,e^{\frac{2i\pi }{5}\times 3}$&\vline     &$110\,e^{\frac{2i\pi }{5}\times 3}$ \\ \hline
\end{tabular}
\caption{
$W_{\Sigma_{1,1}}$ with the word $w_7^{(z)}$ in Eq.\eqref{Words}, based on lattice gauge theory calculation.
}\label{W7compareAppendix}
\end{table}

\subsection{More punctures}
\label{appendix: more punctures}

Now we give a brief sketch on how to generalize to the case of a torus with multiple punctures.
This will be useful if we study the representations of $\text{MCG}(\Sigma_{g,0})$.
For example, in the following is the triangulation for a torus with two punctures:
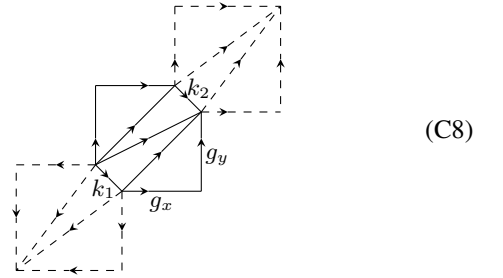
\begin{eqnarray}\label{Genus3Triangle_appendix}
\begin{small}
\begin{tikzpicture}[baseline={(current bounding box.center)}]
\draw (-20pt,20pt)--(-20pt,0pt); \draw [>=stealth,<-] (-20pt, 0pt) -- (-20pt,-10pt);
\draw (20pt,10pt)--(20pt,0pt); \draw [>=stealth,<-] (20pt, 0pt) -- (20pt,-20pt);
\draw [>=stealth,->] (-20pt,20pt)--(0pt,20pt); \draw (0pt,20pt)--(10pt,20pt);
\draw [>=stealth,->] (-10pt,-20pt)--(0pt,-20pt);  \draw (0pt, -20pt) -- (20pt,-20pt);
\draw [>=stealth,->] (-20pt,-10pt)--(-15pt,-15pt); \draw (-15pt,-15pt)--(-10pt,-20pt);

\draw [>=stealth,->] (-20pt,-10pt)--(-5pt, 5pt);\draw (-5pt,5pt)--(10pt,20pt);
\draw [>=stealth,->] (-10pt,-20pt)--(5pt,-5pt);\draw(5pt,-5pt)--(20pt, 10pt);
\draw [>=stealth,->] (-20pt,-10pt)--(0pt, 0pt);\draw (0pt,0pt)--(20pt,10pt);

\draw [dashed][>=stealth,->] (-20pt,-10pt)--(-35pt,-10pt); \draw[dashed](-35pt,-10pt)--(-50pt,-10pt);
\draw [dashed][>=stealth,->] (-10pt,-20pt)--(-10pt,-35pt);\draw[dashed](-10pt,-35pt)--(-10pt,-50pt);
\draw [dashed](-50pt,-50pt)--(-30pt,-50pt); \draw [dashed][>=stealth,<-] (-30pt,-50pt)--(-10pt,-50pt);
\draw [dashed](-50pt,-50pt)--(-50pt,-30pt); \draw[dashed][>=stealth,<-] (-50pt,-30pt)--(-50pt,-10pt);

\draw [dashed][>=stealth,->] (-10pt,-20pt)--(-30pt,-35pt); \draw [dashed](-30pt,-35pt)--(-50pt,-50pt);
\draw [dashed][>=stealth,->] (-20pt,-10pt)--(-35pt,-30pt); \draw [dashed](-35pt,-30pt)--(-50pt,-50pt);


\draw [dashed](-20+30pt,20+30pt)--(-20+30pt,0+30pt); \draw [dashed][>=stealth,<-] (-20+30pt, 0+30pt) -- (-20+30pt,-10+30pt);
\draw [dashed](20+30pt,20+30pt)--(20+30pt,0+30pt); \draw [dashed][>=stealth,<-] (20+30pt, 0+30pt) -- (20+30pt,-20+30pt);
\draw [dashed][>=stealth,->] (-20+30pt,20+30pt)--(0+30pt,20+30pt); \draw [dashed](0+30pt,20+30pt)--(20+30pt,20+30pt);
\draw [dashed][>=stealth,->] (-10+30pt,-20+30pt)--(0+30pt,-20+30pt);  \draw [dashed](0+30pt, -20+30pt) -- (20+30pt,-20+30pt);
\draw [>=stealth,->] (-20+30pt,-10+30pt)--(-15+30pt,-15+30pt); \draw (-15+30pt,-15+30pt)--(-10+30pt,-20+30pt);

\draw [dashed][>=stealth,->] (-20+30pt,-10+30pt)--(0+30pt, 5+30pt);\draw [dashed](0+30pt,5+30pt)--(20+30pt,20+30pt);
\draw [dashed][>=stealth,->] (-10+30pt,-20+30pt)--(5+30pt,0+30pt);\draw [dashed](5+30pt,0+30pt)--(20+30pt, 20+30pt);

\node at (5pt,-25pt){${g_x}$};
\node at (26pt,-7pt){${g_y}$};
\node at (-17pt,-19pt){${k_1}$};
\node at ( 19pt, 19pt){${k_2}$};

\end{tikzpicture}
\end{small}
\end{eqnarray}
Here we embed this twice-punctured torus in a closed manifold $\Sigma_{3,0}$ of genus $g=3$ for simplicity.
In fact, it is straightforward to check it is the building block for $\Sigma_{g,0}$ with $g>2$.

The generalization of the two Dehn twists $T_x$ and $T_y$ [see Eqs.\eqref{Da_Inverse} and \eqref{Db_Inverse}] to this case is straightforward, except that
now $g_x$ and $g_y$ satisfy the following flat-connection condition (note the group elements are applied from right to left):
\be
g_x \, g_y=k_2^{-1}\, g_y\, g_x\, k_1,
\ee
where $k_1$ and $k_2$ denote the holonomy around the two punctures in \eqref{Genus3Triangle_appendix}.
In addition, the $U(1)$ phase associated to $T_{x(y)}$ transformation in \eqref{TxPhasePuncture} and \eqref{TyPhasePuncture}
will be modified accordingly. 
There are now four triangular prisms $Y\times I$ as compared to three $Y\times I$ in \eqref{TxPhasePuncture} and \eqref{TyPhasePuncture}.
Once $T_x$ and $T_y$ are obtained, one can get the twice-punctured $S$ matrix following the
procedure in Sec.\ref{Sec: Lattice_modularTransform}.

\subsection{Twisted quantum double of $G=\mathbb{Z}_p$}
\label{SubSec: Zp}

To illustrate the lattice gauge theory approach in the main text, we consider the example of a twisted quantum double of $G=\mathbb{Z}_p$,
with $\omega\in H^3(\mathbb{Z}_p,U(1))\cong \mathbb{Z}_p$.

There are two motivations for studying this simple example: one is that it mimics the the calculation of modular data
for MS MTCs with $G=\mathbb{Z}_q\rtimes \mathbb{Z}_p$ (as we mentioned in the main text, the modular data of MS MTCs that depend on the 3-cocyle
$\omega^u$ are the same as those for $G=\mathbb{Z}_p$); the other is that one can see clearly 
how the modular transformations act on the genus-1 basis $|g_x,g_y\rangle$ as well as the wavefunction.
It is noted that the modular transformation acts on the basis and the wavefunction in an `opposite' way.

On the manifold $\Sigma_{1,0}$, one can write down the quasi-particle basis in terms of group-element basis. 
Corresponding to the anyon-type in \eqref{AnyonData}, the quasi-particle basis has the form\cite{HuWanWu1211}
\be
|g,\tilde{\chi}_m\rangle=\frac{1}{\sqrt{|G|}}\sum_{h\in G}\tilde{\chi}_m^g(h)\, |g,h\rangle.
\ee
where the  character $\tilde{\chi}_m^g(h)$ is defined in Eq.\eqref{Character_B}, and has the the explicit 
expression 
\be
\small
\tilde{\chi}^g_m(h)=e^{\frac{2\pi i}{p^2} (pm+u[g]_p)\cdot [h]_p},
\ee
here $[x]_p$ means $x$ mod $p$.
Now we consider the effect of Dehn twist $T_{x,y}:=\frac{1}{|G|}\sum_{t\in G}T_{x,y}^t$. With some straightforward algebra, one can find that
\be\small
T_x |g,\tilde{\chi}_m\rangle=\frac{1}{\sqrt{|G|}}\sum_{h\in G}\tilde{\chi}^g_m(h) \cdot \omega(g, g^{-1}h, g) |g, g^{-1}h\rangle.
\ee
By relabelling $g^{-1}h=:k$, one has
\be\small\label{Tx_Zp_appendix}
\begin{split}
T_x |g,\tilde{\chi}_m\rangle=&\frac{1}{\sqrt{|G|}}\sum_{k\in G}\tilde{\chi}^g_m(gk)\omega(g,k,g) |g,k\rangle \\
=&\frac{1}{\sqrt{|G|}}\sum_{k\in G} e^{\frac{2\pi i}{p^2} (pm+u[g]_p)\cdot ([g]_p+[k]_p)}|g,k\rangle\\
=&\frac{1}{\sqrt{|G|}}\sum_{h\in G} e^{\frac{2\pi i}{p^2} (pm+u[g]_p)\cdot ([g]_p+[h]_p)}|g,h\rangle.
\end{split}
\ee
Comparing the above equations, it is noted that $T_x$ acts on the basis as $|g,h\rangle\to|g,g^{-1}h\rangle$, but 
acts on the wavefunction as $(g,h)\to (g, gh)$. It is similar for $T_y$. 
Eq.\eqref{Tx_Zp_appendix} can be rewritten as
\be
\small
\begin{split}
T_x |g,\tilde{\chi}_m\rangle=&e^{\frac{2\pi i}{p^2}(pm+u[g]_p)\cdot [g]_p}\, |g,\tilde{\chi}_m\rangle\\
=:& \theta_{g,\tilde{\chi}_m} |g,\tilde{\chi}_m\rangle.
\end{split}
\ee
That is, the topological spin of anyon $(g,\tilde{\chi}_m)$ is $\theta_{(g,\tilde{\chi}_m)}=e^{\frac{2\pi i}{p^2}(pm+u[g]_p)\cdot [g]_p}$.
Now let us check the effect of modular $S$ transformation defined by $S:=T_y\cdot T_x^{-1}\cdot T_y$.
After some algebra, one can find that
\be
\small
\begin{split}
S |g,\tilde{\chi}_m\rangle=\frac{1}{\sqrt{|G|}}\sum_{h\in G} 
\tilde{\chi}_m^g(h) W(g,h) |h^{-1},g\rangle,
\end{split}
\ee
where one can check explicitly that $W(g,h)=1$. Therefore, we can obtain 
\be
\small
\begin{split}
\langle g',\tilde{\chi}_{m'}|\, S\,|g,\tilde{\chi}_m\rangle=&\frac{1}{|G|}\sum_{h,h'}\delta_{g',h^{-1}}\delta_{h',g} [\chi_{m'}^{g'}(h')]^{\ast}\cdot \chi_m^g(h)\\
=& \frac{1}{p} \, [\chi_{m'}^{g'}(g)]^{\ast}\cdot \chi_m^g((g')^{-1})\\
=& \frac{1}{p} \, [\chi_{m'}^{g'}(g)]^{\ast}\cdot [\chi_m^g((g']^{\ast}\\
=& \frac{1}{p} \, e^{-\frac{2\pi i}{p^2} \left[ p(m[g']_p+m'[g]_p)+2u[g]_p[g']_p\right]},
\end{split}
\ee
where $[g,g']=1$ and $|G|=p$.
One can find the modular data are the same as those for MS MTCs in \eqref{TopologicalSpin_typeB} and \eqref{S_BB}.
But as pointed out in the main text, for a prime number $p>3$, there are only 3 inequivalent categories for $G=\mathbb{Z}_p$,
while $p$ inequivalent MS MTCs.

\subsection{Some path integrals on 3-simplices}
\label{Sec: appendix_PI}

In this part, we present the expressions of some path integrals on 3-simplices for the cases of 
a punctured torus and a closed genus-2 manifold, which are used in Sec.\ref{Sec: Lattice}.
 
The phase associated to \eqref{TxPhasePuncture} is:
\be\label{u_x^t_PuncturedTorus_appendix}
\small
\begin{split}
u_x^t(\Sigma_{1,1})=&Y_{[\bar{1}24],[\bar{1}'4'2^{\ast}]}\cdot Y_{[\bar{1}14],[\bar{1}'1'2^{\ast}]}\cdot Y_{[134],[1'3'2^{\ast}]}\\
&\cdot \omega([\bar{1}'1'],[1'4'],[4'2^{\ast}])\cdot \omega([1'3'],[3'4'],[4'2^{\ast}])
\end{split}
\ee
The first three terms can be expressed in terms of $3\times 3=9$ 3-cocycles as follows
\be\label{YYY_Tx_punctureTorus}
\small
\begin{split}
&Y_{[\bar{1}24],[\bar{1}'4'2^{\ast}]}\cdot Y_{[\bar{1}14],[\bar{1}'1'2^{\ast}]}\cdot Y_{[134],[1'3'2^{\ast}]}\\
=&\big[
\omega([2^{\ast}\bar{1}], [\bar{1}2],[24])^{-1}\cdot \omega([4'2^{\ast}],[2^{\ast}\bar{1}],[\bar{1}2])^{-1}\\
&\cdot \omega([\bar{1}'4'],[4'2^{\ast}],[2^{\ast}\bar{1}])^{-1}
\big]\cdot\big[\omega([2^{\ast}\bar{1}],[\bar{1}1],[14])\\
&\cdot \omega([\bar{1}'2^{\ast}],[2^{\ast}\bar{1}],[\bar{1}1])\cdot\omega([\bar{1}'1'],[1'2^{\ast}],[2^{\ast}1])\big]\\
&\cdot\big[
\omega([2^{\ast}1],[13],[34])\cdot \omega([3'2^{\ast}],[2^{\ast}1],[13])\\
&\cdot \omega([1'3'],[3'2^{\ast}],[2^{\ast}1])\big]\\
=&\big[\omega(g_y^{-1}g_x^{-1}t^{-1}, g_y, g_x)^{-1}\cdot 
\omega(tg_xt^{-1}, g_y^{-1}g_x^{-1}t^{-1}, g_y)^{-1} \\
&\cdot \omega(tg_yt^{-1}, tg_xt^{-1}, g_y^{-1}g_x^{-1}t^{-1})^{-1}\big]\cdot \big[\omega(g_y^{-1}g_x^{-1}t^{-1},k,g_yg_x)\\
&\cdot\omega(tg_xg_yt^{-1},g_y^{-1}g_x^{-1}t^{-1}, k) \cdot \omega(tkt^{-1}, tg_yg_xt^{-1},g_x^{-1}g_y^{-1}t^{-1})\big]\\
&\cdot \big[
\omega(g_x^{-1}g_y^{-1}t^{-1}, g_x,g_y)\cdot \omega(tg_yt^{-1}, g_x^{-1}g_y^{-1}t^{-1},g_x)\\
&\cdot \omega(tg_xt^{-1}, tg_yt^{-1}, g_x^{-1}g_y^{-1}t^{-1})\big]
\end{split}
\ee
The last two terms in Eq.\eqref{u_x^t_PuncturedTorus_appendix} come from the transformation as follows:

\begin{eqnarray}\label{TxPhasePuncture_last2_appendix}
\footnotesize
\begin{tikzpicture}[baseline={(current bounding box.center)}]

\draw [>=stealth,->] (-5pt,-20pt)--(10pt,-20pt);  \draw (10pt, -20pt) -- (20pt,-20pt);

\draw [>=stealth,->] (-10pt,-10pt)--(0pt,0pt); \draw(0pt,0pt)--(10pt,10pt);

\draw [>=stealth,->] (-10pt,-10pt)--(-7.5pt,-15pt); \draw(-7.5pt,-15pt)--(-5pt,-20pt);

\draw [>=stealth,->](-5pt,-20pt)--(22.5pt,-5pt); \draw(22.5pt,-5pt)--(50pt,10pt);

\draw [>=stealth,->] (-10pt,-10pt)--(20pt,0pt); \draw(20pt,0pt)--(50pt,10pt);

\draw [>=stealth,->] (10pt,10pt)--(30pt,10pt); \draw (30pt,10pt)--(50pt,10pt);
\draw [>=stealth,->] (20pt,-20pt)--(35pt,-5pt); \draw (35pt,-5pt)--(50pt,10pt);
\node at (-5pt,-26pt){${1'}$};
\node at (-15pt,-12pt){${\bar{1}'}$};
\node at (22pt,-26pt){${3'}$};

\node at (5pt,12pt){${4'}$};
\node at (55pt,10pt){${2^{\ast}}$};

\end{tikzpicture}
\quad
\to 
\quad
\begin{tikzpicture}[baseline={(current bounding box.center)}]
\draw [>=stealth,->] (-5pt,-20+50pt)--(10pt,-20+50pt);  \draw (10pt, -20+50pt) -- (20pt,-20+50pt);

\draw [>=stealth,->] (-10pt,-10+50pt)--(0pt,0+50pt); \draw(0pt,0+50pt)--(10pt,10+50pt);

\draw [>=stealth,->] (-10pt,-10+50pt)--(-7.5pt,-15+50pt); \draw(-7.5pt,-15+50pt)--(-5pt,-20+50pt);

\draw [>=stealth,->] (-5pt,30pt)--(2.5pt,45pt); \draw(2.5pt,45pt)--(10pt,60pt);

\draw [>=stealth,->] (10pt,10+50pt)--(30pt,10+50pt); \draw (30pt,10+50pt)--(50pt,10+50pt);
\draw [>=stealth,->] (20pt,-20+50pt)--(35pt,-5+50pt); \draw (35pt,-5+50pt)--(50pt,10+50pt);
\draw [>=stealth,->] (20pt,30pt)--(15pt,45pt); \draw(15pt,45pt)--(10pt,60pt);

\node at (-5pt,25pt){${1'}$};
\node at (25pt,25pt){${3'}$};
\node at (-15pt,-12+50pt){${\bar{1}'}$};

\node at (5pt,62pt){${4'}$};
\node at (58pt,60pt){${2^{\ast}}$};
\end{tikzpicture}
\end{eqnarray}
and has the concrete expression
\be\label{LastTwoCocyle_appendix}
\small
\begin{split}
&\omega([\bar{1}'1'],[1'4'],[4'2^{\ast}])\cdot \omega([1'3'],[3'4'],[4'2^{\ast}])\\
=&\omega(tkt^{-1},tg_yk^{-1}t^{-1}, tg_xt^{-1})\cdot \omega(tg_xt^{-1},tg_x^{-1}g_yt^{-1},tg_xt^{-1}).
\end{split}
\ee
The phase $u_y^t(\Sigma_{1,1})$ associated to \eqref{TyPhasePuncture} is
\be\label{u_y^t_PuncturedTorus_appendix}
\small
\begin{split}
u_y^t(\Sigma_{1,1})=&Y_{[\bar{1}24],[\bar{1}'2'3^{\ast}]}\cdot Y_{[\bar{1}14],[\bar{1}'1'3^{\ast}]}\cdot Y_{[134],[1'4'3^{\ast}]}\\
&\cdot \omega([\bar{1}'1'],[1'2'],[2'3^{\ast}])\cdot \omega([1'2'],[2'4'],[4'3^{\ast}])^{-1}.
\end{split}
\ee
The first three terms can be expressed in terms of $3\times 3=9$ 3-cocycles as follows
\be\label{YYY_Ty_punctureTorus}
\small
\begin{split}
&Y_{[\bar{1}24],[\bar{1}'2'3^{\ast}]}\cdot Y_{[\bar{1}14],[\bar{1}'1'3^{\ast}]}\cdot Y_{[134],[1'4'3^{\ast}]}\\
=&\big[\omega([3^{\ast}\bar{1}],[\bar{1}2],[24])^{-1}\cdot\omega([2'3^{\ast}],[3^{\ast}\bar{1}],[\bar{1}2])^{-1}\\
&\cdot\omega([\bar{1}'2'],[2'3^{\ast}],[3^{\ast}\bar{1}])^{-1}\big]\cdot\big[\omega([3^{\ast}\bar{1}],[\bar{1}1],[14])\\
&\cdot \omega([\bar{1}'3^{\ast}],[3^{\ast}\bar{1}],[\bar{1}1])\cdot\omega([\bar{1}'1'],[1'3^{\ast}],[3^{\ast}1])\big]\\
&\cdot\big[
\omega([3^{\ast}1],[13],[34])\cdot \omega([4'3^{\ast}],[3^{\ast}1],[13])\\
&\cdot \omega([1'4'],[4'3^{\ast}],[3^{\ast}1])\big]\\
=&\big[
\omega(g_y^{-1}g_x^{-1}t^{-1},g_y,g_x)^{-1}\cdot \omega(tg_xt^{-1},g_y^{-1}g_x^{-1}t^{-1},g_y)^{-1}\\
&\cdot\omega(tg_yt^{-1},tg_xt^{-1},g_y^{-1}g_x^{-1}t^{-1})^{-1}
\big]\cdot\big[\omega(g_y^{-1}g_x^{-1}t^{-1},k,g_yg_x)\\
&\cdot\omega(tg_xg_yt^{-1},g_y^{-1}g_x^{-1}t^{-1}, k) \cdot \omega(tkt^{-1}, tg_yg_xt^{-1},g_x^{-1}g_y^{-1}t^{-1})\big]\\
&\cdot \big[
\omega(g_x^{-1}g_y^{-1}t^{-1}, g_x,g_y)\cdot \omega(tg_yt^{-1}, g_x^{-1}g_y^{-1}t^{-1},g_x)\\
&\cdot \omega(tg_xt^{-1}, tg_yt^{-1}, g_x^{-1}g_y^{-1}t^{-1})\big].
\end{split}
\ee
Note that the above result is the same as that in Eq.\eqref{YYY_Tx_punctureTorus}.
Next, the last two terms in Eq.\eqref{u_y^t_PuncturedTorus_appendix} come from the the transformation as follows:
\begin{eqnarray}\label{TyPhasePuncture_last2_appendix}
\footnotesize
\begin{tikzpicture}[baseline={(current bounding box.center)}]

\draw [>=stealth,->] (-5pt,-20pt)--(10pt,-20pt);  \draw (10pt, -20pt) -- (20pt,-20pt);

\draw [>=stealth,->] (-10pt,-10pt)--(0pt,0pt); \draw(0pt,0pt)--(10pt,10pt);

\draw [>=stealth,->] (-10pt,-10pt)--(-7.5pt,-15pt); \draw(-7.5pt,-15pt)--(-5pt,-20pt);

\draw [>=stealth,->](-5pt,-20pt)--(22.5pt,-5pt); \draw(22.5pt,-5pt)--(50pt,10pt);

\draw [>=stealth,->] (-10pt,-10pt)--(20pt,0pt); \draw(20pt,0pt)--(50pt,10pt);

\draw [>=stealth,->] (10pt,10pt)--(30pt,10pt); \draw (30pt,10pt)--(50pt,10pt);
\draw [>=stealth,->] (20pt,-20pt)--(35pt,-5pt); \draw (35pt,-5pt)--(50pt,10pt);
\node at (-5pt,-26pt){${1'}$};
\node at (-15pt,-12pt){${\bar{1}'}$};
\node at (22pt,-26pt){${4'}$};

\node at (5pt,12pt){${2'}$};
\node at (55pt,10pt){${3^{\ast}}$};

\end{tikzpicture}
\quad
\to 
\quad
\begin{tikzpicture}[baseline={(current bounding box.center)}]
\draw [>=stealth,->] (-5pt,-20+50pt)--(10pt,-20+50pt);  \draw (10pt, -20+50pt) -- (20pt,-20+50pt);

\draw [>=stealth,->] (-10pt,-10+50pt)--(0pt,0+50pt); \draw(0pt,0+50pt)--(10pt,10+50pt);

\draw [>=stealth,->] (-10pt,-10+50pt)--(-7.5pt,-15+50pt); \draw(-7.5pt,-15+50pt)--(-5pt,-20+50pt);

\draw [>=stealth,->] (-5pt,30pt)--(2.5pt,45pt); \draw(2.5pt,45pt)--(10pt,60pt);

\draw [>=stealth,->] (10pt,10+50pt)--(30pt,10+50pt); \draw (30pt,10+50pt)--(50pt,10+50pt);
\draw [>=stealth,->] (20pt,-20+50pt)--(35pt,-5+50pt); \draw (35pt,-5+50pt)--(50pt,10+50pt);
\draw (15pt,45pt)--(20pt,30pt); \draw[>=stealth,->] (10pt,60pt)--(15pt,45pt);

\node at (-5pt,25pt){${1'}$};
\node at (25pt,25pt){${4'}$};
\node at (-15pt,-12+50pt){${\bar{1}'}$};

\node at (5pt,62pt){${2'}$};
\node at (58pt,60pt){${3^{\ast}}$};
\end{tikzpicture}
\end{eqnarray}
and has the concrete expression as follows:
\be\label{LastTwoCocyle_Ty_appendix}
\small
\begin{split}
&\omega([\bar{1}'1'],[1'2'],[2'3^{\ast}])\cdot \omega([1'2'],[2'4'],[4'3^{\ast}])^{-1}\\
=&\omega(tkt^{-1}, t(g_yk^{-1})t^{-1}, tg_xt^{-1})\\
&\cdot \omega(t(g_yk^{-1})t^{-1}, t(g_y^{-1}g_x)t^{-1}, tg_yt^{-1} )^{-1}.
\end{split}
\ee

The phase associated to the gauge transformation in \eqref{GaugeTransformGenus2} is:
\be\label{PhaseGaugeTransf_genus2_appendix}
\begin{split}
\eta^t(\Sigma_{2,0})=&Y_{[167],[1'6'7']}\cdot Y_{[\bar{1}17],[\bar{1}'1'7']}\cdot Y_{[\bar{1}57],[\bar{1}'5'7']}\\
&\cdot Y_{[134],[1'3'4']}\cdot Y_{[\bar{1}14],[\bar{1}'1'4']}\cdot Y_{[\bar{1}24],[\bar{1}'2'4']},
\end{split}
\ee
The first (last) three terms correspond to the contribution of the left (right) punctured torus. Explicitly, we have
\be\label{GaugeTransf_genus2_A}
\small
\begin{split}
&Y_{[167],[1'6'7']}\cdot Y_{[\bar{1}17],[\bar{1}'1'7']}\cdot Y_{[\bar{1}57],[\bar{1}'5'7']}\\
=&\big[\omega([1'1],[16],[67])^{-1}\cdot \omega([1'6'],[6'6],[67'])\\
&\cdot\omega([1'6],[67'],[7'7])^{-1}\big]\cdot \big[ \omega([\bar{1}'\bar{1}],[\bar{1}1],[17])^{-1}\\
&\cdot \omega([\bar{1}'1'],[1'1],[17])\cdot \omega([\bar{1}'1'],[1'7'],[7'7])^{-1} \big]\\
&\cdot\big[\omega([\bar{1}'\bar{1}],[\bar{1}5],[57])\cdot \omega^{-1}([\bar{1}'5'],[5'5],[57])\\
&\cdot \omega([\bar{1}'5'],[5'7'],[7'7])\big]\\
=&\big[\omega(t^{-1}, g_{2,y}, g_{2,x})^{-1}\cdot \omega(tg_{2,y}t^{-1}, t^{-1}, tg_{2,x})\\
&\cdot \omega(g_{2,y}t^{-1}, tg_{2,x}, t^{-1})^{-1}
\big]\cdot\big[\omega(t^{-1}, k, g_{2,x}g_{2,y})^{-1}\\
&\cdot \omega(tkt^{-1}, t^{-1}, g_{2,x}g_{2,y})\cdot \omega(tkt^{-1}, t g_{2,x}g_{2,y}t^{-1}, t^{-1})^{-1}\big]\\
&\cdot \big[
\omega(t^{-1}, g_{2,x}, g_{2,y})\cdot \omega(tg_{2,x}t^{-1}, t^{-1}, g_{2,y})^{-1}\\
&\cdot \omega(t g_{2,x}t^{-1}, tg_{2,y}t^{-1}, t^{-1})\big].
\end{split}
\ee
and 
\be\label{GaugeTransf_genus2_B}
\small
\begin{split}
&Y_{[134],[1'3'4']}\cdot Y_{[\bar{1}14],[\bar{1}'1'4']}\cdot Y_{[\bar{1}24],[\bar{1}'2'4']}\\
=&\big[
\omega([1'1],[13'],[3'4'])\cdot \omega([13'],[3'3],[34'])^{-1}\\
&\cdot \omega([13],[34'],[4'4])\big]\cdot \big[
\omega([\bar{1}'1'],[1'1],[14'])^{-1}\\
&\cdot \omega([\bar{1}'\bar{1}],[\bar{1}1],[14'])
\cdot \omega([\bar{1}1],[14'],[4'4])\\
&\cdot\big[
\omega([\bar{1}'\bar{1}],[\bar{1}2'],[2'4'])^{-1}\cdot \omega([\bar{1}2'],[2'2],[24'])\\
&\cdot \omega([\bar{1}2],[24'],[4'4])^{-1}
\big]\\
=&\big[\omega(t^{-1}, tg_{1,x}, tg_{1,y}t^{-1})\cdot \omega(tg_{1,x}, t^{-1}, tg_{1,y})^{-1}\\
&\omega(g_{1,x}, tg_{1,y}, t^{-1})\big]\cdot\big[
\omega(tkt^{-1}, t^{-1}, tg_{1,y}g_{1,x})^{-1}\\
&\cdot \omega(t^{-1}, k, tg_{1,y}g_{1,x})
\cdot \omega(k, tg_{1,y}g_{1,x}, t^{-1})
\big]\\
&\cdot\big[
\omega(t^{-1}, tg_{1,y}, tg_{1,x}t^{-1})^{-1}\cdot \omega(tg_{1,y}, t^{-1}, t^{-1}g_{1,x})\\
&\cdot \omega(g_{1,y}, t^{-1}g_{1,x}, t^{-1})^{-1}\big]
\end{split}
\ee

In the following we give the expression of $I_L^t(\Sigma_{1,1})$ in Eq.\eqref{u_Rx_genus2}
[note the difference of ordering from that in \eqref{GaugeTransformGenus2}]:
\be\label{It_L_appendix}
\small
\begin{split}
&Y_{[\bar{1}57],[\bar{1}'5'7']}\cdot Y_{[\bar{1}17],[\bar{1}'1'7']}\cdot  Y_{[167],[1'6'7']}\\
=&\big[
\omega([\bar{1}'\bar{1}],[\bar{1}5],[57])\cdot \omega([\bar{1}'5'],[5'7'],[7'7])\\
&\cdot \omega([\bar{1}'5'],[5'5],[57])^{-1}
\big]\cdot \big[
\omega([\bar{1}'\bar{1}],[\bar{1}1],[17])^{-1}\\
&\cdot \omega([\bar{1}'1'],[1'1],[17]) \cdot\omega([\bar{1}'1'],[1'7'],[7'7])^{-1}
\big]\\
&\cdot\big[
\omega([1'1],[16],[67])^{-1}\cdot \omega([1'6'],[6'7'],[7'6])^{-1}\\
&\cdot \omega([1'7'],[7'6],[67])\big]\\
=&
\big[\omega(t^{-1},g_{2,x}, g_{2,y})\cdot \omega(t g_{2,x} t^{-1}, t g _{2,y} t^{-1}, t^{-1})\\
&\cdot \omega(t g_{2,x} t^{-1}, t^{-1}, t g_{2,y} t^{-1})^{-1}\big]\cdot\big[
\omega(t^{-1},k,g_{2,x}g_{2,y})^{-1}\\
&\cdot \omega(tkt^{-1}, t^{-1}, g_{2,x}g_{2,y})\cdot \omega(tkt^{-1}, tg_{2,x} g_{2,y} t^{-1}, t^{-1})^{-1}
\big]\\
&\cdot\big[
\omega(t^{-1}, g_{2,y}, g_{2,x})^{-1} \cdot \omega^{-1}(tg_{2,y}t^{-1}, tg_{2,x}t^{-1}, g_{2,x}^{-1}t^{-1})\\
&\cdot \omega(tg_{2,x}g_{2,y}t^{-1}, g_{2,x}^{-1}t^{-1}, g_{2,x})\big].
\end{split}
\ee

\section{Galois symmetry in the modular data}
\label{Sec: Galois}

Galois symmetry plays an important role in proving the equivalence of modular data in 
MS MTCs. Here we give a brief review of the Galois symmetry in the modular data\cite{deBoer1991,Coste1994,etingof2005fusion}.

In Refs.\onlinecite{deBoer1991,Coste1994}, it was found that the matrix element $S_{ij}$ of the modular $S$ matrix 
lies in the cyclotomic field $\mathbb{Q}(\zeta_n)$, which is a number field obtained by adjoining a complex primitive root of unity to $\mathbb{Q}$, 
the field of rational numbers.  Here $\zeta_n=e^{\frac{2\pi i}{n}}$ is the $n$-th root of unity.
More explicitly, $\mathbb{Q}(\zeta_n)$ can be thought of complex numbers of the form $a_0+a_1\zeta_n+\cdots +a_k\zeta_n^k$, 
where the coefficients $a_i$ are real.
In other words, each $S_{ij}$ (with $i$, $j\in\Pi_{\mathcal{C}}$) can be written as a polynomial in $\zeta_n$ with rational coefficients.

The proof of the above statement is based on the theorem in 
algebraic number theory that a field extension of $\mathbb{Q}$ is contained in a
cyclotomic field $\mathbb{Q}(\zeta_n)$ if the extension is normal and the corresponding Galois group is abelian.\cite{deBoer1991,marcus1977number}

First, let us consider the extension $L$ of $\mathbb{Q}$ generated by $\lambda_{ij}:=\frac{S_{ij}}{S_{0j}}$, with
$i,j\in \Pi_{\mathcal{C}}$.  It is noted that $\lambda_{ij}$ with a fixed $i$ is the solution of the polynomial equation:
\be
\text{det}(\lambda \mathbf{I}-N_i)=0,
\ee
where $N_i$ is the fusion matrix $N_i=N_{i a}^b$. This indicates the extension of $\mathbb{Q}$ is normal.
Second, we consider the group element in the Galois group $\sigma\in \text{Gal}(L/\mathbb{Q})$.
Since the fusion rules $\lambda_{ai}\lambda_{bi}=N_{ab}^c \lambda_{ci}$ are invariant under the Galois action, 
then one must have
\be\label{g_sijs0j}
\small
\sigma\left(\frac{S_{ij}}{S_{0j}}\right)=\frac{S_{i\hat{\sigma}{(j)}}}{S_{0\hat{\sigma}{(j)}}}.
\ee 
In other words, there is a group morphism from Gal$(L/\mathbb{Q})$ to permutations of anyons $j$ with $j\in \Pi_{\mathcal{C}}$. 
This sends a Galois automorphism $\sigma$ to a bijection $j\to \hat{\sigma}(j)$ of $\Pi_{\mathcal{C}}$.

Next, we need to generalize Eq.\eqref{g_sijs0j} a little bit.
Considering the modular $S$ matrix is unitary and symmetric, and $S_{0j}=S_{0\bar{j}}$, 
one has $(\frac{1}{S_{0j}})^2=\sum_i(\frac{S_{ij}}{S_{0j}})\cdot (\frac{S_{i\bar{j}}}{S_{0\bar{j}}})$.
Applying the automorphism $\sigma$ to this equation, one can obtain
\be
\small
\sigma\left(\frac{1}{S_{0j}^2}\right)=
\sum_i\left(\frac{S_{i\hat{\sigma}(j)}}{S_{0\hat{\sigma}(j)}}\right)\cdot \left(\frac{S_{i\hat{\sigma}(\bar{j})}}{S_{0\hat{\sigma}(\bar{j})}}\right)
=\frac{\delta_{\hat{\sigma}(j),\overline{\hat{\sigma}(\bar{j})}}}{S_{0\hat{\sigma}(j)}\cdot S_{0\hat{\sigma}(\bar{j})} },
\ee
based on which we have $\hat{\sigma}(\bar{j})=\overline{\hat{\sigma}(j)}$, and 
$\sigma (S^2_{0j})=(S_{0\hat{\sigma}(j)})^2=(S_{0\hat{\sigma}(\bar{j})})^2=(S_{0\overline{\hat{\sigma}(j)}})^2$.
Then one has
$
\sigma(S^2_{ij})=\sigma\left(
\frac{S_{ij}^2}{S_{0j}^2}\cdot S_{0j}^2
\right)=(S_{i\hat{\sigma}(j)})^2,
$
and therefore
\be\label{sigma_sij}
\sigma(S_{ij})=\epsilon_{\sigma}(i)\cdot S_{\hat{\sigma}(i),j}=\epsilon_{\sigma}(j)\cdot S_{i,\hat{\sigma}(j)}.
\ee
with $\epsilon_{\sigma}(i),\epsilon_{\sigma}(j)=\pm 1$.
The symmetric property of $S$ matrix is used in Eq.\eqref{sigma_sij}.
One can find the extenston of $\mathbb{Q}$ generated by $S_{ij}$ is finite and normal. 
In addition, considering $\sigma_a,\sigma_b\in \text{Gal}(L/\mathbb{Q})$, it can be checked that 
$\sigma_a\sigma_b(S_{ij})=\sigma_b\sigma_a(S_{ij})$. That is, $\text{Gal}(L/\mathbb{Q})$ is abelian.
Till now, we have seen that the extension of $\mathbb{Q}$ generated by $S_{ij}$ is finite and 
$\text{Gal}(L/\mathbb{Q})$ is abelian. Then based on Kronecker-Weber theorem we arrive at the 
conclusion $S_{ij}\in \mathbb{Q}(\zeta_n)$.\cite{marcus1977number}

The Galois group Gal$(\mathbb{Q}(\zeta_n)/\mathbb{Q})$ is defined to be the automorphisms 
of the field $\mathbb{Q}(\zeta_n)$ which fix $\mathbb{Q}$, and is \textit{isomorphic} to the multiplicative group 
$\mathbb{Z}_n^{\times}$ of integers coprime to $n$. In particular, for any $l\in \mathbb{Z}_n^{\times}$, 
we have an automorphism $\sigma_l\in \text{Gal}(\mathbb{Q}(\zeta_n)/\mathbb{Q})$ sending $\zeta_n$
to $\zeta_n^l$. More explicitly, under the action of $\sigma_l$, 
$a_0+a_1\zeta_n+\cdots +a_k\zeta_n^k$ is sent to $a_0+a_1\zeta^l_n+\cdots +a_k\zeta_n^{lk}$.

Interestingly, for $\sigma\in \text{Gal}(\mathbb{Q}(\zeta_n)/\mathbb{Q})$, it is found that $\sigma$ acts on the modular $T$ matrix as follows\cite{dong2015congruence}
\be\label{sigma2_T}
\sigma^2(T_{ii})=T_{\hat{\sigma}(i)\hat{\sigma}(i)}.
\ee
For the $S$ matrix, in general we have $\sigma^2 (S_{ij})=\epsilon_{\sigma}(i)\cdot \epsilon_{\sigma}(j)\cdot S_{\hat{\sigma}(i),\hat{\sigma}(j) }$,
where $\epsilon_{\sigma}(i)$ and $\epsilon_{\sigma}(j)$ are the same as those in Eq.\eqref{sigma_sij}.
For the (twisted) quantum double of a finite group $G$ as studied in this work, it is found that one always have $\epsilon_{\sigma}(i)=1$ for arbitrary $i\in \Pi_{\mathcal{C}}$.\cite{CGR00}
That is, for a (twisted) quantum double of a finite group, we have
\be\label{sigma2_S}
\sigma^2(S_{ij})=S_{\hat{\sigma}(i)\hat{\sigma}(j)}.
\ee
The Galois action on modular matrices in 
Eqs.\eqref{sigma2_T} and \eqref{sigma2_S} turns out to be useful in proving the equivalence of modular data in MS MTCs
(See also Sec.\ref{Sec: proof_Equiv_ModularData}).

\bibliography{Genus2_Ref}

\end{document}